\documentclass[11pt, a4paper, oneside]{article}
\usepackage[utf8]{inputenc}
\usepackage[british]{babel}
\usepackage[english]{varioref}
\usepackage{hyphenat}
\usepackage[square, numbers]{natbib}
\usepackage[hyphens]{url}
\usepackage{caption}
\usepackage{multicol}
\usepackage{algorithm}
\usepackage{algpseudocode}
\usepackage{amsmath}
\usepackage{amsthm}
\usepackage{amssymb}
\usepackage{mathrsfs}
\usepackage{mathtools}
\usepackage{nicefrac}
\usepackage{etoolbox}
\usepackage{siunitx}
\usepackage[normalem]{ulem}
\usepackage{comment}
\usepackage{fp}
\usepackage{tikz}
\usetikzlibrary{calc, arrows, decorations.markings, decorations.pathmorphing, backgrounds, positioning, fit}
\usepackage[linecolor=red!80!white, backgroundcolor=blue!20!white, textsize=tiny]{todonotes}

\includecomment{comment:figures}

\newcommand*{\LENS}{LENS}
\newcommand*{\NN}{NN}
\newcommand*{\NNS}{NN$_\text{S}$}
\newcommand*{\NNSTwo}{NN$_\text{S}^\text{2}$}
\newcommand*{\NNL}{NN$^\text{L}$}
\newcommand*{\NNSL}{NN$_\text{S}^\text{L}$}
\newcommand*{\CIF}{CIF}
\newcommand*{\CH}{CH}
\newcommand*{\CHC}{CH$_\text{C}$}
\newcommand*{\CHL}{CH$^\text{L}$}
\newcommand*{\CHCL}{CH$_\text{C}^\text{L}$}
\newcommand*{\LPP}{LPP}
\newcommand*{\LPPR}{LPP$^\text{R}$}
\newcommand*{\LPCTwo}{LPC$_2$}
\newcommand*{\LPCTwoR}{LPC$_2^\text{R}$}
\newcommand*{\LPCOne}{LPC$_1$}
\newcommand*{\LPCOneR}{LPC$_1^\text{R}$}
\newcommand*{\TwoO}{+$2$O}
\newcommand*{\ThreeO}{+$3$O}
\newcommand*{\M}[1]{+M$^{#1}$}
\newcommand*{\Lens}[1]{+L}
\newcommand*{\AngleTSP}{AngleTSP}
\newcommand*{\AngleTSPInstances}{{\AngleTSP}-instances}
\newcommand*{\AngleDistanceTSP}{AngleDistanceTSP}
\newcommand*{\AngleDistanceTSPInstances}{{\AngleDistanceTSP}-instances}

\newcommand*{\Ie}{I.\,e.\ }
\newcommand*{\ie}{i.\,e.\ }

\newcommand*{\eg}{e.\,g.\ }
\newcommand*{\etal}{et al.\ }

\newcommand*{\rz}{{\mathbb{R}}}
\newcommand*{\nz}{{\mathbb{N}}}

\newcommand*{\zp}{\bar{z}}
\newcommand*{\zt}{z}
\newcommand*{\lb}{\underline{z}}
\newcommand*{\rg}{\overline{r}\genfrac{}{}{0pt}{}{g}{}}
\newcommand*{\ra}{\overline{t}^{^{\genfrac{}{}{0pt}{}{}{\scalebox{0.8}{$a$}}}}}

\newcommand*{\defeq}{\mathrel{\vcenter{\baselineskip0.5ex \lineskiplimit0pt
			\hbox{\scriptsize.}\hbox{\scriptsize.}}}%
	=}

\newcommand*{\degree}{\ensuremath{^\circ}}

\def\Bigg#1{{\hbox{$\left#1\vbox to17.5\p@{}\right.\n@space$}}}

\newtheorem{axiom}{Axiom}

\newtheorem{example}[axiom]{Example}

\robustify\uline
\def\Decimal{.0000}
\def\Uline#1{\Ulinehelp#1 }
\def\Ulinehelp#1.#2 {%
  #1.#2\setbox0=\hbox{#1\Decimal}\hspace{-\wd0}{\if\relax#2\relax%
  \uline{\phantom{#1.0}}\else\uline{\phantom{#1.#2}}\fi}%
}

\DeclarePairedDelimiterX{\norm}[1]{\lVert}{\rVert}{#1}
\DeclareMathOperator*{\argmin}{argmin}

\newcommand*{\xscale}{0.061}
\newcommand*{\yscale}{2.8}
\newcommand*{\yscaleLarge}{\yscale}
\FPeval\yscaleMiddle{1.666667*\yscale}
\FPeval\yscaleSmall{5*\yscale}
\newcommand*{\axisAdditionalLengthPlus}{0.5}
\newcommand*{\axisAdditionalLengthMinus}{0.15}
\newcommand*{\axisLabel}{0.1}
\newcommand*{\crossSize}{0.075}

\FPeval\xscaleFigureOptimalObjectiveFunctionValuesVsLowerBounds{2*\xscale}
\FPeval\yscaleFigureOptimalObjectiveFunctionValuesVsLowerBounds{15*\yscale}

\FPeval\xscaleEfficiencyStartingHeuristics{4*\xscale}
\FPeval\xscaleEfficiencyImprovementHeuristics{10*\xscale}
\FPeval\yscaleLargeEfficiency{1.67*\yscaleLarge}
\FPeval\yscaleMiddleEfficiency{1.67*\yscaleMiddle}
\FPeval\yscaleSmallEfficiency{1.67*\yscaleSmall}

\newcommand*{\xAxis}{$n$}
\newcommand*{\yAxis}{$\rg$}

\newcommand*{\xAxisEfficiency}{$\ra$}
\newcommand*{\yAxisEfficiency}{$\rg$}

\makeatletter
\newenvironment{figurehere}
{\def\@captype{figure}}
{}
\makeatother



\algrenewcommand\Return{\State \algorithmicreturn{} }%

\algnewcommand\algorithmicsubroutine{\textbf{\large Subroutine: }}
\algnewcommand\Subroutine{\item[\algorithmicsubroutine]}
\algnewcommand\algorithmicmain{\textbf{\large Main: }}
\algnewcommand\Main{\item[\algorithmicmain]}
\algnewcommand\Andoperator{\textbf{and }}
\algrenewcommand\Return{\State \algorithmicreturn{} }
\algnewcommand\algorithmicto{\textbf{to}}
\algrenewtext{For}[2]{\algorithmicfor\ #1 \algorithmicto\ #2 \algorithmicdo}
\algnewcommand\algorithmicdownto{\textbf{downto}}
\algblockdefx[Loop]{Fordownto}{EndFor}[3][]{\algorithmicfor\ #2 \algorithmicdownto\ #3 \algorithmicdo}{\algorithmicend\ \algorithmicfor}
\algnewcommand{\LineComment}[1]{\State \(\triangleright\) #1}

\tikzset{
    arrow/.style={decoration={markings, mark=at position 1 with
    {\arrow[scale=1.5,>=stealth]{>}}}, postaction={decorate}},
    arrow/.default=1
}

\begin{document}
	\sloppy
	
	\title{\bf Geometric and LP-based heuristics for the quadratic travelling salesman problem}
	\author{
		\sc Rostislav Stan\v{e}k{\footnotemark[1]}
		\and
		\sc Peter Greistorfer{\footnotemark[1]}
		\and
		\sc Klaus Ladner{\footnotemark[2]}
		\and
		\sc Ulrich Pferschy{\footnotemark[2]}
	}
	\date{}
	\maketitle
	\renewcommand{\thefootnote}{\fnsymbol{footnote}}
	\footnotetext[1]{
		{\tt \{peter.greistorfer, rostislav.stanek\}@uni-graz.at}.
		Department of Production and Operations Management, University of Graz, Universit\"{a}tsstra{\ss}e 15, 8010 Graz, Austria}
	\footnotetext[2]{
		{\tt \{klaus.ladner, pferschy\}@uni-graz.at}.
		Department of Statistics and Operations Research, University of Graz, Universit\"{a}tsstra{\ss}e 15, 8010 Graz, Austria}
	\renewcommand{\thefootnote}{\arabic{footnote}}
	
	
	\begin{abstract}
		A generalization of the classical TSP is the so-called quadratic travelling salesman problem (QTSP), in which a cost coefficient is associated with the transition in every vertex, i.e.\ with every pair of edges traversed in succession. 
		In this paper we consider two geometrically motivated special cases of the QTSP known from the literature, namely the angular-metric TSP, where transition costs correspond to turning angles in every vertex, and the angular-distance-metric TSP, where a linear combination of turning angles and Euclidean distances is considered.
		
		At first we introduce a wide range of heuristic approaches, motivated by the typical geometric structure of optimal solutions.
		In particular, we exploit lens-shaped neighborhoods of edges and a decomposition of the graph into layers of convex hulls, which are then merged into a tour by a greedy-type procedure or by utilizing an ILP model.
		Secondly, we consider an ILP model for a standard linearization of QTSP and compute fractional solutions of a relaxation.
		By rounding we obtain a collection of subtours, paths and isolated points, which are combined into a tour by various strategies, all of them involving auxiliary ILP models.
		Finally, different improvement heuristics are proposed, most notably a matheuristic which locally reoptimizes the solution for rectangular sectors of the given point set by an ILP approach.
		 
		Extensive computational experiments for benchmark instances from the literature and extensions thereof illustrate the Pareto-efficient frontier of algorithms in a (running time, objective value)-space.
		It turns out that our new methods clearly dominate the previously published heuristics.		
	\end{abstract}
	
	\medskip
	
	\section{Introduction}
		\label{section:introduction}
		The {\em travelling salesman problem} (TSP) is one of the best-known and most widely investigated combinatorial optimization problems with several famous books entirely devoted to its study \cite{LawlerLenstraRinnooyKanShmoys:TheTravelingSalesmanProblemAGuidedTourOfCombinatorialOptimization, Reinelt:TheTravelingSalesmanComputationalSolutionsForTSPApplications, ApplegateBixbyChvatalCook:TheTravelingSalesmanProblemAComputionalStudy, GutinPunnen:TheTravelingSalesmanProblemAndItsVariations}. Plenty of variations of this problem have already been studied (see \eg \cite{GutinPunnen:TheTravelingSalesmanProblemAndItsVariations}).
	
		In the {\em quadratic travelling salesman problem} (QTSP) we are still looking for a {\em tour}, \ie a {\em Hamiltonian cycle}; 
		however, different from the TSP, not only the costs of direct movements from one vertex to the next vertex are considered, but beyond that, costs arising for the successive use of two edges are taken into account. 
		This means that we consider the {\em transition} in each vertex $i$ which depends both on the predecessor and on the successor of $i$ in the tour. 
		Thus, transition costs, such as the effort of turning in path planning \cite{AggarwalCoppersmithKhannaMotwaniSchieber:TheAngularMetricTravelingSalesmanProblem}, changing the equipment in scheduling or the change of transportation means in logistic networks from one edge to another \cite{AmaldiGalbiatiMaffionli:MinimumReloadCostPaths}, can be modelled.
	
		Mathematically, this aspect can be expressed by a quadratic objective function. 
		The resulting optimization problem is known as the {\em (symmetric) quadratic travelling salesman problem} (QTSP)~\cite{FischerHelmberg:TheSymmetricQuadraticTravelingSalesmanProblem}. 
		Note that the quadratic objective can also be extended to include costs for {\em every} pair of edges contained in the tour. 
		Such a definition of quadratic costs was used for the {\em quadratic minimum spanning tree problem} (QMSTP) introduced in~\cite{asxu92} and recently treated in \cite{roma15} and \cite{CusticZhangPunnen:TheQuadraticMinimumSpanningTreeProblemAndItsVariations} (see the references therein). Our version of costs only for successive edges is called {\em adjacent} QMSTP in these papers.
		The same terminology applies for the {\em quadratic shortest path problem} recently considered in \cite{rmfb15} and \cite{huso17}.
		
		In this paper we consider two relevant special cases of the QTSP, where the transition costs correspond to i) {\em turning angles} or to ii) a {\em linear combination of turning angles and Euclidean distances}. 
		In both cases, vertices correspond to points in the Euclidean plane.
		The first problem was introduced by {Aggarwal} \etal \cite{AggarwalCoppersmithKhannaMotwaniSchieber:TheAngularMetricTravelingSalesmanProblem} and is called the {\em angular-metric TSP} (AngleTSP).
		It arises from an application in robotics 
		where changing the driving directions of a robot is more energy consuming for larger turning angles.
		Thus, one would prefer a tour which keeps the movement of the robot as closely as possible to a straight line.
		This is also relevant for steering-constrained robots and vehicles with high weight, were a straight movement is easier to control than narrow curves.
		Another application is the planning of trajectories of high-speed aircraft.
						
		The second problem is the {\em angular-distance-metric TSP} (AngleDistanceTSP) which was first considered by {Savla} \etal \cite{SavlaFrazzoli:TravelingSalespersonProblemsForTheDubinsVehicle} and later in \cite{meur10}.		
		It was introduced for an approximate solution of the {\em TSP for Dubins vehicles}, which also has applications in robotics.
		
		\medskip
		In this paper we follow two lines of research:
		On one hand we focus on geometric properties of ``good'' solutions for both of these problems and derive a number of constructive heuristic solution algorithms. 
		On the other hand we consider relaxations of a linearized ILP formulation for the QTSP studied in~\cite{AichholzerFischerFischerMeierPferschyPilzStanek:MinimizationAndMaximizationVersionsOfTheQuadraticTravellingSalesmanProblem} and derive heuristic methods to generate feasible solutions from them.
		Some of our algorithms are {\em matheuristics} in the sense that they apply ILP models to (exactly) solve subproblems arising in the heuristic approach.
		
%
%

		\subsection{Related literature}
		
		The AngleTSP was introduced by {Aggarwal} \etal \cite{AggarwalCoppersmithKhannaMotwaniSchieber:TheAngularMetricTravelingSalesmanProblem}, who also showed that it is $\mathcal{NP}$-hard. 
		Thus, the AngleDistanceTSP and the QTSP are $\mathcal{NP}$-hard as well, as they are generalizations of the AngleTSP.
		It was also shown that the AngleTSP allows a polynomial time approximation algorithm
		guaranteeing an approximation ratio within $O(\log{n})$.
		
		\smallskip
		Several papers deal with the general QTSP in the context of applications in bioinformatics.
		The first contribution in this area is \citet{JaegerMolitor:AlgorithmsAndExperimentalStudyForTheTravelingSalesmanProblemOfSecondOrder} where the asymmetric version of QTSP is introduced.
		In that case, the cost of traversing vertices may differ between one direction and the other.
		The authors provide seven different heuristics and two exact solution approaches for the asymmetric version. 
		
		More recently, two papers extensively studied the symmetric QTSP and its bioinformatic application.
		The first one, due to {Fischer} \etal \cite{FischerFischerJaegerKeilwagenMolitorGrosse:ExactAlgorithmsAndHeuristicsForTheQuadraticTravelingSalesmanProblemWithAnApplicationInBioinformatics},
		gives three exact approaches and several heuristic algorithms.
		The exact approaches are a transformation to the TSP (which is then solved by an available software package), a Branch-and-Bound scheme and a Branch-and-Cut framework.
		However, these are able to solve only instances of moderate size within reasonable running time.
		Therefore, the authors also develop seven heuristics in 
		\cite{FischerFischerJaegerKeilwagenMolitorGrosse:ExactAlgorithmsAndHeuristicsForTheQuadraticTravelingSalesmanProblemWithAnApplicationInBioinformatics}.
		The computational experiments do not give conclusive results, but suggest that the so-called {\em Cheapest-Insertion Heuristic} (generalizing the corresponding heuristic for the TSP, see Section~\ref{subsection:cheapestInsertionHeuristics})
		can be seen as the best currently known non-exact algorithm for the \AngleTSP{} and the \AngleDistanceTSP{}.
		Thus, it will be used as a reference method for our tests.
		
		The second paper
		\cite{FischerFischerJaegerKeilwagenMolitorGrosse:ComputationalRecognitionOfRNASpliceSitesByExactAlgorithmsForTheQuadraticTravelingSalesmanProblem} 
		by the same group of authors focuses on the bioinformatic details and tries to explore the boundaries of problem sizes that can still be solved to optimality by exact algorithms. 
		To this purpose three different approaches are tested, namely Dynamic Programming, Branch-and-Bound and an improved Branch-and-Cut model, where the latter turns out to be vastly superior.
		
		Polyhedral studies were done for the general symmetric QTSP by \citet{FischerHelmberg:TheSymmetricQuadraticTravelingSalesmanProblem}, who also derived several classes of strengthened subtour elimination constraints and proved that many of them are facet defining. 
		In addition, they provided computational comparisons based on an ILP linearization and on the standard separation approach known from the TSP literature.
		Further polyhedral studies were provided in \cite{FischerFischer2015}.
		Recently, Fischer also analyzed the polytope of the asymmetric QTSP in \cite{Fischer:AnAnalysisOfTheAsymmetricQuadraticTravelingSalesmanPolytope} 
		and presented involved lower bounding procedures based on an extended LP formulation which is solved by column generation in \cite{ejor16}.

	A simple ILP-based approach, based on integral separation of subtours, instead of the previously used standard fractional separation, was presented in~\cite{AichholzerFischerFischerMeierPferschyPilzStanek:MinimizationAndMaximizationVersionsOfTheQuadraticTravellingSalesmanProblem} and performed remarkably well. Also a different linearization with a linear number of additional variables was developed in that paper.
	Moreover, the {\em maximization version} of the QTSP was introduced and some surprising structural properties for the case of the AngleTSP were shown.

		Considering the QTSP from an approximation point of view, only few results are known.
		It is easy to see that the standard TSP with distances $d_{i j}$ can be represented as a special case of QTSP with identical solution value
		by defining costs $\frac 1 2 (d_{ij}+d_{jk})$ for any transition from vertex $i$ via $j$ to $k$.
		Thus, the well-known fact that no constant-ratio approximation can exist for the TSP immediately carries over to the QTSP.
		For the special case of Halin graphs, it was shown in~\cite{halin} that QTSP can be solved in linear time.

	\subsection{Our contribution}\label{subsection:outContribution}

Looking at many optimal solutions of the AngleTSP plotted on the screen, we observed a certain recurrent pattern:
tours tend to consist of large circular arcs trying to avoid sharp corners, resembling in some sense the structure of snail-shells, or the superposition of several circles.
Moreover, solving an ILP model and relaxing the classical subtour elimination constraints, we often obtain a fairly moderate number of subtours, which is very different from the behaviour of the standard TSP.

In this paper we tried to exploit these empirical (although anecdotical) insights in a number of different ways:
\begin{enumerate}
	\item Based on the observation that vertices near to an existing straight line (\ie edge) of a preliminary path may be inserted into the path with little extra cost, the so-called {\em lens procedure} is introduced in Section~\ref{section:lensProcedure}, followed by some standard construction heuristics in Section~\ref{section:trivialConstructionHeuristicsAndRelatedApproaches}.
	\item The snail-shell shape of optimal tours mentioned above gives rise to several solution approaches based on the construction of convex hulls (Section~\ref{section:geometricBasedApproaches}).
	Peeling away convex hulls provides an onion-style set of subtours which can be merged together by different patching procedures, also with the help of ILP models.
	\item Starting from existing ILP models, relaxing both integrality and subtour elimination constraints, but keeping the degree-two-constraint for every vertex, and rounding-up the obtained (fractional) solution, provides us with three types of components: subtours, paths and isolated points. 
	We develop several procedures to transform these building blocks into one tour as described in Section~\ref{section:LPBasedApproaches}.
	Some of the used (sub)procedures utilize again auxiliary ILP models of smaller size.
	\item As for the standard TSP, it often makes sense to apply an improvement heuristic as a postprocessing step. 
	We explore some ideas in this direction in Section~\ref{section:ImprovementApproaches} with particular emphasis on a matheuristic, which locally reoptimizes the solution with respect to all vertices lying in a rectangular sector of the given Euclidean plane; this procedure is then applied on different areas of the graph.
	\item It is obvious that only extensive and carefully performed computational experiments can shed some light on the pros and cons of the individual algorithmic ideas and on the difficult question for their ``best'' combination.
	We pursued many more possibilities than could be reasonable reported in the paper, but tried to narrow down the results to the most promising variants for each concept as reported in Section~\ref{section:computationalResults}.
	All together, it turned out that the LP-based algorithms significantly dominate all other approaches for the \AngleTSP. For the \AngleDistanceTSP{} the situation is more diffuse: apart form the LP-based algorithms, a variant of the well known nearest-neighbour heuristic yields competitive solutions as well. Compared to the heuristics proposed in~\cite{FischerFischerJaegerKeilwagenMolitorGrosse:ExactAlgorithmsAndHeuristicsForTheQuadraticTravelingSalesmanProblemWithAnApplicationInBioinformatics}, 
	our algorithms improve either the objective function values or the running times, and some of them dominate for both parameters.
\end{enumerate}
Although our approaches were mainly motivated by the structure of AngleTSP, it turned out that many of them worked quite well also for the AngleDistanceTSP.
Of course, the actual performances depends heavily on the weighting of the angle- and distance-component.

		\subsection{Formal problem definition}
			\label{subsection:formalProblemDefinitionAndRelatedLiterature}
			Let $G = (V, E)$ be an undirected complete graph with vertex set $V = \{1, 2, \ldots, n\}$ and edge set $E = \big\{\{i, j\} \colon i, j \in V,\, i \neq j\big\}$. 
			A {\em path} $P$ is	an ordered sequence of vertices, \ie $P=(p_1, p_2, \ldots, p_{|P|})$ with $p_i \in V$. 
			Note that we will only consider simple paths, \ie paths containing each vertex at most once.
			A {\em subtour} $T$ is a path with the additional interpretation that all vertices should be visited in the given order and finally the edge from $p_{|T|}$ to $p_1$ is traversed.
			If $T$ contains all vertices of $V$, we simply call $T$ a {\em tour} in the graph, \ie we have a {\em Hamiltonian cycle}. 
			Alternatively, a tour $T$ can be described by $T = \big(\sigma(1), \sigma(2), \ldots, \sigma(n)\big)$ for a permutation $\sigma$ of the vertices $1$, $2$, \ldots, $n$. 
			For a tour $T = \{t_1, t_2, \ldots, t_n\}$ we define the set of all {\em tour edges} as $E(T) \defeq \big\{\{t_1, t_2\}, \{t_2, t_3\}, \ldots, \{t_n, t_1\}\big\}$.
			Finally, for auxiliary graphs $\widehat{G}=(\widehat{V}, \widehat{E})$, we denote the {\em set of all neighbouring vertices} of a vertex $v \in \widehat{V}$ by $\delta(v)$.
			
			In the QTSP we associate costs with every pair of adjacent edges traversed in succession. So, using the (incident) edges $e$ and $f$ one after the other in a tour gives rise to a certain cost value $c_{e f} \in \rz_0^+$, which is assigned to the edge pair $(e, f)$. 
			Note that in this paper we only consider the symmetric case where $c_{e f} = c_{f e}$ for all $e \neq f \in E$. 
			Equivalently, we can state costs for every triple of vertices $(i,j,k) \in V\times V\times V$ by setting $c_{ijk}=c_{ef}$ for $e=\{i,j\}$ and $f=\{j,k\}$.
			Clearly, $c_{ijk}=c_{kji}$.
			The QTSP asks for a {\em tour} $T$ minimizing the {\em objective function}
			\begin{equation}
				\label{equation:ov}
				\zt(G, T) \defeq \left(\sum_{i = 1}^{n - 2}{c_{\sigma(i) \sigma(i + 1) \sigma(i + 2)}}\right) + c_{\sigma(n - 1) \sigma(n) \sigma(1)} + c_{\sigma(n) \sigma(1) \sigma(2)}.
			\end{equation}
			In analogy, the cost of a subtour is defined.
			For convenience, we define the contribution to the objective function of a path $P=(p_1, p_2, \ldots, p_{|P|})$:
			\begin{equation}
				\label{equation:path}
				\zp (G, P) \defeq
				\begin{cases}
					0 & \text {if } |P| \leq 2\\
					\sum_{i = 1}^{|P| - 2}{c_{p_i, p_{i + 1}, p_{i + 2}}} & \text{otherwise}
				\end{cases}
			\end{equation}
						
			Although the costs $c_{ijk} \in \rz_0^+$ can be arbitrary non-negative real numbers in general, in this paper we focus on a geometric scenario where vertices correspond to points in the Euclidean plane.
			However, some of our approaches also work in the general case.
			
			For the AngleTSP we define costs representing the turning angle of a tour based on the coordinates $(x_i, y_i) \in \rz^2$ of a vertex $i \in V$, \ie
			\begin{equation}
				\label{equation:turningAngles}
				c_{i j k} = \alpha_{i j k} \defeq \arccos_{[0, \pi]}{\left(
					\frac{
					\left(\genfrac{}{}{0pt}{}{x_j}{y_j}\right) - \left(\genfrac{}{}{0pt}{}{x_i}{y_i}\right)}
					{\norm*{\left(\genfrac{}{}{0pt}{}{x_j}{y_j}\right) - \left(\genfrac{}{}{0pt}{}{x_i}{y_i}\right)}} \cdot 
					\frac{
						\left(\genfrac{}{}{0pt}{}{x_k}{y_k}\right) - \left(\genfrac{}{}{0pt}{}{x_j}{y_j}\right)}
					{\norm*{\left(\genfrac{}{}{0pt}{}{x_k}{y_k}\right) - \left(\genfrac{}{}{0pt}{}{x_j}{y_j}\right)}}
				\right)}
			\end{equation}
			for all $i \neq j \neq k \in V$. 
			Here $\cdot$ denotes the scalar product.
			For the AngleDistanceTSP a weighted sum of the above turning angles and the Euclidean distances between successive vertices  
			is used in the objective function, \ie for arbitrary non-negative real multipliers $\lambda_1$, $\lambda_2$ we define:
			\begin{equation}\label{equation:turningAnglesDistances}
				c_{i j k} = \lambda_1 \cdot\alpha_{i j k} + \frac{\lambda_2}{2} \left(\norm*{\left(\genfrac{}{}{0pt}{}{x_j}{y_j}\right) - \left(\genfrac{}{}{0pt}{}{x_i}{y_i}\right)} + \norm*{\left(\genfrac{}{}{0pt}{}{x_k}{y_k}\right) - \left(\genfrac{}{}{0pt}{}{x_j}{y_j}\right)} \right)
			\end{equation}

	\section{Lens procedure}
		\label{section:lensProcedure}
		Before dealing with the particular solution approaches, we introduce an extension procedure, which can be applied in different algorithms as a subroutine. 
		Given a partial solution consisting of subtours and/or paths and isolated points, the {\em lens procedure} inserts some of the isolated vertices into the subtours and/or paths and thus moves the partial solution closer to feasibility. At the same time the resulting increase of the objective function value is upper-bounded.		
		
		Let $P = (p_1, p_2, \ldots p_{|P|})$ 
		be a path in $G$ and denote its non-empty complement as $F = V \setminus P \neq \emptyset$.		
		Furthermore, let $p_i$ and $p_{i + 1}$, $1 \leq i < |P|$, be two vertices of $P$ traversed in succession. 
		In the lens heuristic we try to include one or more vertices $f \in F$ into the path $P$ between the vertices $p_i$ and $p_{i + 1}$. 
		In particular, we position a {\em lens} between the two vertices $p_i$ and $p_{i + 1}$ in such a way that their two positive curvatures intersect at these vertices and consider the set of points $L \subseteq F$ contained in this lens. 
		The motivation for this new idea comes from the fact that the inclusion of any point inside such a lens causes an additional turning angle that is upper-bounded by the angle $\gamma$ of the curvatures' tangent in $p_i$ (and $p_{i + 1}$ as well) and the line between $p_i$ and $p_{i + 1}$. 
		Moreover, the distance added by inserting $l \in L$, namely  $\Delta = d_{p_i, l} + d_{l, p_{i + 1}} - d_{p_i, p_{i + 1}}$, which is also part of the objective function value in the AngleDistanceTSP, is relatively small. In fact, one can bound $\Delta \leq d_{p_i, p_{i + 1}} \left(\sqrt{\frac{2}{1 - \cos(\pi - 2 \gamma)}} - 1\right)$. 
		Using parameter $\gamma$, the thickness of a lens can be controlled and thus the number $|L|$ of potential insertion candidates: the higher $\gamma$, the larger $|L|$ can get. 
		
		Now we choose the locally best vertex $l' \in L$ for inclusion in the path by minimizing the increase of the objective function value and selecting
		\begin{equation}
			\label{equation:lensProcedure:selectingRuleForTheVertexWhichShouldBeIncluded}
			l' = \arg\min_{l \in L}\big\{\zp(G, P)\colon P = (p_1, \ldots, p_i, l, p_{i + 1}, \ldots, p_{|P|})\big\},
		\end{equation}
		and set $P = (p_1, \ldots, p_i, l', p_{i + 1}, \ldots, p_{|P|})$.
		This procedure is repeated recursively on the edges $\{p_i, l'\}$ and $\{l', p_{i + 1}\}$ until there exists no more candidate, which could be included. 
		Note that non-recursively including all $l \in L$ at once could lead to an unwelcome increase of the objective function value since not only the total sum of all Euclidean distances, but also the sum of all turning angles has to be taken into account. 
		In fact, not all vertices $l \in L$ are necessarily contained in the lenses, which are stretched out in the further recursion steps.
		
		\begin{example}
			\label{example:lensProcedure}
			Let us consider the graph $G = (V, E)$ with $V = \{1, \ldots, 6\}$ depicted in Figure~\ref{figure:lensProcedure:lensHeuristicIdeaStartedOnTheEdge23Step1}, let $P = (1, 2, 3, 4)$ be a path and let us assume that we are solving the AngleTSP. Starting from the edge $\{2, 3\}$ and choosing 
			$\gamma = 30\degree\ (\approx 0.5236)$ 
			we obtain a lens (dashed in Figure~\ref{figure:lensProcedure:lensHeuristicIdeaStartedOnTheEdge23Step1}) containing both remaining vertices $5$ and $6$, \ie $L = \{5, 6\}$. Since $\zp( G, (1, 2, 6, 3, 4)) \approx 1.4201 < 1.8082 \approx \zp(G, (1, 2, 5, 3, 4))$, vertex $l' = 6$ is chosen and the process starts recursively with the new path $P = (1, 2, 6, 3, 4)$ and the two new edges $\{2,6\}$ and $\{6,3\}$ (see Figure~\ref{figure:lensProcedure:lensHeuristicIdeaStartedOnTheEdge23Step2}). The lens stretched out on the edge $\{2,6\}$ contains just one vertex $5$ and the other lens is empty. Thus the process stops with the new path $P = (1, 2, 5, 6, 3, 4)$.
			\begin{figure}[htb]
				\centering
				\begin{comment:figures}
					\begin{multicols}{2}
						\begin{figurehere}
							\centering
							\tikzset{my label/.style ={fill=red}}
							\begin{tikzpicture}[
									scale=0.66,
									decoration={
										markings,
										mark=at position 1.00 with {\arrow[scale=1.3]{angle 90}}
									}
								]
								\draw[dashed] (8.0, 1.5) arc (61.3896:118.6104:6.26560);
								\draw[dashed] (8.0, 1.5) arc (-61.3896:-118.6104:6.26560);
	
								\node[circle, draw=black!100, fill=black!100, thick, inner sep=0pt, minimum size=0.5mm, label=above:{\color{black} $1$}] (node0) at (0.0, 3.0) {};
								\node[circle, draw=black!100, fill=black!100, thick, inner sep=0pt, minimum size=0.5mm, label=above:{\color{black} $2$}] (node1) at (2.0, 1.5) {};
								\node[circle, draw=black!100, fill=black!100, thick, inner sep=0pt, minimum size=0.5mm, label=above:{\color{black} $3$}] (node2) at (8.0, 1.5) {};
								\node[circle, draw=black!100, fill=black!100, thick, inner sep=0pt, minimum size=0.5mm, label=above:{\color{black} $4$}] (node3) at (10.0, 0.0) {};
								\node[circle, draw=black!100, fill=black!100, thick, inner sep=0pt, minimum size=0.5mm, label=above:{\color{black} 5}] (node4) at (3.5, 1.9) {};
								\node[circle, draw=black!100, fill=black!100, thick, inner sep=0pt, minimum size=0.5mm, label={[above, fill=white]90:6}] (node5) at (5.0, 1.7) {};
								
								\draw[-] (node0) -- (node1) -- (node2) -- (node3);
								
								\draw[dotted] (node1) -- (node4) -- (node2);
								\draw[densely dotted] (node1) -- (node5) -- (node2);
							\end{tikzpicture}
							\caption{lens heuristic idea started from the edge $\{2,3\}$: step $1$}
							\label{figure:lensProcedure:lensHeuristicIdeaStartedOnTheEdge23Step1}
						\end{figurehere}
		
						\begin{figurehere}
							\centering
							\begin{tikzpicture}[
									scale=0.66,
									decoration={
										markings,
										mark=at position 1.00 with {\arrow[scale=1.3]{angle 90}}
									}
								]
								\node[circle, draw=black!100, fill=black!100, thick, inner sep=0pt, minimum size=0.5mm, label=above:{\color{black} $1$}] (node0) at (0.0, 3.0) {};
								\node[circle, draw=black!100, fill=black!100, thick, inner sep=0pt, minimum size=0.5mm, label=above:{\color{black} $2$}] (node1) at (2.0, 1.5) {};
								\node[circle, draw=black!100, fill=black!100, thick, inner sep=0pt, minimum size=0.5mm, label=above:{\color{black} $3$}] (node2) at (8.0, 1.5) {};
								\node[circle, draw=black!100, fill=black!100, thick, inner sep=0pt, minimum size=0.5mm, label=above:{\color{black} $4$}] (node3) at (10.0, 0.0) {};
								\node[circle, draw=black!100, fill=black!100, thick, inner sep=0pt, minimum size=0.5mm, label=above:{\color{black} 5}] (node4) at (3.5, 1.9) {};
								\node[circle, draw=black!100, fill=black!100, thick, inner sep=0pt, minimum size=0.5mm, label=above:{\color{black} 6}] (node5) at (5.0, 1.7) {};
								
								\draw[-] (node0) -- (node1) -- (node5) -- (node2) -- (node3);
								
								\draw[dashed] (5.0, 1.7) arc (65.9476:121.4711:3.2243);
								\draw[dashed] (5.0, 1.7) arc (-58.4242:-113.9477:3.2243);
								
								\draw[dashed] (8.0, 1.5) arc (58.3194:113.8429:3.2243);
								\draw[dashed] (8.0, 1.5) arc (-66.0524:-121.5759:3.2243);
	
								\draw[dotted] (node1) -- (node4) -- (node5);
							\end{tikzpicture}
							\caption{lens heuristic idea started from the edge $\{2,3\}$: step $2$}
							\label{figure:lensProcedure:lensHeuristicIdeaStartedOnTheEdge23Step2}
						\end{figurehere}
					\end{multicols}
				\end{comment:figures}
			\end{figure}
		\end{example}
		
		Although the approach described does not solve the problem itself, it can be used in combination with many other heuristics. Such combinations usually yield significantly better results than the original heuristics alone as documented in Section~\ref{section:computationalResults}.
		We will denote the {\em lens procedure} by \LENS{}.
	
	\section{Simple construction heuristics and related approaches}
		\label{section:trivialConstructionHeuristicsAndRelatedApproaches}
		
		\subsection{Nearest-neighbour heuristics}
			\label{subsection:nearestNeighbourHeuristics}
			The {\em nearest-neighbour heuristic} is one of the most common constructive heuristics used for the TSP (for details see \eg \cite{Reinelt:TheTravelingSalesmanComputationalSolutionsForTSPApplications}). 
			The original TSP-idea is to start from a vertex and move to the nearest neighbour (not yet visited) in every step until no vertex remains unvisited. 
			Finally, the tour is closed. In our problem we cannot start just from a vertex since we need two (incident) edges $e$ and $f$ to be able to estimate the cost value $c_{e f}$ of using them in succession. 
			Thus we start from an arbitrary edge $e = \{i,j\} \in E$ and look for a vertex $k \in V$, which minimizes the additional cost, \ie $k = \argmin_{k \in V \setminus \{i, j\}}\{c_{k i j}, c_{i j k}\}$. 
			This vertex is added either at the beginning or at the end of the path $(i, j)$ depending on the respective costs.
			Similarly to the TSP, this can be repeated until no unconnected vertex remains. In the last step, the tour is closed. This basic variant of the nearest-neighbour heuristic will be denoted by \NN{}. Note that \NN{} always looks for the best next vertex in both path directions; in the literature this variant is sometimes also called the {\em two-directional-nearest-neighbour heuristic}. {J\"ager} and {Molitor} \cite{JaegerMolitor:AlgorithmsAndExperimentalStudyForTheTravelingSalesmanProblemOfSecondOrder} and {Fischer} \etal \cite{FischerFischerJaegerKeilwagenMolitorGrosse:ExactAlgorithmsAndHeuristicsForTheQuadraticTravelingSalesmanProblemWithAnApplicationInBioinformatics}, however, understand a slightly different algorithm under this name.
			
			A common idea is to run \NN{} from all possible starting points, in our case from every start edge $e = \{i,j\} \in E$, and to choose the best solution. This variant will be denoted as \NNS{}.
			
			Based on \NNS{}, one can use an improvement heuristic on every generated \NN{}-solution and then choose the best one. We used the well known $2$-opt heuristic (for details see \eg \cite{Reinelt:TheTravelingSalesmanComputationalSolutionsForTSPApplications}), however, in order to reduce the computation time, we do not start from every possible edge $e = \{i,j\} \in E$, but from every edge $e = \{i,j\} \in E$ for a prespecified vertex $i$. We denote this \NNS{}-variant by \NNSTwo{}.
			
			An extended approach includes the lens procedure \LENS{}, described in Section~\ref{section:lensProcedure}, in the nearest neighbour heuristic:
			Whenever a new vertex is selected, \LENS{} is applied on the new edge to possibly include further vertices into the existing path. 
			We denote the resulting extended approach \NNL{} (resp.\ \NNSL{}) 
			for the underlying basic heuristic \NN{} (resp.\ \NNS{}).
			Because of the high running times of \NNSTwo{} together with \LENS{}, we did not pursue this combination any further. 			
			
			Since \NN{} is highly myopic, a look-ahead strategy might be employed which takes the next few potential iterations into account. 
			In particular, we consider not only the nearest neighbour, but the $w \in \nz$ nearest neighbours; for each such candidate we tentatively select it and repeat the whole process for $h \in \nz$ steps, \ie we evaluate the $h$ next steps of the algorithm and in every step we consider again the $w$ best possible successors. 
			In this way we construct a decision tree with {\em tree width} $w$ and {\em tree depth} $h$. 
			Finally, among all candidates generated in the first step, we choose the vertex, which led to the best solution in the last iteration,
			\ie the best leaf of the decision tree.
			This idea has similarities with the {\em piloting} strategy described by {Duin} and {Vo{\ss}}  \cite{DuinVoss:ThePilotMethodAStrategyForHeuristicRepetitionWithApplicationTotheSteinerProblemInGraphs}, where $w \in \nz$ candidates are considered, but the whole algorithm is evaluated for every such candidate (\ie no decision tree is constructed, but the heuristic is run to the end for every candidate generated in the first step). 
			We tested this {\em look-ahead} strategy for all four variants \NN{}, \NNS{}, \NNL{} and \NNSL{}. 
			However, it turned out that the objective function values were not better than the values produced by other ``good'' algorithms, while -- as can be expected -- the running times increased considerably.
			Thus, we did not follow this idea to the end.
			
		\subsection{Cheapest-insertion heuristic}
			\label{subsection:cheapestInsertionHeuristics}
			The next construction approach, again well-known from the TSP-context, 
			is the {\em cheapest-insertion heuristic} (for more details see \eg \cite{Reinelt:TheTravelingSalesmanComputationalSolutionsForTSPApplications}). A small tour is enlarged step by step until it contains all vertices. In particular, among all not visited vertices and all possible insertion positions the cheapest possibility is chosen in every step. Theoretically, the whole process could be started, similarly to \NN{}, with a random tour just having $3$ vertices, {Fischer} \etal \cite{FischerFischerJaegerKeilwagenMolitorGrosse:ExactAlgorithmsAndHeuristicsForTheQuadraticTravelingSalesmanProblemWithAnApplicationInBioinformatics}, however, offer a smarter rule: they start with the subtour $(u', v', w')$, where
			\begin{align}
				\label{equation:cheapestInsertionHeuristicStartingTourUV}
				(u', v')
					& = \arg\min_{u \neq v \in V}
					\left(\min_{x \in V \setminus \{u, v\}} c_{x u v} + 
					\min_{y \in V \setminus \{u, v\}} c_{u v y}\right) \text{ and}\\
				\label{equation:cheapestInsertionHeuristicStartingTourW}
				w'
					& = \arg\hspace*{-0.26cm}\min_{w \in V \setminus \{u', v'\}}
					\zt(G, (u', v', w)).
			\end{align}
			We will denote this algorithm \CIF{}.
			
			This algorithm exactly corresponds to the 
			{\em Cheapest-Insert Heuristic} (CI) in \cite{FischerFischerJaegerKeilwagenMolitorGrosse:ExactAlgorithmsAndHeuristicsForTheQuadraticTravelingSalesmanProblemWithAnApplicationInBioinformatics},
			where it can be seen as the winner under the presented non-exact algorithms if the instance classes are restricted to the \AngleTSP{} and the \AngleDistanceTSP; thus it will be used as a ``reference method'' in this paper.
	
	\section{Approaches based on convex hulls}
		\label{section:geometricBasedApproaches}
		Although the majority of the test instances used in \cite{FischerHelmberg:TheSymmetricQuadraticTravelingSalesmanProblem, Fischer:APolyhedralStudyOfQuadraticTravelingSalesmanProblems, FischerFischerJaegerKeilwagenMolitorGrosse:ExactAlgorithmsAndHeuristicsForTheQuadraticTravelingSalesmanProblemWithAnApplicationInBioinformatics} are {\AngleTSP}- or {\AngleDistanceTSP}-instances (for exact definitions of these instance-types see Section~\ref{subsubsection:benchmarkInstances}), all heuristics presented in these papers are designed for the general QTSP setting and hardly make use of the geometric properties of AngleTSP and AngleDistanceTSP. 
		Focusing on the AngleTSP we observed that optimal solutions consist of collections of large ``circles'' or ``spiral'' shapes (compare {Aichholzer} \etal \cite{AichholzerFischerFischerMeierPferschyPilzStanek:MinimizationAndMaximizationVersionsOfTheQuadraticTravellingSalesmanProblem}).
		We try to use this observation to design the following approach for the AngleTSP (note that we do not target the AngleDistanceTSP in this section).
		Recall that vertices are points in the Euclidean plane.
		
		If all vertices $v \in V$ of $G$ lay on a convex hull, an optimal AngleTSP-tour is obvious: it simply corresponds to the convex hull. Thus the main idea of our approach can be described as follows. 
		First, the convex hull $H^{1} \subseteq V$ of all vertices in $V$ is computed and defines the first (outermost) subtour $T^{1}$. 
		In the next step, all vertices in $H^{1}$ are removed from the graph and a new convex hull $H^{2} \subseteq V \setminus H^{1}$ is computed on the remaining vertex set, defining a new subtour $T^{2}$.
		This process is repeated until at most two unused vertices remain; 
		these vertices are included into the innermost (last) subtour by the \CIF{}-idea. 
		Finally, all $\tau$ obtained subtours $T^{1}, T^{2}, \ldots, T^{\tau}$ are merged into one single tour. We tested two different merging approaches.

		\subsection{Greedy merging approach}
			\label{subsection:geometricBasedApproaches:GreedyMergingApproach}
			Consider two different arbitrarily chosen subtours $T^{i}$ and $T^{\ell}$ with $1 \leq i \neq \ell \leq \tau$ and an arbitrary edge from each of them denoted by $e = \{u_e, v_e\} \in E\left(T^{i}\right)$ and $f = \{u_f, v_f\} \in E\left(T^{\ell}\right)$.
			There are two possibilities of merging the subtours $T^{i}$ and $T^{\ell}$: 
			either we connect them by the new edges $\{u_e, u_f\}$ and $\{v_e, v_f\}$ or by the new edges $\{u_e, v_f\}$ and $\{u_f, v_e\}$. 
			In our algorithm we try both possibilities for all edge pairs $e \in E\left(T^{i}\right)$ and $f \in E\left(T^{\ell}\right)$ and for all pairs of subtours $T^{i}$ and $T^{\ell}$.
			Among all possibilities, we choose the one with the smallest sum of
			objective function values over all subtours contained in the respective solution.
			The number of subtours decreases by one in this step. This procedure is repeated until a single tour is obtained. The algorithm in this form (\ie convex-hull approach combined with the described Greedy procedure) is abbreviated \CH{}.
			
			This simple Greedy idea can be enhanced by using {\em piloting}, \ie a look-ahead probing strategy, applied in an analogous way as it was done to \NN{}-based approaches in Section~\ref{subsection:nearestNeighbourHeuristics}. 
			However, in most cases the resulting algorithm yields the same tours as the basic {\em Greedy approach} while considerably worsening the computational time, so we did not follow this idea.
			
		\subsection{ILP merging approach}
			\label{subsection:geometricBasedApproaches:ILPMergingApproach}
			Another possibility is to merge all subtours $T^{1}, T^{2}, \ldots, T^{\tau}$, $k \geq 2$, at once by using a single ILP model. 
			The idea is to remove exactly one edge in every subtour and then to connect the resulting paths into a single tour in an optimal way. 
			
			In order to do so, we define an auxiliary graph $\widehat{G} = (\widehat{V}, F)$ where each vertex $i \in V$ is represented by a pair of vertices $a, b \in\widehat{V}$.
			The edge set $F \defeq F_L \cup F_S \cup F_C$ contains three types of edges:
			$F_L$, called {\em long edges}, replaces the original edges in a subtour, 
			$F_S$, called {\em short edges}, joins two vertices $a, b$ arising from one vertex $i \in V$
			and $F_C$, called {\em connecting edges}, joins vertices of different subtours.
			Formally, each subtour $T^{i}$ consists of $\left(v_1^{i}, v_2^{i}, \ldots, v_{\left| T^{i} \right|}^{i}\right)$ for $1 \leq i \leq \tau$
			with $v_j^{i} \in V$. 
			We define for $1 \leq i \leq \tau$:
			\begin{alignat}{2}
				\widehat{V}^{i}
					& = \left\{a_j^{i}, b_j^{i} \colon 1 \leq j \leq \left|T^{i}\right|\right\} 
						\nonumber\\
				F_S^{i}
					& = \left\{ \{a_j^{i}, b_j^{i}\} \colon 1 \leq j \leq \left|T^{i}\right|\right\}
						 \nonumber\\
				F_L^{i}
					& = \left\{ \{b_j^{i}, a_{j+1}^{i} \} \colon 1 \leq j \leq \left|T^{i}\right| - 1\right\}
						&\ \cup\ \left\{ \{ b_{\left|T^{i}\right|}, a_1^{i} \} \right\} \nonumber
			\end{alignat}
			For every pair of subtours $T^{i}$ and $T^{\ell}$ with $1 \leq i \neq \ell \leq \tau$ we define all possible connecting edges:
			\begin{alignat}{2}
				F_C^{i\ell} = \left\{ \{u,v\} \colon u \in \widehat{V}^{i}, v \in \widehat{V}^{\ell} \right\}  \nonumber
			\end{alignat}
			Finally, we have
			\begin{alignat}{2}
				\widehat{V} &\defeq \cup_{i = 1}^\tau \widehat{V}^{i},\\
				F_S &\defeq \cup_{i = 1}^\tau F_S^{i},\
				F_L \defeq \cup_{i = 1}^\tau F_L^{i},\ 
				F_C \defeq \cup_{i = 1}^\tau \cup_{\ell =1,\, \ell\neq i}^\tau F_C^{i\ell}\\
				F & \defeq F_S \cup F_L \cup F_C
			\end{alignat}
			
			\begin{example}
				\label{example:geometricBasedApproaches:ILPMergingApproach:creationOfTheGraphWidehatGWidehatVWidehatE}
				Consider the graph $G = (V, E)$ depicted in Figure~\ref{figure:geometricBasedApproaches:ILPMergingApproach:creationOfTheGraphWidehatGWidehatVWidehatE:graphGVESubtoursT1T2} and let $T^{1} = \left(v_1^{1}, v_2^{1}, v_3^{1}, v_4^{1}\right)$ and $T^{2} = \left(v_1^{2}, v_2^{2}, v_3^{2}, v_4^{2}\right)$ be two subtours. The corresponding graph $\widehat{G} = (\widehat{V}, F)$ is depicted in Figure~\ref{figure:geometricBasedApproaches:ILPMergingApproach:creationOfTheGraphWidehatGWidehatVWidehatE:graphWidehatGWidehatVWidehatE}. The edges $\{a_j^i, b_j^i\} \in F_S^{i}$, $i=1,2$, are depicted densely dotted, the edges in $F_L^i$, $i=1,2$, solid and some of the edges in $F_C$ incident to $a_2^1$ dashed. 
				\begin{figure}[htb]
					\centering
					\begin{comment:figures}
						\begin{multicols}{2}
							\begin{figurehere}
								\centering
								\begin{tikzpicture}[
										scale=0.89,
										decoration={
											markings,
											mark=at position 1.00 with {\arrow[scale=1.3]{angle 90}}
										}
									]
									\node[circle, draw=black!100, fill=black!100, thick, inner sep=0pt, minimum size=0.5mm, label=above:{\color{black} $v_1^{1}$}] (node0l) at (-3.0, 2.0) {};
									\node[circle, draw=black!100, fill=black!100, thick, inner sep=0pt, minimum size=0.5mm, label=above:{\color{black} $v_2^{1}$}] (node1l) at (-1.0, 2.0) {};
									\node[circle, draw=black!100, fill=black!100, thick, inner sep=0pt, minimum size=0.5mm, label=below:{\color{black} $v_3^{1}$}] (node2l) at (-1.0, 0.0) {};
									\node[circle, draw=black!100, fill=black!100, thick, inner sep=0pt, minimum size=0.5mm, label=below:{\color{black} $v_4^{1}$}] (node3l) at (-3.0, 0.0) {};
									
									\draw[-] (node0l) -- (node1l) -- (node2l) -- (node3l) -- (node0l);
									
									\node[circle, draw=black!100, fill=black!100, thick, inner sep=0pt, minimum size=0.5mm, label=above:{\color{black} $v_1^{2}$}] (node0r) at (1.0, 2.0) {};
									\node[circle, draw=black!100, fill=black!100, thick, inner sep=0pt, minimum size=0.5mm, label=above:{\color{black} $v_2^{2}$}] (node1r) at (3.0, 2.0) {};
									\node[circle, draw=black!100, fill=black!100, thick, inner sep=0pt, minimum size=0.5mm, label=below:{\color{black} $v_3^{2}$}] (node2r) at (3.0, 0.0) {};
									\node[circle, draw=black!100, fill=black!100, thick, inner sep=0pt, minimum size=0.5mm, label=below:{\color{black} $v_4^{2}$}] (node3r) at (1.0, 0.0) {};
									
									\draw[-] (node0r) -- (node1r) -- (node2r) -- (node3r) -- (node0r);
									
									\node at (-2.0, 1.0) {$T^{1}$};
									\node at (2.0, 1.0) {$T^{2}$};
								\end{tikzpicture}
								\caption{graph $G = (V, E)$ with subtours $T^{1}$ and $T^{2}$}
								\label{figure:geometricBasedApproaches:ILPMergingApproach:creationOfTheGraphWidehatGWidehatVWidehatE:graphGVESubtoursT1T2}
							\end{figurehere}
							
							\begin{figurehere}
								\centering
								\begin{tikzpicture}[
										scale=0.89,
										decoration={
											markings,
											mark=at position 1.00 with {\arrow[scale=1.3]{angle 90}}
										}
									]
									\node[circle, draw=black!100, fill=black!100, thick, inner sep=0pt, minimum size=0.5mm, label=left:{\color{black} ${a}_1^{1}$}] (node0l) at (-3.0, 1.8) {};
									\node[circle, draw=black!100, fill=black!100, thick, inner sep=0pt, minimum size=0.5mm, label=above:{\color{black} ${b}_1^{1}$}] (node1l) at (-2.8, 2.0) {};
									\node[circle, draw=black!100, fill=black!100, thick, inner sep=0pt, minimum size=0.5mm, label=above:{\color{black} ${a}_2^{1}$}] (node2l) at (-1.2, 2.0) {};
									\node[circle, draw=black!100, fill=black!100, thick, inner sep=0pt, minimum size=0.5mm, label=below left:{\color{black} ${b}_2^{1}$}] (node3l) at (-1.0, 1.8) {};
									\node[circle, draw=black!100, fill=black!100, thick, inner sep=0pt, minimum size=0.5mm, label=above left :{\color{black} ${a}_3^{1}$}] (node4l) at (-1.0, 0.2) {};
									\node[circle, draw=black!100, fill=black!100, thick, inner sep=0pt, minimum size=0.5mm, label=below:{\color{black} ${b}_3^{1}$}] (node5l) at (-1.2, 0.0) {};
									\node[circle, draw=black!100, fill=black!100, thick, inner sep=0pt, minimum size=0.5mm, label=below:{\color{black} ${a}_4^{1}$}] (node6l) at (-2.8, 0.0) {};
									\node[circle, draw=black!100, fill=black!100, thick, inner sep=0pt, minimum size=0.5mm, label=left:{\color{black} ${b}_4^{1}$}] (node7l) at (-3.0, 0.2) {};
									
									\draw[densely dotted] (node0l) -- (node1l) (node2l) -- (node3l) (node4l) -- (node5l) (node6l) -- (node7l);
									\draw[-] (node1l) -- (node2l) (node3l) -- (node4l) (node5l) -- (node6l) (node7l) -- (node0l);
									
									\node[circle, draw=black!100, fill=black!100, thick, inner sep=0pt, minimum size=0.5mm, label=below right:{\color{black} ${a}_1^{2}$}] (node0r) at (1.0, 1.8) {};
									\node[circle, draw=black!100, fill=black!100, thick, inner sep=0pt, minimum size=0.5mm, label=above:{\color{black} ${b}_1^{2}$}] (node1r) at (1.2, 2.0) {};
									\node[circle, draw=black!100, fill=black!100, thick, inner sep=0pt, minimum size=0.5mm, label=above:{\color{black} ${a}_2^{2}$}] (node2r) at (2.8, 2.0) {};
									\node[circle, draw=black!100, fill=black!100, thick, inner sep=0pt, minimum size=0.5mm, label=right:{\color{black} ${b}_2^{2}$}] (node3r) at (3.0, 1.8) {};
									\node[circle, draw=black!100, fill=black!100, thick, inner sep=0pt, minimum size=0.5mm, label=right:{\color{black} ${a}_3^{2}$}] (node4r) at (3.0, 0.2) {};
									\node[circle, draw=black!100, fill=black!100, thick, inner sep=0pt, minimum size=0.5mm, label=below:{\color{black} ${b}_3^{2}$}] (node5r) at (2.8, 0.0) {};
									\node[circle, draw=black!100, fill=black!100, thick, inner sep=0pt, minimum size=0.5mm, label=below:{\color{black} ${a}_4^{2}$}] (node6r) at (1.2, 0.0) {};
									\node[circle, draw=black!100, fill=black!100, thick, inner sep=0pt, minimum size=0.5mm, label=above right:{\color{black} ${b}_4^{2}$}] (node7r) at (1.0, 0.2) {};
									
									\draw[densely dotted] (node0r) -- (node1r) (node2r) -- (node3r) (node4r) -- (node5r) (node6r) -- (node7r);
									\draw[-] (node1r) -- (node2r) (node3r) -- (node4r) (node5r) -- (node6r) (node7r) -- (node0r);
									
									\draw[dashed] (node2l) -- (node1r);
									\draw[dashed] (node2l) -- (node0r);
									\draw[dashed] (node2l) to [out=0, in=135] (node7r);
									\draw[dashed] (node2l) to [out=0, in=180] (node6r);
								\end{tikzpicture}
								\caption{graph $\widehat{G} = (\widehat{V}, F)$ (not all edges $F_C$ are depicted)}
								\label{figure:geometricBasedApproaches:ILPMergingApproach:creationOfTheGraphWidehatGWidehatVWidehatE:graphWidehatGWidehatVWidehatE}
							\end{figurehere}
						\end{multicols}
					\end{comment:figures}
				\end{figure}
			\end{example}
			
			The goal of this auxiliary graph $\widehat{G} = (\widehat{V}, F)$ is to transform the {\em quadratic costs} (depending on two edges) into {\em linear weights} (depending on just one particular edge). 
			Obviously, by opening for each tour $T^i$ one long edge $l^i \in F_L^i$ and its two incident short edges $s_1^i, s_2^i \in F_S^i$, 
			and choosing exactly one connecting edge $r^i \in \cup_{\ell = 1,\, \ell\neq i}^k F_C^{i \ell}$ for each tour $T^i$, $1 \leq i \leq \tau $, we can build one single subtour in the auxiliary graph $\widehat{G} = (\widehat{V}, F)$, which corresponds to a tour in the original graph $G = (V, E)$.
			
			\begin{example}
				\label{example:geometricBasedApproaches:ILPMergingApproach:creationOfATourInTheGraphWidehatGWidehatVWidehatE}
				Let us again consider the graphs depicted in Figures~\ref{figure:geometricBasedApproaches:ILPMergingApproach:creationOfTheGraphWidehatGWidehatVWidehatE:graphGVESubtoursT1T2} and \ref{figure:geometricBasedApproaches:ILPMergingApproach:creationOfTheGraphWidehatGWidehatVWidehatE:graphWidehatGWidehatVWidehatE}. By taking all long and short edges except for the long edges $\left\{b_2^1, a_3^1\right\}$ and $\left\{b_4^2, a_1^2\right\}$ (one for every $i = 1, 2$) and the incident short edges $\left\{a_2^1, b_2^1\right\}$, $\left\{a_3^1, b_3^1\right\}$, $\left\{a_1^2, b_1^2\right\}$ and $\left\{a_4^2, b_4^2\right\}$ (two for every $i = 1, 2$) and by taking the two connecting edges $\left\{a_2^1, b_1^2\right\}$ and $\left\{a_4^2, b_3^1\right\}$ (one for every $i = 1, 2$) we clearly obtain the single subtour $T = \left(a_1^1, b_1^1, a_2^1, b_1^2, a_2^2, b_2^2, a_3^2, b_3^2, a_4^2, b_3^1, a_4^1, b_4^1\right)$ in the auxiliary graph $\widehat{G} = (\widehat{V}, F)$ corresponding to the tour $T = \left(v_1^1, v_2^1, v_1^2, v_2^2, v_3^2, v_4^2, v_3^1, v_4^1\right)$ in the original graph $G = (V, E)$. Note also that the vertices $b_2^1$, $a_3^1$, $b_4^2$ and $a_1^2$ (\ie two vertices for every $i = 1, 2$) remain unconnected in the auxiliary graph $\widehat{G} = (\widehat{V}, F)$ after our merging procedure.
			\end{example}
			
			Now, we assign linear weights $w_e \in \rz_0^+$ to all edges $e \in F$. We also write $w_{u v}$ for $w_e$ if $e = \{u, v\}$. For the short edges $s \in F_S$, we set
				\[
	w_{a_j^i b_j^i} = c_{v_{j-1}^i v_j^i v_{j+1}^i}
	\]
			for all $1 \leq i \leq \tau$ with the obvious cyclic completion around $j=\left|T^i\right|$.
	This means that the weight of a short edge represents the quadratic cost arising from the turn in the corresponding vertex $v_j^i$ in the particular subtour $T^i$ for $1 \leq i \leq \tau$. 
	For the long edges $l \in F_L$ we set $w_l = 0$ and for the connecting edges $F_C$ we set
				\begin{align}
	w_{a_j^i a_k^\ell}
	& = c_{v_{j - 1}^i v_j^i v_k^\ell} + c_{v_j^i v_k^\ell v_{k - 1}^\ell}\nonumber\\
	w_{a_j^i b_k^\ell}
	& = c_{v_{j - 1}^i v_j^i v_k^\ell} + c_{v_j^i v_k^\ell v_{k + 1}^\ell}\nonumber\\
	w_{b_j^i a_k^\ell}
	& = c_{v_{j + 1}^i v_j^i v_k^\ell} + c_{v_j^i v_k^\ell v_{k - 1}^\ell}\nonumber\\
	w_{b_j^i b_k^\ell}
	& = c_{v_{j + 1}^i v_j^i v_k^\ell} + c_{v_j^i v_k^\ell v_{k + 1}^\ell},\nonumber
	\end{align}	
			\ie the weights reflect the additional costs arising by the new connections.
			Again, the wrapping around of indices at the beginning and end of each subtour follows in a natural way.
			
			\begin{example}
				\label{example:geometricBasedApproaches:ILPMergingApproach:settingOfTheLinearWeights}
				Let us again consider the graphs depicted in Figures~\ref{figure:geometricBasedApproaches:ILPMergingApproach:creationOfTheGraphWidehatGWidehatVWidehatE:graphGVESubtoursT1T2} and \ref{figure:geometricBasedApproaches:ILPMergingApproach:creationOfTheGraphWidehatGWidehatVWidehatE:graphWidehatGWidehatVWidehatE}. Then
				\begin{align*}
					& w_{a_2^1 b_2^1} = c_{v_1^1 v_2^1 v_3^1},\, w_{a_3^1 b_3^1} = c_{v_2^1 v_3^1 v_4^1},\, w_{a_4^2 b_4^2} = c_{v_3^2 v_4^2 v_1^2},\ w_{a_1^2 b_1^2} = c_{v_4^2 v_1^2 v_2^2},\ w_{a_2^2 b_2^2} = c_{v_1^2 v_2^2 v_3^2},\\
					& w_{a_2^1 b_1^2} = c_{v_1^1 v_2^1 v_1^2} + c_{v_2^1 v_1^2 v_2^2},\, w_{a_2^1 a_1^2} = c_{v_1^1 v_2^1 v_1^2} + c_{v_2^1 v_1^2 v_4^2},\, \text{etc.}
				\end{align*}
				
				If we would merge the original two subtours $T^1 = \left(v_1^1, v_2^1, v_3^1, v_4^1\right)$ and $T^2 = \left(v_1^2, v_2^2, v_3^2, v_4^2\right)$ to one tour $T = \left(v_1^1, v_2^1, v_1^2, v_2^2, v_3^2, v_4^2, v_3^1, v_4^1\right)$ (in the same way as in Example~\ref{example:geometricBasedApproaches:ILPMergingApproach:creationOfATourInTheGraphWidehatGWidehatVWidehatE}), then we would decrease the total costs by $c_{v_1^1 v_2^1 v_3^1}$, $c_{v_2^1 v_3^1 v_4^1}$, $c_{v_4^2 v_1^2 v_2^2}$ and $c_{v_3^2 v_4^2 v_1^2}$ and increase them by $c_{v_1^1 v_2^1 v_1^2}$, $c_{v_2^1 v_1^2 v_2^2}$, $c_{v_3^2 v_4^2 v_3^1}$ and $c_{v_4^2 v_3^1 v_4^1}$.
				In the auxiliary graph $\widehat{G} = (\widehat{V}, F)$, this exactly corresponds to omitting the long and short edges $\left\{b_2^1, a_3^1\right\}$, $\left\{b_4^2, a_1^2\right\}$, $\left\{a_2^1, b_2^1\right\}$, $\left\{a_3^1, b_3^1\right\}$, $\left\{a_1^2, b_1^2\right\}$ and $\left\{a_4^2, b_4^2\right\}$ and inserting the connecting edges $\left\{a_2^1, b_1^2\right\}$ and $\left\{a_4^2, b_3^1\right\}$.
				Note that we omit exactly $6$ edges (one long and two incident short edges for every $i = 1,2$) and insert $2$ connecting edges (one for every $i = 1, 2$).
			\end{example}
			
			After having defined the weights, for every $1 \leq i \leq \tau$ in the graph $\widehat{G} = (\widehat{V}, F)$, we want to remove one long edge $l^i \in F_L^i$ and its two incident short edges $s_1^i, s_2^i \in F_S^i$, and choose exactly one new connecting edge $r^i \in \cup_{\ell = 1,\, \ell\neq i}^k F_C^{i \ell}$ in order to obtain one single subtour $\widehat{T}$ (which can be subsequently transformed into one single tour $T$ in the original graph $G = (V, E)$). The edges of this process are chosen such that the total costs of the merging are minimized. This is realized by the following ILP.
			Let $x_{f} \in \{0, 1\}$ be a binary variable indicating whether the edge $f \in F$ is included in the tour $\widehat{T}$ or not. Then we can model the described problem as follows.
			\allowdisplaybreaks[1]
			\begin{alignat}{5}
				\label{equation:geometricBasedApproaches:ILPMergingApproach:ILPObjectiveFunction}
				\mbox{min}\ 
					&& \sum_{f \in F} w_f x_f
						&&
							&
								&&\\
				\label{equation:geometricBasedApproaches:ILPMergingApproach:ILPRelaxed2MatchingConstraints}
				\mbox{s.t.}\ 
					&& \sum_{u \in \delta(v)}{x_{u v}} \ 
						&& \leq \ 
							& 2 \quad
								&& \forall\ v \in \widehat{V},\\
				\label{equation:geometricBasedApproaches:ILPMergingApproach:ILPSubtourEliminationConstraints}
					&& \sum_{u \neq v \in S}{x_{u v}} \ 
						&& \leq \ 
							& |S| - 1 \quad
								&&  \forall\ S \subset \widehat{V},\ S \neq \emptyset,\\
				\label{equation:geometricBasedApproaches:ILPMergingApproach:ILPSubtourConstraintsLongEdges}
					&& \sum_{l \in F_L^i}{x_l} \ 
						&& = \ 
							& \left|T^i\right| - 1 \quad
								&&  \forall\ 1 \leq i \leq \tau,\\
				\label{equation:geometricBasedApproaches:ILPMergingApproach:ILPSubtourConstraintsShortEdges}
					&& \sum_{s \in F_S^i}{x_s} \ 
						&& = \ 
							& \left|T^i\right| - 2 \quad
								&&  \forall\ 1 \leq i \leq \tau,\\
				\label{equation:geometricBasedApproaches:ILPMergingApproach:ILPLongEdgeIfTwoShortEdges}
					&& 2 x_l \ 
						&& \geq \ 
							& \hspace*{-0.32cm}\sum_{\substack{\{t, u\} \in F_S \cup F_C\\t \in \delta(u)}}\hspace*{-0.3cm} x_{t u} + \hspace*{-0.3cm}\sum_{\substack{\{v, t\} \in F_S \cup F_C\\t \in \delta(v)}}\hspace*{-0.3cm} x_{v t} \quad
								&& \forall\ l = \left\{u,v \right\} \in F_L,\\
					&& x_l \ 
						&& \leq \ 
							& x_{s_1} + x_{s_2} \quad
								&& \forall\ l \in F_L^i,\ \forall\ 1 \leq i \leq \tau,\nonumber\\
				\label{equation:geometricBasedApproaches:ILPMergingApproach:ILP3NoLongEdgeIfNoShortEdge}
					&& 
						&& 
							& 
								&& s_1 \neq s_2 \in F_S^i,\ \left|s_1 \cap l\right| = \left|s_2 \cap l\right| = 1,\\
				\label{equation:geometricBasedApproaches:ILPMergingApproach:ILPEverySubtourHasToBeConnected}
					&& \hspace*{-0.6cm}\sum_{u \in \widehat{V}^i,\ v \in \widehat{V} \setminus \widehat{V}^i}\hspace*{-0.6cm} x_{u, v} \ 
						&& = \ 
							& 2 \ 
								&& 1 \leq i \leq \tau,\\
				\label{equation:geometricBasedApproaches:ILPMergingApproach:ILPIntegrality}
					&& x_f \ 
						&& \in \ 
							& \{0, 1\} \quad
								&& \forall\ f \in F.
			\end{alignat}
			\allowdisplaybreaks[0]
			The constraints \eqref{equation:geometricBasedApproaches:ILPMergingApproach:ILPRelaxed2MatchingConstraints} ensure that the vertex degree is at most $2$. Note that we cannot enforce equality since we leave some vertices unconnected in graph $\widehat{G}$ (see also Example~\ref{example:geometricBasedApproaches:ILPMergingApproach:creationOfATourInTheGraphWidehatGWidehatVWidehatE}). \eqref{equation:geometricBasedApproaches:ILPMergingApproach:ILPSubtourEliminationConstraints} define the standard subtour elimination constraints while \eqref{equation:geometricBasedApproaches:ILPMergingApproach:ILPSubtourConstraintsLongEdges} and
			 \eqref{equation:geometricBasedApproaches:ILPMergingApproach:ILPSubtourConstraintsShortEdges}			 
			  guarantee that exactly one long edge $l \in F_L^i$ and
			  two short edges $s_1^i \neq s_2^i \in F_S^i$
			  are excluded for every $1 \leq i \leq \tau$.
			  \eqref{equation:geometricBasedApproaches:ILPMergingApproach:ILPLongEdgeIfTwoShortEdges} and \eqref{equation:geometricBasedApproaches:ILPMergingApproach:ILP3NoLongEdgeIfNoShortEdge} imply that each of the two excluded short edges $s_1^i$, $s_2^i$ shares a vertex with the removed long edge $l$.
			   Moreover, the constraints \eqref{equation:geometricBasedApproaches:ILPMergingApproach:ILPLongEdgeIfTwoShortEdges} ensure that all vertices, which are not included in the final tour $\widehat{T}$, are isolated and not connected among each other. 
			The cut constraints \eqref{equation:geometricBasedApproaches:ILPMergingApproach:ILPEverySubtourHasToBeConnected} make sure that every vertex set $\widehat{V}^i$, $ 1 \leq i \leq \tau$, is connected to its complement $\widehat{V} \setminus \widehat{V}^i$ by exactly two connecting edges $e \in F_C$. Finally, the integrality constraints are expressed in \eqref{equation:geometricBasedApproaches:ILPMergingApproach:ILPIntegrality}. Note also that \eqref{equation:geometricBasedApproaches:ILPMergingApproach:ILPEverySubtourHasToBeConnected} does not follow from \eqref{equation:geometricBasedApproaches:ILPMergingApproach:ILPSubtourEliminationConstraints} since we do not ensure equality in \eqref{equation:geometricBasedApproaches:ILPMergingApproach:ILPRelaxed2MatchingConstraints}.
			
			This model has similarities to the ILP known for the TSP (the objective function \eqref{equation:geometricBasedApproaches:ILPMergingApproach:ILPObjectiveFunction} and the constraints \eqref{equation:geometricBasedApproaches:ILPMergingApproach:ILPSubtourEliminationConstraints} and \eqref{equation:geometricBasedApproaches:ILPMergingApproach:ILPIntegrality} are the same and the constraints \eqref{equation:geometricBasedApproaches:ILPMergingApproach:ILPRelaxed2MatchingConstraints} are similar). To solve the above ILP, which has an exponential number of subtour elimination constraints \eqref{equation:geometricBasedApproaches:ILPMergingApproach:ILPSubtourEliminationConstraints}, we adapt the solution approach proposed by { Pferschy} and {Stan\v{e}k} in \cite{PferschyStanek:GeneratingSubtourEliminationConstraintsForTheTSPFromPureIntegerSolutions}. In particular, we relax the constraints \eqref{equation:geometricBasedApproaches:ILPMergingApproach:ILPSubtourEliminationConstraints}, solve the remaining ILP to integrality, add one subtour elimination constraint for every subtour and repeat the whole process until just one tour $\widehat{T}$ exists.
			
			Unfortunately, it turns out that the {\em ILP approach} for merging the subtours $T^1, T^2, \ldots, T^\tau$ representing the collection of convex hulls does not outperform the {\em Greedy approach} (see Section~\ref{subsection:geometricBasedApproaches:GreedyMergingApproach}) since the objective function values are fairly similar and the ILP approach always needs more time. 
			Nevertheless, the ILP model introduced is used later in Section~\ref{subsection:LPBasedApproaches:enhancedLPBasedStrategies} in a different context, where it yields very good results in competitive computational times. 
			Note that the above model contains much more restrictions than the related ILP for the standard TSP and therefore can be solved much faster than a pure TSP of the same size. 
			
		\subsection{Enhanced convex hull strategies}
			\label{subsection:geometricBasedApproaches:enhancedConvexHullStrategies}
			The solutions generated by the merging of convex hulls as described above yield fairly well-structured solutions in general.
			However, this positive aspect deteriorates towards the inner vertices, \ie for the convex hulls computed in the last few convex hull building iterations. 
			It can be observed that the outer convex hulls (from the first iterations) usually contain a larger number of vertices, while the inner convex hulls (from the last iterations) often consist of just a few vertices. 
			Since the sum of all turning angles in a convex polygon is always $2 \pi$, the average contribution to the objective function value of each vertex is usually much larger for these subtours with a small number of vertices.
			
			Therefore we introduce the following matheuristic: 
			We stop the convex hull building process as soon as the number of remaining vertices is at most a prespecified parameter $C \in \nz$. 
			Thus, we obtain at the end of the process the convex hulls $H^1$, $H^2$, \ldots, $H^{\tau^\prime}$, $\tau^\prime \in \nz$ with the corresponding subtours, and a set of remaining vertices $R = V \setminus \{H^1 \cup H^2 \cup \ldots \cup H^{\tau^\prime}\}$.
			Then we compute an optimal subtour $T^{\tau^\prime + 1}$ on the induced subgraph $G[R]$. 
			This is done by the {\em integral ILP based approach} described in \cite{AichholzerFischerFischerMeierPferschyPilzStanek:MinimizationAndMaximizationVersionsOfTheQuadraticTravellingSalesmanProblem}. 
			By choosing the value of parameter $C$ the trade-off between running time and solution quality can be controlled. 
			This approach is denoted by \CHC{}.
			Clearly, for $C = 2$ we obtain \CH{}.
			
			\bigskip
			A second enhanced strategy incorporates \LENS{} procedure: 
			After computing the vertices $H^i$ of the current convex hull and the corresponding subtour $T^i$ in iteration $i$, $1 \leq i \leq \tau$, 
			we run \LENS{} procedure on every edge in $T^i$ and thus possibly enlarge the subtour with an only moderate increase of the objective function value.
			We will denote this variant by \CHL{}. 
			The combined variant including the features of both  \CHC{} and \CHL{} will be denoted by \CHCL{}.
			
	\section{LP-based approaches}
		\label{section:LPBasedApproaches}
		In this section we present a different solution strategy based on the utilization of LP relaxations.
		The general approach can be described as follows. First, QTSP is modelled via an ILP. In the next step, the LP relaxation of this model is solved 
		and the fractional values of variables are rounded to integrality.
		In this way, one obtains either a partial solution
		or an infeasible solution with only limited violations of feasibility.
		Finally, the result of the rounding process is modified to either complete a partial solution or to reach feasibility. 
		This is a standard approach for designing approximation algorithms (see \eg \cite{WilliamsonShmoys:TheDesignOfApproximationAlgorithms, Gonzalez:HandbookOfApproximationAlgorithmsAndMetaheuristics})
		and heuristics.
		
		The QTSP can be written as the following quadratic integer program with binary edge variables $x_{u v}$ for $\{u, v\} \in E$.
		\allowdisplaybreaks[1]
		\begin{alignat}{5}
			\label{equation:LPBasedApproaches:QPObjectiveFunction}
			\mbox{min}\ 
				&& \sum_{u \neq v \neq t \in V}{c_{u v t} x_{u v} x_{v t}}
					&&
						&
							&&\\
			\label{equation:LPBasedApproaches:QP2MatchingConstraints}
			\mbox{s.t.}\ 
				&& \sum_{u \in \delta(v)}{x_{u v}} \ 
					&& = \ 
						& 2 \quad
							&& \forall\ v \in V,\\
			\label{equation:LPBasedApproaches:QPSubtourEliminationConstraints}
				&& \sum_{u \neq v \in S}{x_{u v}} \ 
					&& \leq \ 
						& |S| - 1 \quad
							&&  \forall\ S \subset V,\ S \neq \emptyset,\\
			\label{equation:LPBasedApproaches:QPIntegrality}
				&& x_{u v} \ 
					&& \in \ 
						& \{0, 1\} \quad
							&& \forall\ \{u, v\} \in E.
		\end{alignat}
		\allowdisplaybreaks[0]
		
		In the objective function \eqref{equation:LPBasedApproaches:QPObjectiveFunction} 
		costs $c_{u v t}$ are taken into account for the pair of consecutive edges $\{u, v\},\, \{v, t\} \in E,\, u \neq v \neq t$, if both edges $\{u, v\}$ and $\{v, t\}$ are contained in the tour. Equations \eqref{equation:LPBasedApproaches:QP2MatchingConstraints} are the {\em degree constraints} ensuring that each vertex is visited exactly once, \eqref{equation:LPBasedApproaches:QPSubtourEliminationConstraints} are the well-known {\em subtour elimination constraints} and, finally, \eqref{equation:LPBasedApproaches:QPIntegrality} are the integrality constraints on the edge variables. In comparison to the standard model for the TSP by {Dantzig} \etal \cite{DantzigFulkersonJohnson:SolutionOfALargeScaleTravelingSalesmanProblem} only the objective function is changed.
	
		This quadratic integer program can be linearized as suggested in \cite{FischerHelmberg:TheSymmetricQuadraticTravelingSalesmanProblem}. 
		We introduce a cubic number of additional integer variables $y_{u v t}$ for all pairs of incident edges 
		$\{u, v\},\, \{v, t\} \in E$, $u \neq v \neq t$, where $y_{u v t} = 1$ if and only if the vertices $u$, $v$ and $t$ are visited in the tour in consecutive order. 
			\allowdisplaybreaks[1]
		\begin{alignat}{5}
			\label{equation:LPBasedApproaches:ILPObjectiveFunction}
			\mbox{min}\ 
				&& \sum_{u \neq v \neq t \in V}{c_{u v t} y_{u v t}}
					&&
						&
							&&\\
			\nonumber
			\mbox{s.t.}\ 
				&& \eqref{equation:LPBasedApproaches:QP2MatchingConstraints}, \eqref{equation:LPBasedApproaches:QPSubtourEliminationConstraints}, \eqref{equation:LPBasedApproaches:QPIntegrality},\ \hspace*{-2cm}
					&& \ 
						& \quad
							&& \\
			\label{equation:LPBasedApproaches:ILPConstraintsCoupelingTheXAndYVariables}
				&& x_{u v} \ 
					&& = \ 
						& \sum_{t \in V \backslash \{u, v\}}{y_{u v t}} = \sum_{t \in V \backslash \{u, v\}}{y_{t u v}} \quad
							&& \forall\ \{u, v\} \in E,\\
			\label{equation:LPBasedApproaches:ILPIntegrality}
				&& y_{u v t} \ 
					&& \in \ 
						& \{0, 1\} \quad
							&& \forall\ u \neq v \neq t \in V.
		\end{alignat}
		\allowdisplaybreaks[0]
		
		The $x$-variables have to correspond to a tour due to the constraints \eqref{equation:LPBasedApproaches:QP2MatchingConstraints}--\eqref{equation:LPBasedApproaches:QPIntegrality}. Apart from that, this model has a linear objective function \eqref{equation:LPBasedApproaches:ILPObjectiveFunction}. 
		Constraints \eqref{equation:LPBasedApproaches:ILPConstraintsCoupelingTheXAndYVariables} couple the $x$- and the $y$-variables. 
		Finally, conditions \eqref{equation:LPBasedApproaches:ILPIntegrality} ensure the integrality of the $y$-variables.
		
		Our approach works as follows. 
		Based on this ILP, we relax the integrality constraints \eqref{equation:LPBasedApproaches:QPIntegrality} and \eqref{equation:LPBasedApproaches:ILPIntegrality} and the subtour elimination constraints \eqref{equation:LPBasedApproaches:QPSubtourEliminationConstraints} and solve the obtained LP (obviously having polynomial size) to optimality. 
		Let us denote by $x_e^\prime$ the obtained solution value of the $x$-variable for each edge $e \in E$. 
		To construct a partial solution we first set $x_e = 0$ for all $e \in E$ and choose
		a rounding parameter $\rho$.		
		Then we consider all $x^\prime$-variables with $x_e^\prime > \rho$ in decreasing order and set $x_e = 1$ if and only if the following two conditions are fulfilled:
		\begin{enumerate}
			\item[C1.] The degree of each vertex remains at most $2$, \ie $\sum_{u \in \delta(v)}{x_{u v}} \leq 2$ for all $v \in V$.
			\item[C2.] The rounding does not imply any cycle in the resulting partial solution.
		\end{enumerate}
		By this rounding procedure we obviously obtain a set of paths and isolated points. 
		From these building blocks we construct a tour, \ie a feasible QTSP-solution, with the following two steps. First, all paths are combined into one cycle by using an auxiliary ILP. Secondly, the isolated points are inserted into this cycle by the \CIF{}-idea. 
		The details of the resulting algorithm, denoted by \LPP{}, can be described as follows.
		
		For the merging of paths into a single cycle we proceed similarly to the merging procedure described in Section~\ref{subsection:geometricBasedApproaches:ILPMergingApproach}
		and transfer the quadratic costs depending on two incident edges in the original graph into linear weights depending on one edge in a new auxiliary graph.
		Let $P^1 = \left(v_1^1, v_2^1, \ldots, v_{\left|P^1\right|}^1\right)$, $P^2 = \left(v_1^2, v_2^2, \ldots, v_{\left|P^2\right|}^2\right)$, \ldots, $P^\tau = \left(v_1^\tau, v_2^\tau, \ldots, v_{\left|P^\tau\right|}^\tau\right)$, where $\cup_{i = 1}^\tau P^i \subseteq V$, be the $\tau$ paths obtained from our rounding procedure (putting aside the isolated vertices for the time being).
		We now define an auxiliary graph $\widehat{G} = (\widehat{V}, F)$, where each path $P^i$, $1 \leq i \leq \tau$, is represented by two vertices $a^i, b^i \in \widehat{V}$. 
		The edge set $F \defeq F_P \cup F_C$ contains two types of edges: 
		{\em Path edges} in $F_P$ represent the original paths and connect the $a$-vertices with their $b$-counterparts.
		{\em Connecting edges} in $F_C$ join vertices representing different paths. 
		Formally we define
		\begin{alignat}{2}
		\widehat{V}
		& \defeq \left\{a^i, b^i\colon i=1,\ldots, \tau\right\},\nonumber\\
		F_P
		& \defeq \left\{\{a^i, b^i\}\colon i=1,\ldots, \tau\right\} \nonumber\\
		F_C
		& \defeq \left\{\{u,v\}\colon u \neq v \in \widehat{V} \right\} \setminus F_P \nonumber
		\end{alignat}
%
			
			\begin{example}
				\label{example:LPBasedApproaches}
				Consider the graph $G = (V, E)$ given in Figure~\ref{figure:LPBasedApproaches:graphGVEWithPathsP1P2AndP3} and let $P^1 = \left(v_1^1, v_2^1\right)$, $P^2 = \left(v_1^2, v_2^2, v_3^2\right)$ and $P^3 = \left(v_1^3, v_2^3, v_3^3\right)$ be three paths obtained from  the rounding procedure. The auxiliary graph $\widehat{G} = (\widehat{V}, F)$ is depicted in Figure~\ref{figure:LPBasedApproaches:graphWidehatGWidehatVF}. 
				Edges $F_P$, where $\{a^i, b^i\}$ corresponds to path $P^i$, $i=1,2,3$, are depicted solid, edges $F_C$ incident to $a^1$ dashed and the edges $F_C$ incident to $b^1$ densely dotted (other edges of $F_C$ are not depicted).
	\end{example}
				\begin{figure}[htb]
					\centering
					\begin{comment:figures}
						\begin{multicols}{2}
							\begin{figurehere}
								\centering
								\begin{tikzpicture}[
										scale=0.2,
										decoration={
											markings,
											mark=at position 1.00 with {\arrow[scale=1.3]{angle 90}}
										}
									]
									\node[circle, draw=white!100, fill=white!100, thick, inner sep=0pt, minimum size=0.5mm, label=above:{\color{white} $a^1$}] (node11) at (0.000000000000, 5.000000000000) {};
									\node[circle, draw=white!100, fill=white!100, thick, inner sep=0pt, minimum size=0.5mm, label=above right:{\color{white} $b^1$}] (node12) at (3.535533905933, 3.535533905933) {};
									\node[circle, draw=white!100, fill=white!100, thick, inner sep=0pt, minimum size=0.5mm, label=right:{\color{white} $a^2$}] (node13) at (5.000000000000, 0.000000000000) {};
									\node[circle, draw=white!100, fill=white!100, thick, inner sep=0pt, minimum size=0.5mm, label=below:{\color{white} $b^2$}] (node15) at (0.000000000000, -5.000000000000) {};
									\node[circle, draw=white!100, fill=white!100, thick, inner sep=0pt, minimum size=0.5mm, label=below left:{\color{white} $a^3$}] (node16) at (-3.535533905933, -3.535533905933) {};
									\node[circle, draw=white!100, fill=white!100, thick, inner sep=0pt, minimum size=0.5mm, label=above left:{\color{white} $b^3$}] (node18) at (-3.535533905933, 3.535533905933) {};
									
									\node[circle, draw=black!100, fill=black!100, thick, inner sep=0pt, minimum size=0.5mm, label=above:{\color{black} $v_1^1$}] (node1) at (0.000000000000, 5.000000000000) {};
									\node[circle, draw=black!100, fill=black!100, thick, inner sep=0pt, minimum size=0.5mm, label=above right:{\color{black} $v_2^1$}] (node2) at (3.535533905933, 3.535533905933) {};
									\node[circle, draw=black!100, fill=black!100, thick, inner sep=0pt, minimum size=0.5mm, label=right:{\color{black} $v_1^2$}] (node3) at (5.000000000000, 0.000000000000) {};
									\node[circle, draw=black!100, fill=black!100, thick, inner sep=0pt, minimum size=0.5mm, label=below right:{\color{black} $v_2^2$}] (node4) at (3.535533905933, -3.535533905933) {};
									\node[circle, draw=black!100, fill=black!100, thick, inner sep=0pt, minimum size=0.5mm, label=below:{\color{black} $v_3^2$}] (node5) at (0.000000000000, -5.000000000000) {};
									\node[circle, draw=black!100, fill=black!100, thick, inner sep=0pt, minimum size=0.5mm, label=below left:{\color{black} $v_1^3$}] (node6) at (-3.535533905933, -3.535533905933) {};
									\node[circle, draw=black!100, fill=black!100, thick, inner sep=0pt, minimum size=0.5mm, label=left:{\color{black} $v_2^3$}] (node7) at (-5.000000000000, 0.000000000000) {};
									\node[circle, draw=black!100, fill=black!100, thick, inner sep=0pt, minimum size=0.5mm, label=above left:{\color{black} $v_3^3$}] (node8) at (-3.535533905933, 3.535533905933) {};
									
									\draw (node1) -- (node2) (node3) -- (node4) -- (node5) (node6) -- (node7) -- (node8);
									
									\node at (1.50, 2.78) {$P^1$};
									\node at (2.70, -2.50) {$P^2$};
									\node at (-3.30, 0.00) {$P^3$};
								\end{tikzpicture}
								\caption{LP-based approach: graph $G = (V, E)$ with paths $P^1$, $P^2$ and $P^3$}
								\label{figure:LPBasedApproaches:graphGVEWithPathsP1P2AndP3}
							\end{figurehere}
			
							\begin{figurehere}
								\centering
								\begin{tikzpicture}[
										scale=0.2,
										decoration={
											markings,
											mark=at position 1.00 with {\arrow[scale=1.3]{angle 90}}
										}
									]
									\node[circle, draw=white!100, fill=white!100, thick, inner sep=0pt, minimum size=0.5mm, label=above:{\color{white} $v_1^1$}] (node11) at (0.000000000000, 5.000000000000) {};
									\node[circle, draw=white!100, fill=white!100, thick, inner sep=0pt, minimum size=0.5mm, label=above right:{\color{white} $v_2^1$}] (node12) at (3.535533905933, 3.535533905933) {};
									\node[circle, draw=white!100, fill=white!100, thick, inner sep=0pt, minimum size=0.5mm, label=right:{\color{white} $v_1^2$}] (node13) at (5.000000000000, 0.000000000000) {};
									\node[circle, draw=white!100, fill=white!100, thick, inner sep=0pt, minimum size=0.5mm, label=below right:{\color{white} $v_2^2$}] (node14) at (3.535533905933, -3.535533905933) {};
									\node[circle, draw=white!100, fill=white!100, thick, inner sep=0pt, minimum size=0.5mm, label=below:{\color{white} $v_3^2$}] (node15) at (0.000000000000, -5.000000000000) {};
									\node[circle, draw=white!100, fill=white!100, thick, inner sep=0pt, minimum size=0.5mm, label=below left:{\color{white} $v_1^3$}] (node16) at (-3.535533905933, -3.535533905933) {};
									\node[circle, draw=white!100, fill=white!100, thick, inner sep=0pt, minimum size=0.5mm, label=left:{\color{white} $v_2^3$}] (node17) at (-5.000000000000, 0.000000000000) {};
									\node[circle, draw=white!100, fill=white!100, thick, inner sep=0pt, minimum size=0.5mm, label=above left:{\color{white} $v_3^3$}] (node18) at (-3.535533905933, 3.535533905933) {};
									
									\node[circle, draw=black!100, fill=black!100, thick, inner sep=0pt, minimum size=0.5mm, label=above:{\color{black} $a^1$}] (node1) at (0.000000000000, 5.000000000000) {};
									\node[circle, draw=black!100, fill=black!100, thick, inner sep=0pt, minimum size=0.5mm, label=above right:{\color{black} $b^1$}] (node2) at (3.535533905933, 3.535533905933) {};
									\node[circle, draw=black!100, fill=black!100, thick, inner sep=0pt, minimum size=0.5mm, label=right:{\color{black} $a^2$}] (node3) at (5.000000000000, 0.000000000000) {};
									\node[circle, draw=black!100, fill=black!100, thick, inner sep=0pt, minimum size=0.5mm, label=below:{\color{black} $b^2$}] (node5) at (0.000000000000, -5.000000000000) {};
									\node[circle, draw=black!100, fill=black!100, thick, inner sep=0pt, minimum size=0.5mm, label=below left:{\color{black} $a^3$}] (node6) at (-3.535533905933, -3.535533905933) {};
									\node[circle, draw=black!100, fill=black!100, thick, inner sep=0pt, minimum size=0.5mm, label=above left:{\color{black} $b^3$}] (node8) at (-3.535533905933, 3.535533905933) {};
									
									\draw (node1) -- (node2) (node3) -- (node5) (node6) -- (node8);
									
									\draw[dashed] (node1) -- (node3) (node1) -- (node5) (node1) -- (node6) (node1) -- (node8);
									
									\draw[densely dotted] (node2) -- (node3) (node2) -- (node5) (node2) -- (node6) (node2) -- (node8);
								\end{tikzpicture}
								\caption{LP-based approach: graph $\widehat{G} = (\widehat{V}, F)$ (not all edges $F_C$ are depicted)}
								\label{figure:LPBasedApproaches:graphWidehatGWidehatVF}
							\end{figurehere}
						\end{multicols}
					\end{comment:figures}
				\end{figure}
			Similarly to the auxiliary graph described in Section~\ref{subsection:geometricBasedApproaches:ILPMergingApproach}
			 $\widehat{G} = (\widehat{V}, F)$ transforms the {\em quadratic costs} depending on two edges into {\em linear weights} depending only on one particular edge. Obviously, by choosing one edge $r \in F_C$, one can compute the cost of the new connection in the original graph $G = (V, E)$. 
			 Formally, we set $w_{uv} = 0$ for all $\{u,v\} \in F_P$ and
			\begin{alignat*}{2}
				w_{a^i a^\ell}
					& = c_{v_2^i v_1^i v_1^\ell} + c_{v_1^i v_1^\ell v_2^\ell},\\
				w_{a^i b^\ell}
					& = c_{v_2^i v_1^i v_{\left|P^\ell\right|}^\ell} + c_{v_1^i v_{\left|P^\ell\right|}^\ell v_{\left|P^\ell\right| - 1}^\ell},\\
				w_{b^i a^\ell}
					& = c_{v_{\left|P^i\right| - 1}^i v_{\left|P^i\right|}^i v_1^\ell} + c_{v_{\left|P^i\right|}^i v_1^\ell v_2^\ell},\\
				w_{b^i b^\ell}
					& = c_{v_{\left|P^i\right| - 1}^i v_{\left|P^i\right|}^i v_{\left|P^\ell\right|}^\ell} + c_{v_{\left|P^i\right|}^i v_{\left|P^\ell\right|}^\ell v_{\left|P^\ell\right| - 1}^\ell}
			\end{alignat*}
			for the connecting edges $F_C$, $1 \leq i \neq \ell \leq \tau$.
			Note that in this way, the costs incurred along the paths are not explicitly included in the merging model, since they remain constant anyway.
			\begin{example}
				\label{example:LPBasedApproaches:settingOfTheLinearWeights}
				Let us again consider the graphs depicted in Figures~\ref{figure:LPBasedApproaches:graphGVEWithPathsP1P2AndP3} and \ref{figure:LPBasedApproaches:graphWidehatGWidehatVF}. Then
				\begin{align*}
					& w_{a^1 b^1} = 0,\, w_{a^2 b^2} = 0,\, w_{a^3 b^3} = 0,\, w_{a^1 a^2} = c_{v_2^1 v_1^1 v_1^2} + c_{v_1^1 v_1^2 v_2^2},\, w_{a^1, b^2} = c_{v_2^1 v_1^1 v_3^2} + c_{v_1^1 v_3^2 v_2^2},\\
					& w_{b^1 a^2} = c_{v_1^1 v_2^1 v_1^2} + c_{v_2^1 v_1^2 v_2^2},\, w_{b^1, b^2} = c_{v_1^1 v_2^1 v_3^2} + c_{v_2^1 v_3^2 v_2^2},\, \text{etc.}
				\end{align*}
			\end{example}
			Now, we want to merge the paths into one cycle $\widehat{T}$ in an optimal way. Assume that $\tau \geq 2$ (other cases will be considered below) and let $x_{f} \in \{0, 1\}$ be a binary variable indicating whether edge $f \in F$ is included in the cycle $\widehat{T}$ or not. 
			Then we can model the merging problem as follows.			
			\allowdisplaybreaks[1]
			\begin{alignat}{5}
				\label{equation:LPBasedApproaches:AuxiliaryILPObjectiveFunction}
				\mbox{min} 
					&& \sum_{f \in F} w_f x_f
						&&
							&
								&&\\
				\label{equation:LPBasedApproaches:AuxiliaryILPRelaxed2MatchingConstraints}
				\mbox{s.t.}
					&& \sum_{u \in \delta(v)} x_{uv} \ 
						&& = \ 
							& 2 \quad
								&& \forall\ v \in \widehat{V},\\
				\label{equation:LPBasedApproaches:AuxiliaryILPSubtourEliminationConstraints}
					&& \sum_{u \neq v \in S}{x_{u v}} \ 
						&& \leq \ 
							& |S| - 1 \quad
								&&  \forall\ S \subset \widehat{V},\ S \neq \emptyset,\\
				\label{equation:LPBasedApproaches:AuxiliaryILPFixingOfThePathEdges}
					&& x_f \ 
						&& = \ 
							& 1 \quad
								&&  \forall\ f \in F_P,\\
				\label{equation:LPBasedApproaches:AuxiliaryILPIntegrality}
					&& x_f \ 
						&& \in \ 
							& \{0, 1\} \quad
								&& \forall\ f \in F.
			\end{alignat}
			\allowdisplaybreaks[0]
			The objective function value \eqref{equation:LPBasedApproaches:AuxiliaryILPObjectiveFunction} and the constraints \eqref{equation:LPBasedApproaches:AuxiliaryILPRelaxed2MatchingConstraints}, \eqref{equation:LPBasedApproaches:AuxiliaryILPSubtourEliminationConstraints} and \eqref{equation:LPBasedApproaches:AuxiliaryILPIntegrality} 
			are equivalent to the standard ILP model for the TSP.
			The additional constraints \eqref{equation:LPBasedApproaches:AuxiliaryILPFixingOfThePathEdges} guarantee that all paths in the original graph are included in the resulting cycle. 
			Similarly to Section~\ref{subsection:geometricBasedApproaches:ILPMergingApproach} we adapt the solution approach proposed by {Pferschy} and {Stan\v{e}k} in \cite{PferschyStanek:GeneratingSubtourEliminationConstraintsForTheTSPFromPureIntegerSolutions} for this problem, \ie we relax the subtour elimination constraints \eqref{equation:LPBasedApproaches:AuxiliaryILPSubtourEliminationConstraints}, solve the remaining ILP to integrality, add one subtour elimination constraint for every subtour and repeat the whole process until just one tour $\widehat{T}$ exists. This cycle can then be easily transformed into a subtour $T$ in the original graph $G = (V, E)$.
			
			At the end, the remaining isolated points are inserted into the tour $T$ by the \CIF{}-idea. 
			Now let us briefly consider the special case that the rounding procedure yields only one path ($\tau=1$):
			If $\left|P^1\right| \geq 3$, we just close the path to a cycle instead of solving the auxiliary ILP and then insert the remaining isolated points by \CIF{}.
			If $\left|P^1\right| = 2$, we use this path as a starting edge of \CIF{}
			similar to \eqref{equation:cheapestInsertionHeuristicStartingTourUV},
			choose the next vertex according to \eqref{equation:cheapestInsertionHeuristicStartingTourW} and continue by running \CIF{}. 
			If no path is formed ($\tau=0$), the default version of \CIF{} is performed.
						
			The choice of the rounding parameter $\rho$ influences the number of (undesired) isolated points generated by the rounding procedure.  
			However, there is a trade-off: 
			Decreasing $\rho$, the number of isolated points decreases, but the generated paths tend to be of lower quality, and vice-versa. Therefore a careful setting of this parameter is required.
			
			Finally, let us remark that in theory we only know that the size of the auxiliary graph $\widehat{G} = (\widehat{V}, F)$ is smaller or equal to the size of the original graph $G = (V, E)$, but in practice it is observed to be much smaller. 
			Roughly spoken, this approach transforms the QTSP instances to TSP instances of at most equal size, \ie in the described algorithm we solve the quadratic problem by using its linear version without increasing the instance size.
			Since the size of usual QTSP test instances is perceived as quite small if the data is used for TSP instances, the exact solution of the auxiliary TSP (with one additional set of constraints) does not increase the running time too much.

		\subsection{Enhanced LP-based strategies}
			\label{subsection:LPBasedApproaches:enhancedLPBasedStrategies}
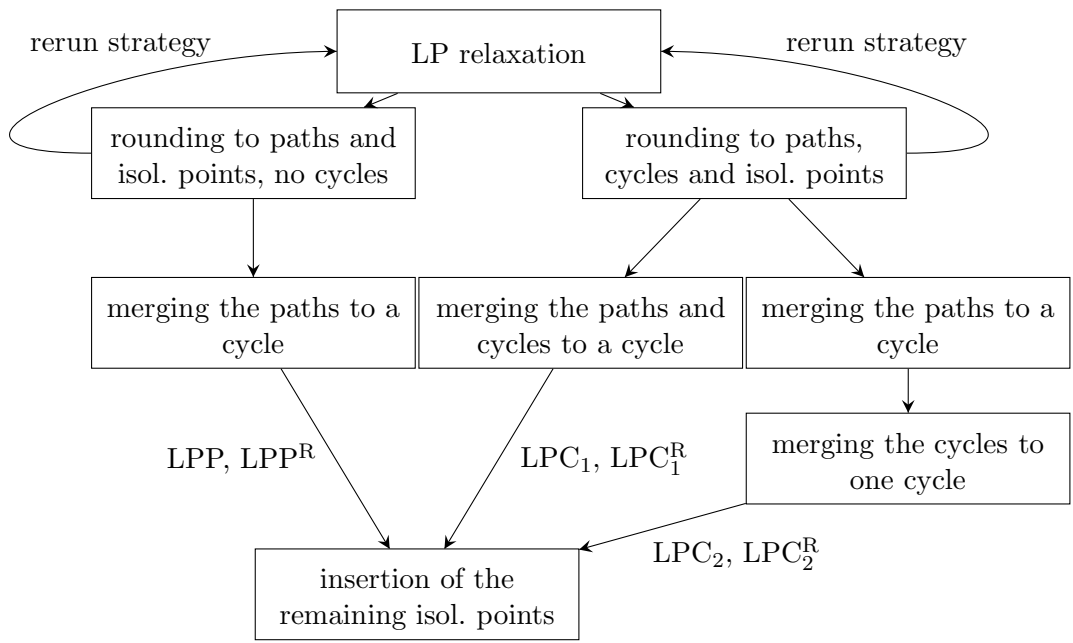
\begin{figure}[htb]
				\centering
				\begin{comment:figures}
					\hspace*{-1.7cm}
					\begin{tikzpicture}[
							xscale=1.025, yscale=0.9,
							node/.style={circle, draw=black!100, fill=white!0, thick, inner sep=0pt, minimum size=25mm}
						]
						\node (node0) at (0.85,0) [black,draw,minimum width=4cm,minimum height=1.1cm,text width=4cm] {\vspace*{-0.5cm}\begin{center}LP relaxation\end{center}};
						\node (node1) at (-2.3,-1.5) [black,draw,minimum width=4cm,minimum height=1.1cm,text width=4cm] {\vspace*{-0.4cm}\begin{center}rounding to paths and isol.~points, no cycles\end{center}};
						\node (node2) at (4,-1.5) [draw,minimum width=4cm,minimum height=1.1cm,text width=4cm] {\vspace*{-0.4cm}\begin{center}rounding to paths, cycles and isol.~points\end{center}};
						\node (node3) at (-2.3,-4) [black,draw,minimum width=4cm,minimum height=1.1cm,text width=4cm] {\vspace*{-0.4cm}\begin{center}merging the paths to a cycle\end{center}};
						\node (node4) at (1.9,-4) [draw,minimum width=4cm,minimum height=1.1cm,text width=4cm] {\vspace*{-0.4cm}\begin{center}merging the paths and cycles to a cycle\end{center}};
						\node (node5) at (6.1,-4) [draw,minimum width=4cm,minimum height=1.1cm,text width=4cm] {\vspace*{-0.4cm}\begin{center}merging the paths to a cycle\end{center}};
						\node (node6) at (6.1,-6) [draw,minimum width=4cm,minimum height=1.1cm,text width=4cm] {\vspace*{-0.4cm}\begin{center}merging the cycles to one cycle\end{center}};
						\node (node7) at (-0.2,-8) [black,draw,minimum width=4cm,minimum height=1.1cm,text width=4cm] {\vspace*{-0.4cm}\begin{center}insertion of the remaining isol.~points\end{center}};
						
	
						\path (node1) edge [loop,looseness=2,out=180, in=180,above] (node0);
						\draw[arrow] (-1.24,0) -- (node0);
						\path (node2) edge [loop,looseness=2,out=0, in=0,above] (node0);
						\draw[arrow] (2.94,0) -- (node0);
						
						\node at (-4, 0.1) {rerun strategy};
						\node at (5.7, 0.1) {rerun strategy};
						
						\draw[arrow] (node0) -- (node1);
						\draw[arrow] (node1) -- (node3);
						\draw[arrow] (node3) -- (node7);
						\draw[arrow] (node2) -- (node4);
						\draw[arrow] (node4) -- (node7);
						\draw[arrow] (node0) -- (node2);
						\draw[arrow] (node2) -- (node5);
						\draw[arrow] (node5) -- (node6);
						\draw[arrow] (node6) -- (node7);
											
						\node at (-2.45, -6) {\LPP, \LPPR};
						\node at (2.2, -6) {\LPCOne, \LPCOneR};
						\node at (3.9, -7.4) {\LPCTwo, \LPCTwoR};
					\end{tikzpicture}
				\end{comment:figures}
				\caption{enhanced LP-based strategies}
				\label{figure:LPBasedApproaches:enhancedLPBasedStrategies:Diagram}
			\end{figure}

			In this section we introduce some enhanced strategies, which lead to better results for the \LPP{}-algorithm. A summary of the different improvement strategies is given in Figure~\ref{figure:LPBasedApproaches:enhancedLPBasedStrategies:Diagram}. 
			\LPP{} corresponds to the left-most possibility, \ie after the {\em rounding to paths and isolated points, no cycles} step, {\em merging the paths to a cycle} yields one cycle, which is enlarged by the remaining isolated points in the last step.
			
			The {\em rerun strategy} is the first possibility for enhancing \LPP{}:  After solving the LP relaxation and performing the rounding procedure, we have $x_e = 1$ for some $e \in E$. Now, we fix $x_e = 1$ for all such edges and resolve the LP with a smaller number of free variables and start the rounding procedure again in order to obtain $x_e = 1$ for more edges $e \in E$. This process is repeated until the number of isolated points remains unchanged. 
			\LPP{} combined with this {\em rerun strategy} will be denote as \LPPR{}.
		
			The next idea focuses on the rounding procedure. In \LPP{} (and in \LPPR{}) $x_e$ is set to $1$ only if the edge $e \in E$ does not close a cycle. Due to this condition we obtain only paths and isolated points in the next step, but we ``lose'' some information provided by the solution of the LP relaxation. Instead, we can also permit to close cycles in the rounding procedure, \ie omit the condition C2, and stick only to the degree condition C1. Then we have to deal with a structure consisting of subtours, paths and isolated points. One obvious possibility is to merge the paths to one cycle in the same way as it is done in \LPP{} and then merge all cycles (\ie the ones directly obtained from the rounding procedure and the one created from the paths) to one subtour. 
			This can be done by subsequently solving two auxiliary ILPs, namely
			\eqref{equation:LPBasedApproaches:AuxiliaryILPObjectiveFunction}--\eqref{equation:LPBasedApproaches:AuxiliaryILPIntegrality}
			and then \eqref{equation:geometricBasedApproaches:ILPMergingApproach:ILPObjectiveFunction}--\eqref{equation:geometricBasedApproaches:ILPMergingApproach:ILPIntegrality} as described in Section~\ref{subsection:geometricBasedApproaches:ILPMergingApproach}.
			As the last step, the isolated points are included by the \CIF{}-idea. This approach corresponds to the right-most possibility in Figure~\ref{figure:LPBasedApproaches:enhancedLPBasedStrategies:Diagram} and can be used with or without the rerun strategy. 
			We will denote the resulting algorithms by \LPCTwoR{} and \LPCTwo{}, respectively.
			
		
			\LPCTwoR{} and \LPCTwo{} use both auxiliary ILPs
			described in Sections~\ref{section:LPBasedApproaches} and \ref{subsection:geometricBasedApproaches:ILPMergingApproach} in succession. 
			A natural idea is to combine these ILPs and to merge all cycles and paths at once. In particular, we deal with all cycles in the same way as in Section~\ref{subsection:geometricBasedApproaches:ILPMergingApproach} and with all paths in the same way as in Section~\ref{section:LPBasedApproaches}. The only difference is that we add {\em connecting edges} $F_C$ also between all transformed cycles and all shrunk paths.
			The weights $w_c$ of these new connections can be computed according to the same principle as it is done for other connecting edges. 
			This variant corresponds to the middle possibility in Figure~\ref{figure:LPBasedApproaches:enhancedLPBasedStrategies:Diagram} and will be denoted by \LPCOneR{} (with rerun) and \LPCOne{} (without rerun).
			
			Note that \LPCOne{} does not necessarily lead to the same or better results than \LPCTwo{}: if the paths and cycles are merged at once (\LPCOne{}), no path edge is allowed to be removed. If, however, the paths are merged to a cycle first and then all cycles are merged together, the possibility of opening a path edge in the second step of \LPCTwo{} is added.
			
			\medskip			
			Finally, we tested algorithms based on {\em randomised rounding} as introduced by {Raghavan} and {Tompson} \cite{RaghavanTompson:RandomizedRoundingATechniqueForProvablyGoodAlgorithmsAndAlgorithmicProofs}.
			Its main principle is to use the optimal solution values of the variables of a LP relaxation as probabilities for rounding up. 
			We tested more variants of this approach applied to \LPP{} and \LPCTwo{} and although the obtained objective function values are good for small values of $n$, the algorithms fail to produce good solutions in a reasonable computational time for larger instances. Thus we did not follow this idea any further.			
%

	\section{Improvement approaches}
		\label{section:ImprovementApproaches}
		The {\em $2$-opt-heuristic} and the {\em $3$-opt-heuristic} belong to the standard improvement algorithms used in the TSP context (for more details see \cite{Reinelt:TheTravelingSalesmanComputationalSolutionsForTSPApplications}). Both algorithms work as follows:
		Basically, for a starting tour $T$, $2$ or $3$ edges are removed and the remaining segments of $T$ are reconnected to obtain a new tour $T^\prime$. If $\zt(T^\prime) < \zt(T)$, an improved tour was found and $T^\prime$ replaces $T$, otherwise $T^\prime$ is discarded. 
		This process is repeated until no improvement is possible. 
		We tested both algorithms and will denote them by \TwoO{} and \ThreeO{} respectively. 
		Note that depending on the precise definition, \ThreeO{} is not always \TwoO{}-inclusive; we decided to use the \TwoO{}-inclusive variant.
		
		In our computational experiments (see Section~\ref{subsection:computationalResults:ImprovementHeuristics}), we made two observations which are in line with the behavior observed for the analogous heuristics applied to the TSP.
		\begin{enumerate}
			\item \TwoO{} has relatively small running times in comparison to all stand-alone heuristics described in Sections~\ref{section:trivialConstructionHeuristicsAndRelatedApproaches}, \ref{section:geometricBasedApproaches}, and \ref{section:LPBasedApproaches} and improves the solutions significantly.
			\item \ThreeO{} has running times comparable to the slower stand-alone heuristics, such as the LP-based ones, and yields significantly better solutions than \TwoO{}.
		\end{enumerate}
		The naive extension of this approach to a {\em $4$-opt-heuristic} can be expected to yield further improvements while the running times would increase dramatically.
		Therefore, we follow a different and more structured approach for removing parts of an existing solution while trying to find improvements.

		\subsection{Magnifying glass matheuristic}
			\label{subsection:ImprovementApproaches:MagnifyingGlassMatheuristic}
	
		The magnifying glass heuristic can be understood as a {\em large neighborhood search}, where a (starting) solution is partially destroyed and repaired by corresponding operators; a process, which is iteratively repeated until a local optimum is reached (see \cite{PisingerRopke:LargeNeighborhoodSearch}). 
		The underlying idea of our approach works as follows: 
		We consider a subset $S\subset V$ of vertices and remove all edges adjacent to $S$ from the tour. 
		Thus, we are left with several paths and isolated points. 
		A new, possibly improved tour containing these paths is computed optimally by means of a quadratic programming model.
		
		Formally, our algorithm works as follows. Let $T = (t_1, t_2, \ldots, t_n)$ be a tour in the graph $G = (V, E)$ and remember that $E(T) = \big\{\{t_1, t_2\}, \{t_2, t_3\}, \ldots, \{t_n, t_1\}\big\}$ is the set of all tour edges.
		Then for a set of vertices $S \subset V$, we remove all tour edges, which have at least one vertex in $S$, \ie all $\{u, v\} \in E(T)$ with $\{u, v\} \cap S \neq \emptyset$. 
		The remaining tour edges
		form a number of $\tau$ paths 
		denoted by $P^1$, $P^2$, \ldots, $P^\tau$, 
		where $P^i = (v_1^i, v_2^i, \ldots, v_{\left|P^i\right|}^i)$,
		$1 \leq i \leq \tau$.
		The idea is to reconnect these paths and the remaining isolated points $I = V \setminus \{P^1 \cup P^2 \cup \ldots \cup P^\tau\}$ 
		(mainly consisting of the vertices in $S$) 
		in an optimal way to build a new tour $T^\prime$ as a possible improvement of $T$.
		This whole process is repeated for different choices of $S$, which constitutes a matheuristic type of improvement procedure.
		
		Although the general procedure is defined for arbitrary sets $S$, the practical motivation is based on the consideration of vertex sets $S$  corresponding to points in a geographical neighbourhood.
		If $S$ contains all vertices of a certain region of the Euclidean plane, it can be expected that the removal of all connections within $S$ yields only a small number of unconnected paths.
		Thus, the optimization should be mainly concerned with the improvement of the tour in a limited part of the given graph.
		This is the reason for using the expression ``magnifying glass'', meaning that we look at a small region of the graph and insert its vertices in a locally optimal way into the existing solution.
		
		Following this intuition, we want to avoid short path fragments arising from the removal process which can be expected to lie close to the region of $S$.
		Therefore, in our implementation we enlarge $S$ by all vertices of paths consisting of at most 3 vertices and reduce $\tau$ accordingly.
		Also all isolated points outside $S$ arising from the removal of edges are added to $S$ thus setting $I$ equal to the enlarged set $S$.
				
		\medskip
		Let us now consider the reconnecting procedure. 
		We have a set of paths $P^i$, $1 \leq i \leq \tau$, and set of isolated points $I$. 
		In order to connect them into a new tour, we define a complete auxiliary graph $\widehat{G} = (\widehat{V}, F)$ with vertex set $\widehat{V}$ representing all isolated points $I$ and containing two vertices $a^i, b^i$ for each path $P^i$, $1 \leq i \leq \tau$.
		We model the reconnecting problem as an instance of QTSP with a smaller number of vertices and the following coefficients:
			\begin{equation*}
				\begin{alignedat}{2}
					w_{u v t}
						& = c_{u v t}
							& \quad
								& \mbox{for all } u \neq v \neq t \in I,\\
					w_{a^i b^i t}
						& = c_{v_{\left|P^i\right| - 1}^i v_{\left|P^i\right|}^i t}
							& \raisebox{-1\normalbaselineskip}[0pt][0pt]{\hspace*{-0.2cm}$\left.{\genfrac{}{}{0pt}{}{\genfrac{}{}{0pt}{}{\genfrac{}{}{0pt}{}{\genfrac{}{}{0pt}{}{}{}}{\genfrac{}{}{0pt}{}{}{}}}{\genfrac{}{}{0pt}{}{}{}}}{\genfrac{}{}{0pt}{}{\genfrac{}{}{0pt}{}{}{}}{\genfrac{}{}{0pt}{}{\genfrac{}{}{0pt}{}{}{}}{\genfrac{}{}{0pt}{}{}{}}}}}\right\}$}
								& \raisebox{-1\normalbaselineskip}[0pt][0pt]{for all $1 \leq i \leq \tau.$ and for all $t \in I$}\\
					w_{t a^i b^i}
						& = c_{t v_1^i v_2^i}
							&
								&\\
				\end{alignedat}
			\end{equation*} 
			Now, an optimal tour can be obviously found by using the standard QTSP integer program, \ie by \eqref{equation:LPBasedApproaches:QPObjectiveFunction}--\eqref{equation:LPBasedApproaches:QPIntegrality} or, as we do, by the linearisation \eqref{equation:LPBasedApproaches:ILPObjectiveFunction}, \eqref{equation:LPBasedApproaches:QP2MatchingConstraints}--\eqref{equation:LPBasedApproaches:QPIntegrality}, \eqref{equation:LPBasedApproaches:ILPConstraintsCoupelingTheXAndYVariables}, \eqref{equation:LPBasedApproaches:ILPIntegrality} described in Section~\ref{section:LPBasedApproaches}, with the additional constraints
			\begin{alignat}{5}
				\label{equation:MagnifyingGlassMatheuristic:AuxiliaryILPFixingOfThePathEdges}
					&& x_{a^i b^i} \ 
						&& = \ 
							& 1 \quad
								&&  \forall\ 1 \leq i \leq \tau.
			\end{alignat}
			These constraints ensure that the edges $\{a^i, b^i\}$, $1 \leq i \leq \tau$, representing the paths $P^i$ are contained in the tour. 
			To solve this ILP to optimality, we adopt the {\em integral approach} proposed by {Aichholzer} et al.\ in \cite{AichholzerFischerFischerMeierPferschyPilzStanek:MinimizationAndMaximizationVersionsOfTheQuadraticTravellingSalesmanProblem}, \ie we relax the subtour elimination constraints \eqref{equation:LPBasedApproaches:QPSubtourEliminationConstraints}, solve the remaining ILP to integrality, add one subtour elimination constraint for every generated subtour and repeat the whole process until the solution consists of just one tour $\widehat{T^\prime}$. 
			This cycle can then be easily transformed into a new tour $T^\prime$ in the original graph with $\zt(T^\prime) \leq \zt(T)$. 
			Again note that in this way, all costs incurred along the paths are not explicitly included in the model, since they remain constant anyway.
			
			\begin{example}
				\label{example:ImprovementApproaches:MagnifyingGlassMatheuristic}
				Consider the graph $G = (V, E)$ with the tour $T$ and with the set $S$ consisting of all vertices inside the square area bounded by the dotted line in Figure~\ref{figure:ImprovementApproaches:MagnifyingGlassMatheuristic:graphGVEWithSubsetSAndWithPathsP1AndP2}. After removing all tour edges having at least one vertex in $S$ (dashed in Figure~\ref{figure:ImprovementApproaches:MagnifyingGlassMatheuristic:graphGVEWithSubsetSAndWithPathsP1AndP2}), we obviously obtain two paths $P^1$ and $P^2$ (solid in Figure~\ref{figure:ImprovementApproaches:MagnifyingGlassMatheuristic:graphGVEWithSubsetSAndWithPathsP1AndP2}) and $8$ isolated vertices.
				
				\begin{figure}[htb]
					\centering
					\begin{comment:figures}
						\begin{multicols}{2}
							\begin{figurehere}
								\centering
								\begin{tikzpicture}[
										scale=4,
										decoration={
											markings,
											mark=at position 1.00 with {\arrow[scale=1.3]{angle 90}}
										}
									]
									\node[circle, draw=black!100, fill=black!100, thick, inner sep=0pt, minimum size=0.5mm, label=above:{\color{black} }] (node0) at (0.664609, 0.882716) {};
									\node[circle, draw=black!100, fill=black!100, thick, inner sep=0pt, minimum size=0.5mm, label=above:{\color{black} }] (node1) at (0.740741, 0.302469) {};
									\node[circle, draw=black!100, fill=black!100, thick, inner sep=0pt, minimum size=0.5mm, label=above:{\color{black} }] (node2) at (0.00617284, 0.878601) {};
									\node[circle, draw=black!100, fill=black!100, thick, inner sep=0pt, minimum size=0.5mm, label=above:{\color{black} }] (node3) at (0.242798, 0.720165) {};
									\node[circle, draw=black!100, fill=black!100, thick, inner sep=0pt, minimum size=0.5mm, label=above:{\color{black} }] (node4) at (0.504115, 0.788066) {};
									\node[circle, draw=black!100, fill=black!100, thick, inner sep=0pt, minimum size=0.5mm, label=above:{\color{black} }] (node5) at (0.376543, 0.0925926) {};
									\node[circle, draw=black!100, fill=black!100, thick, inner sep=0pt, minimum size=0.5mm, label=above:{\color{black} }] (node6) at (0.294239, 0.915638) {};
									\node[circle, draw=black!100, fill=black!100, thick, inner sep=0pt, minimum size=0.5mm, label=above:{\color{black} }] (node7) at (0.890947, 0.86214) {};
									\node[circle, draw=black!100, fill=black!100, thick, inner sep=0pt, minimum size=0.5mm, label=above:{\color{black} }] (node8) at (0.407407, 0.0843621) {};
									\node[circle, draw=black!100, fill=black!100, thick, inner sep=0pt, minimum size=0.5mm, label=above:{\color{black} }] (node9) at (0.541152, 0.100823) {};
									\node[circle, draw=black!100, fill=black!100, thick, inner sep=0pt, minimum size=0.5mm, label=above:{\color{black} }] (node10) at (0.76749, 0.604938) {};
									\node[circle, draw=black!100, fill=black!100, thick, inner sep=0pt, minimum size=0.5mm, label=above:{\color{black} }] (node11) at (0.808642, 0.989712) {};
									\node[circle, draw=black!100, fill=black!100, thick, inner sep=0pt, minimum size=0.5mm, label=above:{\color{black} }] (node12) at (0.878601, 0.337449) {};
									\node[circle, draw=black!100, fill=black!100, thick, inner sep=0pt, minimum size=0.5mm, label=above:{\color{black} }] (node13) at (0.0493827, 0.588477) {};
									\node[circle, draw=black!100, fill=black!100, thick, inner sep=0pt, minimum size=0.5mm, label=above:{\color{black} }] (node14) at (0.335391, 0.983539) {};
									\node[circle, draw=black!100, fill=black!100, thick, inner sep=0pt, minimum size=0.5mm, label=above:{\color{black} }] (node15) at (0.296296, 0.524691) {};
									\node[circle, draw=black!100, fill=black!100, thick, inner sep=0pt, minimum size=0.5mm, label=above:{\color{black} }] (node16) at (0.302469, 0.0514403) {};
									\node[circle, draw=black!100, fill=black!100, thick, inner sep=0pt, minimum size=0.5mm, label=above:{\color{black} }] (node17) at (0.816872, 0.864198) {};
									\node[circle, draw=black!100, fill=black!100, thick, inner sep=0pt, minimum size=0.5mm, label=above:{\color{black} }] (node18) at (0.919753, 0.584362) {};
									\node[circle, draw=black!100, fill=black!100, thick, inner sep=0pt, minimum size=0.5mm, label=above:{\color{black} }] (node19) at (0.54321, 0.94856) {};
									\node[circle, draw=black!100, fill=black!100, thick, inner sep=0pt, minimum size=0.5mm, label=above:{\color{black} }] (node20) at (0.839506, 0.444444) {};
									\node[circle, draw=black!100, fill=black!100, thick, inner sep=0pt, minimum size=0.5mm, label=above:{\color{black} }] (node21) at (0, 0.658436) {};
									\node[circle, draw=black!100, fill=black!100, thick, inner sep=0pt, minimum size=0.5mm, label=above:{\color{black} }] (node22) at (0.31893, 0.41358) {};
									\node[circle, draw=black!100, fill=black!100, thick, inner sep=0pt, minimum size=0.5mm, label=above:{\color{black} }] (node23) at (0.479424, 0.251029) {};
									\node[circle, draw=black!100, fill=black!100, thick, inner sep=0pt, minimum size=0.5mm, label=above:{\color{black} }] (node24) at (0.997942, 0.54321) {};
									\node[circle, draw=black!100, fill=black!100, thick, inner sep=0pt, minimum size=0.5mm, label=above:{\color{black} }] (node25) at (0.341564, 0.779835) {};
									\node[circle, draw=black!100, fill=black!100, thick, inner sep=0pt, minimum size=0.5mm, label=above:{\color{black} }] (node26) at (0.106996, 0.67284) {};
									\node[circle, draw=black!100, fill=black!100, thick, inner sep=0pt, minimum size=0.5mm, label=above:{\color{black} }] (node27) at (0.728395, 0) {};
									\node[circle, draw=black!100, fill=black!100, thick, inner sep=0pt, minimum size=0.5mm, label=above:{\color{black} }] (node28) at (1, 0.302469) {};
									\node[circle, draw=black!100, fill=black!100, thick, inner sep=0pt, minimum size=0.5mm, label=above:{\color{black} }] (node29) at (0.0555556, 0.860082) {};
									
									\draw[-] (node16) -- (node8) -- (node9) -- (node27) -- (node1) -- (node20) -- (node18) -- (node7) -- (node17) -- (node0) -- (node19);
									\draw[-] (node4) -- (node12) -- (node28) -- (node24) -- (node10) -- (node11);
									\draw[-] (node13) -- (node15) -- (node22) -- (node23) -- (node5) -- (node16);
																	
									\draw[thick, dotted] (-0.025, 0.625) -- (0.375, 0.625) -- (0.375, 1.025) -- (-0.025, 1.025) -- (-0.025, 0.625);
									
									\draw[dashed] (node19) -- (node14) -- (node2) -- (node29) -- (node6) -- (node4);
									\draw[dashed] (node11) -- (node25) -- (node3) -- (node26) -- (node21) -- (node13);
									
									\node at (0.185, 0.825) {$S$};
									\node at (0.658, 0.728) {$P^1$};
									\node at (0.6, 0.17) {$P^2$};
								\end{tikzpicture}
								\caption{graph $G = (V, E)$ with subset $S$ and paths $P^1$ and $P^2$}
								\label{figure:ImprovementApproaches:MagnifyingGlassMatheuristic:graphGVEWithSubsetSAndWithPathsP1AndP2}
							\end{figurehere}
							
							\begin{figurehere}
								\centering
								\begin{tikzpicture}[
										scale=4,
										decoration={
											markings,
											mark=at position 1.00 with {\arrow[scale=1.3]{angle 90}}
										}
									]
									\node[circle, draw=white!100, fill=white!100, thick, inner sep=0pt, minimum size=0.5mm, label=above:{\color{black} }] (node0) at (0.664609, 0.882716) {};
									\node[circle, draw=white!100, fill=white!100, thick, inner sep=0pt, minimum size=0.5mm, label=above:{\color{black} }] (node1) at (0.740741, 0.302469) {};
									\node[circle, draw=white!100, fill=white!100, thick, inner sep=0pt, minimum size=0.5mm, label=above:{\color{black} }] (node2x) at (0.00617284, 0.878601) {};
									\node[circle, draw=white!100, fill=white!100, thick, inner sep=0pt, minimum size=0.5mm, label=above:{\color{black} }] (node3x) at (0.242798, 0.720165) {};
									\node[circle, draw=white!100, fill=white!100, thick, inner sep=0pt, minimum size=0.5mm, label=above:{\color{black} }] (node4x) at (0.504115, 0.788066) {};
									\node[circle, draw=white!100, fill=white!100, thick, inner sep=0pt, minimum size=0.5mm, label=above:{\color{black} }] (node5) at (0.376543, 0.0925926) {};
									\node[circle, draw=white!100, fill=white!100, thick, inner sep=0pt, minimum size=0.5mm, label=above:{\color{black} }] (node6x) at (0.294239, 0.915638) {};
									\node[circle, draw=white!100, fill=white!100, thick, inner sep=0pt, minimum size=0.5mm, label=above:{\color{black} }] (node7) at (0.890947, 0.86214) {};
									\node[circle, draw=white!100, fill=white!100, thick, inner sep=0pt, minimum size=0.5mm, label=above:{\color{black} }] (node8) at (0.407407, 0.0843621) {};
									\node[circle, draw=white!100, fill=white!100, thick, inner sep=0pt, minimum size=0.5mm, label=above:{\color{black} }] (node9) at (0.541152, 0.100823) {};
									\node[circle, draw=white!100, fill=white!100, thick, inner sep=0pt, minimum size=0.5mm, label=above:{\color{black} }] (node10) at (0.76749, 0.604938) {};
									\node[circle, draw=white!100, fill=white!100, thick, inner sep=0pt, minimum size=0.5mm, label=above:{\color{black} }] (node11x) at (0.808642, 0.989712) {};
									\node[circle, draw=white!100, fill=white!100, thick, inner sep=0pt, minimum size=0.5mm, label=above:{\color{black} }] (node12) at (0.878601, 0.337449) {};
									\node[circle, draw=white!100, fill=white!100, thick, inner sep=0pt, minimum size=0.5mm, label=above:{\color{black} }] (node13x) at (0.0493827, 0.588477) {};
									\node[circle, draw=white!100, fill=white!100, thick, inner sep=0pt, minimum size=0.5mm, label=above:{\color{black} }] (node14x) at (0.335391, 0.983539) {};
									\node[circle, draw=white!100, fill=white!100, thick, inner sep=0pt, minimum size=0.5mm, label=above:{\color{black} }] (node15) at (0.296296, 0.524691) {};
									\node[circle, draw=white!100, fill=white!100, thick, inner sep=0pt, minimum size=0.5mm, label=above:{\color{black} }] (node16) at (0.302469, 0.0514403) {};
									\node[circle, draw=white!100, fill=white!100, thick, inner sep=0pt, minimum size=0.5mm, label=above:{\color{black} }] (node17) at (0.816872, 0.864198) {};
									\node[circle, draw=white!100, fill=white!100, thick, inner sep=0pt, minimum size=0.5mm, label=above:{\color{black} }] (node18) at (0.919753, 0.584362) {};
									\node[circle, draw=white!100, fill=white!100, thick, inner sep=0pt, minimum size=0.5mm, label=above:{\color{black} }] (node19x) at (0.54321, 0.94856) {};
									\node[circle, draw=white!100, fill=white!100, thick, inner sep=0pt, minimum size=0.5mm, label=above:{\color{black} }] (node20) at (0.839506, 0.444444) {};
									\node[circle, draw=white!100, fill=white!100, thick, inner sep=0pt, minimum size=0.5mm, label=above:{\color{black} }] (node21x) at (0, 0.658436) {};
									\node[circle, draw=white!100, fill=white!100, thick, inner sep=0pt, minimum size=0.5mm, label=above:{\color{black} }] (node22) at (0.31893, 0.41358) {};
									\node[circle, draw=white!100, fill=white!100, thick, inner sep=0pt, minimum size=0.5mm, label=above:{\color{black} }] (node23) at (0.479424, 0.251029) {};
									\node[circle, draw=white!100, fill=white!100, thick, inner sep=0pt, minimum size=0.5mm, label=above:{\color{black} }] (node24) at (0.997942, 0.54321) {};
									\node[circle, draw=white!100, fill=white!100, thick, inner sep=0pt, minimum size=0.5mm, label=above:{\color{black} }] (node25x) at (0.341564, 0.779835) {};
									\node[circle, draw=white!100, fill=white!100, thick, inner sep=0pt, minimum size=0.5mm, label=above:{\color{black} }] (node26x) at (0.106996, 0.67284) {};
									\node[circle, draw=white!100, fill=white!100, thick, inner sep=0pt, minimum size=0.5mm, label=above:{\color{black} }] (node27) at (0.728395, 0) {};
									\node[circle, draw=white!100, fill=white!100, thick, inner sep=0pt, minimum size=0.5mm, label=above:{\color{black} }] (node28) at (1, 0.302469) {};
									\node[circle, draw=white!100, fill=white!100, thick, inner sep=0pt, minimum size=0.5mm, label=above:{\color{black} }] (node29x) at (0.0555556, 0.860082) {};
	
									\begin{scope}[shift={(0.1, -0.25)}]
										\node[circle, draw=black!100, fill=black!100, thick, inner sep=0pt, minimum size=0.5mm, label=above:{\color{black} }] (node2) at (0.00617284, 0.878601) {};
										\node[circle, draw=black!100, fill=black!100, thick, inner sep=0pt, minimum size=0.5mm, label=above:{\color{black} }] (node3) at (0.242798, 0.720165) {};
										\node[circle, draw=black!100, fill=black!100, thick, inner sep=0pt, minimum size=0.5mm, label={[label distance=-0.15cm]-125:{\color{black} $a^1$}}] (node4) at (0.504115, 0.788066) {};
										\node[circle, draw=black!100, fill=black!100, thick, inner sep=0pt, minimum size=0.5mm, label=above:{\color{black} }] (node6) at (0.294239, 0.915638) {};
										\node[circle, draw=black!100, fill=black!100, thick, inner sep=0pt, minimum size=0.5mm, label=above:{\color{black} $b^1$}] (node11) at (0.808642, 0.989712) {};
										\node[circle, draw=black!100, fill=black!100, thick, inner sep=0pt, minimum size=0.5mm, label=below left:{\color{black} $a^2$}] (node13) at (0.0493827, 0.588477) {};
										\node[circle, draw=black!100, fill=black!100, thick, inner sep=0pt, minimum size=0.5mm, label=above:{\color{black} }] (node14) at (0.335391, 0.983539) {};
										\node[circle, draw=black!100, fill=black!100, thick, inner sep=0pt, minimum size=0.5mm, label=above:{\color{black} $b^2$}] (node19) at (0.54321, 0.94856) {};
										\node[circle, draw=black!100, fill=black!100, thick, inner sep=0pt, minimum size=0.5mm, label=above:{\color{black} }] (node21) at (0, 0.658436) {};
										\node[circle, draw=black!100, fill=black!100, thick, inner sep=0pt, minimum size=0.5mm, label=above:{\color{black} }] (node25) at (0.341564, 0.779835) {};
										\node[circle, draw=black!100, fill=black!100, thick, inner sep=0pt, minimum size=0.5mm, label=above:{\color{black} }] (node26) at (0.106996, 0.67284) {};
										\node[circle, draw=black!100, fill=black!100, thick, inner sep=0pt, minimum size=0.5mm, label=above:{\color{black} }] (node29) at (0.0555556, 0.860082) {};
										
										\draw[thick, dotted] (-0.025, 0.625) -- (0.375, 0.625) -- (0.375, 1.025) -- (-0.025, 1.025) -- (-0.025, 0.625);
										
										\draw[-] (node11) -- (node4);
										\draw[-] (node19) to[out = -90, in = 0] (0.375, 0.588477) -- (node13);
									\end{scope}
									
																	
									
								\end{tikzpicture}
								\caption{graph $\widehat{G} = (\widehat{V}, F)$ (depicted are only edges fixed by~\eqref{equation:MagnifyingGlassMatheuristic:AuxiliaryILPFixingOfThePathEdges})}
								\label{figure:ImprovementApproaches:MagnifyingGlassMatheuristic:graphGVF}
							\end{figurehere}
						\end{multicols}
					\end{comment:figures}
				\end{figure}
				
				The auxiliary graph $\widehat{G} = (\widehat{V}, F)$ can be seen in Figure~\ref{figure:ImprovementApproaches:MagnifyingGlassMatheuristic:graphGVF}. 
				It consists of all isolated vertices $I$ and of the vertices $a^1$, $b^1$ and $a^2$, $b^2$ representing the paths $P^1$ and $P^2$, respectively.
				Although $\widehat{G}$ is complete, for a better representation only the edges $\{a^1, b^1\}, \{a^2, b^2\}$ fixed by~\eqref{equation:MagnifyingGlassMatheuristic:AuxiliaryILPFixingOfThePathEdges} are depicted.
			\end{example}
			
			This procedure could be performed for any subset $S$.
			Since we deal with the AngleTSP and the AngleDistanceTSP and all our test instances are based on points in the Euclidean plane, we generate a sequence of sets $S$ systematically by moving a square (our magnifying glass) over the set of points. 
			Given a fixed parameter $s$, we place a square of size $s \times s$ in the left upper corner of the point set of the instance. All vertices inside form the first set $S$. 
			After obtaining a new tour $T^\prime$, we shift the square to the right by $\nicefrac{2}{3}\cdot s$ (rounded to the nearest integer) and find a new tour $T^{\prime\prime}$. 
			This process is repeated until we reach the right upper corner of the instance. 
			Then we proceed with the next row, placed $\nicefrac{2}{3}\cdot s$ (rounded to the nearest integer) below the first one. In this way, row after row, we continue to sweep over the point set until we reach the bottom right corner of the instance.
			Clearly, this approach guarantees that every vertex $v \in V$ is contained in at least one set $S \subset V$.
			To stimulate a consideration of point connections in multiple directions, we use overlapping sets $S$.
			
			Concerning the choice of $s$ we note that all our test instances are based on points in the Euclidean plane, which are uniformly distributed on $\{0, \ldots, 500\}^2$ (for further details see Section~\ref{subsection:TestSetUp}). 
			Thus, by setting 
			$s = 500 \sqrt{\frac{k}{n}}$ (rounded to the nearest integer), we have $k$ vertices in every square of size $s \times s$ in expectation. 
			This improvement heuristic will be denoted \M{k}, where $k$ is the expected number of vertices in each square.

		\subsection{Lens metaheuristic}
			\label{subsection:ImprovementApproaches:LensMetaheuristic}
			A natural improvement strategy is given by the use of {\em metaheuristics}, which offer possibilities of escaping local optima, as they may occur e.g.\ in \TwoO{} and \ThreeO{}.
			While the design options for metaheuristics are unlimited, we decided to restrict ourselves to approaches based on the {\em lens neighbourhood}.
			This neighbourhood can be evaluated quickly and still allows a substantial diversification of the current solution.
			
			\subsubsection{Lens neighbourhood}
				\label{subsubsection:ImprovementApproaches:LensMetaheuristic:LensNeighbourhood}
				Let $T = (t_1, t_2, \ldots, t_n)$ be a tour in $G = (V, E)$. 
				Then, for a fixed thickness parameter $\gamma$, we position a lens on all tour edges $e_t \in E(T)$ in the same way as described for the {\em lens procedure} in Section~\ref{section:lensProcedure}. 
				Let $t_l$ be a vertex in the lens, which is positioned on the tour edge $\{t_i, t_{i+1}\}$ (the wrapping around of indices at the beginning and at the end of the tour $T$ follows in a natural way). 
				Then we can remove the tour edges $\{t_{l - 1}, t_l\}$, $\{t_l, t_{l + 1}\}$ and $\{t_i, t_{i+1}\}$ from the tour $T$ and add the edges $\{t_i, t_l\}$, $\{t_l, t_{i+1}\}$ and $\{t_{l - 1}, t_{l + 1}\}$
				in order to obtain a new tour $T^\prime$.

				\begin{example}
					\label{example:LensNeighbourhood}
					Let us consider Figure~\ref{figure:ImprovementApproaches:LensMetaheuristic:GraphGVEWithTourTAndWithOneLensStretchedOnTheEdge12} depicting the graph $G = (V, E)$ and the tour $T = \{1, 2, 3, 4, 5, 6, 7, 8, 9, 10, 11\}$. Assuming that we are solving the AngleTSP, $\zt(T) \approx 13.0661$. The lens positioned on the edge $\{1, 2\}$ with $\gamma = 30\degree\ (\approx 0.5236)$ is depicted dashed in Figure~\ref{figure:ImprovementApproaches:LensMetaheuristic:GraphGVEWithTourTAndWithOneLensStretchedOnTheEdge12} and contains only the vertex $8$. By removing the edges $\{7, 8\}$, $\{8, 9\}$ and $\{1, 2\}$ and by adding the edges $\{1, 8\}$, $\{8, 2\}$ and $\{7, 9\}$ we obtain the new improved tour $T^\prime = \{1, 8, 2, 3, 4, 5, 6, 7, 9, 10, 11\}$ with $\zt(T^\prime) \approx 12.5664$ depicted in Figure~\ref{figure:ImprovementApproaches:LensMetaheuristic:GraphGVEWithTourT}.
					\begin{figure}[htb]
						\centering
						\begin{comment:figures}
							\begin{multicols}{2}
								\begin{figurehere}
									\centering
									\begin{tikzpicture}[
											scale=4,
											decoration={
												markings,
												mark=at position 1.00 with {\arrow[scale=1.3]{angle 90}}
											}
										]
										\node[circle, draw=black!100, fill=black!100, thick, inner sep=0pt, minimum size=0.5mm, label=left:{\color{black} 1}] (node0) at (0.333333, 0.800000) {};
										\node[circle, draw=black!100, fill=black!100, thick, inner sep=0pt, minimum size=0.5mm, label=below left:{\color{black} 2}] (node1) at (0.333333, 0.200000) {};
										\node[circle, draw=black!100, fill=black!100, thick, inner sep=0pt, minimum size=0.5mm, label=above:{\color{black} 3}] (node2) at (0.666666, 0.111111) {};
										\node[circle, draw=black!100, fill=black!100, thick, inner sep=0pt, minimum size=0.5mm, label=below right:{\color{black} 4}] (node3) at (0.880000, 0.211111) {};
										\node[circle, draw=black!100, fill=black!100, thick, inner sep=0pt, minimum size=0.5mm, label=above right:{\color{black} 5}] (node4) at (1.000000, 1.000000) {};
										\node[circle, draw=black!100, fill=black!100, thick, inner sep=0pt, minimum size=0.5mm, label=above left:{\color{black} 6}] (node5) at (0.100000, 0.900000) {};
										\node[circle, draw=black!100, fill=black!100, thick, inner sep=0pt, minimum size=0.5mm, label=below left:{\color{black} 7}] (node6) at (0.000000, 0.500000) {};
										\node[circle, draw=black!100, fill=black!100, thick, inner sep=0pt, minimum size=0.5mm, label=above left:{\color{black} 8}] (node7) at (0.300000, 0.550000) {};
										\node[circle, draw=black!100, fill=black!100, thick, inner sep=0pt, minimum size=0.5mm, label=below left:{\color{black} 9}] (node8) at (0.888889, 0.500000) {};
										\node[circle, draw=black!100, fill=black!100, thick, inner sep=0pt, minimum size=0.5mm, label=right:{\color{black} 10}] (node9) at (0.800000, 0.888889) {};
										\node[circle, draw=black!100, fill=black!100, thick, inner sep=0pt, minimum size=0.5mm, label=below:{\color{black} 11}] (node10) at (0.666666, 0.888889) {};
										
										\draw[-] (node0) -- (node1) -- (node2) -- (node3) -- (node4) -- (node5) -- (node6) -- (node7) -- (node8) -- (node9) -- (node10) -- (node0);
										
										\node at (0.599999, 0.666666) {$T$};
										
										\draw[dashed] (0.333333, 0.800000) arc (151.352110:208.647890:0.625749);
										\draw[dashed] (0.333333, 0.200000) arc (-28.647890:28.647890:0.625749);
									\end{tikzpicture}
									\caption{graph $G = (V, E)$ with tour $T$ and with a lens on edge $\{1, 2\}$}
									\label{figure:ImprovementApproaches:LensMetaheuristic:GraphGVEWithTourTAndWithOneLensStretchedOnTheEdge12}
								\end{figurehere}
								
								\begin{figurehere}
									\centering
									\begin{tikzpicture}[
											scale=4,
											decoration={
												markings,
												mark=at position 1.00 with {\arrow[scale=1.3]{angle 90}}
											}
										]
										\node[circle, draw=black!100, fill=black!100, thick, inner sep=0pt, minimum size=0.5mm, label=left:{\color{black} 1}] (node0) at (0.333333, 0.800000) {};
										\node[circle, draw=black!100, fill=black!100, thick, inner sep=0pt, minimum size=0.5mm, label=below left:{\color{black} 2}] (node1) at (0.333333, 0.200000) {};
										\node[circle, draw=black!100, fill=black!100, thick, inner sep=0pt, minimum size=0.5mm, label=above:{\color{black} 3}] (node2) at (0.666666, 0.111111) {};
										\node[circle, draw=black!100, fill=black!100, thick, inner sep=0pt, minimum size=0.5mm, label=below right:{\color{black} 4}] (node3) at (0.880000, 0.211111) {};
										\node[circle, draw=black!100, fill=black!100, thick, inner sep=0pt, minimum size=0.5mm, label=above right:{\color{black} 5}] (node4) at (1.000000, 1.000000) {};
										\node[circle, draw=black!100, fill=black!100, thick, inner sep=0pt, minimum size=0.5mm, label=above left:{\color{black} 6}] (node5) at (0.100000, 0.900000) {};
										\node[circle, draw=black!100, fill=black!100, thick, inner sep=0pt, minimum size=0.5mm, label=below left:{\color{black} 7}] (node6) at (0.000000, 0.500000) {};
										\node[circle, draw=black!100, fill=black!100, thick, inner sep=0pt, minimum size=0.5mm, label=above left:{\color{black} 8}] (node7) at (0.300000, 0.550000) {};
										\node[circle, draw=black!100, fill=black!100, thick, inner sep=0pt, minimum size=0.5mm, label=below left:{\color{black} 9}] (node8) at (0.888889, 0.500000) {};
										\node[circle, draw=black!100, fill=black!100, thick, inner sep=0pt, minimum size=0.5mm, label=right:{\color{black} 10}] (node9) at (0.800000, 0.888889) {};
										\node[circle, draw=black!100, fill=black!100, thick, inner sep=0pt, minimum size=0.5mm, label=below:{\color{black} 11}] (node10) at (0.666666, 0.888889) {};
										
										\draw[-] (node0) -- (node7) -- (node1) -- (node2) -- (node3) -- (node4) -- (node5) -- (node6) -- (node8) -- (node9) -- (node10) -- (node0);
										
										\node at (0.599999, 0.666666) {$T^\prime$};
										
									\end{tikzpicture}
									\caption{graph $G = (V, E)$ with tour $T^\prime$}
									\label{figure:ImprovementApproaches:LensMetaheuristic:GraphGVEWithTourT}
								\end{figurehere}
								
								$ $
							\end{multicols}
						\end{comment:figures}
					\end{figure}
				\end{example}
				
				The {\em lens neighbourhood} $L(T)$ of a tour $T$ is defined as the set of all tours $T'$ obtained by considering all edges $e_t \in E(T)$, positioning a lens on $e_t$ and creating a new tour $T'$
				for every vertex $t_l$ contained in such a lens as described above.
				Obviously, the lens neighbourhood is a subset of the {\em $3$-opt-neighbourhood}. 
				Since the metaheuristics we design below are based on the evaluation of a large number of iterations, the $3$-opt-neighbourhood would be impractical, while our lens neighbourhood reduces the $3$-opt-neighbourhood, allows a parametrization of this reduction and keeps ``good'' neighbours in the tour.
			
			\medskip			
			Of course, one could use the lens neighbourhood as a simple local search heuristic most probably yielding worse solutions than \ThreeO{} with lower  running times. 
			However, our goal is to escape from local optima. 
			Thus we allow a worsening of the solution by using the concept of {\em simulated annealing}. 
			This metaheuristic is very well-known and has been widely used during the last decades, thus we only refer to \cite{Reeves:ModernHeuristicTechniquesForCombinatorialProblems} for more details.
			
			Given a tour $T$ in $G = (V, E)$ let $L(T) = \{T_1^\prime, T_2^\prime, \ldots, T_{\left|L(T)\right|}^\prime\}$ be the lens neighbourhood of this tour. Then we define the {\em positive} and the {\em negative neighbourhood} as
			\begin{equation*}
				L^+(T) \defeq \big\{T^\prime \in L(T)\colon \zt(T^\prime) < \zt(T)\big\} \text{ and } L^-(T) \defeq L(T) \setminus L^+(T),
			\end{equation*}
			respectively. We tested three variants of applying the acceptance rule of simulated annealing with the lens neighbourhood:
			\begin{description}
				\item[Random choice:] 
				Choose a random tour $T^\prime \in L(T)$ among all neighbours, accept it if $T^\prime \in L^+(T)$ and use the acceptance rule if $T^\prime \in L^-(T)$.
				\item[Least-cost choice:] 
				Take $T^\prime = \argmin_{t^\prime \in L(T)}{\big\{\zt(t^\prime)\big\}}$ and deal with it in the same way as in the previous case, \ie $T^\prime$ is taken if $T^\prime \in L^+(T)$ and the acceptance rule is applied if $T^\prime \in L^-(T)$.
				\item[Combined choice:] 
				If $L^+(T) \neq \emptyset$, choose and accept a random $T^\prime \in L^+(T)$. Otherwise take $T^\prime = \argmin_{t^\prime \in L^-(T)}{\big\{\zt(t^\prime)\big\}}$ and apply the acceptance rule.
			\end{description}
			The acceptance rule accepts the new tour $T^\prime \in L^-(T)$ with probability
			\begin{equation}
				\label{equation:SimulatedAnnealingProbability}
				P = e^{\frac{\frac{\zt(T) - \zt(T^\prime)}{\zt(T)} n}{t(\ell)}},
			\end{equation}
			where $t(\ell)$ is a so-called temperature parameter depending on the {\em iteration number} $\ell \in \{0,1,\ldots, I\}$ with $I$ denoting the maximum number of iterations.		
		 In our implementation, one {\em iteration} corresponds to one choice of $T^\prime \in L(T)$. 
			We use a linearly decreasing temperature function $t(\ell)$
			 with $t(0) = 1$ and $t(I) = 0$, \ie we set
			\begin{equation}
				\label{equation:SimulatedAnnealingTemperature}
				t(\ell) \defeq \frac{I - \ell}{I}.
			\end{equation}
			Note that the algorithm can stop before reaching $I$ iterations, \eg if the lens neighbourhood is empty.
			
			Our preliminary computational experiments have shown that among these three variants, {\em least-cost choice} leads to slightly better results than the {\em random} and the {\em combined choice}. 
			Therefore we perform only {\em least-cost choice} in the reminder of the paper and denote it by \Lens{}.
			
	
	\section{Computational results}
		\label{section:computationalResults}
		\subsection{Test set-up}
			\label{subsection:TestSetUp}
			We describe the setup of our computational tests in this section.
			
			\subsubsection{Benchmark instances}
				\label{subsubsection:benchmarkInstances}	
				
	Our benchmark instances are based on the instance specification developed in~\cite{Fischer:APolyhedralStudyOfQuadraticTravelingSalesmanProblems} and \cite{FischerHelmberg:TheSymmetricQuadraticTravelingSalesmanProblem}
and also used in \cite{AichholzerFischerFischerMeierPferschyPilzStanek:MinimizationAndMaximizationVersionsOfTheQuadraticTravellingSalesmanProblem} from where we take over the integer point sets.
As in \cite{AichholzerFischerFischerMeierPferschyPilzStanek:MinimizationAndMaximizationVersionsOfTheQuadraticTravellingSalesmanProblem} we do not round the costs to integers.
Furthermore, we add point sets for new sizes $n = 5$ and $85 \leq n \leq 200$ generated by the same algorithm
as used by the previous papers.

				We test two types of quadratic instances\footnote{All instances, the lower bound values, the optimal objective function values (whenever they are available) as well as the complete test results for all combinations of algorithms are available on {\tt arXiv}.}:
				\begin{description}
					\item[\AngleTSPInstances] are based on points in the Euclidean plane: First, we randomly choose $n$ points independently from the discrete uniform distribution $\{0, \ldots, 500\} \times \{0, \ldots, 500\}$.
					Then we compute the {\em turning angles} $\alpha$, multiply by $1000$, in the same way as \cite{Fischer:APolyhedralStudyOfQuadraticTravelingSalesmanProblems} and \cite{AichholzerFischerFischerMeierPferschyPilzStanek:MinimizationAndMaximizationVersionsOfTheQuadraticTravellingSalesmanProblem}, and then we round them to $12$ decimal places.
					
					\item[\AngleDistanceTSPInstances] extend the above \AngleTSPInstances{} by combining the turning angles with the Euclidean distances between the points in a weighted sum. 
					Taking the same point sets in the plane
					we denote the Euclidean distances between vertices $i$ and $j$ as $d_{i j}$. 
					Following the literature, we construct QTSP-distances 
					for a weighting parameter $\rho \in \rz_0^+$, setting $\lambda_1 = 100 \rho$ and $\lambda_2 = 100$, the cost coefficients \eqref{equation:turningAnglesDistances} reduce to:
						\begin{equation}
							\label{equation:angleDistanceInstances}
							c_{i j k} \defeq 100 \left(\rho \cdot \alpha_{i j k} + \frac{d_{i j} + {d_{j k}}}{2}\right)
						\end{equation}
						Again, all distances are rounded to $12$ decimal places.
						
						These instances were introduced in~\cite{SavlaFrazzoli:TravelingSalespersonProblemsForTheDubinsVehicle} for an approximate solution of the {\em TSP for Dubins vehicle}. 
						They lie between \AngleTSPInstances{} (for $\rho \to \infty$) and standard TSP instances (for $\rho = 0$). For compatibility with the literature we set $\rho = 40$ for all our tests as done in~\cite{Fischer:APolyhedralStudyOfQuadraticTravelingSalesmanProblems} and \cite{AichholzerFischerFischerMeierPferschyPilzStanek:MinimizationAndMaximizationVersionsOfTheQuadraticTravellingSalesmanProblem}.
				\end{description}
				For each size $n = 5, 10, \ldots, 200$ we took/generated $10$ point sets used for $10$ {\AngleTSP}- and $10$ {\AngleDistanceTSP}-instances.
				
			\subsubsection{Test environment}
				\label{subsubsection:testEnvironment}
				All tests were run on an {\em Intel(R) Core(TM) i7-4790 CPU @ 3.60GHz with 32 GB RAM} under {\em Linux}\footnote{Precise version: {\em Linux 4.4.0-75-generic \#96-Ubuntu SMP x86\_64 x86\_64 x86\_64 GNU{\slash}Linux}.} and all programs, including the reference algorithm from the literature, were implemented in {\em Python}\footnote{Precise version: {\em Python 3.5.2.}}.
				ILPs were set up using {\em PuLP}\footnote{Precise version: {\em PuLP 1.6.0.}} and solved with {\em Gurobi}\footnote{Precise version: {\em Gurobi 6.5.1 (Linux x86\_64)} in the standard setting (\ie the solver is allowed to use multiple threads if solving an ILP).}. 
				Moreover, in order to guarantee the relative reproducibility of our computational results, we (i) allowed no additional swap memory and (ii) ran all tests separately without other user processes in background.
			
			\subsubsection{Evaluation layout}
				\label{subsubsection:EvaluationLayout}
				As a basis for comparisons for smaller instances with $n \leq 75$ (\AngleTSPInstances) resp.\ $n \leq 100$ (\AngleDistanceTSPInstances) we computed optimal objective function values  by using the {\em integral ILP based approach} described in~\cite{AichholzerFischerFischerMeierPferschyPilzStanek:MinimizationAndMaximizationVersionsOfTheQuadraticTravellingSalesmanProblem}.
				For larger instances (\ie $n \geq 80$ resp.\ $n\geq 105$) 
				running times become prohibitive and we settled for lower bounds.
				These were computed by using the linearisation \eqref{equation:LPBasedApproaches:ILPObjectiveFunction}, \eqref{equation:LPBasedApproaches:QP2MatchingConstraints}--\eqref{equation:LPBasedApproaches:QPIntegrality}, \eqref{equation:LPBasedApproaches:ILPConstraintsCoupelingTheXAndYVariables}, \eqref{equation:LPBasedApproaches:ILPIntegrality} described in Section~\ref{section:LPBasedApproaches}, relaxing the subtour elimination constraints \eqref{equation:LPBasedApproaches:QPSubtourEliminationConstraints} and the integrality constraints \eqref{equation:LPBasedApproaches:QPIntegrality} and \eqref{equation:LPBasedApproaches:ILPIntegrality} and by solving the remaining LP using an LP solver. \Ie we used the same LP relaxation as in Section~\ref{section:LPBasedApproaches} for our LP-based approaches.
				
				For every instance $X$ and every heuristic $H$ we define the {\bf objective function value ratio}
				\begin{equation}
					\label{equation:ration}
					r_H^X \defeq \frac{\zt(H^X)}{\lb(X)},
				\end{equation}
				where $\zt(H^X)$ denotes the objective function value of the solution given by the heuristic $H$ for instance $X$ and $\lb(X)$ is the lower bound for instance $X$. Obviously, $r_H^X \geq 1$ for all $H$ and $X$; $r_H^X - 1$ expresses the relative gap between the objective function value of the heuristic solution and the lower bound. 
				We always report the {\em geometric mean ratio} values for the heuristic $H$ and all instances $X$ of the same type (\AngleTSP{} or \AngleDistanceTSP) and the same size $n$. 
				Such {\em means of objective function value ratios} will be denoted by $\rg$. 
				Recall the jump implied by the change from optimal values to lower bounds as the basis of the comparison when moving from smaller to larger instances.
				
				For the running times,
				we report the {\em arithmetic means} $\ra$
				again for every heuristic over all instances of the same type (\AngleTSP{} or \AngleDistanceTSP) and size $n$.

		\bigskip 
		To estimate the meaning of gaps to the lower bound for larger instances,
		we try to give an estimation for the unkown differences between lower bounds and optimal solution values by computing these values for the smaller instances.
		A certain extrapolation of these values to larger instances seems justified.
		It turns out that (see Table~\ref{table:OptimalObjectiveFunctionValueRatioMeans}):
		\begin{enumerate}
			\item The objective function value ratio means $\rg$ are significantly smaller for the \AngleDistanceTSPInstances{}.
			\item We can expect them to be over $10\%$ for larger \AngleTSPInstances{} and over $6.5\%$ for larger \AngleDistanceTSP.
		\end{enumerate}
		
		\addtolength{\tabcolsep}{-2.4pt}
		\begin{table}[htbp!]
			\footnotesize
			\centering
			\begin{tabular}{l|*{10}{r}}
				\multicolumn{1}{l|}{$n$} & \multicolumn{1}{r}{5} & \multicolumn{1}{r}{10} & \multicolumn{1}{r}{15} & \multicolumn{1}{r}{20} & \multicolumn{1}{r}{25} & \multicolumn{1}{r}{30} & \multicolumn{1}{r}{35} & \multicolumn{1}{r}{40} & \multicolumn{1}{r}{45} & \multicolumn{1}{r}{50}\\
				\hline
				\AngleTSP & 1.0000 & 1.0109 & 1.0296 & 1.0653 & 1.0956 & 1.0924 & 1.0958 & 1.1036 & 1.1122 & 1.1042\\
				\AngleDistanceTSP & 1.0000 & 1.0000 & 1.0000 & 1.0029 & 1.0135 & 1.0211 & 1.0359 & 1.0306 & 1.0430 & 1.0514\\
				\hline\hline
				\multicolumn{1}{l|}{$n$} & \multicolumn{1}{r}{55} & \multicolumn{1}{r}{60} & \multicolumn{1}{r}{65} & \multicolumn{1}{r}{70} & \multicolumn{1}{r}{75} & \multicolumn{1}{r}{80} & \multicolumn{1}{r}{85} & \multicolumn{1}{r}{90} & \multicolumn{1}{r}{95} & \multicolumn{1}{r}{100}\\
				\hline
				\AngleTSP & 1.1029 & 1.1094 & 1.0866 & 1.0997 & 1.1035 & \multicolumn{1}{r}{--} & \multicolumn{1}{r}{--} & \multicolumn{1}{r}{--} & \multicolumn{1}{r}{--} & \multicolumn{1}{r}{--}\\
				\AngleDistanceTSP & 1.0500 & 1.0510 & 1.0498 & 1.0548 & 1.0537 & 1.0549 & 1.0578 & 1.0620 & 1.0632 & 1.0662\\
			\end{tabular}
			\caption{optimal objective function value ratio means $\rg$}
			\label{table:OptimalObjectiveFunctionValueRatioMeans}
		\end{table}
		\addtolength{\tabcolsep}{+2.4pt}
		
				
		\subsection{Stand-alone heuristics}
			\label{subsection:computationalResults:StartingHeuristics}
			The proposed stand-alone heuristics can be divided into three groups: the {\em simple construction and related heuristics} (see Section~\ref{section:trivialConstructionHeuristicsAndRelatedApproaches}), the {\em heuristics based on convex hulls} (see Section~\ref{section:geometricBasedApproaches}) and the {\em LP-based heuristics} (see Section~\ref{section:LPBasedApproaches}). In the following, we first discuss the results with respect to the objective function values and later we put them into the context of running times.
			
			Consider first the simple construction and related heuristics. This group contains three parameter-free heuristics, \NNS{}, \NNSTwo{} and \CIF{}, and one approach, which uses \LENS{} (see Section~\ref{section:lensProcedure}) as sub-procedure, namely \NNSL{}. We tested \NNSL{} with the lens thickness parameter $\gamma = 10\degree, 20\degree, \ldots, 90\degree$ and obtained the best results for $\gamma = 40\degree\ (\approx 0.6981)$; thus we used this value of $\gamma$ for all further experiments. The objective function value ratio means for all four heuristics can be seen in Figure~\ref{figure:SimpleConstructionAndRelatedHeuristics}. 
			
			\begin{comment:figures}
	\begin{figure}[htb!]
		\centering
			\begin{tikzpicture}[xscale=\xscale, yscale=\yscaleLarge]
				\pgfgettransformentries{\xscaleTikz}{\@tempa}{\@tempa}{\yscaleTikz}{\@tempa}{\@tempa}
			
				\draw[very thin, color=gray, xstep=20, ystep=0.5] (0, 0) grid (200, 1.5 );
				
				\begin{scope}[shift={(0, -1)}]
					\def\crossSizeX{\crossSize / \xscaleTikz};
					\def\crossSizeY{\crossSize / \yscaleTikz};
					
					\def\crossOne{(-\crossSizeX,-\crossSizeY) -- (\crossSizeX,\crossSizeY) (-\crossSizeX,\crossSizeY) -- (\crossSizeX,-\crossSizeY)};
					\def\crossTwo{(-\crossSizeX,0) -- (\crossSizeX,0) (0,\crossSizeY) -- (0,-\crossSizeY)};
					\def\crossThree{(0,0) -- (0,\crossSizeY) (0,0) -- (\crossSizeX,-\crossSizeY) (0,0) -- (-\crossSizeX,-\crossSizeY)};
					\def\crossFour{(0,0) -- (0,-\crossSizeY) (0,0) -- (\crossSizeX,\crossSizeY) (0,0) -- (-\crossSizeX,\crossSizeY)};
					\def\crossFive{(0,0) -- (\crossSizeX,0) (0,0) -- (-\crossSizeX,\crossSizeY) (0,0) -- (-\crossSizeX,-\crossSizeY)};
					\def\crossSix{(0,0) -- (-\crossSizeX,0) (0,0) -- (\crossSizeX,\crossSizeY) (0,0) -- (\crossSizeX,-\crossSizeY)};
					
					\draw[black, shift={( 5, 1.00518180757 )}] \crossOne;
					\draw[black, shift={( 10, 1.10031329725 )}] \crossOne;
					\draw[black, shift={( 15, 1.13660142608 )}] \crossOne;
					\draw[black, shift={( 20, 1.18124295096 )}] \crossOne;
					\draw[black, shift={( 25, 1.2267075139 )}] \crossOne;
					\draw[black, shift={( 30, 1.26185659646 )}] \crossOne;
					\draw[black, shift={( 35, 1.3049254852 )}] \crossOne;
					\draw[black, shift={( 40, 1.27032021948 )}] \crossOne;
					\draw[black, shift={( 45, 1.34616774084 )}] \crossOne;
					\draw[black, shift={( 50, 1.3603428863 )}] \crossOne;
					\draw[black, shift={( 55, 1.33523480373 )}] \crossOne;
					\draw[black, shift={( 60, 1.32534297699 )}] \crossOne;
					\draw[black, shift={( 65, 1.40679910196 )}] \crossOne;
					\draw[black, shift={( 70, 1.41881764242 )}] \crossOne;
					\draw[black, shift={( 75, 1.4179708235 )}] \crossOne;
					\draw[black, shift={( 80, 1.59454846764 )}] \crossOne;
					\draw[black, shift={( 85, 1.57389582222 )}] \crossOne;
					\draw[black, shift={( 90, 1.62469134064 )}] \crossOne;
					\draw[black, shift={( 95, 1.64270638055 )}] \crossOne;
					\draw[black, shift={( 100, 1.65138717376 )}] \crossOne;
					\draw[black, shift={( 105, 1.62047918608 )}] \crossOne;
					\draw[black, shift={( 110, 1.61972850384 )}] \crossOne;
					\draw[black, shift={( 115, 1.65552021204 )}] \crossOne;
					\draw[black, shift={( 120, 1.67449331362 )}] \crossOne;
					\draw[black, shift={( 125, 1.61571412254 )}] \crossOne;
					\draw[black, shift={( 130, 1.65314376667 )}] \crossOne;
					\draw[black, shift={( 135, 1.66316896946 )}] \crossOne;
					\draw[black, shift={( 140, 1.71617750021 )}] \crossOne;
					\draw[black, shift={( 145, 1.69577136878 )}] \crossOne;
					\draw[black, shift={( 150, 1.68293201321 )}] \crossOne;
					\draw[black, shift={( 155, 1.70653576983 )}] \crossOne;
					\draw[black, shift={( 160, 1.687173932 )}] \crossOne;
					\draw[black, shift={( 165, 1.70928338068 )}] \crossOne;
					\draw[black, shift={( 170, 1.73691200838 )}] \crossOne;
					\draw[black, shift={( 175, 1.74755341209 )}] \crossOne;
					\draw[black, shift={( 180, 1.68571542699 )}] \crossOne;
					\draw[black, shift={( 185, 1.71309596836 )}] \crossOne;
					\draw[black, shift={( 190, 1.7230869045 )}] \crossOne;
					\draw[black, shift={( 195, 1.74890542295 )}] \crossOne;
					\draw[black, shift={( 200, 1.7277211506 )}] \crossOne;
					
					\draw[black!75, dotted] ( 5, 1.00518180757 ) -- ( 10, 1.10031329725 ) -- ( 15, 1.13660142608 ) -- ( 20, 1.18124295096 ) -- ( 25, 1.2267075139 ) -- ( 30, 1.26185659646 ) -- ( 35, 1.3049254852 ) -- ( 40, 1.27032021948 ) -- ( 45, 1.34616774084 ) -- ( 50, 1.3603428863 ) -- ( 55, 1.33523480373 ) -- ( 60, 1.32534297699 ) -- ( 65, 1.40679910196 ) -- ( 70, 1.41881764242 ) -- ( 75, 1.4179708235 ) ( 80, 1.59454846764 ) -- ( 85, 1.57389582222 ) -- ( 90, 1.62469134064 ) -- ( 95, 1.64270638055 ) -- ( 100, 1.65138717376 ) -- ( 105, 1.62047918608 ) -- ( 110, 1.61972850384 ) -- ( 115, 1.65552021204 ) -- ( 120, 1.67449331362 ) -- ( 125, 1.61571412254 ) -- ( 130, 1.65314376667 ) -- ( 135, 1.66316896946 ) -- ( 140, 1.71617750021 ) -- ( 145, 1.69577136878 ) -- ( 150, 1.68293201321 ) -- ( 155, 1.70653576983 ) -- ( 160, 1.687173932 ) -- ( 165, 1.70928338068 ) -- ( 170, 1.73691200838 ) -- ( 175, 1.74755341209 ) -- ( 180, 1.68571542699 ) -- ( 185, 1.71309596836 ) -- ( 190, 1.7230869045 ) -- ( 195, 1.74890542295 ) -- ( 200, 1.7277211506 );

					\draw[red, shift={( 5, 0.999999908761 )}] \crossTwo;
					\draw[red, shift={( 10, 1.00794058849 )}] \crossTwo;
					\draw[red, shift={( 15, 1.01245631791 )}] \crossTwo;
					\draw[red, shift={( 20, 1.04860718098 )}] \crossTwo;
					\draw[red, shift={( 25, 1.07020659582 )}] \crossTwo;
					\draw[red, shift={( 30, 1.11172907056 )}] \crossTwo;
					\draw[red, shift={( 35, 1.13274021661 )}] \crossTwo;
					\draw[red, shift={( 40, 1.12872694166 )}] \crossTwo;
					\draw[red, shift={( 45, 1.16868013806 )}] \crossTwo;
					\draw[red, shift={( 50, 1.21958061984 )}] \crossTwo;
					\draw[red, shift={( 55, 1.21325004225 )}] \crossTwo;
					\draw[red, shift={( 60, 1.22949191739 )}] \crossTwo;
					\draw[red, shift={( 65, 1.26148912182 )}] \crossTwo;
					\draw[red, shift={( 70, 1.2769433694 )}] \crossTwo;
					\draw[red, shift={( 75, 1.28701764831 )}] \crossTwo;
					\draw[red, shift={( 80, 1.4415662453 )}] \crossTwo;
					\draw[red, shift={( 85, 1.43812573615 )}] \crossTwo;
					\draw[red, shift={( 90, 1.48170615075 )}] \crossTwo;
					\draw[red, shift={( 95, 1.49145461506 )}] \crossTwo;
					\draw[red, shift={( 100, 1.52869225037 )}] \crossTwo;
					\draw[red, shift={( 105, 1.49772397201 )}] \crossTwo;
					\draw[red, shift={( 110, 1.5120556657 )}] \crossTwo;
					\draw[red, shift={( 115, 1.54675808345 )}] \crossTwo;
					\draw[red, shift={( 120, 1.57605329198 )}] \crossTwo;
					\draw[red, shift={( 125, 1.53600981829 )}] \crossTwo;
					\draw[red, shift={( 130, 1.5470794417 )}] \crossTwo;
					\draw[red, shift={( 135, 1.56746871479 )}] \crossTwo;
					\draw[red, shift={( 140, 1.59243995976 )}] \crossTwo;
					\draw[red, shift={( 145, 1.59590523781 )}] \crossTwo;
					\draw[red, shift={( 150, 1.59346156942 )}] \crossTwo;
					\draw[red, shift={( 155, 1.59755315554 )}] \crossTwo;
					\draw[red, shift={( 160, 1.6043631232 )}] \crossTwo;
					\draw[red, shift={( 165, 1.61251406199 )}] \crossTwo;
					\draw[red, shift={( 170, 1.62614000305 )}] \crossTwo;
					\draw[red, shift={( 175, 1.64653789932 )}] \crossTwo;
					\draw[red, shift={( 180, 1.62801046262 )}] \crossTwo;
					\draw[red, shift={( 185, 1.63666615932 )}] \crossTwo;
					\draw[red, shift={( 190, 1.65358081686 )}] \crossTwo;
					\draw[red, shift={( 195, 1.65246898587 )}] \crossTwo;
					\draw[red, shift={( 200, 1.64732581433 )}] \crossTwo;
									
					\draw[red!75, densely dotted]( 5, 0.999999908761 ) -- ( 10, 1.00794058849 ) -- ( 15, 1.01245631791 ) -- ( 20, 1.04860718098 ) -- ( 25, 1.07020659582 ) -- ( 30, 1.11172907056 ) -- ( 35, 1.13274021661 ) -- ( 40, 1.12872694166 ) -- ( 45, 1.16868013806 ) -- ( 50, 1.21958061984 ) -- ( 55, 1.21325004225 ) -- ( 60, 1.22949191739 ) -- ( 65, 1.26148912182 ) -- ( 70, 1.2769433694 ) -- ( 75, 1.28701764831 ) ( 80, 1.4415662453 ) -- ( 85, 1.43812573615 ) -- ( 90, 1.48170615075 ) -- ( 95, 1.49145461506 ) -- ( 100, 1.52869225037 ) -- ( 105, 1.49772397201 ) -- ( 110, 1.5120556657 ) -- ( 115, 1.54675808345 ) -- ( 120, 1.57605329198 ) -- ( 125, 1.53600981829 ) -- ( 130, 1.5470794417 ) -- ( 135, 1.56746871479 ) -- ( 140, 1.59243995976 ) -- ( 145, 1.59590523781 ) -- ( 150, 1.59346156942 ) -- ( 155, 1.59755315554 ) -- ( 160, 1.6043631232 ) -- ( 165, 1.61251406199 ) -- ( 170, 1.62614000305 ) -- ( 175, 1.64653789932 ) -- ( 180, 1.62801046262 ) -- ( 185, 1.63666615932 ) -- ( 190, 1.65358081686 ) -- ( 195, 1.65246898587 ) -- ( 200, 1.64732581433 );

					\draw[blue, shift={( 5, 0.999999908761 )}] \crossThree;
					\draw[blue, shift={( 10, 1.04951882078 )}] \crossThree;
					\draw[blue, shift={( 15, 1.04857556141 )}] \crossThree;
					\draw[blue, shift={( 20, 1.10261853665 )}] \crossThree;
					\draw[blue, shift={( 25, 1.11183102643 )}] \crossThree;
					\draw[blue, shift={( 30, 1.11226127168 )}] \crossThree;
					\draw[blue, shift={( 35, 1.17304243857 )}] \crossThree;
					\draw[blue, shift={( 40, 1.20138789613 )}] \crossThree;
					\draw[blue, shift={( 45, 1.20012932271 )}] \crossThree;
					\draw[blue, shift={( 50, 1.24980293013 )}] \crossThree;
					\draw[blue, shift={( 55, 1.27403284301 )}] \crossThree;
					\draw[blue, shift={( 60, 1.28199423268 )}] \crossThree;
					\draw[blue, shift={( 65, 1.29989375502 )}] \crossThree;
					\draw[blue, shift={( 70, 1.32103327249 )}] \crossThree;
					\draw[blue, shift={( 75, 1.36981445462 )}] \crossThree;
					\draw[blue, shift={( 80, 1.48951986668 )}] \crossThree;
					\draw[blue, shift={( 85, 1.54693799433 )}] \crossThree;
					\draw[blue, shift={( 90, 1.52244948385 )}] \crossThree;
					\draw[blue, shift={( 95, 1.57340261209 )}] \crossThree;
					\draw[blue, shift={( 100, 1.56265015208 )}] \crossThree;
					\draw[blue, shift={( 105, 1.58773488082 )}] \crossThree;
					\draw[blue, shift={( 110, 1.59430342444 )}] \crossThree;
					\draw[blue, shift={( 115, 1.64212214865 )}] \crossThree;
					\draw[blue, shift={( 120, 1.60330048494 )}] \crossThree;
					\draw[blue, shift={( 125, 1.64021068792 )}] \crossThree;
					\draw[blue, shift={( 130, 1.62616069762 )}] \crossThree;
					\draw[blue, shift={( 135, 1.63270652479 )}] \crossThree;
					\draw[blue, shift={( 140, 1.71825185253 )}] \crossThree;
					\draw[blue, shift={( 145, 1.6894968797 )}] \crossThree;
					\draw[blue, shift={( 150, 1.67464063706 )}] \crossThree;
					\draw[blue, shift={( 155, 1.7058550991 )}] \crossThree;
					\draw[blue, shift={( 160, 1.72069304752 )}] \crossThree;
					\draw[blue, shift={( 165, 1.73746861745 )}] \crossThree;
					\draw[blue, shift={( 170, 1.81164632248 )}] \crossThree;
					\draw[blue, shift={( 175, 1.75559720298 )}] \crossThree;
					\draw[blue, shift={( 180, 1.77410022649 )}] \crossThree;
					\draw[blue, shift={( 185, 1.79604228966 )}] \crossThree;
					\draw[blue, shift={( 190, 1.78907062482 )}] \crossThree;
					\draw[blue, shift={( 195, 1.79472919842 )}] \crossThree;
					\draw[blue, shift={( 200, 1.79010945915 )}] \crossThree;
									
					\draw[blue!75, loosely dotted]( 5, 0.999999908761 ) -- ( 10, 1.04951882078 ) -- ( 15, 1.04857556141 ) -- ( 20, 1.10261853665 ) -- ( 25, 1.11183102643 ) -- ( 30, 1.11226127168 ) -- ( 35, 1.17304243857 ) -- ( 40, 1.20138789613 ) -- ( 45, 1.20012932271 ) -- ( 50, 1.24980293013 ) -- ( 55, 1.27403284301 ) -- ( 60, 1.28199423268 ) -- ( 65, 1.29989375502 ) -- ( 70, 1.32103327249 ) -- ( 75, 1.36981445462 ) ( 80, 1.48951986668 ) -- ( 85, 1.54693799433 ) -- ( 90, 1.52244948385 ) -- ( 95, 1.57340261209 ) -- ( 100, 1.56265015208 ) -- ( 105, 1.58773488082 ) -- ( 110, 1.59430342444 ) -- ( 115, 1.64212214865 ) -- ( 120, 1.60330048494 ) -- ( 125, 1.64021068792 ) -- ( 130, 1.62616069762 ) -- ( 135, 1.63270652479 ) -- ( 140, 1.71825185253 ) -- ( 145, 1.6894968797 ) -- ( 150, 1.67464063706 ) -- ( 155, 1.7058550991 ) -- ( 160, 1.72069304752 ) -- ( 165, 1.73746861745 ) -- ( 170, 1.81164632248 ) -- ( 175, 1.75559720298 ) -- ( 180, 1.77410022649 ) -- ( 185, 1.79604228966 ) -- ( 190, 1.78907062482 ) -- ( 195, 1.79472919842 ) -- ( 200, 1.79010945915 );

					\draw[magenta, shift={( 5, 1.01679955335 )}] \crossFour;
					\draw[magenta, shift={( 10, 1.080073611 )}] \crossFour;
					\draw[magenta, shift={( 15, 1.18026215693 )}] \crossFour;
					\draw[magenta, shift={( 20, 1.18458890948 )}] \crossFour;
					\draw[magenta, shift={( 25, 1.2578211909 )}] \crossFour;
					\draw[magenta, shift={( 30, 1.2724051611 )}] \crossFour;
					\draw[magenta, shift={( 35, 1.31643660904 )}] \crossFour;
					\draw[magenta, shift={( 40, 1.32227175042 )}] \crossFour;
					\draw[magenta, shift={( 45, 1.36383073399 )}] \crossFour;
					\draw[magenta, shift={( 50, 1.38227858127 )}] \crossFour;
					\draw[magenta, shift={( 55, 1.44394511387 )}] \crossFour;
					\draw[magenta, shift={( 60, 1.44444994802 )}] \crossFour;
					\draw[magenta, shift={( 65, 1.48274618813 )}] \crossFour;
					\draw[magenta, shift={( 70, 1.53151064666 )}] \crossFour;
					\draw[magenta, shift={( 75, 1.57856850225 )}] \crossFour;
					\draw[magenta, shift={( 80, 1.73661042322 )}] \crossFour;
					\draw[magenta, shift={( 85, 1.76893024566 )}] \crossFour;
					\draw[magenta, shift={( 90, 1.76600815125 )}] \crossFour;
					\draw[magenta, shift={( 95, 1.84730746364 )}] \crossFour;
					\draw[magenta, shift={( 100, 1.76598911975 )}] \crossFour;
					\draw[magenta, shift={( 105, 1.8148101692 )}] \crossFour;
					\draw[magenta, shift={( 110, 1.86547551453 )}] \crossFour;
					\draw[magenta, shift={( 115, 1.86892660118 )}] \crossFour;
					\draw[magenta, shift={( 120, 1.89086698927 )}] \crossFour;
					\draw[magenta, shift={( 125, 1.89799259673 )}] \crossFour;
					\draw[magenta, shift={( 130, 1.9268776868 )}] \crossFour;
					\draw[magenta, shift={( 135, 1.92009230541 )}] \crossFour;
					\draw[magenta, shift={( 140, 1.98111013661 )}] \crossFour;
					\draw[magenta, shift={( 145, 1.99816786397 )}] \crossFour;
					\draw[magenta, shift={( 150, 1.99986084382 )}] \crossFour;
					\draw[magenta, shift={( 155, 1.93550807484 )}] \crossFour;
					\draw[magenta, shift={( 160, 2.0451622926 )}] \crossFour;
					\draw[magenta, shift={( 165, 2.04106065341 )}] \crossFour;
					\draw[magenta, shift={( 170, 2.09322337901 )}] \crossFour;
					\draw[magenta, shift={( 175, 2.09405376175 )}] \crossFour;
					\draw[magenta, shift={( 180, 2.02553173324 )}] \crossFour;
					\draw[magenta, shift={( 185, 2.09054218338 )}] \crossFour;
					\draw[magenta, shift={( 190, 2.06436147465 )}] \crossFour;
					\draw[magenta, shift={( 195, 2.15709525995 )}] \crossFour;
					\draw[magenta, shift={( 200, 2.11769245881 )}] \crossFour;
									
					\draw[magenta!75, dash pattern={on 1pt off 1pt}]( 5, 1.01679955335 ) -- ( 10, 1.080073611 ) -- ( 15, 1.18026215693 ) -- ( 20, 1.18458890948 ) -- ( 25, 1.2578211909 ) -- ( 30, 1.2724051611 ) -- ( 35, 1.31643660904 ) -- ( 40, 1.32227175042 ) -- ( 45, 1.36383073399 ) -- ( 50, 1.38227858127 ) -- ( 55, 1.44394511387 ) -- ( 60, 1.44444994802 ) -- ( 65, 1.48274618813 ) -- ( 70, 1.53151064666 ) -- ( 75, 1.57856850225 ) ( 80, 1.73661042322 ) -- ( 85, 1.76893024566 ) -- ( 90, 1.76600815125 ) -- ( 95, 1.84730746364 ) -- ( 100, 1.76598911975 ) -- ( 105, 1.8148101692 ) -- ( 110, 1.86547551453 ) -- ( 115, 1.86892660118 ) -- ( 120, 1.89086698927 ) -- ( 125, 1.89799259673 ) -- ( 130, 1.9268776868 ) -- ( 135, 1.92009230541 ) -- ( 140, 1.98111013661 ) -- ( 145, 1.99816786397 ) -- ( 150, 1.99986084382 ) -- ( 155, 1.93550807484 ) -- ( 160, 2.0451622926 ) -- ( 165, 2.04106065341 ) -- ( 170, 2.09322337901 ) -- ( 175, 2.09405376175 ) -- ( 180, 2.02553173324 ) -- ( 185, 2.09054218338 ) -- ( 190, 2.06436147465 ) -- ( 195, 2.15709525995 ) -- ( 200, 2.11769245881 );



				\end{scope}
				
				\def\axisAdditionalLengthPlusTikzX{\axisAdditionalLengthPlus / \xscaleTikz}
				\def\axisAdditionalLengthMinusTikzX{\axisAdditionalLengthMinus / \xscaleTikz}
				\draw[arrow] (-\axisAdditionalLengthMinusTikzX, 0) -- (200, 0)  -- +(\axisAdditionalLengthPlusTikzX, 0) node[right] {\xAxis};
				\def\xTotalLengthPlus{200+\axisAdditionalLengthPlusTikzX}
				\draw (\xTotalLengthPlus, 0) node[right] {$\qquad$};
				\def\axisAdditionalLengthPlusTikzY{\axisAdditionalLengthPlus / \yscaleTikz}
				\def\axisAdditionalLengthMinusTikzY{\axisAdditionalLengthMinus / \yscaleTikz}
				\draw[arrow] (0, -\axisAdditionalLengthMinusTikzY) -- (0, 1.5 ) -- +(0, \axisAdditionalLengthPlusTikzY) node[above, yshift=-0.15cm] {\yAxis};
				
				\def\axisLabelTikzY{\axisLabel / \yscaleTikz}
				\foreach \pos in {0, 20, 40, 60, 80, 100, 120, 140, 160, 180, 200} \draw[shift={(\pos, 0)}] (0, \axisLabelTikzY) -- (0, -\axisLabelTikzY) node[below] {$\pos$};
				
				\def\axisLabelTikzX{\axisLabel / \xscaleTikz}
				\begin{scope}[shift={(0, -1)}]
					\draw[shift={(0,1 )}] (\axisLabelTikzX, 0) -- (-\axisLabelTikzX, 0) node[left] {$1$};
					\draw[shift={(0,1.5 )}] (\axisLabelTikzX, 0) -- (-\axisLabelTikzX, 0) node[left] {$1.5$};
					\draw[shift={(0,2 )}] (\axisLabelTikzX, 0) -- (-\axisLabelTikzX, 0) node[left] {$2$};
					\draw[shift={(0,2.5 )}] (\axisLabelTikzX, 0) -- (-\axisLabelTikzX, 0) node[left] {$2.5$};
				\end{scope}
			\end{tikzpicture}
		\AngleTSPInstances
		\vspace{-0.6cm}
		
			\begin{tikzpicture}[xscale=\xscale, yscale=\yscaleSmall]
				\pgfgettransformentries{\xscaleTikz}{\@tempa}{\@tempa}{\yscaleTikz}{\@tempa}{\@tempa}
			
				\draw[very thin, color=gray, xstep=20, ystep=0.1] (0, 0) grid (200, 0.3 );
				
				\begin{scope}[shift={(0, -1)}]
					\def\crossSizeX{\crossSize / \xscaleTikz};
					\def\crossSizeY{\crossSize / \yscaleTikz};
					
					\def\crossOne{(-\crossSizeX,-\crossSizeY) -- (\crossSizeX,\crossSizeY) (-\crossSizeX,\crossSizeY) -- (\crossSizeX,-\crossSizeY)};
					\def\crossTwo{(-\crossSizeX,0) -- (\crossSizeX,0) (0,\crossSizeY) -- (0,-\crossSizeY)};
					\def\crossThree{(0,0) -- (0,\crossSizeY) (0,0) -- (\crossSizeX,-\crossSizeY) (0,0) -- (-\crossSizeX,-\crossSizeY)};
					\def\crossFour{(0,0) -- (0,-\crossSizeY) (0,0) -- (\crossSizeX,\crossSizeY) (0,0) -- (-\crossSizeX,\crossSizeY)};
					\def\crossFive{(0,0) -- (\crossSizeX,0) (0,0) -- (-\crossSizeX,\crossSizeY) (0,0) -- (-\crossSizeX,-\crossSizeY)};
					\def\crossSix{(0,0) -- (-\crossSizeX,0) (0,0) -- (\crossSizeX,\crossSizeY) (0,0) -- (\crossSizeX,-\crossSizeY)};
					
					\draw[black, shift={( 5, 1.00032187456 )}] \crossOne;
					\draw[black, shift={( 10, 1.02320623775 )}] \crossOne;
					\draw[black, shift={( 15, 1.06748501388 )}] \crossOne;
					\draw[black, shift={( 20, 1.07247738909 )}] \crossOne;
					\draw[black, shift={( 25, 1.04342276543 )}] \crossOne;
					\draw[black, shift={( 30, 1.08246012422 )}] \crossOne;
					\draw[black, shift={( 35, 1.09378782694 )}] \crossOne;
					\draw[black, shift={( 40, 1.08957402326 )}] \crossOne;
					\draw[black, shift={( 45, 1.10875492437 )}] \crossOne;
					\draw[black, shift={( 50, 1.10709483842 )}] \crossOne;
					\draw[black, shift={( 55, 1.12648837321 )}] \crossOne;
					\draw[black, shift={( 60, 1.12646455471 )}] \crossOne;
					\draw[black, shift={( 65, 1.12291208167 )}] \crossOne;
					\draw[black, shift={( 70, 1.11958367959 )}] \crossOne;
					\draw[black, shift={( 75, 1.13343482586 )}] \crossOne;
					\draw[black, shift={( 80, 1.12938107913 )}] \crossOne;
					\draw[black, shift={( 85, 1.14344647282 )}] \crossOne;
					\draw[black, shift={( 90, 1.13137958179 )}] \crossOne;
					\draw[black, shift={( 95, 1.142103774 )}] \crossOne;
					\draw[black, shift={( 100, 1.13987743272 )}] \crossOne;
					\draw[black, shift={( 105, 1.2098995837 )}] \crossOne;
					\draw[black, shift={( 110, 1.2275236753 )}] \crossOne;
					\draw[black, shift={( 115, 1.23968709101 )}] \crossOne;
					\draw[black, shift={( 120, 1.2257335457 )}] \crossOne;
					\draw[black, shift={( 125, 1.22516926374 )}] \crossOne;
					\draw[black, shift={( 130, 1.2433770005 )}] \crossOne;
					\draw[black, shift={( 135, 1.25187947213 )}] \crossOne;
					\draw[black, shift={( 140, 1.25112562072 )}] \crossOne;
					\draw[black, shift={( 145, 1.26176209289 )}] \crossOne;
					\draw[black, shift={( 150, 1.25077927241 )}] \crossOne;
					\draw[black, shift={( 155, 1.26551906346 )}] \crossOne;
					\draw[black, shift={( 160, 1.25949107717 )}] \crossOne;
					\draw[black, shift={( 165, 1.25767837863 )}] \crossOne;
					\draw[black, shift={( 170, 1.27682002993 )}] \crossOne;
					\draw[black, shift={( 175, 1.27220027138 )}] \crossOne;
					\draw[black, shift={( 180, 1.27246225232 )}] \crossOne;
					\draw[black, shift={( 185, 1.27207582773 )}] \crossOne;
					\draw[black, shift={( 190, 1.27292905611 )}] \crossOne;
					\draw[black, shift={( 195, 1.28732511812 )}] \crossOne;
					\draw[black, shift={( 200, 1.28254208718 )}] \crossOne;
					
					\draw[black!75, dotted] ( 5, 1.00032187456 ) -- ( 10, 1.02320623775 ) -- ( 15, 1.06748501388 ) -- ( 20, 1.07247738909 ) -- ( 25, 1.04342276543 ) -- ( 30, 1.08246012422 ) -- ( 35, 1.09378782694 ) -- ( 40, 1.08957402326 ) -- ( 45, 1.10875492437 ) -- ( 50, 1.10709483842 ) -- ( 55, 1.12648837321 ) -- ( 60, 1.12646455471 ) -- ( 65, 1.12291208167 ) -- ( 70, 1.11958367959 ) -- ( 75, 1.13343482586 ) -- ( 80, 1.12938107913 ) -- ( 85, 1.14344647282 ) -- ( 90, 1.13137958179 ) -- ( 95, 1.142103774 ) -- ( 100, 1.13987743272 ) ( 105, 1.2098995837 ) -- ( 110, 1.2275236753 ) -- ( 115, 1.23968709101 ) -- ( 120, 1.2257335457 ) -- ( 125, 1.22516926374 ) -- ( 130, 1.2433770005 ) -- ( 135, 1.25187947213 ) -- ( 140, 1.25112562072 ) -- ( 145, 1.26176209289 ) -- ( 150, 1.25077927241 ) -- ( 155, 1.26551906346 ) -- ( 160, 1.25949107717 ) -- ( 165, 1.25767837863 ) -- ( 170, 1.27682002993 ) -- ( 175, 1.27220027138 ) -- ( 180, 1.27246225232 ) -- ( 185, 1.27207582773 ) -- ( 190, 1.27292905611 ) -- ( 195, 1.28732511812 ) -- ( 200, 1.28254208718 );

					\draw[red, shift={( 5, 0.999999996966 )}] \crossTwo;
					\draw[red, shift={( 10, 1.00019681156 )}] \crossTwo;
					\draw[red, shift={( 15, 1.00238222981 )}] \crossTwo;
					\draw[red, shift={( 20, 1.0013673183 )}] \crossTwo;
					\draw[red, shift={( 25, 1.00360602001 )}] \crossTwo;
					\draw[red, shift={( 30, 1.00242693635 )}] \crossTwo;
					\draw[red, shift={( 35, 1.00209861905 )}] \crossTwo;
					\draw[red, shift={( 40, 1.00443077626 )}] \crossTwo;
					\draw[red, shift={( 45, 1.00372480954 )}] \crossTwo;
					\draw[red, shift={( 50, 1.0092330973 )}] \crossTwo;
					\draw[red, shift={( 55, 1.00806492994 )}] \crossTwo;
					\draw[red, shift={( 60, 1.01224260045 )}] \crossTwo;
					\draw[red, shift={( 65, 1.00991157082 )}] \crossTwo;
					\draw[red, shift={( 70, 1.01065344516 )}] \crossTwo;
					\draw[red, shift={( 75, 1.01547710396 )}] \crossTwo;
					\draw[red, shift={( 80, 1.0190876865 )}] \crossTwo;
					\draw[red, shift={( 85, 1.02153683598 )}] \crossTwo;
					\draw[red, shift={( 90, 1.01973366669 )}] \crossTwo;
					\draw[red, shift={( 95, 1.02118143504 )}] \crossTwo;
					\draw[red, shift={( 100, 1.01871101505 )}] \crossTwo;
					\draw[red, shift={( 105, 1.08481991434 )}] \crossTwo;
					\draw[red, shift={( 110, 1.09621501817 )}] \crossTwo;
					\draw[red, shift={( 115, 1.1036713264 )}] \crossTwo;
					\draw[red, shift={( 120, 1.09955661518 )}] \crossTwo;
					\draw[red, shift={( 125, 1.10052412291 )}] \crossTwo;
					\draw[red, shift={( 130, 1.10773679015 )}] \crossTwo;
					\draw[red, shift={( 135, 1.11331854492 )}] \crossTwo;
					\draw[red, shift={( 140, 1.11270128318 )}] \crossTwo;
					\draw[red, shift={( 145, 1.11124549416 )}] \crossTwo;
					\draw[red, shift={( 150, 1.11132040311 )}] \crossTwo;
					\draw[red, shift={( 155, 1.12247158465 )}] \crossTwo;
					\draw[red, shift={( 160, 1.1224752582 )}] \crossTwo;
					\draw[red, shift={( 165, 1.1228393803 )}] \crossTwo;
					\draw[red, shift={( 170, 1.13290241899 )}] \crossTwo;
					\draw[red, shift={( 175, 1.13002191952 )}] \crossTwo;
					\draw[red, shift={( 180, 1.13382231911 )}] \crossTwo;
					\draw[red, shift={( 185, 1.13518698703 )}] \crossTwo;
					\draw[red, shift={( 190, 1.1362229858 )}] \crossTwo;
					\draw[red, shift={( 195, 1.14024029521 )}] \crossTwo;
					\draw[red, shift={( 200, 1.1422386282 )}] \crossTwo;
									
					\draw[red!75, densely dotted]( 5, 0.999999996966 ) -- ( 10, 1.00019681156 ) -- ( 15, 1.00238222981 ) -- ( 20, 1.0013673183 ) -- ( 25, 1.00360602001 ) -- ( 30, 1.00242693635 ) -- ( 35, 1.00209861905 ) -- ( 40, 1.00443077626 ) -- ( 45, 1.00372480954 ) -- ( 50, 1.0092330973 ) -- ( 55, 1.00806492994 ) -- ( 60, 1.01224260045 ) -- ( 65, 1.00991157082 ) -- ( 70, 1.01065344516 ) -- ( 75, 1.01547710396 ) -- ( 80, 1.0190876865 ) -- ( 85, 1.02153683598 ) -- ( 90, 1.01973366669 ) -- ( 95, 1.02118143504 ) -- ( 100, 1.01871101505 ) ( 105, 1.08481991434 ) -- ( 110, 1.09621501817 ) -- ( 115, 1.1036713264 ) -- ( 120, 1.09955661518 ) -- ( 125, 1.10052412291 ) -- ( 130, 1.10773679015 ) -- ( 135, 1.11331854492 ) -- ( 140, 1.11270128318 ) -- ( 145, 1.11124549416 ) -- ( 150, 1.11132040311 ) -- ( 155, 1.12247158465 ) -- ( 160, 1.1224752582 ) -- ( 165, 1.1228393803 ) -- ( 170, 1.13290241899 ) -- ( 175, 1.13002191952 ) -- ( 180, 1.13382231911 ) -- ( 185, 1.13518698703 ) -- ( 190, 1.1362229858 ) -- ( 195, 1.14024029521 ) -- ( 200, 1.1422386282 );

					\draw[blue, shift={( 5, 1.00032187456 )}] \crossThree;
					\draw[blue, shift={( 10, 1.02320623775 )}] \crossThree;
					\draw[blue, shift={( 15, 1.0695809027 )}] \crossThree;
					\draw[blue, shift={( 20, 1.06694548399 )}] \crossThree;
					\draw[blue, shift={( 25, 1.04455402439 )}] \crossThree;
					\draw[blue, shift={( 30, 1.0813939774 )}] \crossThree;
					\draw[blue, shift={( 35, 1.0920689125 )}] \crossThree;
					\draw[blue, shift={( 40, 1.08514300226 )}] \crossThree;
					\draw[blue, shift={( 45, 1.11331554378 )}] \crossThree;
					\draw[blue, shift={( 50, 1.10388820625 )}] \crossThree;
					\draw[blue, shift={( 55, 1.1117416571 )}] \crossThree;
					\draw[blue, shift={( 60, 1.11621661695 )}] \crossThree;
					\draw[blue, shift={( 65, 1.12000840169 )}] \crossThree;
					\draw[blue, shift={( 70, 1.10406271574 )}] \crossThree;
					\draw[blue, shift={( 75, 1.13021553329 )}] \crossThree;
					\draw[blue, shift={( 80, 1.12432301373 )}] \crossThree;
					\draw[blue, shift={( 85, 1.131045704 )}] \crossThree;
					\draw[blue, shift={( 90, 1.1280030015 )}] \crossThree;
					\draw[blue, shift={( 95, 1.13334925838 )}] \crossThree;
					\draw[blue, shift={( 100, 1.13941555219 )}] \crossThree;
					\draw[blue, shift={( 105, 1.20210701894 )}] \crossThree;
					\draw[blue, shift={( 110, 1.21888746086 )}] \crossThree;
					\draw[blue, shift={( 115, 1.23064208562 )}] \crossThree;
					\draw[blue, shift={( 120, 1.2181804359 )}] \crossThree;
					\draw[blue, shift={( 125, 1.21960638143 )}] \crossThree;
					\draw[blue, shift={( 130, 1.23044957569 )}] \crossThree;
					\draw[blue, shift={( 135, 1.23530855155 )}] \crossThree;
					\draw[blue, shift={( 140, 1.24240311837 )}] \crossThree;
					\draw[blue, shift={( 145, 1.24418623102 )}] \crossThree;
					\draw[blue, shift={( 150, 1.23779555408 )}] \crossThree;
					\draw[blue, shift={( 155, 1.24866148302 )}] \crossThree;
					\draw[blue, shift={( 160, 1.245987192 )}] \crossThree;
					\draw[blue, shift={( 165, 1.25369818659 )}] \crossThree;
					\draw[blue, shift={( 170, 1.25255189744 )}] \crossThree;
					\draw[blue, shift={( 175, 1.25378070661 )}] \crossThree;
					\draw[blue, shift={( 180, 1.26008574811 )}] \crossThree;
					\draw[blue, shift={( 185, 1.26081877912 )}] \crossThree;
					\draw[blue, shift={( 190, 1.25899665221 )}] \crossThree;
					\draw[blue, shift={( 195, 1.26718117383 )}] \crossThree;
					\draw[blue, shift={( 200, 1.26466297 )}] \crossThree;
									
					\draw[blue!75, loosely dotted]( 5, 1.00032187456 ) -- ( 10, 1.02320623775 ) -- ( 15, 1.0695809027 ) -- ( 20, 1.06694548399 ) -- ( 25, 1.04455402439 ) -- ( 30, 1.0813939774 ) -- ( 35, 1.0920689125 ) -- ( 40, 1.08514300226 ) -- ( 45, 1.11331554378 ) -- ( 50, 1.10388820625 ) -- ( 55, 1.1117416571 ) -- ( 60, 1.11621661695 ) -- ( 65, 1.12000840169 ) -- ( 70, 1.10406271574 ) -- ( 75, 1.13021553329 ) -- ( 80, 1.12432301373 ) -- ( 85, 1.131045704 ) -- ( 90, 1.1280030015 ) -- ( 95, 1.13334925838 ) -- ( 100, 1.13941555219 ) ( 105, 1.20210701894 ) -- ( 110, 1.21888746086 ) -- ( 115, 1.23064208562 ) -- ( 120, 1.2181804359 ) -- ( 125, 1.21960638143 ) -- ( 130, 1.23044957569 ) -- ( 135, 1.23530855155 ) -- ( 140, 1.24240311837 ) -- ( 145, 1.24418623102 ) -- ( 150, 1.23779555408 ) -- ( 155, 1.24866148302 ) -- ( 160, 1.245987192 ) -- ( 165, 1.25369818659 ) -- ( 170, 1.25255189744 ) -- ( 175, 1.25378070661 ) -- ( 180, 1.26008574811 ) -- ( 185, 1.26081877912 ) -- ( 190, 1.25899665221 ) -- ( 195, 1.26718117383 ) -- ( 200, 1.26466297 );

					\draw[magenta, shift={( 5, 1.01621836567 )}] \crossFour;
					\draw[magenta, shift={( 10, 1.02307834716 )}] \crossFour;
					\draw[magenta, shift={( 15, 1.08933505347 )}] \crossFour;
					\draw[magenta, shift={( 20, 1.08923633668 )}] \crossFour;
					\draw[magenta, shift={( 25, 1.10650094514 )}] \crossFour;
					\draw[magenta, shift={( 30, 1.08579003228 )}] \crossFour;
					\draw[magenta, shift={( 35, 1.10261012461 )}] \crossFour;
					\draw[magenta, shift={( 40, 1.09714221862 )}] \crossFour;
					\draw[magenta, shift={( 45, 1.11820898616 )}] \crossFour;
					\draw[magenta, shift={( 50, 1.12561139511 )}] \crossFour;
					\draw[magenta, shift={( 55, 1.11390941749 )}] \crossFour;
					\draw[magenta, shift={( 60, 1.11282100775 )}] \crossFour;
					\draw[magenta, shift={( 65, 1.11366411323 )}] \crossFour;
					\draw[magenta, shift={( 70, 1.1179444663 )}] \crossFour;
					\draw[magenta, shift={( 75, 1.13776897016 )}] \crossFour;
					\draw[magenta, shift={( 80, 1.11805604225 )}] \crossFour;
					\draw[magenta, shift={( 85, 1.12346748642 )}] \crossFour;
					\draw[magenta, shift={( 90, 1.12943016792 )}] \crossFour;
					\draw[magenta, shift={( 95, 1.12318951161 )}] \crossFour;
					\draw[magenta, shift={( 100, 1.12921756103 )}] \crossFour;
					\draw[magenta, shift={( 105, 1.21014849876 )}] \crossFour;
					\draw[magenta, shift={( 110, 1.19826574024 )}] \crossFour;
					\draw[magenta, shift={( 115, 1.20243973668 )}] \crossFour;
					\draw[magenta, shift={( 120, 1.2151962699 )}] \crossFour;
					\draw[magenta, shift={( 125, 1.21800187535 )}] \crossFour;
					\draw[magenta, shift={( 130, 1.21909143906 )}] \crossFour;
					\draw[magenta, shift={( 135, 1.23286550407 )}] \crossFour;
					\draw[magenta, shift={( 140, 1.23669162428 )}] \crossFour;
					\draw[magenta, shift={( 145, 1.22720897655 )}] \crossFour;
					\draw[magenta, shift={( 150, 1.23741073746 )}] \crossFour;
					\draw[magenta, shift={( 155, 1.2449347436 )}] \crossFour;
					\draw[magenta, shift={( 160, 1.242253607 )}] \crossFour;
					\draw[magenta, shift={( 165, 1.24205538851 )}] \crossFour;
					\draw[magenta, shift={( 170, 1.25139775524 )}] \crossFour;
					\draw[magenta, shift={( 175, 1.2477592213 )}] \crossFour;
					\draw[magenta, shift={( 180, 1.24013567424 )}] \crossFour;
					\draw[magenta, shift={( 185, 1.27042979812 )}] \crossFour;
					\draw[magenta, shift={( 190, 1.26190076831 )}] \crossFour;
					\draw[magenta, shift={( 195, 1.27352826319 )}] \crossFour;
					\draw[magenta, shift={( 200, 1.26411228051 )}] \crossFour;
									
					\draw[magenta!75, dash pattern={on 1pt off 1pt}]( 5, 1.01621836567 ) -- ( 10, 1.02307834716 ) -- ( 15, 1.08933505347 ) -- ( 20, 1.08923633668 ) -- ( 25, 1.10650094514 ) -- ( 30, 1.08579003228 ) -- ( 35, 1.10261012461 ) -- ( 40, 1.09714221862 ) -- ( 45, 1.11820898616 ) -- ( 50, 1.12561139511 ) -- ( 55, 1.11390941749 ) -- ( 60, 1.11282100775 ) -- ( 65, 1.11366411323 ) -- ( 70, 1.1179444663 ) -- ( 75, 1.13776897016 ) -- ( 80, 1.11805604225 ) -- ( 85, 1.12346748642 ) -- ( 90, 1.12943016792 ) -- ( 95, 1.12318951161 ) -- ( 100, 1.12921756103 ) ( 105, 1.21014849876 ) -- ( 110, 1.19826574024 ) -- ( 115, 1.20243973668 ) -- ( 120, 1.2151962699 ) -- ( 125, 1.21800187535 ) -- ( 130, 1.21909143906 ) -- ( 135, 1.23286550407 ) -- ( 140, 1.23669162428 ) -- ( 145, 1.22720897655 ) -- ( 150, 1.23741073746 ) -- ( 155, 1.2449347436 ) -- ( 160, 1.242253607 ) -- ( 165, 1.24205538851 ) -- ( 170, 1.25139775524 ) -- ( 175, 1.2477592213 ) -- ( 180, 1.24013567424 ) -- ( 185, 1.27042979812 ) -- ( 190, 1.26190076831 ) -- ( 195, 1.27352826319 ) -- ( 200, 1.26411228051 );



				\end{scope}
				
				\def\axisAdditionalLengthPlusTikzX{\axisAdditionalLengthPlus / \xscaleTikz}
				\def\axisAdditionalLengthMinusTikzX{\axisAdditionalLengthMinus / \xscaleTikz}
				\draw[arrow] (-\axisAdditionalLengthMinusTikzX, 0) -- (200, 0)  -- +(\axisAdditionalLengthPlusTikzX, 0) node[right] {\xAxis};
				\def\xTotalLengthPlus{200+\axisAdditionalLengthPlusTikzX}
				\draw (\xTotalLengthPlus, 0) node[right] {$\qquad$};
				\def\axisAdditionalLengthPlusTikzY{\axisAdditionalLengthPlus / \yscaleTikz}
				\def\axisAdditionalLengthMinusTikzY{\axisAdditionalLengthMinus / \yscaleTikz}
				\draw[arrow] (0, -\axisAdditionalLengthMinusTikzY) -- (0, 0.3 ) -- +(0, \axisAdditionalLengthPlusTikzY) node[above, yshift=-0.15cm] {\yAxis};
				
				\def\axisLabelTikzY{\axisLabel / \yscaleTikz}
				\foreach \pos in {0, 20, 40, 60, 80, 100, 120, 140, 160, 180, 200} \draw[shift={(\pos, 0)}] (0, \axisLabelTikzY) -- (0, -\axisLabelTikzY) node[below] {$\pos$};
				
				\def\axisLabelTikzX{\axisLabel / \xscaleTikz}
				\begin{scope}[shift={(0, -1)}]
					\draw[shift={(0,1 )}] (\axisLabelTikzX, 0) -- (-\axisLabelTikzX, 0) node[left] {$1$};
					\draw[shift={(0,1.1 )}] (\axisLabelTikzX, 0) -- (-\axisLabelTikzX, 0) node[left] {$1.1$};
					\draw[shift={(0,1.2 )}] (\axisLabelTikzX, 0) -- (-\axisLabelTikzX, 0) node[left] {$1.2$};
					\draw[shift={(0,1.3 )}] (\axisLabelTikzX, 0) -- (-\axisLabelTikzX, 0) node[left] {$1.3$};
				\end{scope}
			\end{tikzpicture}
		\AngleDistanceTSPInstances
		\vspace*{-0.1cm}
		\caption[simple construction and related heuristics: \CIF{}, \NNS{}, \NNSL{}, \NNSTwo{}]{
			simple construction and related heuristics:\hspace*{-10cm}\\
			\begin{tikzpicture}[xscale=\xscale, yscale=2.0]
				\pgfgettransformentries{\xscaleTikz}{\@tempa}{\@tempa}{\yscaleTikz}{\@tempa}{\@tempa}
				\def\crossSizeX{\crossSize / \xscaleTikz};
				\def\crossSizeY{\crossSize / \yscaleTikz};
				
				\def\crossOne{(-\crossSizeX,-\crossSizeY) -- (\crossSizeX,\crossSizeY) (-\crossSizeX,\crossSizeY) -- (\crossSizeX,-\crossSizeY)};
				\def\crossTwo{(-\crossSizeX,0) -- (\crossSizeX,0) (0,\crossSizeY) -- (0,-\crossSizeY)};
				\def\crossThree{(0,0) -- (0,\crossSizeY) (0,0) -- (\crossSizeX,-\crossSizeY) (0,0) -- (-\crossSizeX,-\crossSizeY)};
				\def\crossFour{(0,0) -- (0,-\crossSizeY) (0,0) -- (\crossSizeX,\crossSizeY) (0,0) -- (-\crossSizeX,\crossSizeY)};
				\def\crossFive{(0,0) -- (\crossSizeX,0) (0,0) -- (-\crossSizeX,\crossSizeY) (0,0) -- (-\crossSizeX,-\crossSizeY)};
				\def\crossSix{(0,0) -- (-\crossSizeX,0) (0,0) -- (\crossSizeX,\crossSizeY) (0,0) -- (\crossSizeX,-\crossSizeY)};
				
				\draw[black, shift={(0, 0)}] \crossOne;
				\draw[black, shift={(5, 0)}] \crossOne;
				\draw[black, shift={(10, 0)}] \crossOne;
				\draw[black, shift={(15, 0)}] \crossOne;
				\draw[black!75, dotted] (-2.5, 0) -- (17.5, 0);
				\node at (40, 0) [minimum width=3cm] {\parbox{1.1cm}\NNS{}};

				\draw[red, shift={(80, 0)}] \crossTwo;
				\draw[red, shift={(85, 0)}] \crossTwo;
				\draw[red, shift={(90, 0)}] \crossTwo;
				\draw[red, shift={(95, 0)}] \crossTwo;
				\draw[red!75, densely dotted] (77.5, 0) -- (97.5, 0);
				\node at (120, 0) [minimum width=3cm] {\parbox{1.1cm}\NNSTwo{}};
				
				\draw[blue, shift={(160, 0)}] \crossThree;
				\draw[blue, shift={(165, 0)}] \crossThree;
				\draw[blue, shift={(170, 0)}] \crossThree;
				\draw[blue, shift={(175, 0)}] \crossThree;
				\draw[blue!75, loosely dotted] (157.5, 0) -- (177.5, 0);
				\node at (200, 0) [minimum width=3cm] {\parbox{1.1cm}\NNSL{}};
				
				\draw[magenta, shift={(0, -0.25)}] \crossFour;
				\draw[magenta, shift={(5, -0.25)}] \crossFour;
				\draw[magenta, shift={(10, -0.25)}] \crossFour;
				\draw[magenta, shift={(15, -0.25)}] \crossFour;
				\draw[magenta!75, dash pattern={on 1pt off 1pt}] (-2.5, -0.25) -- (17.5, -0.25);
				\node at (40, -0.25) [minimum width=3cm] {\parbox{1.1cm}\CIF{}};
			\end{tikzpicture}
		}
		\label{figure:SimpleConstructionAndRelatedHeuristics}
	\end{figure}
			\end{comment:figures}
			
			If we consider \CIF{}, which is our ``reference method'' from \cite{FischerFischerJaegerKeilwagenMolitorGrosse:ExactAlgorithmsAndHeuristicsForTheQuadraticTravelingSalesmanProblemWithAnApplicationInBioinformatics}), we can see that it is outperformed by all other heuristics of this group for \AngleTSPInstances{} and that it yields results similar to \NNS{} and \NNSL{} for the \AngleDistanceTSPInstances.
			A comparison of the nearest-neighbour-based heuristics (\NNS{}, \NNSTwo{} and \NNSL{}) gives us \NNSTwo{} as a clear winner. Furthermore, comparing to \NNS{}, we can see that \NNSL{} performs slightly worse for {\AngleTSP}- and slightly better for {\AngleDistanceTSP}-instances. 
				
			\begin{comment:figures}
	\begin{figure}[htb!]
		\centering
		\begin{tikzpicture}[xscale=\xscale, yscale=\yscaleMiddle]
			\pgfgettransformentries{\xscaleTikz}{\@tempa}{\@tempa}{\yscaleTikz}{\@tempa}{\@tempa}
		
			\draw[very thin, color=gray, xstep=20, ystep=0.3] (0, 0) grid (200, 0.9 );
			
			\begin{scope}[shift={(0, -1)}]
				\def\crossSizeX{\crossSize / \xscaleTikz};
				\def\crossSizeY{\crossSize / \yscaleTikz};
				
				\def\crossOne{(-\crossSizeX,-\crossSizeY) -- (\crossSizeX,\crossSizeY) (-\crossSizeX,\crossSizeY) -- (\crossSizeX,-\crossSizeY)};
				\def\crossTwo{(-\crossSizeX,0) -- (\crossSizeX,0) (0,\crossSizeY) -- (0,-\crossSizeY)};
				\def\crossThree{(0,0) -- (0,\crossSizeY) (0,0) -- (\crossSizeX,-\crossSizeY) (0,0) -- (-\crossSizeX,-\crossSizeY)};
				\def\crossFour{(0,0) -- (0,-\crossSizeY) (0,0) -- (\crossSizeX,\crossSizeY) (0,0) -- (-\crossSizeX,\crossSizeY)};
				\def\crossFive{(0,0) -- (\crossSizeX,0) (0,0) -- (-\crossSizeX,\crossSizeY) (0,0) -- (-\crossSizeX,-\crossSizeY)};
				\def\crossSix{(0,0) -- (-\crossSizeX,0) (0,0) -- (\crossSizeX,\crossSizeY) (0,0) -- (\crossSizeX,-\crossSizeY)};
				
				\draw[black, shift={( 5, 0.999999908761 )}] \crossOne;
				\draw[black, shift={( 10, 1.07443362989 )}] \crossOne;
				\draw[black, shift={( 15, 1.10634236399 )}] \crossOne;
				\draw[black, shift={( 20, 1.17647291586 )}] \crossOne;
				\draw[black, shift={( 25, 1.23132086225 )}] \crossOne;
				\draw[black, shift={( 30, 1.28147444639 )}] \crossOne;
				\draw[black, shift={( 35, 1.28184320416 )}] \crossOne;
				\draw[black, shift={( 40, 1.30964766137 )}] \crossOne;
				\draw[black, shift={( 45, 1.30962886515 )}] \crossOne;
				\draw[black, shift={( 50, 1.39626702925 )}] \crossOne;
				\draw[black, shift={( 55, 1.36407967977 )}] \crossOne;
				\draw[black, shift={( 60, 1.372174589 )}] \crossOne;
				\draw[black, shift={( 65, 1.42441283667 )}] \crossOne;
				\draw[black, shift={( 70, 1.45614444513 )}] \crossOne;
				\draw[black, shift={( 75, 1.42214292475 )}] \crossOne;
				\draw[black, shift={( 80, 1.61951229747 )}] \crossOne;
				\draw[black, shift={( 85, 1.5822009781 )}] \crossOne;
				\draw[black, shift={( 90, 1.64785677243 )}] \crossOne;
				\draw[black, shift={( 95, 1.64110731546 )}] \crossOne;
				\draw[black, shift={( 100, 1.66931952084 )}] \crossOne;
				\draw[black, shift={( 105, 1.64698940523 )}] \crossOne;
				\draw[black, shift={( 110, 1.64035255033 )}] \crossOne;
				\draw[black, shift={( 115, 1.67373579449 )}] \crossOne;
				\draw[black, shift={( 120, 1.69722571914 )}] \crossOne;
				\draw[black, shift={( 125, 1.64410473976 )}] \crossOne;
				\draw[black, shift={( 130, 1.67183406497 )}] \crossOne;
				\draw[black, shift={( 135, 1.68970775792 )}] \crossOne;
				\draw[black, shift={( 140, 1.75006550348 )}] \crossOne;
				\draw[black, shift={( 145, 1.72589981327 )}] \crossOne;
				\draw[black, shift={( 150, 1.70345487361 )}] \crossOne;
				\draw[black, shift={( 155, 1.72823556163 )}] \crossOne;
				\draw[black, shift={( 160, 1.70932249861 )}] \crossOne;
				\draw[black, shift={( 165, 1.74166363875 )}] \crossOne;
				\draw[black, shift={( 170, 1.76854372719 )}] \crossOne;
				\draw[black, shift={( 175, 1.78470009443 )}] \crossOne;
				\draw[black, shift={( 180, 1.72592616491 )}] \crossOne;
				\draw[black, shift={( 185, 1.72842221335 )}] \crossOne;
				\draw[black, shift={( 190, 1.73911385175 )}] \crossOne;
				\draw[black, shift={( 195, 1.75547435511 )}] \crossOne;
				\draw[black, shift={( 200, 1.74171721214 )}] \crossOne;
				
				\draw[black!75, dotted] ( 5, 0.999999908761 ) -- ( 10, 1.07443362989 ) -- ( 15, 1.10634236399 ) -- ( 20, 1.17647291586 ) -- ( 25, 1.23132086225 ) -- ( 30, 1.28147444639 ) -- ( 35, 1.28184320416 ) -- ( 40, 1.30964766137 ) -- ( 45, 1.30962886515 ) -- ( 50, 1.39626702925 ) -- ( 55, 1.36407967977 ) -- ( 60, 1.372174589 ) -- ( 65, 1.42441283667 ) -- ( 70, 1.45614444513 ) -- ( 75, 1.42214292475 ) ( 80, 1.61951229747 ) -- ( 85, 1.5822009781 ) -- ( 90, 1.64785677243 ) -- ( 95, 1.64110731546 ) -- ( 100, 1.66931952084 ) -- ( 105, 1.64698940523 ) -- ( 110, 1.64035255033 ) -- ( 115, 1.67373579449 ) -- ( 120, 1.69722571914 ) -- ( 125, 1.64410473976 ) -- ( 130, 1.67183406497 ) -- ( 135, 1.68970775792 ) -- ( 140, 1.75006550348 ) -- ( 145, 1.72589981327 ) -- ( 150, 1.70345487361 ) -- ( 155, 1.72823556163 ) -- ( 160, 1.70932249861 ) -- ( 165, 1.74166363875 ) -- ( 170, 1.76854372719 ) -- ( 175, 1.78470009443 ) -- ( 180, 1.72592616491 ) -- ( 185, 1.72842221335 ) -- ( 190, 1.73911385175 ) -- ( 195, 1.75547435511 ) -- ( 200, 1.74171721214 );

				\draw[red, shift={( 5, 0.999999908761 )}] \crossTwo;
				\draw[red, shift={( 10, 1.00000003264 )}] \crossTwo;
				\draw[red, shift={( 15, 1.00000001335 )}] \crossTwo;
				\draw[red, shift={( 20, 0.999999994143 )}] \crossTwo;
				\draw[red, shift={( 25, 1.10866693194 )}] \crossTwo;
				\draw[red, shift={( 30, 1.15779806878 )}] \crossTwo;
				\draw[red, shift={( 35, 1.20311569565 )}] \crossTwo;
				\draw[red, shift={( 40, 1.22901843439 )}] \crossTwo;
				\draw[red, shift={( 45, 1.2521595481 )}] \crossTwo;
				\draw[red, shift={( 50, 1.29705237623 )}] \crossTwo;
				\draw[red, shift={( 55, 1.29836890869 )}] \crossTwo;
				\draw[red, shift={( 60, 1.31725150599 )}] \crossTwo;
				\draw[red, shift={( 65, 1.39302759213 )}] \crossTwo;
				\draw[red, shift={( 70, 1.38071782958 )}] \crossTwo;
				\draw[red, shift={( 75, 1.35726326264 )}] \crossTwo;
				\draw[red, shift={( 80, 1.55504855998 )}] \crossTwo;
				\draw[red, shift={( 85, 1.55410873392 )}] \crossTwo;
				\draw[red, shift={( 90, 1.59463518026 )}] \crossTwo;
				\draw[red, shift={( 95, 1.5782809934 )}] \crossTwo;
				\draw[red, shift={( 100, 1.59957213207 )}] \crossTwo;
				\draw[red, shift={( 105, 1.61392741111 )}] \crossTwo;
				\draw[red, shift={( 110, 1.58790144567 )}] \crossTwo;
				\draw[red, shift={( 115, 1.61376377786 )}] \crossTwo;
				\draw[red, shift={( 120, 1.65603685368 )}] \crossTwo;
				\draw[red, shift={( 125, 1.60503805691 )}] \crossTwo;
				\draw[red, shift={( 130, 1.62379059936 )}] \crossTwo;
				\draw[red, shift={( 135, 1.65430995243 )}] \crossTwo;
				\draw[red, shift={( 140, 1.72581738468 )}] \crossTwo;
				\draw[red, shift={( 145, 1.68827717686 )}] \crossTwo;
				\draw[red, shift={( 150, 1.68056866934 )}] \crossTwo;
				\draw[red, shift={( 155, 1.69538249597 )}] \crossTwo;
				\draw[red, shift={( 160, 1.6784893776 )}] \crossTwo;
				\draw[red, shift={( 165, 1.71935429183 )}] \crossTwo;
				\draw[red, shift={( 170, 1.72291235488 )}] \crossTwo;
				\draw[red, shift={( 175, 1.73884359693 )}] \crossTwo;
				\draw[red, shift={( 180, 1.70182784051 )}] \crossTwo;
				\draw[red, shift={( 185, 1.69665674139 )}] \crossTwo;
				\draw[red, shift={( 190, 1.72186536792 )}] \crossTwo;
				\draw[red, shift={( 195, 1.73253005107 )}] \crossTwo;
				\draw[red, shift={( 200, 1.72116832433 )}] \crossTwo;
								
				\draw[red!75, densely dotted]( 5, 0.999999908761 ) -- ( 10, 1.00000003264 ) -- ( 15, 1.00000001335 ) -- ( 20, 0.999999994143 ) -- ( 25, 1.10866693194 ) -- ( 30, 1.15779806878 ) -- ( 35, 1.20311569565 ) -- ( 40, 1.22901843439 ) -- ( 45, 1.2521595481 ) -- ( 50, 1.29705237623 ) -- ( 55, 1.29836890869 ) -- ( 60, 1.31725150599 ) -- ( 65, 1.39302759213 ) -- ( 70, 1.38071782958 ) -- ( 75, 1.35726326264 ) ( 80, 1.55504855998 ) -- ( 85, 1.55410873392 ) -- ( 90, 1.59463518026 ) -- ( 95, 1.5782809934 ) -- ( 100, 1.59957213207 ) -- ( 105, 1.61392741111 ) -- ( 110, 1.58790144567 ) -- ( 115, 1.61376377786 ) -- ( 120, 1.65603685368 ) -- ( 125, 1.60503805691 ) -- ( 130, 1.62379059936 ) -- ( 135, 1.65430995243 ) -- ( 140, 1.72581738468 ) -- ( 145, 1.68827717686 ) -- ( 150, 1.68056866934 ) -- ( 155, 1.69538249597 ) -- ( 160, 1.6784893776 ) -- ( 165, 1.71935429183 ) -- ( 170, 1.72291235488 ) -- ( 175, 1.73884359693 ) -- ( 180, 1.70182784051 ) -- ( 185, 1.69665674139 ) -- ( 190, 1.72186536792 ) -- ( 195, 1.73253005107 ) -- ( 200, 1.72116832433 );

				\draw[blue, shift={( 5, 0.999999908761 )}] \crossThree;
				\draw[blue, shift={( 10, 1.00000003264 )}] \crossThree;
				\draw[blue, shift={( 15, 1.00000001335 )}] \crossThree;
				\draw[blue, shift={( 20, 0.999999994143 )}] \crossThree;
				\draw[blue, shift={( 25, 1.10314965774 )}] \crossThree;
				\draw[blue, shift={( 30, 1.09738999027 )}] \crossThree;
				\draw[blue, shift={( 35, 1.12829486245 )}] \crossThree;
				\draw[blue, shift={( 40, 1.18066153615 )}] \crossThree;
				\draw[blue, shift={( 45, 1.16234301115 )}] \crossThree;
				\draw[blue, shift={( 50, 1.21452430522 )}] \crossThree;
				\draw[blue, shift={( 55, 1.2436662087 )}] \crossThree;
				\draw[blue, shift={( 60, 1.27491708486 )}] \crossThree;
				\draw[blue, shift={( 65, 1.30809842225 )}] \crossThree;
				\draw[blue, shift={( 70, 1.28969615016 )}] \crossThree;
				\draw[blue, shift={( 75, 1.33023793492 )}] \crossThree;
				\draw[blue, shift={( 80, 1.48424158583 )}] \crossThree;
				\draw[blue, shift={( 85, 1.50848815202 )}] \crossThree;
				\draw[blue, shift={( 90, 1.49146484253 )}] \crossThree;
				\draw[blue, shift={( 95, 1.54106763346 )}] \crossThree;
				\draw[blue, shift={( 100, 1.51220970532 )}] \crossThree;
				\draw[blue, shift={( 105, 1.57000596798 )}] \crossThree;
				\draw[blue, shift={( 110, 1.56713937801 )}] \crossThree;
				\draw[blue, shift={( 115, 1.60779663985 )}] \crossThree;
				\draw[blue, shift={( 120, 1.54462678532 )}] \crossThree;
				\draw[blue, shift={( 125, 1.62779899396 )}] \crossThree;
				\draw[blue, shift={( 130, 1.58210356427 )}] \crossThree;
				\draw[blue, shift={( 135, 1.5971705538 )}] \crossThree;
				\draw[blue, shift={( 140, 1.65791941868 )}] \crossThree;
				\draw[blue, shift={( 145, 1.66226687833 )}] \crossThree;
				\draw[blue, shift={( 150, 1.66286151729 )}] \crossThree;
				\draw[blue, shift={( 155, 1.67419343054 )}] \crossThree;
				\draw[blue, shift={( 160, 1.69520726804 )}] \crossThree;
				\draw[blue, shift={( 165, 1.69755024213 )}] \crossThree;
				\draw[blue, shift={( 170, 1.76098855355 )}] \crossThree;
				\draw[blue, shift={( 175, 1.72244463261 )}] \crossThree;
				\draw[blue, shift={( 180, 1.72868872251 )}] \crossThree;
				\draw[blue, shift={( 185, 1.77045922662 )}] \crossThree;
				\draw[blue, shift={( 190, 1.72899680627 )}] \crossThree;
				\draw[blue, shift={( 195, 1.7514014867 )}] \crossThree;
				\draw[blue, shift={( 200, 1.72197936516 )}] \crossThree;
								
				\draw[blue!75, loosely dotted]( 5, 0.999999908761 ) -- ( 10, 1.00000003264 ) -- ( 15, 1.00000001335 ) -- ( 20, 0.999999994143 ) -- ( 25, 1.10314965774 ) -- ( 30, 1.09738999027 ) -- ( 35, 1.12829486245 ) -- ( 40, 1.18066153615 ) -- ( 45, 1.16234301115 ) -- ( 50, 1.21452430522 ) -- ( 55, 1.2436662087 ) -- ( 60, 1.27491708486 ) -- ( 65, 1.30809842225 ) -- ( 70, 1.28969615016 ) -- ( 75, 1.33023793492 ) ( 80, 1.48424158583 ) -- ( 85, 1.50848815202 ) -- ( 90, 1.49146484253 ) -- ( 95, 1.54106763346 ) -- ( 100, 1.51220970532 ) -- ( 105, 1.57000596798 ) -- ( 110, 1.56713937801 ) -- ( 115, 1.60779663985 ) -- ( 120, 1.54462678532 ) -- ( 125, 1.62779899396 ) -- ( 130, 1.58210356427 ) -- ( 135, 1.5971705538 ) -- ( 140, 1.65791941868 ) -- ( 145, 1.66226687833 ) -- ( 150, 1.66286151729 ) -- ( 155, 1.67419343054 ) -- ( 160, 1.69520726804 ) -- ( 165, 1.69755024213 ) -- ( 170, 1.76098855355 ) -- ( 175, 1.72244463261 ) -- ( 180, 1.72868872251 ) -- ( 185, 1.77045922662 ) -- ( 190, 1.72899680627 ) -- ( 195, 1.7514014867 ) -- ( 200, 1.72197936516 );




			\end{scope}
			
			\def\axisAdditionalLengthPlusTikzX{\axisAdditionalLengthPlus / \xscaleTikz}
			\def\axisAdditionalLengthMinusTikzX{\axisAdditionalLengthMinus / \xscaleTikz}
			\draw[arrow] (-\axisAdditionalLengthMinusTikzX, 0) -- (200, 0)  -- +(\axisAdditionalLengthPlusTikzX, 0) node[right] {\xAxis};
			\def\xTotalLengthPlus{200+\axisAdditionalLengthPlusTikzX}
			\draw (\xTotalLengthPlus, 0) node[right] {$\qquad$};
			\def\axisAdditionalLengthPlusTikzY{\axisAdditionalLengthPlus / \yscaleTikz}
			\def\axisAdditionalLengthMinusTikzY{\axisAdditionalLengthMinus / \yscaleTikz}
			\draw[arrow] (0, -\axisAdditionalLengthMinusTikzY) -- (0, 0.9 ) -- +(0, \axisAdditionalLengthPlusTikzY) node[above, yshift=-0.15cm] {\yAxis};
			
			\def\axisLabelTikzY{\axisLabel / \yscaleTikz}
			\foreach \pos in {0, 20, 40, 60, 80, 100, 120, 140, 160, 180, 200} \draw[shift={(\pos, 0)}] (0, \axisLabelTikzY) -- (0, -\axisLabelTikzY) node[below] {$\pos$};
			
			\def\axisLabelTikzX{\axisLabel / \xscaleTikz}
			\begin{scope}[shift={(0, -1)}]
				\draw[shift={(0,1 )}] (\axisLabelTikzX, 0) -- (-\axisLabelTikzX, 0) node[left] {$1$};
				\draw[shift={(0,1.3 )}] (\axisLabelTikzX, 0) -- (-\axisLabelTikzX, 0) node[left] {$1.3$};
				\draw[shift={(0,1.6 )}] (\axisLabelTikzX, 0) -- (-\axisLabelTikzX, 0) node[left] {$1.6$};
				\draw[shift={(0,1.9 )}] (\axisLabelTikzX, 0) -- (-\axisLabelTikzX, 0) node[left] {$1.9$};
			\end{scope}
		\end{tikzpicture}
		\AngleTSPInstances
		\vspace*{-0.1cm}
		\caption[heuristics based on convex hulls: \CH{}, \CHC{}, \CHCL{}]{
			heuristics based on convex hulls:\hspace*{-8cm}\\
			\begin{tikzpicture}[xscale=\xscale, yscale=2.0]
				\pgfgettransformentries{\xscaleTikz}{\@tempa}{\@tempa}{\yscaleTikz}{\@tempa}{\@tempa}
				\def\crossSizeX{\crossSize / \xscaleTikz};
				\def\crossSizeY{\crossSize / \yscaleTikz};
				
				\def\crossOne{(-\crossSizeX,-\crossSizeY) -- (\crossSizeX,\crossSizeY) (-\crossSizeX,\crossSizeY) -- (\crossSizeX,-\crossSizeY)};
				\def\crossTwo{(-\crossSizeX,0) -- (\crossSizeX,0) (0,\crossSizeY) -- (0,-\crossSizeY)};
				\def\crossThree{(0,0) -- (0,\crossSizeY) (0,0) -- (\crossSizeX,-\crossSizeY) (0,0) -- (-\crossSizeX,-\crossSizeY)};
				\def\crossFour{(0,0) -- (0,-\crossSizeY) (0,0) -- (\crossSizeX,\crossSizeY) (0,0) -- (-\crossSizeX,\crossSizeY)};
				\def\crossFive{(0,0) -- (\crossSizeX,0) (0,0) -- (-\crossSizeX,\crossSizeY) (0,0) -- (-\crossSizeX,-\crossSizeY)};
				\def\crossSix{(0,0) -- (-\crossSizeX,0) (0,0) -- (\crossSizeX,\crossSizeY) (0,0) -- (\crossSizeX,-\crossSizeY)};
				
				\draw[black, shift={(0, 0)}] \crossOne;
				\draw[black, shift={(5, 0)}] \crossOne;
				\draw[black, shift={(10, 0)}] \crossOne;
				\draw[black, shift={(15, 0)}] \crossOne;
				\draw[black!75, dotted] (-2.5, 0) -- (17.5, 0);
				\node at (40, 0) [minimum width=3cm] {\parbox{1.1cm}\CH{}};

				\draw[red, shift={(80, 0)}] \crossTwo;
				\draw[red, shift={(85, 0)}] \crossTwo;
				\draw[red, shift={(90, 0)}] \crossTwo;
				\draw[red, shift={(95, 0)}] \crossTwo;
				\draw[red!75, densely dotted] (77.5, 0) -- (97.5, 0);
				\node at (120, 0) [minimum width=3cm] {\parbox{1.1cm}\CHC{}};
				
				\draw[blue, shift={(160, 0)}] \crossThree;
				\draw[blue, shift={(165, 0)}] \crossThree;
				\draw[blue, shift={(170, 0)}] \crossThree;
				\draw[blue, shift={(175, 0)}] \crossThree;
				\draw[blue!75, loosely dotted] (157.5, 0) -- (177.5, 0);
				\node at (200, 0) [minimum width=3cm] {\parbox{1.1cm}\CHCL{}};
			\end{tikzpicture}
		}
		\label{figure:HeuristicsBasedOnConvexHulls}
	\end{figure}
			\end{comment:figures}
			
			
			Figure~\ref{figure:HeuristicsBasedOnConvexHulls} sums up results for \CH{}, \CHC{} and \CHCL{}, which were tested only for the \AngleTSPInstances. Again, after making preliminary tests, we fixed the lens thickness parameter $\gamma = 40\degree\ (\approx 0.6981)$ for \CHCL{}. The choice of parameter $C$ for \CHC{} and \CHCL{} always entails a trade-off: for larger values of $C$, we obviously obtain better objective function values but larger computation times. 
			Our preliminary tests have shown that there are relatively large objective function value differences between $C=2$, $C=10$ and $C=20$; the step to $C=30$, however, only slightly improves the objective function values while the running times significantly increase. Thus we set $C=20$ for all our further tests.

			\begin{comment:figures}
	\begin{figure}[htb!]
		\centering
			\begin{tikzpicture}[xscale=\xscale, yscale=\yscaleMiddle]
				\pgfgettransformentries{\xscaleTikz}{\@tempa}{\@tempa}{\yscaleTikz}{\@tempa}{\@tempa}
			
				\draw[very thin, color=gray, xstep=20, ystep=0.3] (0, 0) grid (200, 0.9 );
				
				\begin{scope}[shift={(0, -1)}]
					\def\crossSizeX{\crossSize / \xscaleTikz};
					\def\crossSizeY{\crossSize / \yscaleTikz};
					
					\def\crossOne{(-\crossSizeX,-\crossSizeY) -- (\crossSizeX,\crossSizeY) (-\crossSizeX,\crossSizeY) -- (\crossSizeX,-\crossSizeY)};
					\def\crossTwo{(-\crossSizeX,0) -- (\crossSizeX,0) (0,\crossSizeY) -- (0,-\crossSizeY)};
					\def\crossThree{(0,0) -- (0,\crossSizeY) (0,0) -- (\crossSizeX,-\crossSizeY) (0,0) -- (-\crossSizeX,-\crossSizeY)};
					\def\crossFour{(0,0) -- (0,-\crossSizeY) (0,0) -- (\crossSizeX,\crossSizeY) (0,0) -- (-\crossSizeX,\crossSizeY)};
					\def\crossFive{(0,0) -- (\crossSizeX,0) (0,0) -- (-\crossSizeX,\crossSizeY) (0,0) -- (-\crossSizeX,-\crossSizeY)};
					\def\crossSix{(0,0) -- (-\crossSizeX,0) (0,0) -- (\crossSizeX,\crossSizeY) (0,0) -- (\crossSizeX,-\crossSizeY)};
					
					\draw[black, shift={( 5, 0.999999908761 )}] \crossOne;
					\draw[black, shift={( 10, 1.07315624855 )}] \crossOne;
					\draw[black, shift={( 15, 1.06377418611 )}] \crossOne;
					\draw[black, shift={( 20, 1.09479164244 )}] \crossOne;
					\draw[black, shift={( 25, 1.09135857196 )}] \crossOne;
					\draw[black, shift={( 30, 1.11083836395 )}] \crossOne;
					\draw[black, shift={( 35, 1.14004142239 )}] \crossOne;
					\draw[black, shift={( 40, 1.14780059055 )}] \crossOne;
					\draw[black, shift={( 45, 1.16345643091 )}] \crossOne;
					\draw[black, shift={( 50, 1.20690890278 )}] \crossOne;
					\draw[black, shift={( 55, 1.17538174481 )}] \crossOne;
					\draw[black, shift={( 60, 1.21953316992 )}] \crossOne;
					\draw[black, shift={( 65, 1.19762322498 )}] \crossOne;
					\draw[black, shift={( 70, 1.19898200766 )}] \crossOne;
					\draw[black, shift={( 75, 1.2259327387 )}] \crossOne;
					\draw[black, shift={( 80, 1.38471643165 )}] \crossOne;
					\draw[black, shift={( 85, 1.36220239938 )}] \crossOne;
					\draw[black, shift={( 90, 1.37166750068 )}] \crossOne;
					\draw[black, shift={( 95, 1.39707554132 )}] \crossOne;
					\draw[black, shift={( 100, 1.45482533162 )}] \crossOne;
					\draw[black, shift={( 105, 1.45632277458 )}] \crossOne;
					\draw[black, shift={( 110, 1.44812640726 )}] \crossOne;
					\draw[black, shift={( 115, 1.4314196322 )}] \crossOne;
					\draw[black, shift={( 120, 1.48785956689 )}] \crossOne;
					\draw[black, shift={( 125, 1.46063191257 )}] \crossOne;
					\draw[black, shift={( 130, 1.52410355022 )}] \crossOne;
					\draw[black, shift={( 135, 1.5250054176 )}] \crossOne;
					\draw[black, shift={( 140, 1.52283672987 )}] \crossOne;
					\draw[black, shift={( 145, 1.5384805059 )}] \crossOne;
					\draw[black, shift={( 150, 1.5754997671 )}] \crossOne;
					\draw[black, shift={( 155, 1.51667668374 )}] \crossOne;
					\draw[black, shift={( 160, 1.5484863103 )}] \crossOne;
					\draw[black, shift={( 165, 1.53958405389 )}] \crossOne;
					\draw[black, shift={( 170, 1.5553527341 )}] \crossOne;
					\draw[black, shift={( 175, 1.58869261472 )}] \crossOne;
					\draw[black, shift={( 180, 1.59657225257 )}] \crossOne;
					\draw[black, shift={( 185, 1.620865054 )}] \crossOne;
					\draw[black, shift={( 190, 1.62135141354 )}] \crossOne;
					\draw[black, shift={( 195, 1.6079222616 )}] \crossOne;
					\draw[black, shift={( 200, 1.64383947017 )}] \crossOne;
					
					\draw[black!75, dotted] ( 5, 0.999999908761 ) -- ( 10, 1.07315624855 ) -- ( 15, 1.06377418611 ) -- ( 20, 1.09479164244 ) -- ( 25, 1.09135857196 ) -- ( 30, 1.11083836395 ) -- ( 35, 1.14004142239 ) -- ( 40, 1.14780059055 ) -- ( 45, 1.16345643091 ) -- ( 50, 1.20690890278 ) -- ( 55, 1.17538174481 ) -- ( 60, 1.21953316992 ) -- ( 65, 1.19762322498 ) -- ( 70, 1.19898200766 ) -- ( 75, 1.2259327387 ) ( 80, 1.38471643165 ) -- ( 85, 1.36220239938 ) -- ( 90, 1.37166750068 ) -- ( 95, 1.39707554132 ) -- ( 100, 1.45482533162 ) -- ( 105, 1.45632277458 ) -- ( 110, 1.44812640726 ) -- ( 115, 1.4314196322 ) -- ( 120, 1.48785956689 ) -- ( 125, 1.46063191257 ) -- ( 130, 1.52410355022 ) -- ( 135, 1.5250054176 ) -- ( 140, 1.52283672987 ) -- ( 145, 1.5384805059 ) -- ( 150, 1.5754997671 ) -- ( 155, 1.51667668374 ) -- ( 160, 1.5484863103 ) -- ( 165, 1.53958405389 ) -- ( 170, 1.5553527341 ) -- ( 175, 1.58869261472 ) -- ( 180, 1.59657225257 ) -- ( 185, 1.620865054 ) -- ( 190, 1.62135141354 ) -- ( 195, 1.6079222616 ) -- ( 200, 1.64383947017 );

					\draw[red, shift={( 5, 0.999999908761 )}] \crossTwo;
					\draw[red, shift={( 10, 1.0840622574 )}] \crossTwo;
					\draw[red, shift={( 15, 1.0602328131 )}] \crossTwo;
					\draw[red, shift={( 20, 1.05583576999 )}] \crossTwo;
					\draw[red, shift={( 25, 1.10268091273 )}] \crossTwo;
					\draw[red, shift={( 30, 1.08640058202 )}] \crossTwo;
					\draw[red, shift={( 35, 1.1395596618 )}] \crossTwo;
					\draw[red, shift={( 40, 1.12048030411 )}] \crossTwo;
					\draw[red, shift={( 45, 1.12565295241 )}] \crossTwo;
					\draw[red, shift={( 50, 1.13396965695 )}] \crossTwo;
					\draw[red, shift={( 55, 1.10427579677 )}] \crossTwo;
					\draw[red, shift={( 60, 1.10022896893 )}] \crossTwo;
					\draw[red, shift={( 65, 1.1242943644 )}] \crossTwo;
					\draw[red, shift={( 70, 1.13012155224 )}] \crossTwo;
					\draw[red, shift={( 75, 1.12573532165 )}] \crossTwo;
					\draw[red, shift={( 80, 1.24911570955 )}] \crossTwo;
					\draw[red, shift={( 85, 1.22737098691 )}] \crossTwo;
					\draw[red, shift={( 90, 1.23289518389 )}] \crossTwo;
					\draw[red, shift={( 95, 1.26751368563 )}] \crossTwo;
					\draw[red, shift={( 100, 1.2584498721 )}] \crossTwo;
					\draw[red, shift={( 105, 1.23655779339 )}] \crossTwo;
					\draw[red, shift={( 110, 1.23187118062 )}] \crossTwo;
					\draw[red, shift={( 115, 1.26486348605 )}] \crossTwo;
					\draw[red, shift={( 120, 1.28236238547 )}] \crossTwo;
					\draw[red, shift={( 125, 1.24074339799 )}] \crossTwo;
					\draw[red, shift={( 130, 1.24807996328 )}] \crossTwo;
					\draw[red, shift={( 135, 1.26434586252 )}] \crossTwo;
					\draw[red, shift={( 140, 1.28082894178 )}] \crossTwo;
					\draw[red, shift={( 145, 1.25628614279 )}] \crossTwo;
					\draw[red, shift={( 150, 1.26384726539 )}] \crossTwo;
					\draw[red, shift={( 155, 1.2899283107 )}] \crossTwo;
					\draw[red, shift={( 160, 1.25185943171 )}] \crossTwo;
					\draw[red, shift={( 165, 1.28905925542 )}] \crossTwo;
					\draw[red, shift={( 170, 1.24749150955 )}] \crossTwo;
					\draw[red, shift={( 175, 1.24744819376 )}] \crossTwo;
					\draw[red, shift={( 180, 1.25697065149 )}] \crossTwo;
					\draw[red, shift={( 185, 1.26772713732 )}] \crossTwo;
					\draw[red, shift={( 190, 1.29514094466 )}] \crossTwo;
					\draw[red, shift={( 195, 1.30298815406 )}] \crossTwo;
					\draw[red, shift={( 200, 1.28034457849 )}] \crossTwo;
									
					\draw[red!75, densely dotted]( 5, 0.999999908761 ) -- ( 10, 1.0840622574 ) -- ( 15, 1.0602328131 ) -- ( 20, 1.05583576999 ) -- ( 25, 1.10268091273 ) -- ( 30, 1.08640058202 ) -- ( 35, 1.1395596618 ) -- ( 40, 1.12048030411 ) -- ( 45, 1.12565295241 ) -- ( 50, 1.13396965695 ) -- ( 55, 1.10427579677 ) -- ( 60, 1.10022896893 ) -- ( 65, 1.1242943644 ) -- ( 70, 1.13012155224 ) -- ( 75, 1.12573532165 ) ( 80, 1.24911570955 ) -- ( 85, 1.22737098691 ) -- ( 90, 1.23289518389 ) -- ( 95, 1.26751368563 ) -- ( 100, 1.2584498721 ) -- ( 105, 1.23655779339 ) -- ( 110, 1.23187118062 ) -- ( 115, 1.26486348605 ) -- ( 120, 1.28236238547 ) -- ( 125, 1.24074339799 ) -- ( 130, 1.24807996328 ) -- ( 135, 1.26434586252 ) -- ( 140, 1.28082894178 ) -- ( 145, 1.25628614279 ) -- ( 150, 1.26384726539 ) -- ( 155, 1.2899283107 ) -- ( 160, 1.25185943171 ) -- ( 165, 1.28905925542 ) -- ( 170, 1.24749150955 ) -- ( 175, 1.24744819376 ) -- ( 180, 1.25697065149 ) -- ( 185, 1.26772713732 ) -- ( 190, 1.29514094466 ) -- ( 195, 1.30298815406 ) -- ( 200, 1.28034457849 );

					\draw[blue, shift={( 5, 0.999999908761 )}] \crossThree;
					\draw[blue, shift={( 10, 1.07941551675 )}] \crossThree;
					\draw[blue, shift={( 15, 1.054668787 )}] \crossThree;
					\draw[blue, shift={( 20, 1.05851034214 )}] \crossThree;
					\draw[blue, shift={( 25, 1.05482126858 )}] \crossThree;
					\draw[blue, shift={( 30, 1.08011399486 )}] \crossThree;
					\draw[blue, shift={( 35, 1.09051946806 )}] \crossThree;
					\draw[blue, shift={( 40, 1.09643469797 )}] \crossThree;
					\draw[blue, shift={( 45, 1.09845939601 )}] \crossThree;
					\draw[blue, shift={( 50, 1.12061933957 )}] \crossThree;
					\draw[blue, shift={( 55, 1.08813492464 )}] \crossThree;
					\draw[blue, shift={( 60, 1.08970152829 )}] \crossThree;
					\draw[blue, shift={( 65, 1.09321172064 )}] \crossThree;
					\draw[blue, shift={( 70, 1.12687935504 )}] \crossThree;
					\draw[blue, shift={( 75, 1.10316301465 )}] \crossThree;
					\draw[blue, shift={( 80, 1.22297116854 )}] \crossThree;
					\draw[blue, shift={( 85, 1.22883224294 )}] \crossThree;
					\draw[blue, shift={( 90, 1.23504273997 )}] \crossThree;
					\draw[blue, shift={( 95, 1.24468922779 )}] \crossThree;
					\draw[blue, shift={( 100, 1.25148609454 )}] \crossThree;
					\draw[blue, shift={( 105, 1.22598137777 )}] \crossThree;
					\draw[blue, shift={( 110, 1.22940907017 )}] \crossThree;
					\draw[blue, shift={( 115, 1.24747254749 )}] \crossThree;
					\draw[blue, shift={( 120, 1.22679297102 )}] \crossThree;
					\draw[blue, shift={( 125, 1.21659590156 )}] \crossThree;
					\draw[blue, shift={( 130, 1.21766853385 )}] \crossThree;
					\draw[blue, shift={( 135, 1.25260883446 )}] \crossThree;
					\draw[blue, shift={( 140, 1.25368956022 )}] \crossThree;
					\draw[blue, shift={( 145, 1.24543093713 )}] \crossThree;
					\draw[blue, shift={( 150, 1.23934058866 )}] \crossThree;
					\draw[blue, shift={( 155, 1.23102857472 )}] \crossThree;
					\draw[blue, shift={( 160, 1.25694525677 )}] \crossThree;
					\draw[blue, shift={( 165, 1.2571923187 )}] \crossThree;
					\draw[blue, shift={( 170, 1.24154251296 )}] \crossThree;
					\draw[blue, shift={( 175, 1.23704176609 )}] \crossThree;
					\draw[blue, shift={( 180, 1.23551992264 )}] \crossThree;
					\draw[blue, shift={( 185, 1.24437342033 )}] \crossThree;
					\draw[blue, shift={( 190, 1.23886939437 )}] \crossThree;
					\draw[blue, shift={( 195, 1.27991152912 )}] \crossThree;
					\draw[blue, shift={( 200, 1.24383424319 )}] \crossThree;
									
					\draw[blue!75, loosely dotted]( 5, 0.999999908761 ) -- ( 10, 1.07941551675 ) -- ( 15, 1.054668787 ) -- ( 20, 1.05851034214 ) -- ( 25, 1.05482126858 ) -- ( 30, 1.08011399486 ) -- ( 35, 1.09051946806 ) -- ( 40, 1.09643469797 ) -- ( 45, 1.09845939601 ) -- ( 50, 1.12061933957 ) -- ( 55, 1.08813492464 ) -- ( 60, 1.08970152829 ) -- ( 65, 1.09321172064 ) -- ( 70, 1.12687935504 ) -- ( 75, 1.10316301465 ) ( 80, 1.22297116854 ) -- ( 85, 1.22883224294 ) -- ( 90, 1.23504273997 ) -- ( 95, 1.24468922779 ) -- ( 100, 1.25148609454 ) -- ( 105, 1.22598137777 ) -- ( 110, 1.22940907017 ) -- ( 115, 1.24747254749 ) -- ( 120, 1.22679297102 ) -- ( 125, 1.21659590156 ) -- ( 130, 1.21766853385 ) -- ( 135, 1.25260883446 ) -- ( 140, 1.25368956022 ) -- ( 145, 1.24543093713 ) -- ( 150, 1.23934058866 ) -- ( 155, 1.23102857472 ) -- ( 160, 1.25694525677 ) -- ( 165, 1.2571923187 ) -- ( 170, 1.24154251296 ) -- ( 175, 1.23704176609 ) -- ( 180, 1.23551992264 ) -- ( 185, 1.24437342033 ) -- ( 190, 1.23886939437 ) -- ( 195, 1.27991152912 ) -- ( 200, 1.24383424319 );

					\draw[magenta, shift={( 5, 0.999999908761 )}] \crossFour;
					\draw[magenta, shift={( 10, 1.07941551675 )}] \crossFour;
					\draw[magenta, shift={( 15, 1.054668787 )}] \crossFour;
					\draw[magenta, shift={( 20, 1.05851034214 )}] \crossFour;
					\draw[magenta, shift={( 25, 1.0513043869 )}] \crossFour;
					\draw[magenta, shift={( 30, 1.0771571731 )}] \crossFour;
					\draw[magenta, shift={( 35, 1.08162222356 )}] \crossFour;
					\draw[magenta, shift={( 40, 1.09744870968 )}] \crossFour;
					\draw[magenta, shift={( 45, 1.09228634084 )}] \crossFour;
					\draw[magenta, shift={( 50, 1.12061933957 )}] \crossFour;
					\draw[magenta, shift={( 55, 1.08748371008 )}] \crossFour;
					\draw[magenta, shift={( 60, 1.08397928398 )}] \crossFour;
					\draw[magenta, shift={( 65, 1.09837426566 )}] \crossFour;
					\draw[magenta, shift={( 70, 1.12280570636 )}] \crossFour;
					\draw[magenta, shift={( 75, 1.1002486545 )}] \crossFour;
					\draw[magenta, shift={( 80, 1.21806683419 )}] \crossFour;
					\draw[magenta, shift={( 85, 1.22683947592 )}] \crossFour;
					\draw[magenta, shift={( 90, 1.23247831073 )}] \crossFour;
					\draw[magenta, shift={( 95, 1.244231978 )}] \crossFour;
					\draw[magenta, shift={( 100, 1.25201170136 )}] \crossFour;
					\draw[magenta, shift={( 105, 1.22578972065 )}] \crossFour;
					\draw[magenta, shift={( 110, 1.22862377684 )}] \crossFour;
					\draw[magenta, shift={( 115, 1.2380757942 )}] \crossFour;
					\draw[magenta, shift={( 120, 1.22645508594 )}] \crossFour;
					\draw[magenta, shift={( 125, 1.21540256452 )}] \crossFour;
					\draw[magenta, shift={( 130, 1.21766853385 )}] \crossFour;
					\draw[magenta, shift={( 135, 1.24898991894 )}] \crossFour;
					\draw[magenta, shift={( 140, 1.24967554585 )}] \crossFour;
					\draw[magenta, shift={( 145, 1.24436410079 )}] \crossFour;
					\draw[magenta, shift={( 150, 1.23715331441 )}] \crossFour;
					\draw[magenta, shift={( 155, 1.22826838516 )}] \crossFour;
					\draw[magenta, shift={( 160, 1.25711740161 )}] \crossFour;
					\draw[magenta, shift={( 165, 1.25235007488 )}] \crossFour;
					\draw[magenta, shift={( 170, 1.24455398981 )}] \crossFour;
					\draw[magenta, shift={( 175, 1.23409672882 )}] \crossFour;
					\draw[magenta, shift={( 180, 1.23090988036 )}] \crossFour;
					\draw[magenta, shift={( 185, 1.24197708506 )}] \crossFour;
					\draw[magenta, shift={( 190, 1.23558088696 )}] \crossFour;
					\draw[magenta, shift={( 195, 1.27941662751 )}] \crossFour;
					\draw[magenta, shift={( 200, 1.24301197132 )}] \crossFour;
									
					\draw[magenta!75, dash pattern={on 1pt off 1pt}]( 5, 0.999999908761 ) -- ( 10, 1.07941551675 ) -- ( 15, 1.054668787 ) -- ( 20, 1.05851034214 ) -- ( 25, 1.0513043869 ) -- ( 30, 1.0771571731 ) -- ( 35, 1.08162222356 ) -- ( 40, 1.09744870968 ) -- ( 45, 1.09228634084 ) -- ( 50, 1.12061933957 ) -- ( 55, 1.08748371008 ) -- ( 60, 1.08397928398 ) -- ( 65, 1.09837426566 ) -- ( 70, 1.12280570636 ) -- ( 75, 1.1002486545 ) ( 80, 1.21806683419 ) -- ( 85, 1.22683947592 ) -- ( 90, 1.23247831073 ) -- ( 95, 1.244231978 ) -- ( 100, 1.25201170136 ) -- ( 105, 1.22578972065 ) -- ( 110, 1.22862377684 ) -- ( 115, 1.2380757942 ) -- ( 120, 1.22645508594 ) -- ( 125, 1.21540256452 ) -- ( 130, 1.21766853385 ) -- ( 135, 1.24898991894 ) -- ( 140, 1.24967554585 ) -- ( 145, 1.24436410079 ) -- ( 150, 1.23715331441 ) -- ( 155, 1.22826838516 ) -- ( 160, 1.25711740161 ) -- ( 165, 1.25235007488 ) -- ( 170, 1.24455398981 ) -- ( 175, 1.23409672882 ) -- ( 180, 1.23090988036 ) -- ( 185, 1.24197708506 ) -- ( 190, 1.23558088696 ) -- ( 195, 1.27941662751 ) -- ( 200, 1.24301197132 );



				\end{scope}
				
				\def\axisAdditionalLengthPlusTikzX{\axisAdditionalLengthPlus / \xscaleTikz}
				\def\axisAdditionalLengthMinusTikzX{\axisAdditionalLengthMinus / \xscaleTikz}
				\draw[arrow] (-\axisAdditionalLengthMinusTikzX, 0) -- (200, 0)  -- +(\axisAdditionalLengthPlusTikzX, 0) node[right] {\xAxis};
				\def\xTotalLengthPlus{200+\axisAdditionalLengthPlusTikzX}
				\draw (\xTotalLengthPlus, 0) node[right] {$\qquad$};
				\def\axisAdditionalLengthPlusTikzY{\axisAdditionalLengthPlus / \yscaleTikz}
				\def\axisAdditionalLengthMinusTikzY{\axisAdditionalLengthMinus / \yscaleTikz}
				\draw[arrow] (0, -\axisAdditionalLengthMinusTikzY) -- (0, 0.9 ) -- +(0, \axisAdditionalLengthPlusTikzY) node[above, yshift=-0.15cm] {\yAxis};
				
				\def\axisLabelTikzY{\axisLabel / \yscaleTikz}
				\foreach \pos in {0, 20, 40, 60, 80, 100, 120, 140, 160, 180, 200} \draw[shift={(\pos, 0)}] (0, \axisLabelTikzY) -- (0, -\axisLabelTikzY) node[below] {$\pos$};
				
				\def\axisLabelTikzX{\axisLabel / \xscaleTikz}
				\begin{scope}[shift={(0, -1)}]
					\draw[shift={(0,1 )}] (\axisLabelTikzX, 0) -- (-\axisLabelTikzX, 0) node[left] {$1$};
					\draw[shift={(0,1.3 )}] (\axisLabelTikzX, 0) -- (-\axisLabelTikzX, 0) node[left] {$1.3$};
					\draw[shift={(0,1.6 )}] (\axisLabelTikzX, 0) -- (-\axisLabelTikzX, 0) node[left] {$1.6$};
					\draw[shift={(0,1.9 )}] (\axisLabelTikzX, 0) -- (-\axisLabelTikzX, 0) node[left] {$1.9$};
				\end{scope}
			\end{tikzpicture}
		\AngleTSPInstances
		\vspace{-0.6cm}
		
			\begin{tikzpicture}[xscale=\xscale, yscale=\yscaleSmall]
				\pgfgettransformentries{\xscaleTikz}{\@tempa}{\@tempa}{\yscaleTikz}{\@tempa}{\@tempa}
			
				\draw[very thin, color=gray, xstep=20, ystep=0.1] (0, 0) grid (200, 0.3 );
				
				\begin{scope}[shift={(0, -1)}]
					\def\crossSizeX{\crossSize / \xscaleTikz};
					\def\crossSizeY{\crossSize / \yscaleTikz};
					
					\def\crossOne{(-\crossSizeX,-\crossSizeY) -- (\crossSizeX,\crossSizeY) (-\crossSizeX,\crossSizeY) -- (\crossSizeX,-\crossSizeY)};
					\def\crossTwo{(-\crossSizeX,0) -- (\crossSizeX,0) (0,\crossSizeY) -- (0,-\crossSizeY)};
					\def\crossThree{(0,0) -- (0,\crossSizeY) (0,0) -- (\crossSizeX,-\crossSizeY) (0,0) -- (-\crossSizeX,-\crossSizeY)};
					\def\crossFour{(0,0) -- (0,-\crossSizeY) (0,0) -- (\crossSizeX,\crossSizeY) (0,0) -- (-\crossSizeX,\crossSizeY)};
					\def\crossFive{(0,0) -- (\crossSizeX,0) (0,0) -- (-\crossSizeX,\crossSizeY) (0,0) -- (-\crossSizeX,-\crossSizeY)};
					\def\crossSix{(0,0) -- (-\crossSizeX,0) (0,0) -- (\crossSizeX,\crossSizeY) (0,0) -- (\crossSizeX,-\crossSizeY)};
					
					\draw[black, shift={( 5, 0.999999996966 )}] \crossOne;
					\draw[black, shift={( 10, 1.02306355979 )}] \crossOne;
					\draw[black, shift={( 15, 1.00882743483 )}] \crossOne;
					\draw[black, shift={( 20, 1.03616926105 )}] \crossOne;
					\draw[black, shift={( 25, 1.09121859036 )}] \crossOne;
					\draw[black, shift={( 30, 1.06654911596 )}] \crossOne;
					\draw[black, shift={( 35, 1.05265872403 )}] \crossOne;
					\draw[black, shift={( 40, 1.05894015913 )}] \crossOne;
					\draw[black, shift={( 45, 1.03601477263 )}] \crossOne;
					\draw[black, shift={( 50, 1.06675925061 )}] \crossOne;
					\draw[black, shift={( 55, 1.05089914851 )}] \crossOne;
					\draw[black, shift={( 60, 1.05850009601 )}] \crossOne;
					\draw[black, shift={( 65, 1.05710214749 )}] \crossOne;
					\draw[black, shift={( 70, 1.05744937546 )}] \crossOne;
					\draw[black, shift={( 75, 1.04958900944 )}] \crossOne;
					\draw[black, shift={( 80, 1.04915946965 )}] \crossOne;
					\draw[black, shift={( 85, 1.05879358052 )}] \crossOne;
					\draw[black, shift={( 90, 1.05918878172 )}] \crossOne;
					\draw[black, shift={( 95, 1.06152098391 )}] \crossOne;
					\draw[black, shift={( 100, 1.05353569159 )}] \crossOne;
					\draw[black, shift={( 105, 1.11982864903 )}] \crossOne;
					\draw[black, shift={( 110, 1.12703993066 )}] \crossOne;
					\draw[black, shift={( 115, 1.12600956285 )}] \crossOne;
					\draw[black, shift={( 120, 1.13210532725 )}] \crossOne;
					\draw[black, shift={( 125, 1.13515083769 )}] \crossOne;
					\draw[black, shift={( 130, 1.14817110726 )}] \crossOne;
					\draw[black, shift={( 135, 1.15036332178 )}] \crossOne;
					\draw[black, shift={( 140, 1.14583821704 )}] \crossOne;
					\draw[black, shift={( 145, 1.13952383541 )}] \crossOne;
					\draw[black, shift={( 150, 1.14587967102 )}] \crossOne;
					\draw[black, shift={( 155, 1.15913220668 )}] \crossOne;
					\draw[black, shift={( 160, 1.15553999629 )}] \crossOne;
					\draw[black, shift={( 165, 1.15082022932 )}] \crossOne;
					\draw[black, shift={( 170, 1.15344277265 )}] \crossOne;
					\draw[black, shift={( 175, 1.17263786879 )}] \crossOne;
					\draw[black, shift={( 180, 1.16603962566 )}] \crossOne;
					\draw[black, shift={( 185, 1.16326483212 )}] \crossOne;
					\draw[black, shift={( 190, 1.16991669245 )}] \crossOne;
					\draw[black, shift={( 195, 1.16706282817 )}] \crossOne;
					\draw[black, shift={( 200, 1.17484712931 )}] \crossOne;
					
					\draw[black!75, dotted] ( 5, 0.999999996966 ) -- ( 10, 1.02306355979 ) -- ( 15, 1.00882743483 ) -- ( 20, 1.03616926105 ) -- ( 25, 1.09121859036 ) -- ( 30, 1.06654911596 ) -- ( 35, 1.05265872403 ) -- ( 40, 1.05894015913 ) -- ( 45, 1.03601477263 ) -- ( 50, 1.06675925061 ) -- ( 55, 1.05089914851 ) -- ( 60, 1.05850009601 ) -- ( 65, 1.05710214749 ) -- ( 70, 1.05744937546 ) -- ( 75, 1.04958900944 ) -- ( 80, 1.04915946965 ) -- ( 85, 1.05879358052 ) -- ( 90, 1.05918878172 ) -- ( 95, 1.06152098391 ) -- ( 100, 1.05353569159 ) ( 105, 1.11982864903 ) -- ( 110, 1.12703993066 ) -- ( 115, 1.12600956285 ) -- ( 120, 1.13210532725 ) -- ( 125, 1.13515083769 ) -- ( 130, 1.14817110726 ) -- ( 135, 1.15036332178 ) -- ( 140, 1.14583821704 ) -- ( 145, 1.13952383541 ) -- ( 150, 1.14587967102 ) -- ( 155, 1.15913220668 ) -- ( 160, 1.15553999629 ) -- ( 165, 1.15082022932 ) -- ( 170, 1.15344277265 ) -- ( 175, 1.17263786879 ) -- ( 180, 1.16603962566 ) -- ( 185, 1.16326483212 ) -- ( 190, 1.16991669245 ) -- ( 195, 1.16706282817 ) -- ( 200, 1.17484712931 );

					\draw[red, shift={( 5, 0.999999996966 )}] \crossTwo;
					\draw[red, shift={( 10, 1.02306355979 )}] \crossTwo;
					\draw[red, shift={( 15, 1.00882743483 )}] \crossTwo;
					\draw[red, shift={( 20, 1.03616926105 )}] \crossTwo;
					\draw[red, shift={( 25, 1.08702629776 )}] \crossTwo;
					\draw[red, shift={( 30, 1.07501278798 )}] \crossTwo;
					\draw[red, shift={( 35, 1.0554180775 )}] \crossTwo;
					\draw[red, shift={( 40, 1.06993011307 )}] \crossTwo;
					\draw[red, shift={( 45, 1.04106948871 )}] \crossTwo;
					\draw[red, shift={( 50, 1.07932240847 )}] \crossTwo;
					\draw[red, shift={( 55, 1.05888875794 )}] \crossTwo;
					\draw[red, shift={( 60, 1.07855314717 )}] \crossTwo;
					\draw[red, shift={( 65, 1.06982449125 )}] \crossTwo;
					\draw[red, shift={( 70, 1.07129844548 )}] \crossTwo;
					\draw[red, shift={( 75, 1.07333947593 )}] \crossTwo;
					\draw[red, shift={( 80, 1.0732839007 )}] \crossTwo;
					\draw[red, shift={( 85, 1.06985953022 )}] \crossTwo;
					\draw[red, shift={( 90, 1.0590113688 )}] \crossTwo;
					\draw[red, shift={( 95, 1.07469928932 )}] \crossTwo;
					\draw[red, shift={( 100, 1.06926775685 )}] \crossTwo;
					\draw[red, shift={( 105, 1.13065256822 )}] \crossTwo;
					\draw[red, shift={( 110, 1.14074402312 )}] \crossTwo;
					\draw[red, shift={( 115, 1.14175171225 )}] \crossTwo;
					\draw[red, shift={( 120, 1.14685074429 )}] \crossTwo;
					\draw[red, shift={( 125, 1.14391408567 )}] \crossTwo;
					\draw[red, shift={( 130, 1.15181074169 )}] \crossTwo;
					\draw[red, shift={( 135, 1.15830247396 )}] \crossTwo;
					\draw[red, shift={( 140, 1.15085154892 )}] \crossTwo;
					\draw[red, shift={( 145, 1.1512844842 )}] \crossTwo;
					\draw[red, shift={( 150, 1.16113980998 )}] \crossTwo;
					\draw[red, shift={( 155, 1.15705661228 )}] \crossTwo;
					\draw[red, shift={( 160, 1.14090529955 )}] \crossTwo;
					\draw[red, shift={( 165, 1.14321114416 )}] \crossTwo;
					\draw[red, shift={( 170, 1.15657940925 )}] \crossTwo;
					\draw[red, shift={( 175, 1.15546935884 )}] \crossTwo;
					\draw[red, shift={( 180, 1.16743734287 )}] \crossTwo;
					\draw[red, shift={( 185, 1.15844727146 )}] \crossTwo;
					\draw[red, shift={( 190, 1.15928974213 )}] \crossTwo;
					\draw[red, shift={( 195, 1.17148112649 )}] \crossTwo;
					\draw[red, shift={( 200, 1.16582534666 )}] \crossTwo;
									
					\draw[red!75, densely dotted]( 5, 0.999999996966 ) -- ( 10, 1.02306355979 ) -- ( 15, 1.00882743483 ) -- ( 20, 1.03616926105 ) -- ( 25, 1.08702629776 ) -- ( 30, 1.07501278798 ) -- ( 35, 1.0554180775 ) -- ( 40, 1.06993011307 ) -- ( 45, 1.04106948871 ) -- ( 50, 1.07932240847 ) -- ( 55, 1.05888875794 ) -- ( 60, 1.07855314717 ) -- ( 65, 1.06982449125 ) -- ( 70, 1.07129844548 ) -- ( 75, 1.07333947593 ) -- ( 80, 1.0732839007 ) -- ( 85, 1.06985953022 ) -- ( 90, 1.0590113688 ) -- ( 95, 1.07469928932 ) -- ( 100, 1.06926775685 ) ( 105, 1.13065256822 ) -- ( 110, 1.14074402312 ) -- ( 115, 1.14175171225 ) -- ( 120, 1.14685074429 ) -- ( 125, 1.14391408567 ) -- ( 130, 1.15181074169 ) -- ( 135, 1.15830247396 ) -- ( 140, 1.15085154892 ) -- ( 145, 1.1512844842 ) -- ( 150, 1.16113980998 ) -- ( 155, 1.15705661228 ) -- ( 160, 1.14090529955 ) -- ( 165, 1.14321114416 ) -- ( 170, 1.15657940925 ) -- ( 175, 1.15546935884 ) -- ( 180, 1.16743734287 ) -- ( 185, 1.15844727146 ) -- ( 190, 1.15928974213 ) -- ( 195, 1.17148112649 ) -- ( 200, 1.16582534666 );

					\draw[blue, shift={( 5, 0.999999996966 )}] \crossThree;
					\draw[blue, shift={( 10, 1.01391233295 )}] \crossThree;
					\draw[blue, shift={( 15, 1.00881520308 )}] \crossThree;
					\draw[blue, shift={( 20, 1.03146824918 )}] \crossThree;
					\draw[blue, shift={( 25, 1.05969641962 )}] \crossThree;
					\draw[blue, shift={( 30, 1.05393963049 )}] \crossThree;
					\draw[blue, shift={( 35, 1.04534484807 )}] \crossThree;
					\draw[blue, shift={( 40, 1.07335220966 )}] \crossThree;
					\draw[blue, shift={( 45, 1.037693167 )}] \crossThree;
					\draw[blue, shift={( 50, 1.06772463216 )}] \crossThree;
					\draw[blue, shift={( 55, 1.05286738994 )}] \crossThree;
					\draw[blue, shift={( 60, 1.06763782741 )}] \crossThree;
					\draw[blue, shift={( 65, 1.0588725696 )}] \crossThree;
					\draw[blue, shift={( 70, 1.06357279568 )}] \crossThree;
					\draw[blue, shift={( 75, 1.05743700022 )}] \crossThree;
					\draw[blue, shift={( 80, 1.06493138494 )}] \crossThree;
					\draw[blue, shift={( 85, 1.05843817966 )}] \crossThree;
					\draw[blue, shift={( 90, 1.05659440868 )}] \crossThree;
					\draw[blue, shift={( 95, 1.06936235052 )}] \crossThree;
					\draw[blue, shift={( 100, 1.06800804128 )}] \crossThree;
					\draw[blue, shift={( 105, 1.1186696999 )}] \crossThree;
					\draw[blue, shift={( 110, 1.13760979406 )}] \crossThree;
					\draw[blue, shift={( 115, 1.14102142218 )}] \crossThree;
					\draw[blue, shift={( 120, 1.13793465958 )}] \crossThree;
					\draw[blue, shift={( 125, 1.13708261779 )}] \crossThree;
					\draw[blue, shift={( 130, 1.14563642337 )}] \crossThree;
					\draw[blue, shift={( 135, 1.14970476612 )}] \crossThree;
					\draw[blue, shift={( 140, 1.14509316381 )}] \crossThree;
					\draw[blue, shift={( 145, 1.13978668343 )}] \crossThree;
					\draw[blue, shift={( 150, 1.14975873972 )}] \crossThree;
					\draw[blue, shift={( 155, 1.15000657759 )}] \crossThree;
					\draw[blue, shift={( 160, 1.13868440642 )}] \crossThree;
					\draw[blue, shift={( 165, 1.13648338391 )}] \crossThree;
					\draw[blue, shift={( 170, 1.14033156978 )}] \crossThree;
					\draw[blue, shift={( 175, 1.1400224272 )}] \crossThree;
					\draw[blue, shift={( 180, 1.15096327342 )}] \crossThree;
					\draw[blue, shift={( 185, 1.14779109816 )}] \crossThree;
					\draw[blue, shift={( 190, 1.15326828036 )}] \crossThree;
					\draw[blue, shift={( 195, 1.16114905789 )}] \crossThree;
					\draw[blue, shift={( 200, 1.15457270027 )}] \crossThree;
									
					\draw[blue!75, loosely dotted]( 5, 0.999999996966 ) -- ( 10, 1.01391233295 ) -- ( 15, 1.00881520308 ) -- ( 20, 1.03146824918 ) -- ( 25, 1.05969641962 ) -- ( 30, 1.05393963049 ) -- ( 35, 1.04534484807 ) -- ( 40, 1.07335220966 ) -- ( 45, 1.037693167 ) -- ( 50, 1.06772463216 ) -- ( 55, 1.05286738994 ) -- ( 60, 1.06763782741 ) -- ( 65, 1.0588725696 ) -- ( 70, 1.06357279568 ) -- ( 75, 1.05743700022 ) -- ( 80, 1.06493138494 ) -- ( 85, 1.05843817966 ) -- ( 90, 1.05659440868 ) -- ( 95, 1.06936235052 ) -- ( 100, 1.06800804128 ) ( 105, 1.1186696999 ) -- ( 110, 1.13760979406 ) -- ( 115, 1.14102142218 ) -- ( 120, 1.13793465958 ) -- ( 125, 1.13708261779 ) -- ( 130, 1.14563642337 ) -- ( 135, 1.14970476612 ) -- ( 140, 1.14509316381 ) -- ( 145, 1.13978668343 ) -- ( 150, 1.14975873972 ) -- ( 155, 1.15000657759 ) -- ( 160, 1.13868440642 ) -- ( 165, 1.13648338391 ) -- ( 170, 1.14033156978 ) -- ( 175, 1.1400224272 ) -- ( 180, 1.15096327342 ) -- ( 185, 1.14779109816 ) -- ( 190, 1.15326828036 ) -- ( 195, 1.16114905789 ) -- ( 200, 1.15457270027 );

					\draw[magenta, shift={( 5, 0.999999996966 )}] \crossFour;
					\draw[magenta, shift={( 10, 1.01391233295 )}] \crossFour;
					\draw[magenta, shift={( 15, 1.00881520308 )}] \crossFour;
					\draw[magenta, shift={( 20, 1.03146824918 )}] \crossFour;
					\draw[magenta, shift={( 25, 1.05039554264 )}] \crossFour;
					\draw[magenta, shift={( 30, 1.04405797764 )}] \crossFour;
					\draw[magenta, shift={( 35, 1.04079768206 )}] \crossFour;
					\draw[magenta, shift={( 40, 1.05515541727 )}] \crossFour;
					\draw[magenta, shift={( 45, 1.03186188615 )}] \crossFour;
					\draw[magenta, shift={( 50, 1.06363144435 )}] \crossFour;
					\draw[magenta, shift={( 55, 1.05175046553 )}] \crossFour;
					\draw[magenta, shift={( 60, 1.05941438013 )}] \crossFour;
					\draw[magenta, shift={( 65, 1.05349187697 )}] \crossFour;
					\draw[magenta, shift={( 70, 1.05794694026 )}] \crossFour;
					\draw[magenta, shift={( 75, 1.05621568569 )}] \crossFour;
					\draw[magenta, shift={( 80, 1.06226705602 )}] \crossFour;
					\draw[magenta, shift={( 85, 1.05071823969 )}] \crossFour;
					\draw[magenta, shift={( 90, 1.054855128 )}] \crossFour;
					\draw[magenta, shift={( 95, 1.05948707276 )}] \crossFour;
					\draw[magenta, shift={( 100, 1.0629786039 )}] \crossFour;
					\draw[magenta, shift={( 105, 1.11460820363 )}] \crossFour;
					\draw[magenta, shift={( 110, 1.13471727725 )}] \crossFour;
					\draw[magenta, shift={( 115, 1.14030567583 )}] \crossFour;
					\draw[magenta, shift={( 120, 1.13462140301 )}] \crossFour;
					\draw[magenta, shift={( 125, 1.13742202024 )}] \crossFour;
					\draw[magenta, shift={( 130, 1.14244128585 )}] \crossFour;
					\draw[magenta, shift={( 135, 1.14387493793 )}] \crossFour;
					\draw[magenta, shift={( 140, 1.14134867608 )}] \crossFour;
					\draw[magenta, shift={( 145, 1.12869331269 )}] \crossFour;
					\draw[magenta, shift={( 150, 1.1463404607 )}] \crossFour;
					\draw[magenta, shift={( 155, 1.14811127933 )}] \crossFour;
					\draw[magenta, shift={( 160, 1.13884695012 )}] \crossFour;
					\draw[magenta, shift={( 165, 1.13527131869 )}] \crossFour;
					\draw[magenta, shift={( 170, 1.14019248118 )}] \crossFour;
					\draw[magenta, shift={( 175, 1.13665984568 )}] \crossFour;
					\draw[magenta, shift={( 180, 1.14935697024 )}] \crossFour;
					\draw[magenta, shift={( 185, 1.14779109816 )}] \crossFour;
					\draw[magenta, shift={( 190, 1.15200367131 )}] \crossFour;
					\draw[magenta, shift={( 195, 1.15923643254 )}] \crossFour;
					\draw[magenta, shift={( 200, 1.15457270027 )}] \crossFour;
									
					\draw[magenta!75, dash pattern={on 1pt off 1pt}]( 5, 0.999999996966 ) -- ( 10, 1.01391233295 ) -- ( 15, 1.00881520308 ) -- ( 20, 1.03146824918 ) -- ( 25, 1.05039554264 ) -- ( 30, 1.04405797764 ) -- ( 35, 1.04079768206 ) -- ( 40, 1.05515541727 ) -- ( 45, 1.03186188615 ) -- ( 50, 1.06363144435 ) -- ( 55, 1.05175046553 ) -- ( 60, 1.05941438013 ) -- ( 65, 1.05349187697 ) -- ( 70, 1.05794694026 ) -- ( 75, 1.05621568569 ) -- ( 80, 1.06226705602 ) -- ( 85, 1.05071823969 ) -- ( 90, 1.054855128 ) -- ( 95, 1.05948707276 ) -- ( 100, 1.0629786039 ) ( 105, 1.11460820363 ) -- ( 110, 1.13471727725 ) -- ( 115, 1.14030567583 ) -- ( 120, 1.13462140301 ) -- ( 125, 1.13742202024 ) -- ( 130, 1.14244128585 ) -- ( 135, 1.14387493793 ) -- ( 140, 1.14134867608 ) -- ( 145, 1.12869331269 ) -- ( 150, 1.1463404607 ) -- ( 155, 1.14811127933 ) -- ( 160, 1.13884695012 ) -- ( 165, 1.13527131869 ) -- ( 170, 1.14019248118 ) -- ( 175, 1.13665984568 ) -- ( 180, 1.14935697024 ) -- ( 185, 1.14779109816 ) -- ( 190, 1.15200367131 ) -- ( 195, 1.15923643254 ) -- ( 200, 1.15457270027 );



				\end{scope}
				
				\def\axisAdditionalLengthPlusTikzX{\axisAdditionalLengthPlus / \xscaleTikz}
				\def\axisAdditionalLengthMinusTikzX{\axisAdditionalLengthMinus / \xscaleTikz}
				\draw[arrow] (-\axisAdditionalLengthMinusTikzX, 0) -- (200, 0)  -- +(\axisAdditionalLengthPlusTikzX, 0) node[right] {\xAxis};
				\def\xTotalLengthPlus{200+\axisAdditionalLengthPlusTikzX}
				\draw (\xTotalLengthPlus, 0) node[right] {$\qquad$};
				\def\axisAdditionalLengthPlusTikzY{\axisAdditionalLengthPlus / \yscaleTikz}
				\def\axisAdditionalLengthMinusTikzY{\axisAdditionalLengthMinus / \yscaleTikz}
				\draw[arrow] (0, -\axisAdditionalLengthMinusTikzY) -- (0, 0.3 ) -- +(0, \axisAdditionalLengthPlusTikzY) node[above, yshift=-0.15cm] {\yAxis};
				
				\def\axisLabelTikzY{\axisLabel / \yscaleTikz}
				\foreach \pos in {0, 20, 40, 60, 80, 100, 120, 140, 160, 180, 200} \draw[shift={(\pos, 0)}] (0, \axisLabelTikzY) -- (0, -\axisLabelTikzY) node[below] {$\pos$};
				
				\def\axisLabelTikzX{\axisLabel / \xscaleTikz}
				\begin{scope}[shift={(0, -1)}]
					\draw[shift={(0,1 )}] (\axisLabelTikzX, 0) -- (-\axisLabelTikzX, 0) node[left] {$1$};
					\draw[shift={(0,1.1 )}] (\axisLabelTikzX, 0) -- (-\axisLabelTikzX, 0) node[left] {$1.1$};
					\draw[shift={(0,1.2 )}] (\axisLabelTikzX, 0) -- (-\axisLabelTikzX, 0) node[left] {$1.2$};
					\draw[shift={(0,1.3 )}] (\axisLabelTikzX, 0) -- (-\axisLabelTikzX, 0) node[left] {$1.3$};
				\end{scope}
			\end{tikzpicture}
		\AngleDistanceTSPInstances
		\vspace*{-0.1cm}
		\caption[LP-based heuristics: \LPP{}, \LPPR{}, \LPCOneR{}, \LPCTwoR{}]{
			LP-based heuristics:\hspace*{-5cm}\\
			\begin{tikzpicture}[xscale=\xscale, yscale=2.0]
				\pgfgettransformentries{\xscaleTikz}{\@tempa}{\@tempa}{\yscaleTikz}{\@tempa}{\@tempa}
				\def\crossSizeX{\crossSize / \xscaleTikz};
				\def\crossSizeY{\crossSize / \yscaleTikz};
				
				\def\crossOne{(-\crossSizeX,-\crossSizeY) -- (\crossSizeX,\crossSizeY) (-\crossSizeX,\crossSizeY) -- (\crossSizeX,-\crossSizeY)};
				\def\crossTwo{(-\crossSizeX,0) -- (\crossSizeX,0) (0,\crossSizeY) -- (0,-\crossSizeY)};
				\def\crossThree{(0,0) -- (0,\crossSizeY) (0,0) -- (\crossSizeX,-\crossSizeY) (0,0) -- (-\crossSizeX,-\crossSizeY)};
				\def\crossFour{(0,0) -- (0,-\crossSizeY) (0,0) -- (\crossSizeX,\crossSizeY) (0,0) -- (-\crossSizeX,\crossSizeY)};
				\def\crossFive{(0,0) -- (\crossSizeX,0) (0,0) -- (-\crossSizeX,\crossSizeY) (0,0) -- (-\crossSizeX,-\crossSizeY)};
				\def\crossSix{(0,0) -- (-\crossSizeX,0) (0,0) -- (\crossSizeX,\crossSizeY) (0,0) -- (\crossSizeX,-\crossSizeY)};
				
				\draw[black, shift={(0, 0)}] \crossOne;
				\draw[black, shift={(5, 0)}] \crossOne;
				\draw[black, shift={(10, 0)}] \crossOne;
				\draw[black, shift={(15, 0)}] \crossOne;
				\draw[black!75, dotted] (-2.5, 0) -- (17.5, 0);
				\node at (40, 0) [minimum width=3cm] {\parbox{1.1cm}\LPP{}};

				\draw[red, shift={(80, 0)}] \crossTwo;
				\draw[red, shift={(85, 0)}] \crossTwo;
				\draw[red, shift={(90, 0)}] \crossTwo;
				\draw[red, shift={(95, 0)}] \crossTwo;
				\draw[red!75, densely dotted] (77.5, 0) -- (97.5, 0);
				\node at (120, 0) [minimum width=3cm] {\parbox{1.1cm}\LPPR{}};
				
				\draw[blue, shift={(160, 0)}] \crossThree;
				\draw[blue, shift={(165, 0)}] \crossThree;
				\draw[blue, shift={(170, 0)}] \crossThree;
				\draw[blue, shift={(175, 0)}] \crossThree;
				\draw[blue!75, loosely dotted] (157.5, 0) -- (177.5, 0);
				\node at (200, 0) [minimum width=3cm] {\parbox{1.1cm}\LPCOneR{}};

				\draw[magenta, shift={(0, -0.25)}] \crossFour;
				\draw[magenta, shift={(5, -0.25)}] \crossFour;
				\draw[magenta, shift={(10, -0.25)}] \crossFour;
				\draw[magenta, shift={(15, -0.25)}] \crossFour;
				\draw[magenta!75, dash pattern={on 1pt off 1pt}] (-2.5, -0.25) -- (17.5, -0.25);
				\node at (40, -0.25) [minimum width=3cm] {\parbox{1.1cm}\LPCTwoR{}};
			\end{tikzpicture}
		}
		\label{figure:LPBasedHeuristics}
	\end{figure}
			\end{comment:figures}
			
			Figure~\ref{figure:HeuristicsBasedOnConvexHulls} gives us a clear ranking: \CHCL{} yields  the best objective function values followed by \CHC{} and \CH{}.
			This effect, however, seems to be less significant for larger values of $n$.


		The third group of stand-alone heuristics consists of \LPP{}, \LPPR{}, \LPCOneR{} and \LPCTwoR{}. 
		Again, preliminary tests gave us clear evidence for the value of the rounding parameter $\rho$, which we fixed to  $\rho = 0.5$ in all four heuristics.

			The ranking, shown in Figure~\ref{figure:LPBasedHeuristics}, is the same for {\AngleTSP}- and {\AngleDistanceTSP}-instances: ordered from the worst to the best heuristic we get \LPP{}, \LPPR{}, \LPCOneR{} and \LPCTwoR{}, where the last two approaches yield very similar results. The step between \LPP{} and \LPPR{} (\ie between not using vs.\ using the rerun strategy) is significant, the differences between \LPCOneR{} and \LPCTwoR{} are rather marginal.
			
			\medskip
			
			Let us now consider the efficiency of the particular algorithms shown as trade-off between the running time and the solution quality. Therefore we plot for every heuristic the objective function value ratio $\rg$ against the running time $\ra$ taken as means over all instances with $105 \leq n \leq 200$ in Figure~\ref{figure:StartingHeuristicEfficienciesForInstancesWith105LessOrEqualTonLessOrEqual200} (the exact values can be found in Tables~\ref{table:ObjectiveFunctionValueRatioMeansForAllTestInstancesWith105LessOrEqualTonLessOrEqual200} and \ref{table:RunningTimeMeansForAllTestInstancesWith105LessOrEqualTonLessOrEqual200} in the Appendix). 
			Thus, each heuristic is represented by a single point in the (running time, objective function ratio)-space. We concentrated on the larger instance sizes $105 \leq n \leq 200$, because
			\begin{enumerate}
				\item then, as mentioned, all objective function values are compared with the lower bounds (and not with the optima) and
				\item larger values can be expected to give a better indication for the general behavior
				(note that \eg \CHC{} always yields an optimum if $n \leq C$).
			\end{enumerate}
			
			\begin{comment:figures}
	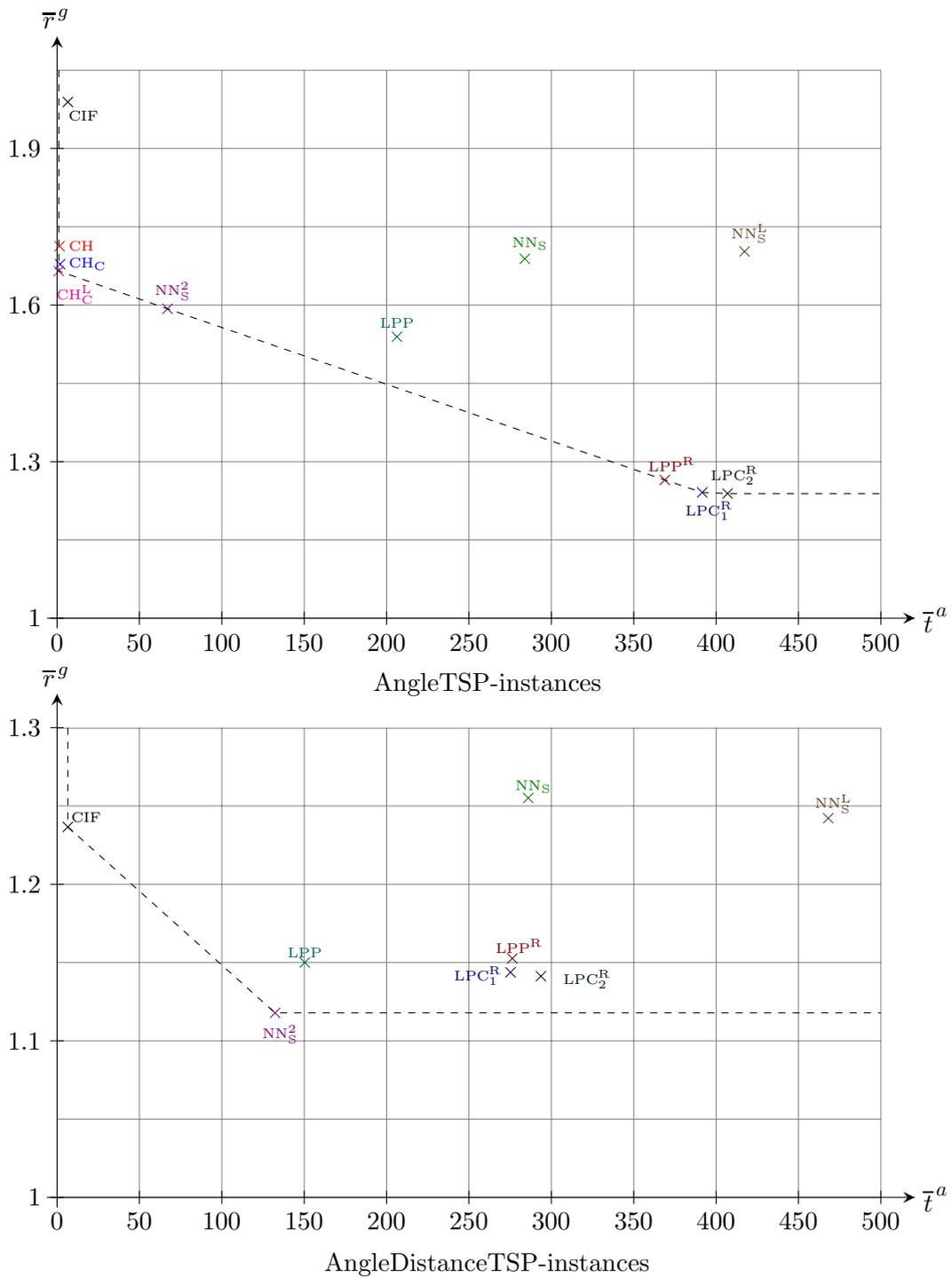
\begin{figure}[htbp!]
		\centering
			\begin{tikzpicture}[xscale=\xscaleEfficiencyStartingHeuristics, yscale=\yscaleMiddleEfficiency]
				\pgfgettransformentries{\xscaleTikz}{\@tempa}{\@tempa}{\yscaleTikz}{\@tempa}{\@tempa}
			
				\draw[very thin, color=gray, xstep=5, ystep=0.15] (0, 0) grid (50, 1.05 );
				
				\begin{scope}[shift={(0, -1)}]
					\def\crossSizeX{\crossSize / \xscaleTikz};
					\def\crossSizeY{\crossSize / \yscaleTikz};
					
					\def\crossOne{(-\crossSizeX,-\crossSizeY) -- (\crossSizeX,\crossSizeY) (-\crossSizeX,\crossSizeY) -- (\crossSizeX,-\crossSizeY)};
					\def\crossTwo{(-\crossSizeX,0) -- (\crossSizeX,0) (0,\crossSizeY) -- (0,-\crossSizeY)};
					\def\crossThree{(0,0) -- (0,\crossSizeY) (0,0) -- (\crossSizeX,-\crossSizeY) (0,0) -- (-\crossSizeX,-\crossSizeY)};
					\def\crossFour{(0,0) -- (0,-\crossSizeY) (0,0) -- (\crossSizeX,\crossSizeY) (0,0) -- (-\crossSizeX,\crossSizeY)};
					\def\crossFive{(0,0) -- (\crossSizeX,0) (0,0) -- (-\crossSizeX,\crossSizeY) (0,0) -- (-\crossSizeX,-\crossSizeY)};
					\def\crossSix{(0,0) -- (-\crossSizeX,0) (0,0) -- (\crossSizeX,\crossSizeY) (0,0) -- (\crossSizeX,-\crossSizeY)};
					
					\draw[black!50!green, shift={( 28.38884851682, 1.6886750798)}] \crossOne;
					\node[label={[label distance=-0.14cm]+90:{\color{black!50!green} \tiny \; \NNS{}}}] at ( 28.38884851682, 1.6886750798) {};
	
					\draw[violet, shift={( 6.70159488237, 1.59284252)}] \crossOne;
					\node[label={[label distance=-0.14cm]+90:{\color{violet} \tiny \; \NNSTwo{}}}] at ( 6.70159488237, 1.59284252) {};
	
					\draw[black!50!brown, shift={( 41.72670590806, 1.7026226113)}] \crossOne;
					\node[label={[label distance=-0.14cm]+90:{\color{black!50!brown} \tiny \; \NNSL{}}}] at ( 41.72670590806, 1.7026226113) {};
					
					\draw[black, shift={( 0.65365459931, 1.9891574343)}] \crossOne;
					\node[label={[label distance=-0.25cm]-3:{\color{black} \tiny \hspace*{-0.0101cm}\CIF{}}}] at (0.65365459931, 1.9891574343) {};
					
					\draw[red, shift={( 0.150280828, 1.7128395721)}] \crossOne;
					\node[label={[label distance=-0.132cm]+1:{\color{red} \tiny \CH{}}}] at ( 0.150280828, 1.7128395721) {};
	
					\draw[blue, shift={( 0.18191259646, 1.6782692782)}] \crossOne;
					\node[label={[label distance=-0.14cm]+1:{\color{blue} \tiny \CHC{}}}] at ( 0.18191259646, 1.6782692782) {};
	
					\draw[magenta, shift={( 0.10235768318, 1.6651628557)}] \crossOne;
					\node[label={[label distance=-0.1cm]-90:{\color{magenta} \tiny \CHCL{}}}] at ( 1.1, 1.66) {};
					
					\draw[black!25!teal, shift={( 20.62801763701, 1.5392871641)}] \crossOne;
					\node[label={[label distance=-0.14cm]+90:{\color{black!25!teal} \tiny \LPP{}}}] at ( 20.62801763701, 1.5392871641) {};
					
					\draw[black!50!red, shift={( 36.88421900809, 1.264778304)}] \crossOne;
					\node[label={[label distance=-0.14cm]+90:{\color{black!50!red} \tiny \;\, \LPPR{}}}] at ( 36.88421900809, 1.264778304) {};
					
					\draw[black!50!blue, shift={( 39.15620841944, 1.2409769694)}] \crossOne;
					\node[label={[label distance=-0.14cm]-90:{\color{black!50!blue} \tiny \;\, \LPCOneR{}}}] at ( 39.15620841944, 1.2409769694) {};
					
					\draw[black!90!green, shift={( 40.6945596118, 1.2388916232)}] \crossOne;
					\node[label={[label distance=-0.14cm]+90:{\color{black!90!green} \tiny \;\, \LPCTwoR{}}}] at ( 40.6945596118, 1.2388916232) {};

					\draw[dashed] ( 0.10235768318, 2.05) -- ( 0.10235768318, 1.6651628557) -- ( 6.70159488237, 1.59284252) -- ( 36.88421900809, 1.264778304) -- ( 39.15620841944, 1.2409769694) -- ( 40.6945596118, 1.2388916232) -- ( 50, 1.2388916232);
				\end{scope}
				
				\def\axisAdditionalLengthPlusTikzX{\axisAdditionalLengthPlus / \xscaleTikz}
				\def\axisAdditionalLengthMinusTikzX{\axisAdditionalLengthMinus / \xscaleTikz}
				\draw[arrow] (-\axisAdditionalLengthMinusTikzX, 0) -- (50, 0)  -- +(\axisAdditionalLengthPlusTikzX, 0) node[right] {\xAxisEfficiency};
				\def\xTotalLengthPlus{50+\axisAdditionalLengthPlusTikzX}
				\draw (\xTotalLengthPlus, 0) node[right] {$\qquad$};
				\def\axisAdditionalLengthPlusTikzY{\axisAdditionalLengthPlus / \yscaleTikz}
				\def\axisAdditionalLengthMinusTikzY{\axisAdditionalLengthMinus / \yscaleTikz}
				\draw[arrow] (0, -\axisAdditionalLengthMinusTikzY) -- (0, 1.05 ) -- +(0, \axisAdditionalLengthPlusTikzY) node[above, yshift=-0.15cm] {\yAxisEfficiency};
				
				\def\axisLabelTikzY{\axisLabel / \yscaleTikz}
				
				\draw[shift={(0, 0)}] (0, \axisLabelTikzY) -- (0, -\axisLabelTikzY) node[below] {$0$};
				\draw[shift={(5, 0)}] (0, \axisLabelTikzY) -- (0, -\axisLabelTikzY) node[below] {$50$};
				\draw[shift={(10, 0)}] (0, \axisLabelTikzY) -- (0, -\axisLabelTikzY) node[below] {$100$};
				\draw[shift={(15, 0)}] (0, \axisLabelTikzY) -- (0, -\axisLabelTikzY) node[below] {$150$};
				\draw[shift={(20, 0)}] (0, \axisLabelTikzY) -- (0, -\axisLabelTikzY) node[below] {$200$};
				\draw[shift={(25, 0)}] (0, \axisLabelTikzY) -- (0, -\axisLabelTikzY) node[below] {$250$};
				\draw[shift={(30, 0)}] (0, \axisLabelTikzY) -- (0, -\axisLabelTikzY) node[below] {$300$};
				\draw[shift={(35, 0)}] (0, \axisLabelTikzY) -- (0, -\axisLabelTikzY) node[below] {$350$};
				\draw[shift={(40, 0)}] (0, \axisLabelTikzY) -- (0, -\axisLabelTikzY) node[below] {$400$};
				\draw[shift={(45, 0)}] (0, \axisLabelTikzY) -- (0, -\axisLabelTikzY) node[below] {$450$};
				\draw[shift={(50, 0)}] (0, \axisLabelTikzY) -- (0, -\axisLabelTikzY) node[below] {$500$};
				
				\def\axisLabelTikzX{\axisLabel / \xscaleTikz}
				\begin{scope}[shift={(0, -1)}]
					\draw[shift={(0,1 )}] (\axisLabelTikzX, 0) -- (-\axisLabelTikzX, 0) node[left] {$1$};
					\draw[shift={(0,1.3 )}] (\axisLabelTikzX, 0) -- (-\axisLabelTikzX, 0) node[left] {$1.3$};
					\draw[shift={(0,1.6 )}] (\axisLabelTikzX, 0) -- (-\axisLabelTikzX, 0) node[left] {$1.6$};
					\draw[shift={(0,1.9 )}] (\axisLabelTikzX, 0) -- (-\axisLabelTikzX, 0) node[left] {$1.9$};
				\end{scope}
			\end{tikzpicture}
		AngleTSP-instances
		\vspace{-0.6cm}
		
			\begin{tikzpicture}[xscale=\xscaleEfficiencyStartingHeuristics, yscale=\yscaleSmallEfficiency]
				\pgfgettransformentries{\xscaleTikz}{\@tempa}{\@tempa}{\yscaleTikz}{\@tempa}{\@tempa}
			
				\draw[very thin, color=gray, xstep=5, ystep=0.05] (0, 0) grid (50, 0.3 );
				
				\begin{scope}[shift={(0, -1)}]
					\def\crossSizeX{\crossSize / \xscaleTikz};
					\def\crossSizeY{\crossSize / \yscaleTikz};
					
					\def\crossOne{(-\crossSizeX,-\crossSizeY) -- (\crossSizeX,\crossSizeY) (-\crossSizeX,\crossSizeY) -- (\crossSizeX,-\crossSizeY)};
					\def\crossTwo{(-\crossSizeX,0) -- (\crossSizeX,0) (0,\crossSizeY) -- (0,-\crossSizeY)};
					\def\crossThree{(0,0) -- (0,\crossSizeY) (0,0) -- (\crossSizeX,-\crossSizeY) (0,0) -- (-\crossSizeX,-\crossSizeY)};
					\def\crossFour{(0,0) -- (0,-\crossSizeY) (0,0) -- (\crossSizeX,\crossSizeY) (0,0) -- (-\crossSizeX,\crossSizeY)};
					\def\crossFive{(0,0) -- (\crossSizeX,0) (0,0) -- (-\crossSizeX,\crossSizeY) (0,0) -- (-\crossSizeX,-\crossSizeY)};
					\def\crossSix{(0,0) -- (-\crossSizeX,0) (0,0) -- (\crossSizeX,\crossSizeY) (0,0) -- (\crossSizeX,-\crossSizeY)};
					
					\draw[black!50!green, shift={( 28.59435007989, 1.2551262705)}] \crossOne;
					\node[label={[label distance=-0.14cm]+90:{\color{black!50!green} \tiny \; \NNS{}}}] at ( 28.59435007989, 1.2551262705) {};
	
					\draw[violet, shift={( 13.23109778368, 1.1178631211)}] \crossOne;
					\node[label={[label distance=-0.14cm]-90:{\color{violet} \tiny \; \NNSTwo{}}}] at ( 13.23109778368, 1.1178631211) {};
	
					\draw[black!50!brown, shift={( 46.80508613992, 1.2421763869)}] \crossOne;
					\node[label={[label distance=-0.14cm]+90:{\color{black!50!brown} \tiny \; \NNSL{}}}] at ( 46.80508613992, 1.2421763869) {};
					
					\draw[black, shift={( 0.6546783278, 1.2366105023)}] \crossOne;
					\node[label={[label distance=-0.25cm]+3:{\color{black} \tiny \hspace*{-0.06cm} \CIF{}}}] at ( 0.6546783278, 1.2366105023) {};
					
	
	
					
					\draw[black!25!teal, shift={( 15.03578750372, 1.1500200958)}] \crossOne;
					\node[label={[label distance=-0.14cm]+90:{\color{black!25!teal} \tiny \LPP{}}}] at ( 15.03578750372, 1.1500200958) {};
					
					\draw[black!50!red, shift={( 27.63158933294, 1.1526055484)}] \crossOne;
					\node[label={[label distance=-0.14cm]+90:{\color{black!50!red} \tiny \;\, \LPPR{}}}] at ( 27.63158933294, 1.1526055484) {};
					
					\draw[black!50!blue, shift={( 27.52135980511, 1.1437441692)}] \crossOne;
					\node[label={[label distance=-0.14cm]+180:{\color{black!50!blue} \tiny \;\, \LPCOneR{}}}] at ( 27.52135980511, 1.1437441692) {};
					
					\draw[black!90!green, shift={( 29.36623909736, 1.1412803501)}] \crossOne;
					\node[label={[label distance=-0.14cm]+0:{\color{black!90!green} \tiny \;\, \LPCTwoR{}}}] at ( 29.36623909736, 1.1412803501) {};

					\draw[dashed] ( 0.6546783278, 1.3) -- ( 0.6546783278, 1.2366105023) -- ( 13.23109778368, 1.1178631211) -- ( 50, 1.1178631211);
				\end{scope}
				
				\def\axisAdditionalLengthPlusTikzX{\axisAdditionalLengthPlus / \xscaleTikz}
				\def\axisAdditionalLengthMinusTikzX{\axisAdditionalLengthMinus / \xscaleTikz}
				\draw[arrow] (-\axisAdditionalLengthMinusTikzX, 0) -- (50, 0)  -- +(\axisAdditionalLengthPlusTikzX, 0) node[right] {\xAxisEfficiency};
				\def\xTotalLengthPlus{50+\axisAdditionalLengthPlusTikzX}
				\draw (\xTotalLengthPlus, 0) node[right] {$\qquad$};
				\def\axisAdditionalLengthPlusTikzY{\axisAdditionalLengthPlus / \yscaleTikz}
				\def\axisAdditionalLengthMinusTikzY{\axisAdditionalLengthMinus / \yscaleTikz}
				\draw[arrow] (0, -\axisAdditionalLengthMinusTikzY) -- (0, 0.3 ) -- +(0, \axisAdditionalLengthPlusTikzY) node[above, yshift=-0.15cm] {\yAxisEfficiency};
				
				\def\axisLabelTikzY{\axisLabel / \yscaleTikz}
				
				\draw[shift={(0, 0)}] (0, \axisLabelTikzY) -- (0, -\axisLabelTikzY) node[below] {$0$};
				\draw[shift={(5, 0)}] (0, \axisLabelTikzY) -- (0, -\axisLabelTikzY) node[below] {$50$};
				\draw[shift={(10, 0)}] (0, \axisLabelTikzY) -- (0, -\axisLabelTikzY) node[below] {$100$};
				\draw[shift={(15, 0)}] (0, \axisLabelTikzY) -- (0, -\axisLabelTikzY) node[below] {$150$};
				\draw[shift={(20, 0)}] (0, \axisLabelTikzY) -- (0, -\axisLabelTikzY) node[below] {$200$};
				\draw[shift={(25, 0)}] (0, \axisLabelTikzY) -- (0, -\axisLabelTikzY) node[below] {$250$};
				\draw[shift={(30, 0)}] (0, \axisLabelTikzY) -- (0, -\axisLabelTikzY) node[below] {$300$};
				\draw[shift={(35, 0)}] (0, \axisLabelTikzY) -- (0, -\axisLabelTikzY) node[below] {$350$};
				\draw[shift={(40, 0)}] (0, \axisLabelTikzY) -- (0, -\axisLabelTikzY) node[below] {$400$};
				\draw[shift={(45, 0)}] (0, \axisLabelTikzY) -- (0, -\axisLabelTikzY) node[below] {$450$};
				\draw[shift={(50, 0)}] (0, \axisLabelTikzY) -- (0, -\axisLabelTikzY) node[below] {$500$};
				
				\def\axisLabelTikzX{\axisLabel / \xscaleTikz}
				\begin{scope}[shift={(0, -1)}]
					\draw[shift={(0,1 )}] (\axisLabelTikzX, 0) -- (-\axisLabelTikzX, 0) node[left] {$1$};
					\draw[shift={(0,1.1 )}] (\axisLabelTikzX, 0) -- (-\axisLabelTikzX, 0) node[left] {$1.1$};
					\draw[shift={(0,1.2 )}] (\axisLabelTikzX, 0) -- (-\axisLabelTikzX, 0) node[left] {$1.2$};
					\draw[shift={(0,1.3 )}] (\axisLabelTikzX, 0) -- (-\axisLabelTikzX, 0) node[left] {$1.3$};
				\end{scope}
			\end{tikzpicture}
		AngleDistanceTSP-instances
		\vspace*{-0.1cm}
		\caption{stand-alone heuristic trade-offs for instances with $105 \leq n \leq 200$; Pareto-frontier is plotted dashed}
		\label{figure:StartingHeuristicEfficienciesForInstancesWith105LessOrEqualTonLessOrEqual200}
	\end{figure}
			\end{comment:figures}
			
			Consider now the trade-offs for the \AngleTSPInstances. Among the ``fast'' heuristics (\CIF{}, \CH{}, \CHC{} and \CHCL{}), \CHCL{} clearly Pareto-dominates the other ones.
			In this context it is interesting that it is superior to \CH{} in both the objective function values and the running times. 
			\NNSTwo{} can be seen as a good trade-off between the tour quality and the time investment. 
			If better objective function values are sought, \LPPR{}, \LPCOneR{} or \LPCTwoR{} would be the obvious choice, \LPCTwoR{}, however, brings only a marginal improvement in comparison to \LPCOneR{}. The results for the \AngleDistanceTSPInstances{} are in many aspects similar to the results for the \AngleTSPInstances, there is, however, one main difference: \NNSTwo{} significantly outperforms all other heuristics, although its running time advantage decreases.
		
		\subsection{Improvement heuristics}
			\label{subsection:computationalResults:ImprovementHeuristics}
			Concerning improvement heuristics, it would not make sense to present results for all possible combinations of stand-alone and improvement heuristics in detail.
			Instead, we discuss the results for the best stand-alone heuristic combined with various improvement approaches; in particular, we took \LPCTwoR{} for the \AngleTSPInstances{} and \NNSTwo{} for the \AngleDistanceTSPInstances{} as the stand-alone heuristic (results for all other combinations are summed up in Tables~\ref{table:ObjectiveFunctionValueRatioMeansForAllTestInstancesWith105LessOrEqualTonLessOrEqual200} and \ref{table:RunningTimeMeansForAllTestInstancesWith105LessOrEqualTonLessOrEqual200} in the Appendix).
			
			Again, we must fix some parameters first. Similar to \CHC{} and \CHCL{} we have a trade-off in \M{}: $k$ expresses the expected number of vertices in our magnifying glass, thus larger values of $k$ lead to better objective function values, but worse running times. After various tests we decided to use two variants, namely \M{15} and \M{20}; the first one is roughly half-way between \TwoO{} and \ThreeO{} with respect to the computation times and the latter one yields the overall best results with respect to the objective function values. 
			In our metaheuristic \Lens{}, we have just two parameters: after preliminary tests we could set the lens thickness to $\gamma = 20\degree\ (\approx 0.3491)$ and the maximum number of iterations to $I = 2000$.
			
			\begin{comment:figures}
	\begin{figure}[htb!]
		\centering
			\begin{tikzpicture}[xscale=\xscale, yscale=\yscaleSmall]
				\pgfgettransformentries{\xscaleTikz}{\@tempa}{\@tempa}{\yscaleTikz}{\@tempa}{\@tempa}
			
				\draw[very thin, color=gray, xstep=20, ystep=0.1] (0, 0) grid (200, 0.3 );
				
				\begin{scope}[shift={(0, -1)}]
					\def\crossSizeX{\crossSize / \xscaleTikz};
					\def\crossSizeY{\crossSize / \yscaleTikz};
					
					\def\crossOne{(-\crossSizeX,-\crossSizeY) -- (\crossSizeX,\crossSizeY) (-\crossSizeX,\crossSizeY) -- (\crossSizeX,-\crossSizeY)};
					\def\crossTwo{(-\crossSizeX,0) -- (\crossSizeX,0) (0,\crossSizeY) -- (0,-\crossSizeY)};
					\def\crossThree{(0,0) -- (0,\crossSizeY) (0,0) -- (\crossSizeX,-\crossSizeY) (0,0) -- (-\crossSizeX,-\crossSizeY)};
					\def\crossFour{(0,0) -- (0,-\crossSizeY) (0,0) -- (\crossSizeX,\crossSizeY) (0,0) -- (-\crossSizeX,\crossSizeY)};
					\def\crossFive{(0,0) -- (\crossSizeX,0) (0,0) -- (-\crossSizeX,\crossSizeY) (0,0) -- (-\crossSizeX,-\crossSizeY)};
					\def\crossSix{(0,0) -- (-\crossSizeX,0) (0,0) -- (\crossSizeX,\crossSizeY) (0,0) -- (\crossSizeX,-\crossSizeY)};
					
					\draw[black, shift={( 5, 0.999999908761 )}] \crossOne;
					\draw[black, shift={( 10, 1.07941551675 )}] \crossOne;
					\draw[black, shift={( 15, 1.054668787 )}] \crossOne;
					\draw[black, shift={( 20, 1.05851034214 )}] \crossOne;
					\draw[black, shift={( 25, 1.0513043869 )}] \crossOne;
					\draw[black, shift={( 30, 1.0771571731 )}] \crossOne;
					\draw[black, shift={( 35, 1.08162222356 )}] \crossOne;
					\draw[black, shift={( 40, 1.09744870968 )}] \crossOne;
					\draw[black, shift={( 45, 1.09228634084 )}] \crossOne;
					\draw[black, shift={( 50, 1.12061933957 )}] \crossOne;
					\draw[black, shift={( 55, 1.08748371008 )}] \crossOne;
					\draw[black, shift={( 60, 1.08397928398 )}] \crossOne;
					\draw[black, shift={( 65, 1.09837426566 )}] \crossOne;
					\draw[black, shift={( 70, 1.12280570636 )}] \crossOne;
					\draw[black, shift={( 75, 1.1002486545 )}] \crossOne;
					\draw[black, shift={( 80, 1.21806683419 )}] \crossOne;
					\draw[black, shift={( 85, 1.22683947592 )}] \crossOne;
					\draw[black, shift={( 90, 1.23247831073 )}] \crossOne;
					\draw[black, shift={( 95, 1.244231978 )}] \crossOne;
					\draw[black, shift={( 100, 1.25201170136 )}] \crossOne;
					\draw[black, shift={( 105, 1.22578972065 )}] \crossOne;
					\draw[black, shift={( 110, 1.22862377684 )}] \crossOne;
					\draw[black, shift={( 115, 1.2380757942 )}] \crossOne;
					\draw[black, shift={( 120, 1.22645508594 )}] \crossOne;
					\draw[black, shift={( 125, 1.21540256452 )}] \crossOne;
					\draw[black, shift={( 130, 1.21766853385 )}] \crossOne;
					\draw[black, shift={( 135, 1.24898991894 )}] \crossOne;
					\draw[black, shift={( 140, 1.24967554585 )}] \crossOne;
					\draw[black, shift={( 145, 1.24436410079 )}] \crossOne;
					\draw[black, shift={( 150, 1.23715331441 )}] \crossOne;
					\draw[black, shift={( 155, 1.22826838516 )}] \crossOne;
					\draw[black, shift={( 160, 1.25711740161 )}] \crossOne;
					\draw[black, shift={( 165, 1.25235007488 )}] \crossOne;
					\draw[black, shift={( 170, 1.24455398981 )}] \crossOne;
					\draw[black, shift={( 175, 1.23409672882 )}] \crossOne;
					\draw[black, shift={( 180, 1.23090988036 )}] \crossOne;
					\draw[black, shift={( 185, 1.24197708506 )}] \crossOne;
					\draw[black, shift={( 190, 1.23558088696 )}] \crossOne;
					\draw[black, shift={( 195, 1.27941662751 )}] \crossOne;
					\draw[black, shift={( 200, 1.24301197132 )}] \crossOne;
					
					\draw[black!75, dotted] ( 5, 0.999999908761 ) -- ( 10, 1.07941551675 ) -- ( 15, 1.054668787 ) -- ( 20, 1.05851034214 ) -- ( 25, 1.0513043869 ) -- ( 30, 1.0771571731 ) -- ( 35, 1.08162222356 ) -- ( 40, 1.09744870968 ) -- ( 45, 1.09228634084 ) -- ( 50, 1.12061933957 ) -- ( 55, 1.08748371008 ) -- ( 60, 1.08397928398 ) -- ( 65, 1.09837426566 ) -- ( 70, 1.12280570636 ) -- ( 75, 1.1002486545 ) ( 80, 1.21806683419 ) -- ( 85, 1.22683947592 ) -- ( 90, 1.23247831073 ) -- ( 95, 1.244231978 ) -- ( 100, 1.25201170136 ) -- ( 105, 1.22578972065 ) -- ( 110, 1.22862377684 ) -- ( 115, 1.2380757942 ) -- ( 120, 1.22645508594 ) -- ( 125, 1.21540256452 ) -- ( 130, 1.21766853385 ) -- ( 135, 1.24898991894 ) -- ( 140, 1.24967554585 ) -- ( 145, 1.24436410079 ) -- ( 150, 1.23715331441 ) -- ( 155, 1.22826838516 ) -- ( 160, 1.25711740161 ) -- ( 165, 1.25235007488 ) -- ( 170, 1.24455398981 ) -- ( 175, 1.23409672882 ) -- ( 180, 1.23090988036 ) -- ( 185, 1.24197708506 ) -- ( 190, 1.23558088696 ) -- ( 195, 1.27941662751 ) -- ( 200, 1.24301197132 );

					\draw[red, shift={( 5, 0.999999908761 )}] \crossTwo;
					\draw[red, shift={( 10, 1.03781576641 )}] \crossTwo;
					\draw[red, shift={( 15, 1.01934010913 )}] \crossTwo;
					\draw[red, shift={( 20, 1.03625000004 )}] \crossTwo;
					\draw[red, shift={( 25, 1.02882019208 )}] \crossTwo;
					\draw[red, shift={( 30, 1.05145200977 )}] \crossTwo;
					\draw[red, shift={( 35, 1.07093665823 )}] \crossTwo;
					\draw[red, shift={( 40, 1.08083549518 )}] \crossTwo;
					\draw[red, shift={( 45, 1.07350322924 )}] \crossTwo;
					\draw[red, shift={( 50, 1.08936016318 )}] \crossTwo;
					\draw[red, shift={( 55, 1.07079634016 )}] \crossTwo;
					\draw[red, shift={( 60, 1.06953656621 )}] \crossTwo;
					\draw[red, shift={( 65, 1.07965973301 )}] \crossTwo;
					\draw[red, shift={( 70, 1.09344065724 )}] \crossTwo;
					\draw[red, shift={( 75, 1.08596529288 )}] \crossTwo;
					\draw[red, shift={( 80, 1.19782473443 )}] \crossTwo;
					\draw[red, shift={( 85, 1.20392161223 )}] \crossTwo;
					\draw[red, shift={( 90, 1.1969538853 )}] \crossTwo;
					\draw[red, shift={( 95, 1.21502345696 )}] \crossTwo;
					\draw[red, shift={( 100, 1.20880534074 )}] \crossTwo;
					\draw[red, shift={( 105, 1.22100814812 )}] \crossTwo;
					\draw[red, shift={( 110, 1.21163180979 )}] \crossTwo;
					\draw[red, shift={( 115, 1.22272865337 )}] \crossTwo;
					\draw[red, shift={( 120, 1.20857577395 )}] \crossTwo;
					\draw[red, shift={( 125, 1.20987225526 )}] \crossTwo;
					\draw[red, shift={( 130, 1.20369827108 )}] \crossTwo;
					\draw[red, shift={( 135, 1.22288103261 )}] \crossTwo;
					\draw[red, shift={( 140, 1.22448585081 )}] \crossTwo;
					\draw[red, shift={( 145, 1.22739803268 )}] \crossTwo;
					\draw[red, shift={( 150, 1.22048892577 )}] \crossTwo;
					\draw[red, shift={( 155, 1.20776211796 )}] \crossTwo;
					\draw[red, shift={( 160, 1.22062115225 )}] \crossTwo;
					\draw[red, shift={( 165, 1.23944354083 )}] \crossTwo;
					\draw[red, shift={( 170, 1.21605477131 )}] \crossTwo;
					\draw[red, shift={( 175, 1.2200115625 )}] \crossTwo;
					\draw[red, shift={( 180, 1.20801431702 )}] \crossTwo;
					\draw[red, shift={( 185, 1.2236547461 )}] \crossTwo;
					\draw[red, shift={( 190, 1.21405916268 )}] \crossTwo;
					\draw[red, shift={( 195, 1.25648480355 )}] \crossTwo;
					\draw[red, shift={( 200, 1.22551752034 )}] \crossTwo;
									
					\draw[red!75, densely dotted]( 5, 0.999999908761 ) -- ( 10, 1.03781576641 ) -- ( 15, 1.01934010913 ) -- ( 20, 1.03625000004 ) -- ( 25, 1.02882019208 ) -- ( 30, 1.05145200977 ) -- ( 35, 1.07093665823 ) -- ( 40, 1.08083549518 ) -- ( 45, 1.07350322924 ) -- ( 50, 1.08936016318 ) -- ( 55, 1.07079634016 ) -- ( 60, 1.06953656621 ) -- ( 65, 1.07965973301 ) -- ( 70, 1.09344065724 ) -- ( 75, 1.08596529288 ) ( 80, 1.19782473443 ) -- ( 85, 1.20392161223 ) -- ( 90, 1.1969538853 ) -- ( 95, 1.21502345696 ) -- ( 100, 1.20880534074 ) -- ( 105, 1.22100814812 ) -- ( 110, 1.21163180979 ) -- ( 115, 1.22272865337 ) -- ( 120, 1.20857577395 ) -- ( 125, 1.20987225526 ) -- ( 130, 1.20369827108 ) -- ( 135, 1.22288103261 ) -- ( 140, 1.22448585081 ) -- ( 145, 1.22739803268 ) -- ( 150, 1.22048892577 ) -- ( 155, 1.20776211796 ) -- ( 160, 1.22062115225 ) -- ( 165, 1.23944354083 ) -- ( 170, 1.21605477131 ) -- ( 175, 1.2200115625 ) -- ( 180, 1.20801431702 ) -- ( 185, 1.2236547461 ) -- ( 190, 1.21405916268 ) -- ( 195, 1.25648480355 ) -- ( 200, 1.22551752034 );

					\draw[blue, shift={( 5, 0.999999908761 )}] \crossThree;
					\draw[blue, shift={( 10, 1.00691641152 )}] \crossThree;
					\draw[blue, shift={( 15, 1.00740536496 )}] \crossThree;
					\draw[blue, shift={( 20, 1.0275869019 )}] \crossThree;
					\draw[blue, shift={( 25, 1.01781063015 )}] \crossThree;
					\draw[blue, shift={( 30, 1.02194863555 )}] \crossThree;
					\draw[blue, shift={( 35, 1.026377895 )}] \crossThree;
					\draw[blue, shift={( 40, 1.05259411188 )}] \crossThree;
					\draw[blue, shift={( 45, 1.03979661758 )}] \crossThree;
					\draw[blue, shift={( 50, 1.0587165705 )}] \crossThree;
					\draw[blue, shift={( 55, 1.0352089113 )}] \crossThree;
					\draw[blue, shift={( 60, 1.03887748681 )}] \crossThree;
					\draw[blue, shift={( 65, 1.0535980836 )}] \crossThree;
					\draw[blue, shift={( 70, 1.05225900334 )}] \crossThree;
					\draw[blue, shift={( 75, 1.0463456191 )}] \crossThree;
					\draw[blue, shift={( 80, 1.16293292358 )}] \crossThree;
					\draw[blue, shift={( 85, 1.1809613864 )}] \crossThree;
					\draw[blue, shift={( 90, 1.1719938778 )}] \crossThree;
					\draw[blue, shift={( 95, 1.17793479145 )}] \crossThree;
					\draw[blue, shift={( 100, 1.19043796953 )}] \crossThree;
					\draw[blue, shift={( 105, 1.18459917955 )}] \crossThree;
					\draw[blue, shift={( 110, 1.18318123982 )}] \crossThree;
					\draw[blue, shift={( 115, 1.18235186487 )}] \crossThree;
					\draw[blue, shift={( 120, 1.18447611579 )}] \crossThree;
					\draw[blue, shift={( 125, 1.17775189614 )}] \crossThree;
					\draw[blue, shift={( 130, 1.17055269871 )}] \crossThree;
					\draw[blue, shift={( 135, 1.18485016753 )}] \crossThree;
					\draw[blue, shift={( 140, 1.18888672386 )}] \crossThree;
					\draw[blue, shift={( 145, 1.1929594469 )}] \crossThree;
					\draw[blue, shift={( 150, 1.19506684666 )}] \crossThree;
					\draw[blue, shift={( 155, 1.18726519043 )}] \crossThree;
					\draw[blue, shift={( 160, 1.19058915527 )}] \crossThree;
					\draw[blue, shift={( 165, 1.19360802093 )}] \crossThree;
					\draw[blue, shift={( 170, 1.1903302 )}] \crossThree;
					\draw[blue, shift={( 175, 1.1871565849 )}] \crossThree;
					\draw[blue, shift={( 180, 1.17842720399 )}] \crossThree;
					\draw[blue, shift={( 185, 1.19399276319 )}] \crossThree;
					\draw[blue, shift={( 190, 1.19475220208 )}] \crossThree;
					\draw[blue, shift={( 195, 1.21773108189 )}] \crossThree;
					\draw[blue, shift={( 200, 1.1977138871 )}] \crossThree;
									
					\draw[blue!75, loosely dotted]( 5, 0.999999908761 ) -- ( 10, 1.00691641152 ) -- ( 15, 1.00740536496 ) -- ( 20, 1.0275869019 ) -- ( 25, 1.01781063015 ) -- ( 30, 1.02194863555 ) -- ( 35, 1.026377895 ) -- ( 40, 1.05259411188 ) -- ( 45, 1.03979661758 ) -- ( 50, 1.0587165705 ) -- ( 55, 1.0352089113 ) -- ( 60, 1.03887748681 ) -- ( 65, 1.0535980836 ) -- ( 70, 1.05225900334 ) -- ( 75, 1.0463456191 ) ( 80, 1.16293292358 ) -- ( 85, 1.1809613864 ) -- ( 90, 1.1719938778 ) -- ( 95, 1.17793479145 ) -- ( 100, 1.19043796953 ) -- ( 105, 1.18459917955 ) -- ( 110, 1.18318123982 ) -- ( 115, 1.18235186487 ) -- ( 120, 1.18447611579 ) -- ( 125, 1.17775189614 ) -- ( 130, 1.17055269871 ) -- ( 135, 1.18485016753 ) -- ( 140, 1.18888672386 ) -- ( 145, 1.1929594469 ) -- ( 150, 1.19506684666 ) -- ( 155, 1.18726519043 ) -- ( 160, 1.19058915527 ) -- ( 165, 1.19360802093 ) -- ( 170, 1.1903302 ) -- ( 175, 1.1871565849 ) -- ( 180, 1.17842720399 ) -- ( 185, 1.19399276319 ) -- ( 190, 1.19475220208 ) -- ( 195, 1.21773108189 ) -- ( 200, 1.1977138871 );

					\draw[magenta, shift={( 5, 0.999999908761 )}] \crossFour;
					\draw[magenta, shift={( 10, 1.00000003264 )}] \crossFour;
					\draw[magenta, shift={( 15, 1.00000001335 )}] \crossFour;
					\draw[magenta, shift={( 20, 0.999999994143 )}] \crossFour;
					\draw[magenta, shift={( 25, 1.00026906241 )}] \crossFour;
					\draw[magenta, shift={( 30, 1.02370674299 )}] \crossFour;
					\draw[magenta, shift={( 35, 1.01712475142 )}] \crossFour;
					\draw[magenta, shift={( 40, 1.00672966166 )}] \crossFour;
					\draw[magenta, shift={( 45, 1.01458265175 )}] \crossFour;
					\draw[magenta, shift={( 50, 1.02926017559 )}] \crossFour;
					\draw[magenta, shift={( 55, 1.01559159764 )}] \crossFour;
					\draw[magenta, shift={( 60, 1.02103884039 )}] \crossFour;
					\draw[magenta, shift={( 65, 1.02934726032 )}] \crossFour;
					\draw[magenta, shift={( 70, 1.03098051482 )}] \crossFour;
					\draw[magenta, shift={( 75, 1.02879335113 )}] \crossFour;
					\draw[magenta, shift={( 80, 1.1376646549 )}] \crossFour;
					\draw[magenta, shift={( 85, 1.14396735364 )}] \crossFour;
					\draw[magenta, shift={( 90, 1.14336832725 )}] \crossFour;
					\draw[magenta, shift={( 95, 1.14420165654 )}] \crossFour;
					\draw[magenta, shift={( 100, 1.14960078139 )}] \crossFour;
					\draw[magenta, shift={( 105, 1.16093569147 )}] \crossFour;
					\draw[magenta, shift={( 110, 1.14534201049 )}] \crossFour;
					\draw[magenta, shift={( 115, 1.16012594285 )}] \crossFour;
					\draw[magenta, shift={( 120, 1.16358683506 )}] \crossFour;
					\draw[magenta, shift={( 125, 1.15400693601 )}] \crossFour;
					\draw[magenta, shift={( 130, 1.1445409843 )}] \crossFour;
					\draw[magenta, shift={( 135, 1.16094824611 )}] \crossFour;
					\draw[magenta, shift={( 140, 1.16169921851 )}] \crossFour;
					\draw[magenta, shift={( 145, 1.16444059898 )}] \crossFour;
					\draw[magenta, shift={( 150, 1.16850686158 )}] \crossFour;
					\draw[magenta, shift={( 155, 1.1669263966 )}] \crossFour;
					\draw[magenta, shift={( 160, 1.17406390018 )}] \crossFour;
					\draw[magenta, shift={( 165, 1.16727735763 )}] \crossFour;
					\draw[magenta, shift={( 170, 1.17087284465 )}] \crossFour;
					\draw[magenta, shift={( 175, 1.16672100124 )}] \crossFour;
					\draw[magenta, shift={( 180, 1.15967557426 )}] \crossFour;
					\draw[magenta, shift={( 185, 1.17777886357 )}] \crossFour;
					\draw[magenta, shift={( 190, 1.17588250922 )}] \crossFour;
					\draw[magenta, shift={( 195, 1.18517337369 )}] \crossFour;
					\draw[magenta, shift={( 200, 1.17606072963 )}] \crossFour;
									
					\draw[magenta!75, dash pattern={on 1pt off 1pt}]( 5, 0.999999908761 ) -- ( 10, 1.00000003264 ) -- ( 15, 1.00000001335 ) -- ( 20, 0.999999994143 ) -- ( 25, 1.00026906241 ) -- ( 30, 1.02370674299 ) -- ( 35, 1.01712475142 ) -- ( 40, 1.00672966166 ) -- ( 45, 1.01458265175 ) -- ( 50, 1.02926017559 ) -- ( 55, 1.01559159764 ) -- ( 60, 1.02103884039 ) -- ( 65, 1.02934726032 ) -- ( 70, 1.03098051482 ) -- ( 75, 1.02879335113 ) ( 80, 1.1376646549 ) -- ( 85, 1.14396735364 ) -- ( 90, 1.14336832725 ) -- ( 95, 1.14420165654 ) -- ( 100, 1.14960078139 ) -- ( 105, 1.16093569147 ) -- ( 110, 1.14534201049 ) -- ( 115, 1.16012594285 ) -- ( 120, 1.16358683506 ) -- ( 125, 1.15400693601 ) -- ( 130, 1.1445409843 ) -- ( 135, 1.16094824611 ) -- ( 140, 1.16169921851 ) -- ( 145, 1.16444059898 ) -- ( 150, 1.16850686158 ) -- ( 155, 1.1669263966 ) -- ( 160, 1.17406390018 ) -- ( 165, 1.16727735763 ) -- ( 170, 1.17087284465 ) -- ( 175, 1.16672100124 ) -- ( 180, 1.15967557426 ) -- ( 185, 1.17777886357 ) -- ( 190, 1.17588250922 ) -- ( 195, 1.18517337369 ) -- ( 200, 1.17606072963 );

					\draw[black!50!brown, shift={( 5, 0.999999908761 )}] \crossFive;
					\draw[black!50!brown, shift={( 10, 1.00000003264 )}] \crossFive;
					\draw[black!50!brown, shift={( 15, 1.00000001335 )}] \crossFive;
					\draw[black!50!brown, shift={( 20, 0.999999994143 )}] \crossFive;
					\draw[black!50!brown, shift={( 25, 1.00000004341 )}] \crossFive;
					\draw[black!50!brown, shift={( 30, 1.00289226341 )}] \crossFive;
					\draw[black!50!brown, shift={( 35, 1.00410109084 )}] \crossFive;
					\draw[black!50!brown, shift={( 40, 1.0038812887 )}] \crossFive;
					\draw[black!50!brown, shift={( 45, 1.01187541742 )}] \crossFive;
					\draw[black!50!brown, shift={( 50, 1.03051744266 )}] \crossFive;
					\draw[black!50!brown, shift={( 55, 1.01420830899 )}] \crossFive;
					\draw[black!50!brown, shift={( 60, 1.01103997131 )}] \crossFive;
					\draw[black!50!brown, shift={( 65, 1.0159065612 )}] \crossFive;
					\draw[black!50!brown, shift={( 70, 1.01810381119 )}] \crossFive;
					\draw[black!50!brown, shift={( 75, 1.01922997434 )}] \crossFive;
					\draw[black!50!brown, shift={( 80, 1.13424432364 )}] \crossFive;
					\draw[black!50!brown, shift={( 85, 1.14250738343 )}] \crossFive;
					\draw[black!50!brown, shift={( 90, 1.14036613917 )}] \crossFive;
					\draw[black!50!brown, shift={( 95, 1.1268905656 )}] \crossFive;
					\draw[black!50!brown, shift={( 100, 1.14677932684 )}] \crossFive;
					\draw[black!50!brown, shift={( 105, 1.15752258091 )}] \crossFive;
					\draw[black!50!brown, shift={( 110, 1.14530981519 )}] \crossFive;
					\draw[black!50!brown, shift={( 115, 1.1480448308 )}] \crossFive;
					\draw[black!50!brown, shift={( 120, 1.14617891166 )}] \crossFive;
					\draw[black!50!brown, shift={( 125, 1.1510133606 )}] \crossFive;
					\draw[black!50!brown, shift={( 130, 1.14036629543 )}] \crossFive;
					\draw[black!50!brown, shift={( 135, 1.16150843278 )}] \crossFive;
					\draw[black!50!brown, shift={( 140, 1.14585359503 )}] \crossFive;
					\draw[black!50!brown, shift={( 145, 1.15096527021 )}] \crossFive;
					\draw[black!50!brown, shift={( 150, 1.15646444818 )}] \crossFive;
					\draw[black!50!brown, shift={( 155, 1.16134873787 )}] \crossFive;
					\draw[black!50!brown, shift={( 160, 1.16308017702 )}] \crossFive;
					\draw[black!50!brown, shift={( 165, 1.16329107937 )}] \crossFive;
					\draw[black!50!brown, shift={( 170, 1.16786481788 )}] \crossFive;
					\draw[black!50!brown, shift={( 175, 1.15946105787 )}] \crossFive;
					\draw[black!50!brown, shift={( 180, 1.15869969967 )}] \crossFive;
					\draw[black!50!brown, shift={( 185, 1.16240466628 )}] \crossFive;
					\draw[black!50!brown, shift={( 190, 1.1715538122 )}] \crossFive;
					\draw[black!50!brown, shift={( 195, 1.17357139912 )}] \crossFive;
					\draw[black!50!brown, shift={( 200, 1.16980729677 )}] \crossFive;
									
					\draw[black!50!brown!75, dash pattern={on 1pt off 2pt}]( 5, 0.999999908761 ) -- ( 10, 1.00000003264 ) -- ( 15, 1.00000001335 ) -- ( 20, 0.999999994143 ) -- ( 25, 1.00000004341 ) -- ( 30, 1.00289226341 ) -- ( 35, 1.00410109084 ) -- ( 40, 1.0038812887 ) -- ( 45, 1.01187541742 ) -- ( 50, 1.03051744266 ) -- ( 55, 1.01420830899 ) -- ( 60, 1.01103997131 ) -- ( 65, 1.0159065612 ) -- ( 70, 1.01810381119 ) -- ( 75, 1.01922997434 ) ( 80, 1.13424432364 ) -- ( 85, 1.14250738343 ) -- ( 90, 1.14036613917 ) -- ( 95, 1.1268905656 ) -- ( 100, 1.14677932684 ) -- ( 105, 1.15752258091 ) -- ( 110, 1.14530981519 ) -- ( 115, 1.1480448308 ) -- ( 120, 1.14617891166 ) -- ( 125, 1.1510133606 ) -- ( 130, 1.14036629543 ) -- ( 135, 1.16150843278 ) -- ( 140, 1.14585359503 ) -- ( 145, 1.15096527021 ) -- ( 150, 1.15646444818 ) -- ( 155, 1.16134873787 ) -- ( 160, 1.16308017702 ) -- ( 165, 1.16329107937 ) -- ( 170, 1.16786481788 ) -- ( 175, 1.15946105787 ) -- ( 180, 1.15869969967 ) -- ( 185, 1.16240466628 ) -- ( 190, 1.1715538122 ) -- ( 195, 1.17357139912 ) -- ( 200, 1.16980729677 );

					\draw[black!25!teal, shift={( 5, 0.999999908761 )}] \crossSix;
					\draw[black!25!teal, shift={( 10, 1.07231990758 )}] \crossSix;
					\draw[black!25!teal, shift={( 15, 1.04855638347 )}] \crossSix;
					\draw[black!25!teal, shift={( 20, 1.05649610152 )}] \crossSix;
					\draw[black!25!teal, shift={( 25, 1.04690776966 )}] \crossSix;
					\draw[black!25!teal, shift={( 30, 1.06682104282 )}] \crossSix;
					\draw[black!25!teal, shift={( 35, 1.06885313171 )}] \crossSix;
					\draw[black!25!teal, shift={( 40, 1.08375334806 )}] \crossSix;
					\draw[black!25!teal, shift={( 45, 1.07904442174 )}] \crossSix;
					\draw[black!25!teal, shift={( 50, 1.10203612144 )}] \crossSix;
					\draw[black!25!teal, shift={( 55, 1.07605241581 )}] \crossSix;
					\draw[black!25!teal, shift={( 60, 1.07189388285 )}] \crossSix;
					\draw[black!25!teal, shift={( 65, 1.09019426897 )}] \crossSix;
					\draw[black!25!teal, shift={( 70, 1.10200907491 )}] \crossSix;
					\draw[black!25!teal, shift={( 75, 1.08498701218 )}] \crossSix;
					\draw[black!25!teal, shift={( 80, 1.19748265322 )}] \crossSix;
					\draw[black!25!teal, shift={( 85, 1.21106225023 )}] \crossSix;
					\draw[black!25!teal, shift={( 90, 1.20387596 )}] \crossSix;
					\draw[black!25!teal, shift={( 95, 1.22783599535 )}] \crossSix;
					\draw[black!25!teal, shift={( 100, 1.23687040859 )}] \crossSix;
					\draw[black!25!teal, shift={( 105, 1.21082077131 )}] \crossSix;
					\draw[black!25!teal, shift={( 110, 1.20853362602 )}] \crossSix;
					\draw[black!25!teal, shift={( 115, 1.22253137475 )}] \crossSix;
					\draw[black!25!teal, shift={( 120, 1.20708866019 )}] \crossSix;
					\draw[black!25!teal, shift={( 125, 1.19904544028 )}] \crossSix;
					\draw[black!25!teal, shift={( 130, 1.2042950876 )}] \crossSix;
					\draw[black!25!teal, shift={( 135, 1.23026884653 )}] \crossSix;
					\draw[black!25!teal, shift={( 140, 1.22777004184 )}] \crossSix;
					\draw[black!25!teal, shift={( 145, 1.22810403821 )}] \crossSix;
					\draw[black!25!teal, shift={( 150, 1.22022127571 )}] \crossSix;
					\draw[black!25!teal, shift={( 155, 1.20874285013 )}] \crossSix;
					\draw[black!25!teal, shift={( 160, 1.23768225969 )}] \crossSix;
					\draw[black!25!teal, shift={( 165, 1.23750368052 )}] \crossSix;
					\draw[black!25!teal, shift={( 170, 1.2247328549 )}] \crossSix;
					\draw[black!25!teal, shift={( 175, 1.21510689373 )}] \crossSix;
					\draw[black!25!teal, shift={( 180, 1.21377280087 )}] \crossSix;
					\draw[black!25!teal, shift={( 185, 1.22527131171 )}] \crossSix;
					\draw[black!25!teal, shift={( 190, 1.21761029053 )}] \crossSix;
					\draw[black!25!teal, shift={( 195, 1.25481140411 )}] \crossSix;
					\draw[black!25!teal, shift={( 200, 1.22774714568 )}] \crossSix;
									
					\draw[black!25!teal!75, dash pattern={on 1pt off 3pt}]( 5, 0.999999908761 ) -- ( 10, 1.07231990758 ) -- ( 15, 1.04855638347 ) -- ( 20, 1.05649610152 ) -- ( 25, 1.04690776966 ) -- ( 30, 1.06682104282 ) -- ( 35, 1.06885313171 ) -- ( 40, 1.08375334806 ) -- ( 45, 1.07904442174 ) -- ( 50, 1.10203612144 ) -- ( 55, 1.07605241581 ) -- ( 60, 1.07189388285 ) -- ( 65, 1.09019426897 ) -- ( 70, 1.10200907491 ) -- ( 75, 1.08498701218 ) ( 80, 1.19748265322 ) -- ( 85, 1.21106225023 ) -- ( 90, 1.20387596 ) -- ( 95, 1.22783599535 ) -- ( 100, 1.23687040859 ) -- ( 105, 1.21082077131 ) -- ( 110, 1.20853362602 ) -- ( 115, 1.22253137475 ) -- ( 120, 1.20708866019 ) -- ( 125, 1.19904544028 ) -- ( 130, 1.2042950876 ) -- ( 135, 1.23026884653 ) -- ( 140, 1.22777004184 ) -- ( 145, 1.22810403821 ) -- ( 150, 1.22022127571 ) -- ( 155, 1.20874285013 ) -- ( 160, 1.23768225969 ) -- ( 165, 1.23750368052 ) -- ( 170, 1.2247328549 ) -- ( 175, 1.21510689373 ) -- ( 180, 1.21377280087 ) -- ( 185, 1.22527131171 ) -- ( 190, 1.21761029053 ) -- ( 195, 1.25481140411 ) -- ( 200, 1.22774714568 );

				\end{scope}
				
				\def\axisAdditionalLengthPlusTikzX{\axisAdditionalLengthPlus / \xscaleTikz}
				\def\axisAdditionalLengthMinusTikzX{\axisAdditionalLengthMinus / \xscaleTikz}
				\draw[arrow] (-\axisAdditionalLengthMinusTikzX, 0) -- (200, 0)  -- +(\axisAdditionalLengthPlusTikzX, 0) node[right] {\xAxis};
				\def\xTotalLengthPlus{200+\axisAdditionalLengthPlusTikzX}
				\draw (\xTotalLengthPlus, 0) node[right] {$\qquad$};
				\def\axisAdditionalLengthPlusTikzY{\axisAdditionalLengthPlus / \yscaleTikz}
				\def\axisAdditionalLengthMinusTikzY{\axisAdditionalLengthMinus / \yscaleTikz}
				\draw[arrow] (0, -\axisAdditionalLengthMinusTikzY) -- (0, 0.3 ) -- +(0, \axisAdditionalLengthPlusTikzY) node[above, yshift=-0.15cm] {\yAxis};
				
				\def\axisLabelTikzY{\axisLabel / \yscaleTikz}
				\foreach \pos in {0, 20, 40, 60, 80, 100, 120, 140, 160, 180, 200} \draw[shift={(\pos, 0)}] (0, \axisLabelTikzY) -- (0, -\axisLabelTikzY) node[below] {$\pos$};
				
				\def\axisLabelTikzX{\axisLabel / \xscaleTikz}
				\begin{scope}[shift={(0, -1)}]
					\draw[shift={(0,1 )}] (\axisLabelTikzX, 0) -- (-\axisLabelTikzX, 0) node[left] {$1$};
					\draw[shift={(0,1.1 )}] (\axisLabelTikzX, 0) -- (-\axisLabelTikzX, 0) node[left] {$1.1$};
					\draw[shift={(0,1.2 )}] (\axisLabelTikzX, 0) -- (-\axisLabelTikzX, 0) node[left] {$1.2$};
					\draw[shift={(0,1.3 )}] (\axisLabelTikzX, 0) -- (-\axisLabelTikzX, 0) node[left] {$1.3$};
				\end{scope}
			\end{tikzpicture}
		\AngleTSPInstances
		\vspace{-0.6cm}
		
			\begin{tikzpicture}[xscale=\xscale, yscale=\yscaleSmall]
				\pgfgettransformentries{\xscaleTikz}{\@tempa}{\@tempa}{\yscaleTikz}{\@tempa}{\@tempa}
			
				\draw[very thin, color=gray, xstep=20, ystep=0.1] (0, 0) grid (200, 0.3 );
				
				\begin{scope}[shift={(0, -1)}]
					\def\crossSizeX{\crossSize / \xscaleTikz};
					\def\crossSizeY{\crossSize / \yscaleTikz};
					
					\def\crossOne{(-\crossSizeX,-\crossSizeY) -- (\crossSizeX,\crossSizeY) (-\crossSizeX,\crossSizeY) -- (\crossSizeX,-\crossSizeY)};
					\def\crossTwo{(-\crossSizeX,0) -- (\crossSizeX,0) (0,\crossSizeY) -- (0,-\crossSizeY)};
					\def\crossThree{(0,0) -- (0,\crossSizeY) (0,0) -- (\crossSizeX,-\crossSizeY) (0,0) -- (-\crossSizeX,-\crossSizeY)};
					\def\crossFour{(0,0) -- (0,-\crossSizeY) (0,0) -- (\crossSizeX,\crossSizeY) (0,0) -- (-\crossSizeX,\crossSizeY)};
					\def\crossFive{(0,0) -- (\crossSizeX,0) (0,0) -- (-\crossSizeX,\crossSizeY) (0,0) -- (-\crossSizeX,-\crossSizeY)};
					\def\crossSix{(0,0) -- (-\crossSizeX,0) (0,0) -- (\crossSizeX,\crossSizeY) (0,0) -- (\crossSizeX,-\crossSizeY)};
					
					\draw[black, shift={( 5, 0.999999996966 )}] \crossOne;
					\draw[black, shift={( 10, 1.00019681156 )}] \crossOne;
					\draw[black, shift={( 15, 1.00238222981 )}] \crossOne;
					\draw[black, shift={( 20, 1.0013673183 )}] \crossOne;
					\draw[black, shift={( 25, 1.00360602001 )}] \crossOne;
					\draw[black, shift={( 30, 1.00242693635 )}] \crossOne;
					\draw[black, shift={( 35, 1.00209861905 )}] \crossOne;
					\draw[black, shift={( 40, 1.00443077626 )}] \crossOne;
					\draw[black, shift={( 45, 1.00372480954 )}] \crossOne;
					\draw[black, shift={( 50, 1.0092330973 )}] \crossOne;
					\draw[black, shift={( 55, 1.00806492994 )}] \crossOne;
					\draw[black, shift={( 60, 1.01224260045 )}] \crossOne;
					\draw[black, shift={( 65, 1.00991157082 )}] \crossOne;
					\draw[black, shift={( 70, 1.01065344516 )}] \crossOne;
					\draw[black, shift={( 75, 1.01547710396 )}] \crossOne;
					\draw[black, shift={( 80, 1.0190876865 )}] \crossOne;
					\draw[black, shift={( 85, 1.02153683598 )}] \crossOne;
					\draw[black, shift={( 90, 1.01973366669 )}] \crossOne;
					\draw[black, shift={( 95, 1.02118143504 )}] \crossOne;
					\draw[black, shift={( 100, 1.01871101505 )}] \crossOne;
					\draw[black, shift={( 105, 1.08481991434 )}] \crossOne;
					\draw[black, shift={( 110, 1.09621501817 )}] \crossOne;
					\draw[black, shift={( 115, 1.1036713264 )}] \crossOne;
					\draw[black, shift={( 120, 1.09955661518 )}] \crossOne;
					\draw[black, shift={( 125, 1.10052412291 )}] \crossOne;
					\draw[black, shift={( 130, 1.10773679015 )}] \crossOne;
					\draw[black, shift={( 135, 1.11331854492 )}] \crossOne;
					\draw[black, shift={( 140, 1.11270128318 )}] \crossOne;
					\draw[black, shift={( 145, 1.11124549416 )}] \crossOne;
					\draw[black, shift={( 150, 1.11132040311 )}] \crossOne;
					\draw[black, shift={( 155, 1.12247158465 )}] \crossOne;
					\draw[black, shift={( 160, 1.1224752582 )}] \crossOne;
					\draw[black, shift={( 165, 1.1228393803 )}] \crossOne;
					\draw[black, shift={( 170, 1.13290241899 )}] \crossOne;
					\draw[black, shift={( 175, 1.13002191952 )}] \crossOne;
					\draw[black, shift={( 180, 1.13382231911 )}] \crossOne;
					\draw[black, shift={( 185, 1.13518698703 )}] \crossOne;
					\draw[black, shift={( 190, 1.1362229858 )}] \crossOne;
					\draw[black, shift={( 195, 1.14024029521 )}] \crossOne;
					\draw[black, shift={( 200, 1.1422386282 )}] \crossOne;
					
					\draw[black!75, dotted] ( 5, 0.999999996966 ) -- ( 10, 1.00019681156 ) -- ( 15, 1.00238222981 ) -- ( 20, 1.0013673183 ) -- ( 25, 1.00360602001 ) -- ( 30, 1.00242693635 ) -- ( 35, 1.00209861905 ) -- ( 40, 1.00443077626 ) -- ( 45, 1.00372480954 ) -- ( 50, 1.0092330973 ) -- ( 55, 1.00806492994 ) -- ( 60, 1.01224260045 ) -- ( 65, 1.00991157082 ) -- ( 70, 1.01065344516 ) -- ( 75, 1.01547710396 ) -- ( 80, 1.0190876865 ) -- ( 85, 1.02153683598 ) -- ( 90, 1.01973366669 ) -- ( 95, 1.02118143504 ) -- ( 100, 1.01871101505 ) ( 105, 1.08481991434 ) -- ( 110, 1.09621501817 ) -- ( 115, 1.1036713264 ) -- ( 120, 1.09955661518 ) -- ( 125, 1.10052412291 ) -- ( 130, 1.10773679015 ) -- ( 135, 1.11331854492 ) -- ( 140, 1.11270128318 ) -- ( 145, 1.11124549416 ) -- ( 150, 1.11132040311 ) -- ( 155, 1.12247158465 ) -- ( 160, 1.1224752582 ) -- ( 165, 1.1228393803 ) -- ( 170, 1.13290241899 ) -- ( 175, 1.13002191952 ) -- ( 180, 1.13382231911 ) -- ( 185, 1.13518698703 ) -- ( 190, 1.1362229858 ) -- ( 195, 1.14024029521 ) -- ( 200, 1.1422386282 );

					\draw[blue, shift={( 5, 0.999999996966 )}] \crossThree;
					\draw[blue, shift={( 10, 1.00019681156 )}] \crossThree;
					\draw[blue, shift={( 15, 1.0000000041 )}] \crossThree;
					\draw[blue, shift={( 20, 1.00106623853 )}] \crossThree;
					\draw[blue, shift={( 25, 1.00099046477 )}] \crossThree;
					\draw[blue, shift={( 30, 1.00056041579 )}] \crossThree;
					\draw[blue, shift={( 35, 1.0020546197 )}] \crossThree;
					\draw[blue, shift={( 40, 1.00337365408 )}] \crossThree;
					\draw[blue, shift={( 45, 1.00289725686 )}] \crossThree;
					\draw[blue, shift={( 50, 1.00562542991 )}] \crossThree;
					\draw[blue, shift={( 55, 1.00575346663 )}] \crossThree;
					\draw[blue, shift={( 60, 1.00974085804 )}] \crossThree;
					\draw[blue, shift={( 65, 1.00768776734 )}] \crossThree;
					\draw[blue, shift={( 70, 1.00702395608 )}] \crossThree;
					\draw[blue, shift={( 75, 1.01172349268 )}] \crossThree;
					\draw[blue, shift={( 80, 1.01405984592 )}] \crossThree;
					\draw[blue, shift={( 85, 1.01702281655 )}] \crossThree;
					\draw[blue, shift={( 90, 1.0166689565 )}] \crossThree;
					\draw[blue, shift={( 95, 1.01602720138 )}] \crossThree;
					\draw[blue, shift={( 100, 1.0133965117 )}] \crossThree;
					\draw[blue, shift={( 105, 1.07472405787 )}] \crossThree;
					\draw[blue, shift={( 110, 1.0883989483 )}] \crossThree;
					\draw[blue, shift={( 115, 1.0955750328 )}] \crossThree;
					\draw[blue, shift={( 120, 1.09191320979 )}] \crossThree;
					\draw[blue, shift={( 125, 1.09238291565 )}] \crossThree;
					\draw[blue, shift={( 130, 1.09665632201 )}] \crossThree;
					\draw[blue, shift={( 135, 1.09637532785 )}] \crossThree;
					\draw[blue, shift={( 140, 1.10028406631 )}] \crossThree;
					\draw[blue, shift={( 145, 1.09649663868 )}] \crossThree;
					\draw[blue, shift={( 150, 1.10165442489 )}] \crossThree;
					\draw[blue, shift={( 155, 1.10908697059 )}] \crossThree;
					\draw[blue, shift={( 160, 1.11004268736 )}] \crossThree;
					\draw[blue, shift={( 165, 1.11045478849 )}] \crossThree;
					\draw[blue, shift={( 170, 1.12282281429 )}] \crossThree;
					\draw[blue, shift={( 175, 1.11518036655 )}] \crossThree;
					\draw[blue, shift={( 180, 1.12103306588 )}] \crossThree;
					\draw[blue, shift={( 185, 1.12309417788 )}] \crossThree;
					\draw[blue, shift={( 190, 1.12131237547 )}] \crossThree;
					\draw[blue, shift={( 195, 1.12465173354 )}] \crossThree;
					\draw[blue, shift={( 200, 1.12402658213 )}] \crossThree;
									
					\draw[blue!75, loosely dotted]( 5, 0.999999996966 ) -- ( 10, 1.00019681156 ) -- ( 15, 1.0000000041 ) -- ( 20, 1.00106623853 ) -- ( 25, 1.00099046477 ) -- ( 30, 1.00056041579 ) -- ( 35, 1.0020546197 ) -- ( 40, 1.00337365408 ) -- ( 45, 1.00289725686 ) -- ( 50, 1.00562542991 ) -- ( 55, 1.00575346663 ) -- ( 60, 1.00974085804 ) -- ( 65, 1.00768776734 ) -- ( 70, 1.00702395608 ) -- ( 75, 1.01172349268 ) -- ( 80, 1.01405984592 ) -- ( 85, 1.01702281655 ) -- ( 90, 1.0166689565 ) -- ( 95, 1.01602720138 ) -- ( 100, 1.0133965117 ) ( 105, 1.07472405787 ) -- ( 110, 1.0883989483 ) -- ( 115, 1.0955750328 ) -- ( 120, 1.09191320979 ) -- ( 125, 1.09238291565 ) -- ( 130, 1.09665632201 ) -- ( 135, 1.09637532785 ) -- ( 140, 1.10028406631 ) -- ( 145, 1.09649663868 ) -- ( 150, 1.10165442489 ) -- ( 155, 1.10908697059 ) -- ( 160, 1.11004268736 ) -- ( 165, 1.11045478849 ) -- ( 170, 1.12282281429 ) -- ( 175, 1.11518036655 ) -- ( 180, 1.12103306588 ) -- ( 185, 1.12309417788 ) -- ( 190, 1.12131237547 ) -- ( 195, 1.12465173354 ) -- ( 200, 1.12402658213 );

					\draw[magenta, shift={( 5, 0.999999996966 )}] \crossFour;
					\draw[magenta, shift={( 10, 1.00000000532 )}] \crossFour;
					\draw[magenta, shift={( 15, 1.0000000041 )}] \crossFour;
					\draw[magenta, shift={( 20, 1.00000000398 )}] \crossFour;
					\draw[magenta, shift={( 25, 0.99999999396 )}] \crossFour;
					\draw[magenta, shift={( 30, 0.99999999756 )}] \crossFour;
					\draw[magenta, shift={( 35, 1.00047848067 )}] \crossFour;
					\draw[magenta, shift={( 40, 1.00059450144 )}] \crossFour;
					\draw[magenta, shift={( 45, 1.00001452246 )}] \crossFour;
					\draw[magenta, shift={( 50, 1.00165329996 )}] \crossFour;
					\draw[magenta, shift={( 55, 1.00391794269 )}] \crossFour;
					\draw[magenta, shift={( 60, 1.0052015694 )}] \crossFour;
					\draw[magenta, shift={( 65, 1.00251910598 )}] \crossFour;
					\draw[magenta, shift={( 70, 1.00199760288 )}] \crossFour;
					\draw[magenta, shift={( 75, 1.00651813857 )}] \crossFour;
					\draw[magenta, shift={( 80, 1.00562529032 )}] \crossFour;
					\draw[magenta, shift={( 85, 1.00521328622 )}] \crossFour;
					\draw[magenta, shift={( 90, 1.00784830824 )}] \crossFour;
					\draw[magenta, shift={( 95, 1.00613276551 )}] \crossFour;
					\draw[magenta, shift={( 100, 1.00642168627 )}] \crossFour;
					\draw[magenta, shift={( 105, 1.06678729025 )}] \crossFour;
					\draw[magenta, shift={( 110, 1.07507000584 )}] \crossFour;
					\draw[magenta, shift={( 115, 1.08044303722 )}] \crossFour;
					\draw[magenta, shift={( 120, 1.08134616319 )}] \crossFour;
					\draw[magenta, shift={( 125, 1.07999369763 )}] \crossFour;
					\draw[magenta, shift={( 130, 1.08622156911 )}] \crossFour;
					\draw[magenta, shift={( 135, 1.08654182759 )}] \crossFour;
					\draw[magenta, shift={( 140, 1.08496314147 )}] \crossFour;
					\draw[magenta, shift={( 145, 1.08426385441 )}] \crossFour;
					\draw[magenta, shift={( 150, 1.08595417069 )}] \crossFour;
					\draw[magenta, shift={( 155, 1.08994176489 )}] \crossFour;
					\draw[magenta, shift={( 160, 1.08775064297 )}] \crossFour;
					\draw[magenta, shift={( 165, 1.09535503584 )}] \crossFour;
					\draw[magenta, shift={( 170, 1.09725832171 )}] \crossFour;
					\draw[magenta, shift={( 175, 1.09485124907 )}] \crossFour;
					\draw[magenta, shift={( 180, 1.09688361073 )}] \crossFour;
					\draw[magenta, shift={( 185, 1.09629238514 )}] \crossFour;
					\draw[magenta, shift={( 190, 1.09878459709 )}] \crossFour;
					\draw[magenta, shift={( 195, 1.09743760962 )}] \crossFour;
					\draw[magenta, shift={( 200, 1.10033384032 )}] \crossFour;
									
					\draw[magenta!75, dash pattern={on 1pt off 1pt}]( 5, 0.999999996966 ) -- ( 10, 1.00000000532 ) -- ( 15, 1.0000000041 ) -- ( 20, 1.00000000398 ) -- ( 25, 0.99999999396 ) -- ( 30, 0.99999999756 ) -- ( 35, 1.00047848067 ) -- ( 40, 1.00059450144 ) -- ( 45, 1.00001452246 ) -- ( 50, 1.00165329996 ) -- ( 55, 1.00391794269 ) -- ( 60, 1.0052015694 ) -- ( 65, 1.00251910598 ) -- ( 70, 1.00199760288 ) -- ( 75, 1.00651813857 ) -- ( 80, 1.00562529032 ) -- ( 85, 1.00521328622 ) -- ( 90, 1.00784830824 ) -- ( 95, 1.00613276551 ) -- ( 100, 1.00642168627 ) ( 105, 1.06678729025 ) -- ( 110, 1.07507000584 ) -- ( 115, 1.08044303722 ) -- ( 120, 1.08134616319 ) -- ( 125, 1.07999369763 ) -- ( 130, 1.08622156911 ) -- ( 135, 1.08654182759 ) -- ( 140, 1.08496314147 ) -- ( 145, 1.08426385441 ) -- ( 150, 1.08595417069 ) -- ( 155, 1.08994176489 ) -- ( 160, 1.08775064297 ) -- ( 165, 1.09535503584 ) -- ( 170, 1.09725832171 ) -- ( 175, 1.09485124907 ) -- ( 180, 1.09688361073 ) -- ( 185, 1.09629238514 ) -- ( 190, 1.09878459709 ) -- ( 195, 1.09743760962 ) -- ( 200, 1.10033384032 );

					\draw[black!50!brown, shift={( 5, 0.999999996966 )}] \crossFive;
					\draw[black!50!brown, shift={( 10, 1.00000000532 )}] \crossFive;
					\draw[black!50!brown, shift={( 15, 1.0000000041 )}] \crossFive;
					\draw[black!50!brown, shift={( 20, 1.00000000398 )}] \crossFive;
					\draw[black!50!brown, shift={( 25, 0.99999999396 )}] \crossFive;
					\draw[black!50!brown, shift={( 30, 0.99999999756 )}] \crossFive;
					\draw[black!50!brown, shift={( 35, 1.00000000181 )}] \crossFive;
					\draw[black!50!brown, shift={( 40, 1.00059261866 )}] \crossFive;
					\draw[black!50!brown, shift={( 45, 1.00001452246 )}] \crossFive;
					\draw[black!50!brown, shift={( 50, 1.00051423513 )}] \crossFive;
					\draw[black!50!brown, shift={( 55, 1.0037041988 )}] \crossFive;
					\draw[black!50!brown, shift={( 60, 1.00583965902 )}] \crossFive;
					\draw[black!50!brown, shift={( 65, 1.00251910598 )}] \crossFive;
					\draw[black!50!brown, shift={( 70, 1.00251257912 )}] \crossFive;
					\draw[black!50!brown, shift={( 75, 1.00421315752 )}] \crossFive;
					\draw[black!50!brown, shift={( 80, 1.00389265857 )}] \crossFive;
					\draw[black!50!brown, shift={( 85, 1.00492737808 )}] \crossFive;
					\draw[black!50!brown, shift={( 90, 1.00673410935 )}] \crossFive;
					\draw[black!50!brown, shift={( 95, 1.00473037509 )}] \crossFive;
					\draw[black!50!brown, shift={( 100, 1.00477024773 )}] \crossFive;
					\draw[black!50!brown, shift={( 105, 1.0646088256 )}] \crossFive;
					\draw[black!50!brown, shift={( 110, 1.07396896863 )}] \crossFive;
					\draw[black!50!brown, shift={( 115, 1.0808320693 )}] \crossFive;
					\draw[black!50!brown, shift={( 120, 1.07975905236 )}] \crossFive;
					\draw[black!50!brown, shift={( 125, 1.07860851761 )}] \crossFive;
					\draw[black!50!brown, shift={( 130, 1.08231793952 )}] \crossFive;
					\draw[black!50!brown, shift={( 135, 1.08224494322 )}] \crossFive;
					\draw[black!50!brown, shift={( 140, 1.08049761078 )}] \crossFive;
					\draw[black!50!brown, shift={( 145, 1.0814380764 )}] \crossFive;
					\draw[black!50!brown, shift={( 150, 1.08374104015 )}] \crossFive;
					\draw[black!50!brown, shift={( 155, 1.0892154885 )}] \crossFive;
					\draw[black!50!brown, shift={( 160, 1.08737914275 )}] \crossFive;
					\draw[black!50!brown, shift={( 165, 1.08881174499 )}] \crossFive;
					\draw[black!50!brown, shift={( 170, 1.09235482101 )}] \crossFive;
					\draw[black!50!brown, shift={( 175, 1.09101330748 )}] \crossFive;
					\draw[black!50!brown, shift={( 180, 1.09420002075 )}] \crossFive;
					\draw[black!50!brown, shift={( 185, 1.09327696782 )}] \crossFive;
					\draw[black!50!brown, shift={( 190, 1.09489436178 )}] \crossFive;
					\draw[black!50!brown, shift={( 195, 1.09785197882 )}] \crossFive;
					\draw[black!50!brown, shift={( 200, 1.0953503197 )}] \crossFive;
									
					\draw[black!50!brown!75, dash pattern={on 1pt off 2pt}]( 5, 0.999999996966 ) -- ( 10, 1.00000000532 ) -- ( 15, 1.0000000041 ) -- ( 20, 1.00000000398 ) -- ( 25, 0.99999999396 ) -- ( 30, 0.99999999756 ) -- ( 35, 1.00000000181 ) -- ( 40, 1.00059261866 ) -- ( 45, 1.00001452246 ) -- ( 50, 1.00051423513 ) -- ( 55, 1.0037041988 ) -- ( 60, 1.00583965902 ) -- ( 65, 1.00251910598 ) -- ( 70, 1.00251257912 ) -- ( 75, 1.00421315752 ) -- ( 80, 1.00389265857 ) -- ( 85, 1.00492737808 ) -- ( 90, 1.00673410935 ) -- ( 95, 1.00473037509 ) -- ( 100, 1.00477024773 ) ( 105, 1.0646088256 ) -- ( 110, 1.07396896863 ) -- ( 115, 1.0808320693 ) -- ( 120, 1.07975905236 ) -- ( 125, 1.07860851761 ) -- ( 130, 1.08231793952 ) -- ( 135, 1.08224494322 ) -- ( 140, 1.08049761078 ) -- ( 145, 1.0814380764 ) -- ( 150, 1.08374104015 ) -- ( 155, 1.0892154885 ) -- ( 160, 1.08737914275 ) -- ( 165, 1.08881174499 ) -- ( 170, 1.09235482101 ) -- ( 175, 1.09101330748 ) -- ( 180, 1.09420002075 ) -- ( 185, 1.09327696782 ) -- ( 190, 1.09489436178 ) -- ( 195, 1.09785197882 ) -- ( 200, 1.0953503197 );

					\draw[black!25!teal, shift={( 5, 0.999999996966 )}] \crossSix;
					\draw[black!25!teal, shift={( 10, 1.00019681156 )}] \crossSix;
					\draw[black!25!teal, shift={( 15, 1.00238222981 )}] \crossSix;
					\draw[black!25!teal, shift={( 20, 1.0013673183 )}] \crossSix;
					\draw[black!25!teal, shift={( 25, 1.00360602001 )}] \crossSix;
					\draw[black!25!teal, shift={( 30, 1.00242693635 )}] \crossSix;
					\draw[black!25!teal, shift={( 35, 1.0020546197 )}] \crossSix;
					\draw[black!25!teal, shift={( 40, 1.00388876967 )}] \crossSix;
					\draw[black!25!teal, shift={( 45, 1.00372480954 )}] \crossSix;
					\draw[black!25!teal, shift={( 50, 1.00804182954 )}] \crossSix;
					\draw[black!25!teal, shift={( 55, 1.00705869328 )}] \crossSix;
					\draw[black!25!teal, shift={( 60, 1.01200409298 )}] \crossSix;
					\draw[black!25!teal, shift={( 65, 1.00917345345 )}] \crossSix;
					\draw[black!25!teal, shift={( 70, 1.00956058109 )}] \crossSix;
					\draw[black!25!teal, shift={( 75, 1.0135894177 )}] \crossSix;
					\draw[black!25!teal, shift={( 80, 1.01725539539 )}] \crossSix;
					\draw[black!25!teal, shift={( 85, 1.01930793568 )}] \crossSix;
					\draw[black!25!teal, shift={( 90, 1.01879957283 )}] \crossSix;
					\draw[black!25!teal, shift={( 95, 1.01948700475 )}] \crossSix;
					\draw[black!25!teal, shift={( 100, 1.01772688424 )}] \crossSix;
					\draw[black!25!teal, shift={( 105, 1.08134251074 )}] \crossSix;
					\draw[black!25!teal, shift={( 110, 1.09455913125 )}] \crossSix;
					\draw[black!25!teal, shift={( 115, 1.10033711615 )}] \crossSix;
					\draw[black!25!teal, shift={( 120, 1.09770265867 )}] \crossSix;
					\draw[black!25!teal, shift={( 125, 1.09789582739 )}] \crossSix;
					\draw[black!25!teal, shift={( 130, 1.10182892405 )}] \crossSix;
					\draw[black!25!teal, shift={( 135, 1.10979387592 )}] \crossSix;
					\draw[black!25!teal, shift={( 140, 1.10988963158 )}] \crossSix;
					\draw[black!25!teal, shift={( 145, 1.10944948634 )}] \crossSix;
					\draw[black!25!teal, shift={( 150, 1.10728385995 )}] \crossSix;
					\draw[black!25!teal, shift={( 155, 1.11690099872 )}] \crossSix;
					\draw[black!25!teal, shift={( 160, 1.11935375586 )}] \crossSix;
					\draw[black!25!teal, shift={( 165, 1.11915952672 )}] \crossSix;
					\draw[black!25!teal, shift={( 170, 1.12885002631 )}] \crossSix;
					\draw[black!25!teal, shift={( 175, 1.12429571324 )}] \crossSix;
					\draw[black!25!teal, shift={( 180, 1.13038745366 )}] \crossSix;
					\draw[black!25!teal, shift={( 185, 1.13134795424 )}] \crossSix;
					\draw[black!25!teal, shift={( 190, 1.13077497567 )}] \crossSix;
					\draw[black!25!teal, shift={( 195, 1.13463387683 )}] \crossSix;
					\draw[black!25!teal, shift={( 200, 1.13707105631 )}] \crossSix;
									
					\draw[black!25!teal!75, dash pattern={on 1pt off 3pt}]( 5, 0.999999996966 ) -- ( 10, 1.00019681156 ) -- ( 15, 1.00238222981 ) -- ( 20, 1.0013673183 ) -- ( 25, 1.00360602001 ) -- ( 30, 1.00242693635 ) -- ( 35, 1.0020546197 ) -- ( 40, 1.00388876967 ) -- ( 45, 1.00372480954 ) -- ( 50, 1.00804182954 ) -- ( 55, 1.00705869328 ) -- ( 60, 1.01200409298 ) -- ( 65, 1.00917345345 ) -- ( 70, 1.00956058109 ) -- ( 75, 1.0135894177 ) -- ( 80, 1.01725539539 ) -- ( 85, 1.01930793568 ) -- ( 90, 1.01879957283 ) -- ( 95, 1.01948700475 ) -- ( 100, 1.01772688424 ) ( 105, 1.08134251074 ) -- ( 110, 1.09455913125 ) -- ( 115, 1.10033711615 ) -- ( 120, 1.09770265867 ) -- ( 125, 1.09789582739 ) -- ( 130, 1.10182892405 ) -- ( 135, 1.10979387592 ) -- ( 140, 1.10988963158 ) -- ( 145, 1.10944948634 ) -- ( 150, 1.10728385995 ) -- ( 155, 1.11690099872 ) -- ( 160, 1.11935375586 ) -- ( 165, 1.11915952672 ) -- ( 170, 1.12885002631 ) -- ( 175, 1.12429571324 ) -- ( 180, 1.13038745366 ) -- ( 185, 1.13134795424 ) -- ( 190, 1.13077497567 ) -- ( 195, 1.13463387683 ) -- ( 200, 1.13707105631 );

				\end{scope}
				
				\def\axisAdditionalLengthPlusTikzX{\axisAdditionalLengthPlus / \xscaleTikz}
				\def\axisAdditionalLengthMinusTikzX{\axisAdditionalLengthMinus / \xscaleTikz}
				\draw[arrow] (-\axisAdditionalLengthMinusTikzX, 0) -- (200, 0)  -- +(\axisAdditionalLengthPlusTikzX, 0) node[right] {\xAxis};
				\def\xTotalLengthPlus{200+\axisAdditionalLengthPlusTikzX}
				\draw (\xTotalLengthPlus, 0) node[right] {$\qquad$};
				\def\axisAdditionalLengthPlusTikzY{\axisAdditionalLengthPlus / \yscaleTikz}
				\def\axisAdditionalLengthMinusTikzY{\axisAdditionalLengthMinus / \yscaleTikz}
				\draw[arrow] (0, -\axisAdditionalLengthMinusTikzY) -- (0, 0.3 ) -- +(0, \axisAdditionalLengthPlusTikzY) node[above, yshift=-0.15cm] {\yAxis};
				
				\def\axisLabelTikzY{\axisLabel / \yscaleTikz}
				\foreach \pos in {0, 20, 40, 60, 80, 100, 120, 140, 160, 180, 200} \draw[shift={(\pos, 0)}] (0, \axisLabelTikzY) -- (0, -\axisLabelTikzY) node[below] {$\pos$};
				
				\def\axisLabelTikzX{\axisLabel / \xscaleTikz}
				\begin{scope}[shift={(0, -1)}]
					\draw[shift={(0,1 )}] (\axisLabelTikzX, 0) -- (-\axisLabelTikzX, 0) node[left] {$1$};
					\draw[shift={(0,1.1 )}] (\axisLabelTikzX, 0) -- (-\axisLabelTikzX, 0) node[left] {$1.1$};
					\draw[shift={(0,1.2 )}] (\axisLabelTikzX, 0) -- (-\axisLabelTikzX, 0) node[left] {$1.2$};
					\draw[shift={(0,1.3 )}] (\axisLabelTikzX, 0) -- (-\axisLabelTikzX, 0) node[left] {$1.3$};
								\end{scope}
			\end{tikzpicture}
		\AngleDistanceTSPInstances
		\vspace*{-0.1cm}
		\caption[improvement heuristics: \LPCTwoR{}, \LPCTwoR{}\TwoO{}, \LPCTwoR{}\ThreeO{}, \LPCTwoR{}\M{15}, \LPCTwoR{}\M{20}, \LPCTwoR{}\Lens{}]{
			improvement heuristics:\hspace*{-10cm}\\
			\begin{tikzpicture}[xscale=\xscale, yscale=2.0]
				\pgfgettransformentries{\xscaleTikz}{\@tempa}{\@tempa}{\yscaleTikz}{\@tempa}{\@tempa}
				\def\crossSizeX{\crossSize / \xscaleTikz};
				\def\crossSizeY{\crossSize / \yscaleTikz};
				
				\def\crossOne{(-\crossSizeX,-\crossSizeY) -- (\crossSizeX,\crossSizeY) (-\crossSizeX,\crossSizeY) -- (\crossSizeX,-\crossSizeY)};
				\def\crossTwo{(-\crossSizeX,0) -- (\crossSizeX,0) (0,\crossSizeY) -- (0,-\crossSizeY)};
				\def\crossThree{(0,0) -- (0,\crossSizeY) (0,0) -- (\crossSizeX,-\crossSizeY) (0,0) -- (-\crossSizeX,-\crossSizeY)};
				\def\crossFour{(0,0) -- (0,-\crossSizeY) (0,0) -- (\crossSizeX,\crossSizeY) (0,0) -- (-\crossSizeX,\crossSizeY)};
				\def\crossFive{(0,0) -- (\crossSizeX,0) (0,0) -- (-\crossSizeX,\crossSizeY) (0,0) -- (-\crossSizeX,-\crossSizeY)};
				\def\crossSix{(0,0) -- (-\crossSizeX,0) (0,0) -- (\crossSizeX,\crossSizeY) (0,0) -- (\crossSizeX,-\crossSizeY)};
				
				\draw[black, shift={(0, 0)}] \crossOne;
				\draw[black, shift={(5, 0)}] \crossOne;
				\draw[black, shift={(10, 0)}] \crossOne;
				\draw[black, shift={(15, 0)}] \crossOne;
				\draw[black!75, dotted] (-2.5, 0) -- (17.5, 0);
				\node at (40, 0) [minimum width=3cm] {\LPCTwoR{}};

				\draw[red, shift={(80, 0)}] \crossTwo;
				\draw[red, shift={(85, 0)}] \crossTwo;
				\draw[red, shift={(90, 0)}] \crossTwo;
				\draw[red, shift={(95, 0)}] \crossTwo;
				\draw[red!75, densely dotted] (77.5, 0) -- (97.5, 0);
				\node at (126.218, 0) [minimum width=3cm] {\LPCTwoR{}\TwoO{}};
				
				\draw[blue, shift={(160, 0)}] \crossThree;
				\draw[blue, shift={(165, 0)}] \crossThree;
				\draw[blue, shift={(170, 0)}] \crossThree;
				\draw[blue, shift={(175, 0)}] \crossThree;
				\draw[blue!75, loosely dotted] (157.5, 0) -- (177.5, 0);
				\node at (206.28, 0) [minimum width=2cm] {\LPCTwoR{}\ThreeO{}};

				\draw[magenta, shift={(0, -0.25)}] \crossFour;
				\draw[magenta, shift={(5, -0.25)}] \crossFour;
				\draw[magenta, shift={(10, -0.25)}] \crossFour;
				\draw[magenta, shift={(15, -0.25)}] \crossFour;
				\draw[magenta!75, dash pattern={on 1pt off 1pt}] (-2.5, -0.25) -- (17.5, -0.25);
				\node at (47.885, -0.25) [minimum width=3cm] {\LPCTwoR{}\M{15}};
				
				\draw[black!50!brown, shift={(80, -0.25)}] \crossFive;
				\draw[black!50!brown, shift={(85, -0.25)}] \crossFive;
				\draw[black!50!brown, shift={(90, -0.25)}] \crossFive;
				\draw[black!50!brown, shift={(95, -0.25)}] \crossFive;
				\draw[black!50!brown!75, dash pattern={on 1pt off 2pt}] (77.5, -0.25) -- (97.5, -0.25);
				\node at (127.6, -0.25) [minimum width=3cm] {\LPCTwoR{}\M{20}};
				
				\draw[black!25!teal, shift={(160, -0.25)}] \crossSix;
				\draw[black!25!teal, shift={(165, -0.25)}] \crossSix;
				\draw[black!25!teal, shift={(170, -0.25)}] \crossSix;
				\draw[black!25!teal, shift={(175, -0.25)}] \crossSix;
				\draw[black!25!teal!75, dash pattern={on 1pt off 3pt}] (157.5, -0.25) -- (177.5, -0.25);
				\node at (204.35, -0.25) [minimum width=3cm] {\LPCTwoR{}\Lens{15}};
			\end{tikzpicture}
		}
		\label{figure:FigureImprovementHeuristics}
	\end{figure}
			\end{comment:figures}
	
			Next, consider Figure~\ref{figure:FigureImprovementHeuristics}. We can see that the ranking is the same for the {\AngleTSP}- and {\AngleDistanceTSP}-instances: ordered from the worst to the best improvement heuristic we get \Lens{}, \TwoO{}, \ThreeO{}, \M{15} and \M{20}, where \Lens{} and \TwoO{} yield similar objective function values (note that \TwoO{} makes no sense for \NNSTwo{}; see Section~\ref{subsection:nearestNeighbourHeuristics} for more details). By considering concrete relations between the particular improvement heuristics, \ThreeO{} performs better for \AngleTSPInstances{} and can be placed between \TwoO{} and \M{15} as well as \M{20} for this group, for \AngleDistanceTSPInstances, however, \M{15} and \M{20} clearly beat all other approaches. 
					
			\begin{comment:figures}
	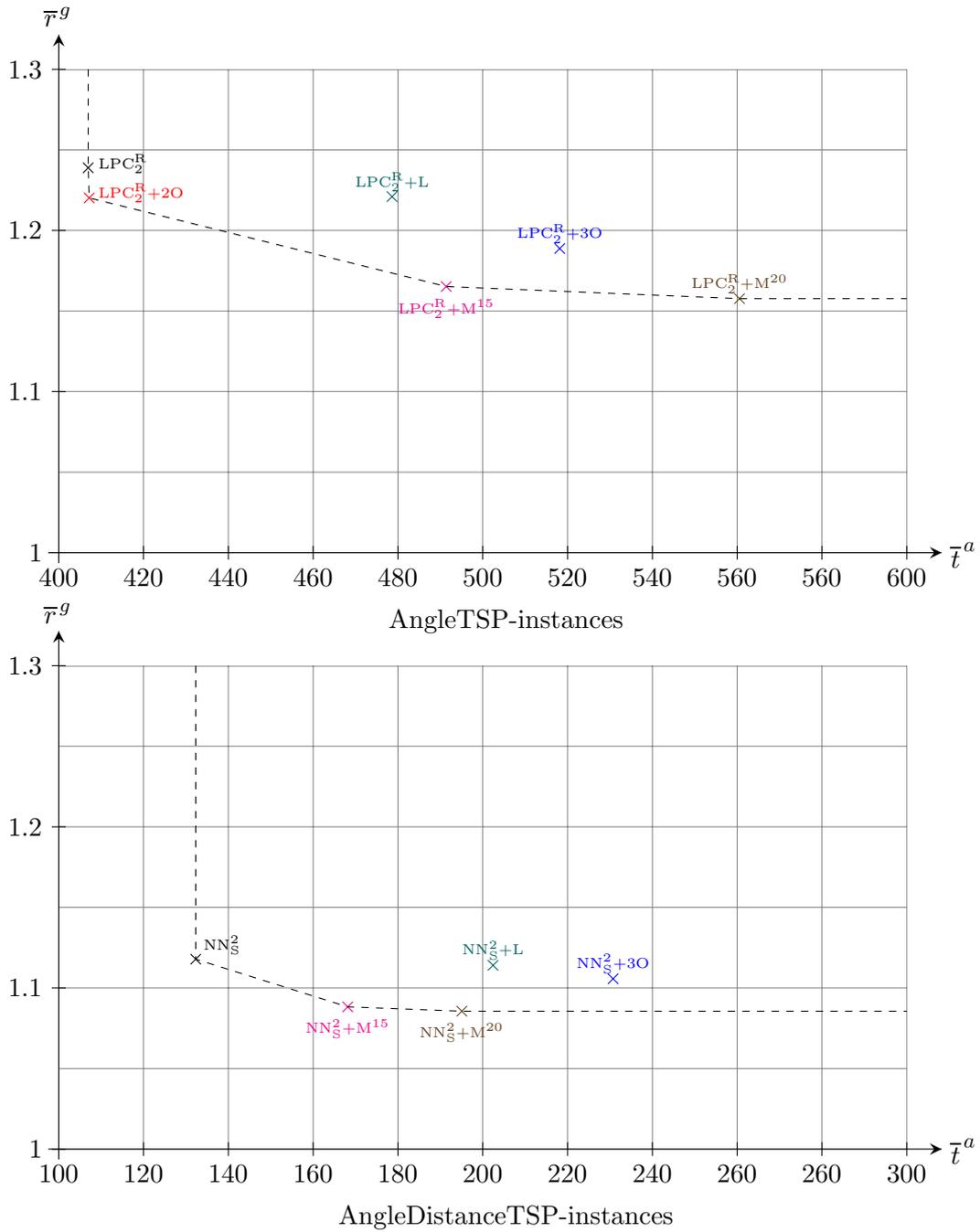
\begin{figure}[htbp!]
		\centering
			\begin{tikzpicture}[xscale=\xscaleEfficiencyImprovementHeuristics, yscale=\yscaleSmallEfficiency]
				\pgfgettransformentries{\xscaleTikz}{\@tempa}{\@tempa}{\yscaleTikz}{\@tempa}{\@tempa}
			
				\draw[very thin, color=gray, xstep=2, ystep=0.05] (0, 0) grid (20, 0.3 );
				
				\begin{scope}[shift={(-40, -1)}]
					\def\crossSizeX{\crossSize / \xscaleTikz};
					\def\crossSizeY{\crossSize / \yscaleTikz};
					
					\def\crossOne{(-\crossSizeX,-\crossSizeY) -- (\crossSizeX,\crossSizeY) (-\crossSizeX,\crossSizeY) -- (\crossSizeX,-\crossSizeY)};
					\def\crossTwo{(-\crossSizeX,0) -- (\crossSizeX,0) (0,\crossSizeY) -- (0,-\crossSizeY)};
					\def\crossThree{(0,0) -- (0,\crossSizeY) (0,0) -- (\crossSizeX,-\crossSizeY) (0,0) -- (-\crossSizeX,-\crossSizeY)};
					\def\crossFour{(0,0) -- (0,-\crossSizeY) (0,0) -- (\crossSizeX,\crossSizeY) (0,0) -- (-\crossSizeX,\crossSizeY)};
					\def\crossFive{(0,0) -- (\crossSizeX,0) (0,0) -- (-\crossSizeX,\crossSizeY) (0,0) -- (-\crossSizeX,-\crossSizeY)};
					\def\crossSix{(0,0) -- (-\crossSizeX,0) (0,0) -- (\crossSizeX,\crossSizeY) (0,0) -- (\crossSizeX,-\crossSizeY)};
					
					\draw[black, shift={( 40.6945596118, 1.2388916232)}] \crossOne;
					\node[label={[label distance=-0.14cm]+10:{\color{black} \tiny \LPCTwoR{}}}] at ( 40.71874207985, 1.232) {};
					
					\draw[red, shift={( 40.71874207985, 1.2201636038)}] \crossOne;
					\node[label={[label distance=-0.14cm]+10:{\color{red} \tiny \LPCTwoR{}\TwoO{}}}] at ( 40.71874207985, 1.214) {};
					
					\draw[blue, shift={( 51.81517047179, 1.1887753658)}] \crossOne;
					\node[label={[label distance=-0.14cm]+90:{\color{blue} \tiny \LPCTwoR{}\ThreeO{}}}] at ( 51.81517047179, 1.1887753658) {};
					
					\draw[magenta, shift={( 49.13908631945, 1.1651858532)}] \crossOne;
					\node[label={[label distance=-0.14cm]-90:{\color{magenta} \tiny \LPCTwoR{}\M{15}}}] at ( 49.13908631945, 1.1651858532) {};
					
					\draw[black!50!brown, shift={( 56.0558237406, 1.1576783496)}] \crossOne;
					\node[label={[label distance=-0.14cm]+90:{\color{black!50!brown} \tiny \LPCTwoR{}\M{20}}}] at ( 56.0558237406, 1.1576783496) {};
					
					\draw[black!25!teal, shift={( 47.86038049161, 1.2210133091)}] \crossOne;
					\node[label={[label distance=-0.14cm]+90:{\color{black!25!teal} \tiny \LPCTwoR{}\Lens{}}}] at ( 47.86038049161, 1.2210133091) {};
					
					\draw[dashed] ( 40.6945596118, 1.3) -- ( 40.6945596118, 1.2388916232) -- ( 40.71874207985, 1.2201636038) -- ( 49.13908631945, 1.1651858532) -- ( 56.0558237406, 1.1576783496) -- ( 60, 1.1576783496);
				\end{scope}
				
				\def\axisAdditionalLengthPlusTikzX{\axisAdditionalLengthPlus / \xscaleTikz}
				\def\axisAdditionalLengthMinusTikzX{\axisAdditionalLengthMinus / \xscaleTikz}
				\draw[arrow] (-\axisAdditionalLengthMinusTikzX, 0) -- (20, 0)  -- +(\axisAdditionalLengthPlusTikzX, 0) node[right] {\xAxisEfficiency};
				\def\xTotalLengthPlus{20+\axisAdditionalLengthPlusTikzX}
				\draw (\xTotalLengthPlus, 0) node[right] {$\qquad$};
				\def\axisAdditionalLengthPlusTikzY{\axisAdditionalLengthPlus / \yscaleTikz}
				\def\axisAdditionalLengthMinusTikzY{\axisAdditionalLengthMinus / \yscaleTikz}
				\draw[arrow] (0, -\axisAdditionalLengthMinusTikzY) -- (0, 0.3 ) -- +(0, \axisAdditionalLengthPlusTikzY) node[above, yshift=-0.15cm] {\yAxisEfficiency};
				
				\def\axisLabelTikzY{\axisLabel / \yscaleTikz}
				
				\draw[shift={(0, 0)}] (0, \axisLabelTikzY) -- (0, -\axisLabelTikzY) node[below] {$400$};
				\draw[shift={(2, 0)}] (0, \axisLabelTikzY) -- (0, -\axisLabelTikzY) node[below] {$420$};
				\draw[shift={(4, 0)}] (0, \axisLabelTikzY) -- (0, -\axisLabelTikzY) node[below] {$440$};
				\draw[shift={(6, 0)}] (0, \axisLabelTikzY) -- (0, -\axisLabelTikzY) node[below] {$460$};
				\draw[shift={(8, 0)}] (0, \axisLabelTikzY) -- (0, -\axisLabelTikzY) node[below] {$480$};
				\draw[shift={(10, 0)}] (0, \axisLabelTikzY) -- (0, -\axisLabelTikzY) node[below] {$500$};
				\draw[shift={(12, 0)}] (0, \axisLabelTikzY) -- (0, -\axisLabelTikzY) node[below] {$520$};
				\draw[shift={(14, 0)}] (0, \axisLabelTikzY) -- (0, -\axisLabelTikzY) node[below] {$540$};
				\draw[shift={(16, 0)}] (0, \axisLabelTikzY) -- (0, -\axisLabelTikzY) node[below] {$560$};
				\draw[shift={(18, 0)}] (0, \axisLabelTikzY) -- (0, -\axisLabelTikzY) node[below] {$560$};
				\draw[shift={(20, 0)}] (0, \axisLabelTikzY) -- (0, -\axisLabelTikzY) node[below] {$600$};
				
				\def\axisLabelTikzX{\axisLabel / \xscaleTikz}
				\begin{scope}[shift={(0, -1)}]
					\draw[shift={(0,1 )}] (\axisLabelTikzX, 0) -- (-\axisLabelTikzX, 0) node[left] {$1$};
					\draw[shift={(0,1.1 )}] (\axisLabelTikzX, 0) -- (-\axisLabelTikzX, 0) node[left] {$1.1$};
					\draw[shift={(0,1.2 )}] (\axisLabelTikzX, 0) -- (-\axisLabelTikzX, 0) node[left] {$1.2$};
					\draw[shift={(0,1.3 )}] (\axisLabelTikzX, 0) -- (-\axisLabelTikzX, 0) node[left] {$1.3$};
				\end{scope}
			\end{tikzpicture}
		\AngleTSPInstances
		\vspace{-0.6cm}
		
			\begin{tikzpicture}[xscale=\xscaleEfficiencyImprovementHeuristics, yscale=\yscaleSmallEfficiency]
				\pgfgettransformentries{\xscaleTikz}{\@tempa}{\@tempa}{\yscaleTikz}{\@tempa}{\@tempa}
			
				\draw[very thin, color=gray, xstep=2, ystep=0.05] (0, 0) grid (20, 0.3 );
				
				\begin{scope}[shift={(-10, -1)}]
					\def\crossSizeX{\crossSize / \xscaleTikz};
					\def\crossSizeY{\crossSize / \yscaleTikz};
					
					\def\crossOne{(-\crossSizeX,-\crossSizeY) -- (\crossSizeX,\crossSizeY) (-\crossSizeX,\crossSizeY) -- (\crossSizeX,-\crossSizeY)};
					\def\crossTwo{(-\crossSizeX,0) -- (\crossSizeX,0) (0,\crossSizeY) -- (0,-\crossSizeY)};
					\def\crossThree{(0,0) -- (0,\crossSizeY) (0,0) -- (\crossSizeX,-\crossSizeY) (0,0) -- (-\crossSizeX,-\crossSizeY)};
					\def\crossFour{(0,0) -- (0,-\crossSizeY) (0,0) -- (\crossSizeX,\crossSizeY) (0,0) -- (-\crossSizeX,\crossSizeY)};
					\def\crossFive{(0,0) -- (\crossSizeX,0) (0,0) -- (-\crossSizeX,\crossSizeY) (0,0) -- (-\crossSizeX,-\crossSizeY)};
					\def\crossSix{(0,0) -- (-\crossSizeX,0) (0,0) -- (\crossSizeX,\crossSizeY) (0,0) -- (\crossSizeX,-\crossSizeY)};
					
					\draw[black, shift={( 13.23109778368, 1.1178631211)}] \crossOne;
					\node[label={[label distance=-0.22cm]+45:{\color{black} \tiny \NNSTwo{}}}] at ( 13.23109778368, 1.1178631211) {};
					
					
					\draw[blue, shift={( 23.07379904044, 1.1057190581)}] \crossOne;
					\node[label={[label distance=-0.14cm]+90:{\color{blue} \tiny \NNSTwo{}\ThreeO{}}}] at ( 23.07379904044, 1.1057190581) {};
					
					\draw[magenta, shift={( 16.81560547066, 1.0882889164)}] \crossOne;
					\node[label={[label distance=-0.14cm]-90:{\color{magenta} \tiny \NNSTwo{}\M{15}}}] at ( 16.81560547066, 1.0882889164) {};
					
					\draw[black!50!brown, shift={( 19.50741522515, 1.0855878491)}] \crossOne;
					\node[label={[label distance=-0.14cm]-90:{\color{black!50!brown} \tiny \NNSTwo{}\M{20}}}] at ( 19.50741522515, 1.0855878491) {};
					
					\draw[black!25!teal, shift={( 20.24255752695, 1.1140393739)}] \crossOne;
					\node[label={[label distance=-0.14cm]+90:{\color{black!25!teal} \tiny \NNSTwo{}\Lens{}}}] at ( 20.24255752695, 1.1140393739) {};
					
					\draw[dashed] ( 13.23109778368, 1.3) -- ( 13.23109778368, 1.1178631211) -- ( 16.81560547066, 1.0882889164) -- ( 19.50741522515, 1.0855878491) -- ( 30, 1.0855878491);
				\end{scope}
				
				\def\axisAdditionalLengthPlusTikzX{\axisAdditionalLengthPlus / \xscaleTikz}
				\def\axisAdditionalLengthMinusTikzX{\axisAdditionalLengthMinus / \xscaleTikz}
				\draw[arrow] (-\axisAdditionalLengthMinusTikzX, 0) -- (20, 0)  -- +(\axisAdditionalLengthPlusTikzX, 0) node[right] {\xAxisEfficiency};
				\def\xTotalLengthPlus{20+\axisAdditionalLengthPlusTikzX}
				\draw (\xTotalLengthPlus, 0) node[right] {$\qquad$};
				\def\axisAdditionalLengthPlusTikzY{\axisAdditionalLengthPlus / \yscaleTikz}
				\def\axisAdditionalLengthMinusTikzY{\axisAdditionalLengthMinus / \yscaleTikz}
				\draw[arrow] (0, -\axisAdditionalLengthMinusTikzY) -- (0, 0.3 ) -- +(0, \axisAdditionalLengthPlusTikzY) node[above, yshift=-0.15cm] {\yAxisEfficiency};
				
				\def\axisLabelTikzY{\axisLabel / \yscaleTikz}
				
				\draw[shift={(0, 0)}] (0, \axisLabelTikzY) -- (0, -\axisLabelTikzY) node[below] {$100$};
				\draw[shift={(2, 0)}] (0, \axisLabelTikzY) -- (0, -\axisLabelTikzY) node[below] {$120$};
				\draw[shift={(4, 0)}] (0, \axisLabelTikzY) -- (0, -\axisLabelTikzY) node[below] {$140$};
				\draw[shift={(6, 0)}] (0, \axisLabelTikzY) -- (0, -\axisLabelTikzY) node[below] {$160$};
				\draw[shift={(8, 0)}] (0, \axisLabelTikzY) -- (0, -\axisLabelTikzY) node[below] {$180$};
				\draw[shift={(10, 0)}] (0, \axisLabelTikzY) -- (0, -\axisLabelTikzY) node[below] {$200$};
				\draw[shift={(12, 0)}] (0, \axisLabelTikzY) -- (0, -\axisLabelTikzY) node[below] {$220$};
				\draw[shift={(14, 0)}] (0, \axisLabelTikzY) -- (0, -\axisLabelTikzY) node[below] {$240$};
				\draw[shift={(16, 0)}] (0, \axisLabelTikzY) -- (0, -\axisLabelTikzY) node[below] {$260$};
				\draw[shift={(18, 0)}] (0, \axisLabelTikzY) -- (0, -\axisLabelTikzY) node[below] {$260$};
				\draw[shift={(20, 0)}] (0, \axisLabelTikzY) -- (0, -\axisLabelTikzY) node[below] {$300$};
				
				\def\axisLabelTikzX{\axisLabel / \xscaleTikz}
				\begin{scope}[shift={(0, -1)}]
					\draw[shift={(0,1 )}] (\axisLabelTikzX, 0) -- (-\axisLabelTikzX, 0) node[left] {$1$};
					\draw[shift={(0,1.1 )}] (\axisLabelTikzX, 0) -- (-\axisLabelTikzX, 0) node[left] {$1.1$};
					\draw[shift={(0,1.2 )}] (\axisLabelTikzX, 0) -- (-\axisLabelTikzX, 0) node[left] {$1.2$};
					\draw[shift={(0,1.3 )}] (\axisLabelTikzX, 0) -- (-\axisLabelTikzX, 0) node[left] {$1.3$};
				\end{scope}
			\end{tikzpicture}
		\AngleDistanceTSPInstances
		\vspace*{-0.1cm}
		\caption{improvement heuristic trade-offs for instances with $105 \leq n \leq 200$; Pareto-frontier is plotted dashed}
		\label{figure:ImprovementHeuristicEfficienciesForInstancesWith105LessOrEqualTonLessOrEqual200}
	\end{figure}
			\end{comment:figures}
			
			\medskip
			
			Figure~\ref{figure:ImprovementHeuristicEfficienciesForInstancesWith105LessOrEqualTonLessOrEqual200} sums up the trade-offs of the particular improvement heuristics.
			Obviously, \TwoO{} performs much better than \Lens{}, because it reaches similar objective function values faster. 
			Very interesting is the performance of \M{15} and \M{20}:
			\M{15} beats \ThreeO{} both in solution qualities and running times. 
			The step from \M{15} to \M{20} costs a significant amount of computation time while the objective function value improvement is rather marginal. 
			Summing up, \TwoO{} and \M{15} offer the best quality/cost ratio.
			
			\bigskip
			Finally, by considering Table~\ref{table:ObjectiveFunctionValueRatioMeansForAllTestInstancesWith105LessOrEqualTonLessOrEqual200} in the Appendix, we can see that the overall best mean objective function value ratio $\rg$ could be obtained by the combinations \LPCOneR\M{20} for the {\AngleTSP}- and \LPCTwoR\M{20} for the {\AngleDistanceTSP}-instances, \ie not by improving the outcome of the best stand-alone heuristics respectively.
			Of course, we cannot estimate how far the solutions are above the optimum in these cases. 
			However, Table~\ref{table:OptimalObjectiveFunctionValueRatioMeans}, allows a rough intuition that the difference between the lower bound gap and the optimal objective function gap is around 10 percentage points for the larger {\AngleTSP}- and 6.5 percentage points for the larger {\AngleDistanceTSP}-instances. Thus the best obtained objective function value ratios $\rg$ of 15.73\% and 8.26\% can be roughly translated into 5\% and 1.5\% gaps to the unknown optimal solution values for the {\AngleTSP}- and {\AngleDistanceTSP}-instances, respectively.

	\section{Conclusions}
		\label{section:conclusions}
		
	Quadratic variants of classical combinatorial optimization problems have received increasing attention in recent years.	
	They permit the inclusion of interdependency effects in the objective function while the standard problems only add up the costs of elements in a feasible solution subset selected from a given ground set.
	In this paper we consider two quadratic variants of the famous TSP corresponding to the following geometric cases:
	Given points in the Euclidean plane, which should be connected by a Hamiltonian cycle, the cost of such a tour is either the sum of {\em turning angles} in every point (AngleTSP)
	or a linear combination of turning angles and Euclidean distances (AngleDistanceTSP).
	Both problems have applications in robotics and other planning problems for moving objects of high inertia.
	
	One main contribution of the paper is a number of constructive heuristics and tour improvement algorithms, which are based on geometric properties of good solutions, particularly encouraging the exploitation of the convex hull for the given point set and subsets thereof.
	A second contribution utilizes the impressive performance of todays ILP solvers.
	We use ILP models to patch together paths and subtours arising either from constructive procedures or from relaxations of more general ILP models.
	Finally, we also solve ILP models as an improvement step by  iteratively selecting a subregion of the point set and removing all solution edges from this area. 
	Then the resulting subproblem is solved to optimality while keeping fixed all tour connections outside this area.
	This matheuristic can be thought of as a magnifying glass which is successively placed on certain parts of the Euclidean plane and rewires the tour in this area in an optimal way. The described principle has similarities to other well-known meta- and matheuristic ideas, such as large neighbourhood search.
	
	Extensive computational experiments on test instances previously used in the literature, but also on larger sized instances, were performed and reported in a condensed way.
	We managed to take a major step forward compared to previously reported heuristics.
	The obtained gaps to the lower bound of roughly 15\% for AngleTSP (8\% for AngleDistanceTSP) for instances up to $n=200$ points 
	could be translated in a cursory way to an expected gap of 5\% (1.5\%) to the unknown optimal solution values.
	Note for comparison that the previously best performing heuristic reported in the literature (corresponding to \CIF{}\ThreeO) yields average gaps to the lower bound of 31\% (12\%) and requires the running times for the $3$-opt heuristic, which rapidly increases for larger instances.
	Running times to reach our best solution values obtained on a standard PC average from 5 to 10 minutes, but with a moderate growth rate for even larger instances.
		
	\medskip
	For future research the still existing optimality gaps of our approaches suggest opportunities for further progress. 
	In particular, there seems to be plenty of potential for developing sophisticated metaheuristics.
	Moreover, our approaches are motivated by the geometry of the point set and the implied cost coefficients in the objective function.
	Moving away from this structure to more general cost values, possibly still preserving some weaker geometric relationship, 
	could give rise to challenging new variants of the QTSP. In this context, note that our LP-based approaches could be used for the general QTSP as well.
	Finally, the underlying idea of our magnifying glass matheuristic, which performed very well in our QTSP setting, can be rather easily applied to other combinatorial optimization problems with a spatial structure of the given objects and decomposable solution sets.
	
	\section*{Acknowledgements}
		\label{section:acknowledgements}
		We would like to thank Anja Fischer for valuable discussions in the early stages of our research.
		Furthermore, we thank Andreas Tramper for his preliminary studies on this problem. Ulrich Pferschy was supported by the COLIBRI Initiative of the University of Graz. 

	
	\newpage
	
	\bibliographystyle{IEEEtranSN}
	\bibliography{HeuristicApproachesForTheQuadraticTravellingSalesmanProblem}

	\medskip
	
	\appendix
	
	\newpage
	
	\section*{Appendix}
		\label{section:appendix}
		In the following two tables the best value in every row is marked by an asterisk and the best value in every column is highlighted by underlying.
		
		\vspace*{-0.2cm}
				
		\renewcommand{\arraystretch}{1.3}
		\begin{table}[H]
			\tiny
			\centering
			\begin{tabular}{p{1.25cm}||*{6}{S[table-format=1.4, detect-weight, mode=text, table-text-alignment=right, table-number-alignment=right, table-space-text-post={*}, table-column-width=1.25cm]}|S[table-format=1.4, detect-weight,mode=text, table-text-alignment=right, table-number-alignment=right, table-space-text-post={*}, table-column-width=1.25cm]}
				\multicolumn{1}{l||}{} & \multicolumn{1}{r}{no improv.} & \multicolumn{1}{r}{\TwoO{}} & \multicolumn{1}{r}{\ThreeO{}} & \multicolumn{1}{r}{\M{15}} & \multicolumn{1}{r}{\M{20}} & \multicolumn{1}{r|}{\Lens{}} & \multicolumn{1}{r}{min}\\
				\hline\hline
				\NNS{} & 1.6887 & 1.6560 & 1.5466 & 1.3637 & 1.3026* & 1.6811 & 1.3026*\\
				\NNSTwo{} & 1.5928 & 1.5928 & 1.5254 & 1.3446 & 1.2867* & 1.5824 & 1.2867*\\
				\NNSL{} & 1.7026 & 1.4266 & 1.2726 & 1.2077 & 1.1886* & 1.6406 & 1.1886*\\
				\CIF{} & 1.9892 & 1.4630 & 1.3121 & 1.2697 & 1.2336* & 1.8689 & 1.2336*\\
				\CH{} & 1.7128 & 1.6999 & 1.5798 & 1.3645 & 1.3069* & 1.7057 & 1.3069*\\
				\CHC{} & 1.6783 & 1.6741 & 1.6026 & 1.3736 & 1.3098* & 1.6748 & 1.3098*\\
				\CHCL{} & 1.6652 & 1.4703 & 1.2757 & 1.2096 & 1.1882* & 1.6083 & 1.1882*\\
				\LPP{} & 1.5393 & 1.3339 & 1.2376 & 1.2118 & 1.1887* & 1.4771 & 1.1887*\\
				\LPPR{} & 1.2648 & 1.2313 & 1.1975 & 1.1718 & 1.1615* & 1.2426 & 1.1615*\\
				\LPCOneR{} & 1.2410 & 1.2210 & \Uline{1.1884} & 1.1655 & \Uline{1.1573}* & 1.2231 & \Uline{1.1573}*\\
				\LPCTwoR{} & \Uline{1.2389} & \Uline{1.2202} & 1.1888 & \Uline{1.1652} & 1.1577* & \Uline{1.2210} & 1.1577*\\[0.03cm]
				\hline
				{min} & \Uline{1.2389} & \Uline{1.2202} & \Uline{1.1884} & \Uline{1.1652} & \Uline{1.1573}* & \Uline{1.2210} & \Uline{1.1573}*\\[0.1cm]
				\multicolumn{8}{c}{\AngleTSPInstances}\\[0.24cm]
				\NNS{} & 1.2551 & 1.1570 & 1.1254 & 1.0993 & 1.0934* & 1.2387 & 1.0934*\\
				\NNSTwo{} & \Uline{1.1179} & 1.1179 & 1.1057 & 1.0883 & 1.0856* & \Uline{1.1140} & 1.0856*\\
				\NNSL{} & 1.2422 & 1.1484 & 1.1201 & 1.0975 & 1.0920* & 1.2244 & 1.0920*\\
				\CIF{} & 1.2366 & 1.1504 & 1.1209 & 1.1035 & 1.0953* & 1.2303 & 1.0953*\\
				\LPP{} & 1.1500 & 1.1134 & 1.0990 & 1.0880 & 1.0857* & 1.1432 & 1.0857*\\
				\LPPR{} & 1.1526 & 1.1077 & 1.0944 & 1.0860 & 1.0836* & 1.1396 & 1.0836*\\
				\LPCOneR{} & 1.1437 & 1.1064 & 1.0936 & 1.0858 & 1.0828* & 1.1333 & 1.0828*\\
				\LPCTwoR{} & 1.1413 & \Uline{1.1058} & \Uline{1.0931} & \Uline{1.0854} & \Uline{1.0826}* & 1.1310 & \Uline{1.0826}*\\[0.03cm]
				\hline
				{min} & \Uline{1.1179} & \Uline{1.1058} & \Uline{1.0931} & \Uline{1.0854} & \Uline{1.0826}* & \Uline{1.1140} & \Uline{1.0826}*\\[0.1cm]
				\multicolumn{8}{c}{\AngleDistanceTSPInstances}\vspace{-0.175cm}
			\end{tabular}
			\caption{objective function value ratio means for all test instances with $105 \leq n \leq 200$}
			\label{table:ObjectiveFunctionValueRatioMeansForAllTestInstancesWith105LessOrEqualTonLessOrEqual200}
		\end{table}
		
		\vspace*{-0.39cm}
		
		\begin{table}[H]
			\tiny
			\centering
			\begin{tabular}{p{1.25cm}||*{6}{S[table-format=3.4, detect-weight, mode=text, table-text-alignment=right, table-number-alignment=right, table-space-text-post={*}, table-column-width=1.25cm]}|S[table-format=3.4, detect-weight,mode=text, table-text-alignment=right, table-number-alignment=right, table-space-text-post={*}, table-column-width=1.25cm]}
				\multicolumn{1}{l||}{} & \multicolumn{1}{r}{no improv.} & \multicolumn{1}{r}{\TwoO{}} & \multicolumn{1}{r}{\ThreeO{}} & \multicolumn{1}{r}{\M{15}} & \multicolumn{1}{r}{\M{20}} & \multicolumn{1}{r|}{\Lens{}} & \multicolumn{1}{r}{min}\\
				\hline\hline
				\NNS{} & 283.8885* & 284.1263 & 413.1111 & 465.3341 & 569.1563 & 356.7633 & 283.8885*\\
				\NNSTwo{} & 67.0159* & 67.1282 & 193.0359 & 240.8241 & 345.4880 & 139.7709 & 67.0159*\\
				\NNSL{} & 417.2671* & 417.8017 & 571.5596 & 517.4951 & 605.8565 & 487.5652 & 417.2671*\\
				\CIF{} & 6.5365* & 7.2117 & 168.7839 & 110.9418 & \Uline{172.3227} & 77.9529 & 6.5365*\\
				\CH{} & 1.5028* & 1.7001 & 137.9146 & 179.3441 & 298.8040 & 74.5875 & 1.5028*\\
				\CHC{} & 1.8191* & 1.9635 & \Uline{111.6876} & 187.8074 & 303.5915 & 74.9145 & 1.8191*\\
				\CHCL{} & \Uline{1.0236}* & \Uline{1.4739} & 165.6817 & \Uline{108.0078} & 201.2742 & \Uline{71.9514} & \Uline{1.0236}*\\
				\LPP{} & 206.2802* & 206.7912 & 340.4795 & 295.4575 & 380.8795 & 277.6503 & 206.2802*\\
				\LPPR{} & 368.8422* & 369.1380 & 483.0896 & 455.4174 & 532.4050 & 440.6636 & 368.8422*\\
				\LPCOneR{} & 391.5621* & 391.8200 & 501.6059 & 475.4146 & 548.8180 & 463.3042 & 391.5621*\\
				\LPCTwoR{} & 406.9456* & 407.1874 & 518.1517 & 491.3909 & 560.5582 & 478.6038 & 406.9456*\\[0.03cm]
				\hline
				{min} & \Uline{1.0236}* & \Uline{1.4739} & \Uline{111.6876} & \Uline{108.0078} & \Uline{172.3227} & \Uline{71.9514} & \Uline{1.0236}*\\[0.1cm]
				\multicolumn{8}{c}{\AngleTSPInstances}\\[0.24cm]
				\NNS{} & 285.9435* & 286.5523 & 412.0495 & 332.5851 & 367.3309 & 357.0333 & 285.9435*\\
				\NNSTwo{} & 132.3110* & 132.4236 & 230.7380 & 168.1561 & 195.0742 & 202.4256 & 132.3110*\\
				\NNSL{} & 468.0509* & 468.6615 & 592.9431 & 513.0854 & 548.2855 & 539.1046 & 468.0509*\\
				\CIF{} & \Uline{6.5468}* & \Uline{7.0853} & \Uline{127.9475} & \Uline{41.8982} & \Uline{71.4263} & \Uline{73.8265} & \Uline{6.5468}*\\
				\LPP{} & 150.3579* & 150.7859 & 252.0784 & 183.6876 & 211.1502 & 220.0040 & 150.3579*\\
				\LPPR{} & 276.3159* & 276.7537 & 376.4522 & 310.7960 & 338.4974 & 347.1968 & 276.3159*\\
				\LPCOneR{} & 275.2136* & 275.6351 & 375.8346 & 309.4856 & 335.4789 & 345.7596 & 275.2136*\\
				\LPCTwoR{} & 293.6624* & 294.0777 & 392.0432 & 327.9660 & 354.4542 & 364.1617 & 293.6624*\\[0.03cm]
				\hline
				{min} & \Uline{6.5468}* & \Uline{7.0853} & \Uline{127.9475} & \Uline{41.8982} & \Uline{71.4263} & \Uline{73.8265} & \Uline{6.5468}*\\[0.1cm]
				\multicolumn{8}{c}{\AngleDistanceTSPInstances}\vspace{-0.175cm}
			\end{tabular}
			\caption{running time means for all test instances with $105 \leq n \leq 200$}
			\label{table:RunningTimeMeansForAllTestInstancesWith105LessOrEqualTonLessOrEqual200}
		\end{table}
\end{document}